%% file: densematter2.tex
\documentclass[graybox,envcountchap,sectrefs]{svmono}

\usepackage{mathptmx}
\usepackage{helvet}
\usepackage{courier}
\usepackage{type1cm}         

\usepackage{makeidx}         
\usepackage{graphicx}        
\usepackage{multicol}        
\usepackage[bottom]{footmisc}

\usepackage{amssymb}  
\usepackage{amsmath, amsfonts}

\newcommand\g{\gamma}

\newcommand\e{\epsilon}

\newcommand{\non}{\nonumber\\}

\newcommand{\be}{\begin{equation}}
\newcommand{\ee}{\end{equation}}
\newcommand{\bea}{\begin{eqnarray}}
\newcommand{\eea}{\end{eqnarray}}
\newcommand{\ba}[1]{\begin{array}{#1}}
\newcommand{\ea}{\end{array}}

\newcommand{\bm}[1]{\mbox{\boldmath${#1}$}}
\newcommand{\uq}{\hat{\mathbf{q}}} 
\newcommand{\uk}{\hat{\mathbf{k}}}
\newcommand{\up}{\hat{\mathbf{p}}}

\newcommand{\vg}{\bm{\gamma}}

\newcommand{\Tr}{{\rm Tr}}

\makeindex             

\begin{document}

\author{Andreas Schmitt}
\title{Dense matter in compact stars}
\subtitle{-- A pedagogical introduction --}
\date{June 25, 2010}
\maketitle

\frontmatter

\include{preface2}

\tableofcontents

\mainmatter

\include{dense2}

\backmatter
\include{appendix2}
\include{glossary2}

\printindex


\end{document}

%% file: preface2.tex
%
%

\preface

The purpose and motivation of these lectures can be summarized in the following two questions:
\begin{itemize}
\item What is the ground state (and its properties) of dense matter?

\item What is the matter composition of a compact star?
\end{itemize}
The two questions are, of course, strongly coupled to each other. Depending on your point of view, you can either consider the 
first as the main question and the second as a consequence or application of the first, or vice versa. 

If you are interested
in fundamental questions in particle physics you may take the former point of view: you ask the question what happens to 
matter if you squeeze it more and more. This leads to fundamental questions 
because at some level of sufficient squeezing you expect to reach the
point where the fundamental degrees of freedom and their interactions become important. That is, at some point 
you will reach a form of matter where not molecules or atoms, but the constituents of an atom, namely neutrons, protons, and electrons, 
are the relevant degrees of freedom. This form of matter, and its variants, constitute one important topic of these lectures and 
is termed nuclear matter.
If you squeeze further, you might reach a level where the constituents of neutrons and protons, namely 
quarks and gluons, become relevant degrees of freedom. This form of matter, termed quark matter or strange quark matter, is the second 
important subject we shall discuss. By studying dense matter, we shall thus learn a lot about the 
fundamental theories and interactions of elementary particles. When trying to understand this kind of dense matter, we would like
to perform experiments and check whether our fundamental theories work or whether there are new phenomena, or maybe even new theories,
that we have not included into our description. Unfortunately, there are currently no experiments on earth which can produce
matter at such ultra-high densities we are talking about. However, this does not mean that this kind of matter does not exist in nature.
On the contrary, we are pretty sure that we have observed objects that contain matter at ultra-high density, namely compact stars.
We may thus consider compact stars as our ``laboratory''. Thinking about the first question has therefore led us to the 
second.

If you are primarily interested in phenomenology, or if you are an astrophysicist, you may start from the second question:
you observe a compact star in nature and would like to understand its properties. 
In this case you start from observations like the rotation frequency, the temperature of the star etc.\ and ask, 
why does the star rotate so slow/so fast, why does
it cool down so slow/so fast? And these questions will inevitably lead you to the attempt to figure out the microscopic structure of the 
star, although you have started from macroscopic observables. You need to know whether the star contains nuclear matter or quark matter or both,
in which phase the respective matter is present, and which properties these phases have. 
It is thus very natural, also from the astrophysicist's point of view, to study the first question.

In any case, we see that both questions are closely related and we don't have to decide which of the two points of view we take. 
If I have to characterize what awaits you in these lectures I would nevertheless say that we shall lean a bit more towards the fundamental aspects. 
In other words, we shall neglect many complications that arise from considering a realistic compact star. A star is a finite system, 
it is inhomogeneous, it underlies the laws of general relativity etc. Although our discussions are always motivated by 
the astrophysical application, we mostly discuss infinite, homogeneous systems and do not elaborate on general relativistic effects. 
Only in discussing the consequences of our microscopic calculations we shall, on a qualitative level, discuss the more realistic setting. 

So what kind of physics will we discuss and which theoretical tools do we need? Since our focus is on nuclear and quark matter,
the dominant interaction that governs the states of matter we are interested in is the strong interaction. The underlying 
theory for this interaction is Quantum Chromodynamics (QCD). Although this theory is uniquely determined by very simple 
symmetry principles, it is extremely hard to solve for most applications. Unfortunately (or fortunately, because this makes it interesting
and challenging) matter at compact star densities eludes rigorous first-principle calculations. 
Therefore, we often have to retreat to simple phenomenological models or have to perform rigorous QCD 
calculations at asymptotically large densities and then extrapolate the results down to the density regime we are interested in.

In the physics of compact stars also the weak interaction plays an important role. 
We shall see that it is responsible for the chemical equilibration of the system, i.e., it fixes
the various chemical potentials. It is also important for the understanding of cooling mechanisms of the star or for transport
properties of nuclear and quark matter. 
Furthermore, our (mostly field-theoretical) treatment always includes nonzero
chemical potentials and sometimes nonzero temperature (for many applications the zero-temperature approximation is sufficient). 
In this sense it goes beyond the standard vacuum field theory formalism.
Basic elements of thermal quantum field theory at finite chemical potential are therefore explained in the appendix. 

All this may sound exciting on the one hand, because it shows that the physics of compact stars is extremely rich (due to the 
diversity of involved physics I found it helpful to include a glossary of important terms at the end of these lecture notes). 
But on the other hand it may also sound like a big challenge for you if you are not familiar with advanced field theory. Nevertheless, 
these lecture notes are not primarily intended as a review for researchers (although they might find it useful too) but as a pedagogical 
introduction for graduate students and advanced undergraduate students. 
For some of our discussions all you need as a prerequisite is some knowledge in thermodynamics
and statistical physics, for instance in chapter \ref{sec:massradius}, which deals almost exclusively with noninteracting systems. 
Some other sections, for instance the calculation of the neutrino emissivity in chapter \ref{sec:cooling} indeed makes use of advanced 
field-theoretical methods at finite temperature. It is not the intention of these lectures to develop the theoretical tools in all 
details before we use them. More importantly, all calculations are physically motivated, thus by understanding the physics 
behind the results and calculations, these lectures aim at making you familiar with the theories and technicalities via ``learning by doing''. 
So at the end of these lectures you will have heard about the basic phenomena and possible microscopic explanations of the physics of 
compact stars, but also will be prepared to start theoretical research in this exciting field yourself, to possibly contribute to the answers to 
the two questions we have started with.

These lectures are based on a course given in the summer semester 2009 at the Vienna University of Technology. I thank all participants of this 
course for the lively discussions and the numerous questions and comments that helped improve these lecture notes. I am also grateful to
M.\ Alford, P.\ Jaikumar, P.\ van Nieuwenhuizen, F.\ Preis, A.\ Rebhan, T.\ Sch\"{a}fer, I.\ Shovkovy, S.\ Stricker, 
N.-O.\ Walliser, Q.\ Wang, and F.\ Weber for helpful comments and discussions.

\vspace{\baselineskip}
\begin{flushright}\noindent
Vienna, January 2010 \hfill {\it Andreas Schmitt}\\
\end{flushright}

%% file: dense2.tex
\chapter{Introduction}
\label{sec:intro}

\section{What is dense matter?}
\label{sec:what2}

\begin{figure}[t]
\begin{center}
\includegraphics[width=0.75\textwidth]{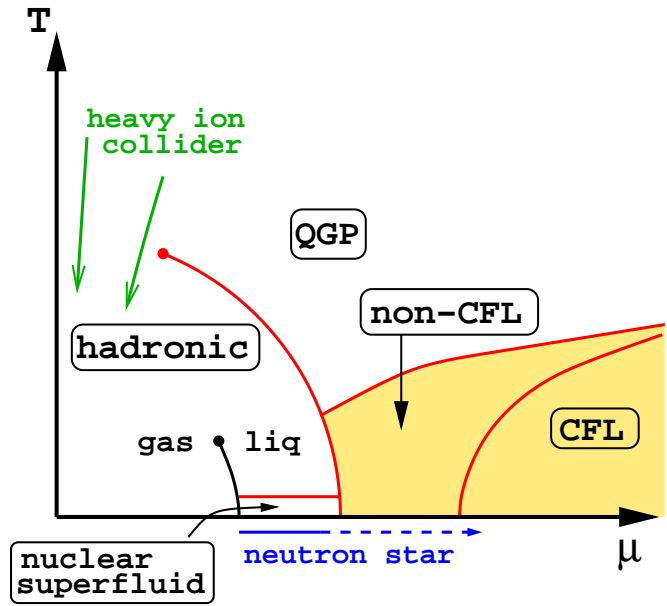}
\caption{Conjectured phase diagram of QCD in the plane of quark chemical potential $\mu$ and temperature $T$. While matter at low
density and high temperature is probed in heavy-ion collisions, cold and dense matter can only be found in neutron stars (compact stars).
We may find (superfluid) nuclear matter and/or deconfined quark matter inside a star. Deconfined quark matter is, at high temperatures, termed
quark-gluon plasma (QGP) and is, at low temperatures, expected to be in a color-superconducting 
state, here labelled by CFL (color-flavor locking), discussed in Sec.\ \ref{sec:CFL}, and non-CFL (some color superconductor other than CFL).}  
\label{figQCDpd}
\end{center}
\end{figure}

The QCD phase diagram collects the equilibrium phases of QCD in the plane of 
quark (or baryon) chemical potential $\mu$ and temperature $T$. We show a sketch of this phase diagram in Fig. \ref{figQCDpd}. 
In this introduction, we are not concerned with the details of this diagram. 
We observe that compact stars, on the scales of this diagram, live in the region of small temperatures
and intermediate densities. They may live in the region where quarks are confined, i.e., in the hadronic phase. This would imply
that they are neutron stars. They may also live in the deconfined region which would make them quark stars.
A compact star may also contain both deconfined and confined quark matter because the star actually has a density profile rather than a homogeneous 
density. In the interior, we expect the density to be larger than at the surface. Therefore, the third possibility is a hybrid star
with a quark core and a nuclear mantle. 

We do currently not know the exact location of most of the phase transition lines in Fig. \ref{figQCDpd}. Therefore, we do not know
the ground state of strongly-interacting quark (or nuclear) matter at the relevant density. As a consequence, we can to some extent only
speculate about the matter composition of the star. 
The reason is, simply speaking, that QCD is notoriously hard to solve for temperatures and densities present in a compact
star. With the help of the phase diagram we can put this statement in a wider context: QCD is {\it asymptotically free}, which means that for large 
momentum exchange the interaction becomes weak. Hence, at sufficiently large temperatures and/or densities, we deal with
weakly interacting quarks and gluons. In the case of large densities (or large chemical potentials) this can be understood from the
uncertainty principle which relates small distances (the interacting particles are very close to each other) to large momenta. As a result
of asymptotic freedom, regions in the phase diagram where $\mu$ and/or $T$ are sufficiently large can be understood from rigorous 
first-principle calculations. These regions, although theoretically under control, are far from being experimentally (even astrophysically) 
accessible. 

If we now go to lower temperatures and densities we have to cross a large unknown territory. Only at small temperatures and densities,
when we are deep in the hadronic phase we have reached an area which again is under control, at least to some extent. 
Theoretically, it is more complicated
than the perturbatively treatable asymptotic regions. After all, hadrons are quite complicated objects once we try to describe them in 
terms of their constituents. However, we can use effective descriptions which can be supported, confirmed, and improved by experiments in 
the lab. Furthermore, at least for vanishing chemical potentials, we can perform brute-force QCD calculations on the computer which gives
us solid theoretical knowledge for certain quantities (at nonvanishing chemical potentials these calculations are problematic due to the 
so-called {\it sign problem}). 

We thus see that compact stars (as well as the quark-gluon plasma created in heavy-ion collisions) reside in a
region of the phase diagram which is hard to access. More positively speaking, this region is interesting and challenging 
because exciting and unknown physics may be discovered and new theoretical tools may need to be developed. Or, in other words, the cold and 
dense matter we talk about in these lectures is interesting from the theoretical point of view because, on the characteristic scale of QCD, 
it is only moderately, not extremely, dense. 

The theoretical tools used in current research to describe cold and dense matter are based on the above observations: 
if we describe quark matter 
we may use perturbative methods which are valid at asymptotically large densities and extrapolate the results down to intermediate densities.
We shall do so for instance in chapter \ref{sec:massradius} where we treat quarks as noninteracting or in Sec.\ \ref{sec:QCDgap} where 
we calculate the color-superconducting gap within perturbative QCD. However, we have to be aware that the extrapolation of the results 
pushes the calculations out of their range of validity by many orders of magnitude. On the other hand, we may use models for nuclear matter 
which are established at low densities by experimental data. We do so for instance in chapter \ref{sec:internuclear}. This time 
we have to extrapolate to {\it larger} densities. Again, the extrapolation is in principle uncontrolled. 

These theoretical challenges emphasize the significance of 
astrophysical observations: we do not simply like to {\it confirm} the results of our calculations by using astrophysical data, we 
need astrophysical input to {\it understand} the theory which we believe to be the underlying theory of strongly interacting matter, 
namely QCD. Therefore, the connection between astrophysical observables and microscopic calculations is one of the main
subjects of these lectures.

\section{What is a compact star?}
\label{sec:what1}
  
Only beaten by black holes, compact stars are the second-densest objects in nature. They have masses of the order of the mass of the sun, 
$M\sim 1.4\,M_\odot$, but radii 
of only about ten kilometers, $R\sim 10\, {\rm km}$. Thus the mass of the sun $M_\odot=1.989\cdot 10^{33}\,{\rm g}$ is concentrated 
in a sphere with a radius $10^5$ times smaller than that of the sun, $R_\odot=6.96\cdot 10^5\,{\rm km}$. We thus estimate the average mass
density in a compact star to be
\be
\rho \simeq 7\cdot 10^{14}\,{\rm g}\,{\rm cm}^{-3} \, .
\ee
This is a few times larger than the density present in heavy nuclei, the nuclear ground state density
\be
\rho_0 = 2.5 \cdot 10^{14}\,{\rm g}\,{\rm cm}^{-3} \, ,
\ee
which corresponds to a baryon number density of $n_0\simeq 0.15\,{\rm fm}^{-3}$. Mass and radius of the star are determined 
by the equation(s) of state of the matter phase(s) inside the star. This is the subject of chapter \ref{sec:massradius}.

In the traditional picture of a compact star, the star is made out of neutron-rich nuclear matter. 
Hence the traditional name is actually {\it neutron star}. This is sometimes still the preferred term, 
even if one talks about a star that contains a quark matter core (which then might be called ``exotic neutron star''). 
Here we shall always use the more general term {\it compact star} to include the possibilities of more exotic matter; after all, 
a significant part of these lectures is about this exotic matter. The term compact star will thus be used in these lectures for an 
object with characteristic mass, radius etc.\ as given in this subsection. It can either be made of nuclear matter or variants thereof 
({\it neutron star}), of a quark matter core with a surrounding mantle of nuclear matter ({\it hybrid star}) or exclusively of (strange) quark matter 
({\it quark star} or {\it strange star}).\footnote{The term compact star is in general 
also used to include {\it white dwarfs}, stars which are less dense 
than neutron stars, hybrid stars, or quark stars, and sometimes even to include black holes. Since we are not concerned with either of
these objects here, we can reserve the term compact star as explained in the text.} Here are a few more basic properties of compact stars:

\begin{itemize}

\item Compact stars are born in a supernova, a spectacular explosion of a giant or supergiant star due to the gravitational collapse of its core. 
Supernovae are very complex, nonequilibrium processes which astrophysicists try to understand with  
hydrodynamic simulations. We shall not be concerned with supernovae in these lectures but should keep in mind that some properties
of the star may be a result of these violent explosions. A possible example is the high velocity with which some of the compact stars 
travel through space.

\item Compact stars are not only extreme with respect to their density. Some of them also rotate very fast with rotation periods in the millisecond
regime, such that their frequencies are 
\be
\nu \lesssim 1\,{\rm ms}^{-1} \, .
\ee
To see that this is really fast, notice that a point on the equator has a velocity of $2\pi R/1{\rm ms} \simeq  0.2\, c$, i.e., it 
moves with 20\% of the speed of light. The current record holder is the star PSR J1748-2446ad,\footnote{The label of the star says that it is 
a ``Pulsating Source of Radio emission'' (PSR) located on the celestial sphere at right ascension 17\,h~48\,min with $-24^\circ$ 46' 
declination; the `J' indicates
the use of the J2000 coordinate system, the suffix `ad' is used to distinguish the object from other pulsars in the same globular cluster Terzan 5.} 
rotating with a period of 1.39 ms.  
Several observations are related to the rotation frequency. First of all, compact stars have been discovered as {\it pulsars}, by observing
pulsating radio signals, for the first time in 1967. These periodic signals are due to the lighthouse effect, i.e., radio emission 
is aligned in a beam along the magnetic axis of the pulsar which spins around the rotation axis, crossing the earth's telescopes
periodically. More interestingly for our purpose, the pure
fact that the rotation of some compact stars can be so fast requires some explanation. 
From the microscopic point of view, this is related to transport properties such 
as viscosity of the matter inside the star, see Sec.\ \ref{sec:misc}. Also {\it pulsar glitches}, sudden jumps in the rotation frequency, 
must find an explanation in the properties of dense matter.

\item Compact stars also have huge magnetic fields,
\be
B\sim 10^{12}\, {\rm G} \, .
\ee
Even larger surface magnetic fields of the order of $B\sim 10^{15}\,{\rm G}$ have been observed (the magnetic field in the core of the star 
possibly being even higher). Such highly magnetized stars are also termed {\it magnetars}. Compare these magnetic fields 
for instance to the earth's magnetic field, $B\sim 0.6\,{\rm G}$, a common hand-held magnet, $B\sim 100\,{\rm G}$, or the strongest steady 
magnetic fields in the laboratory, $B\sim 4.5\cdot 10^5\,{\rm G}$. 

\item Compact stars are cold. This may sound odd, given their temperatures which,
right after they are born in a supernova explosion, can be as high as $T\sim 10^{11}\,{\rm K}$. This corresponds, in units where the 
Boltzmann constant is one, $k_B=1$, to $T\sim 10\,{\rm MeV}$. During the evolution of the star, the temperature 
decreases down to temperatures in the keV range. The dominant cooling mechanism is neutrino emission which we discuss in chapter
\ref{sec:cooling}. There are two reasons why in our context it is appropriate to call compact stars cold, in spite of the apparently large 
temperatures. First, temperatures in the keV range are small compared to the scale set by QCD, for instance the deconfinement transition 
at vanishing quark chemical potential of about $T_c\simeq 170\, {\rm MeV}$. This means compact stars are located basically on the horizontal
axis in the QCD phase diagram in Fig.\ \ref{figQCDpd}. Second, temperatures in compact stars are small compared to the quark (or baryon) 
chemical potential, $T\ll\mu$. This is important for our calculations since it implies that $T=0$ is a good approximation in many cases.  

\end{itemize}

\section{Further reading}

Before we start with the main part, let's mention some literature. Extensive textbooks about compact stars are Refs.\ \cite{weber1,glendenning1,haensel1}. 
A shorter introduction to compact stars and dense matter can be found in the 
review article \cite{Page:2006ud1}. 
Similar reviews are Refs.\ \cite{Madsen:1998uh1,Weber:2004kj1}, with emphasis on quark matter,
and Ref.\ \cite{Lattimer:2006xb1}, with emphasis on astrophysical observations. 
A more theoretical review about quark matter (more precisely, about color-superconducting quark matter), with 
a section about compact star applications is Ref.\ \cite{Alford:2007xm1}.
For an introduction to thermal field theory see the textbooks \cite{kapusta1} and \cite{lebellac1}, on which the appendix of these lecture notes
is partially based. As became clear above, in this course we shall deal with questions which are under debate in current research. 
Therefore, some of the material included here has so far only been available in research papers. The respective references will 
be given in the various chapters. I will not try to be exhaustive in the reference list but rather point out selected references which are 
useful for a deeper understanding of what we discuss in these lectures. If you are interested in more references you can find plenty in the 
quoted papers and textbooks.

\chapter{Mass and radius of the star}
\label{sec:massradius}

In this chapter, we will discuss the most basic properties of a compact star, its mass and radius. We have already given typical values for these
quantities above. Below we shall connect them with microscopic properties of nuclear and quark matter. This connection 
is made by the equation of state from which, in particular, an estimate for the maximum mass of the star can be obtained. Let us begin with a simple
estimate of mass and radius from general relativity. For the stability of the star we need $R>R_s$ where $R$ is the radius of the star, and
$R_s=2MG$ the Schwarzschild radius, with the mass of the star $M$ and the gravitational constant $G=6.672\cdot 10^{-11}\,{\rm m}^3{\rm kg}^{-1}
{\rm s}^{-2}=6.707\cdot 10^{-39}\,{\rm GeV}^{-2}$. (We shall mostly use units common in nuclear and particle physics, $\hbar=c=k_B=1$, although 
astrophysicists often use different units.) For $R<R_s$ the star becomes unstable with respect to the collapse into a black hole. Let us build a simple star made out of a number of nucleons $A$ with mass $m\simeq 939\,{\rm MeV}$ and a distance $r_0\simeq 0.5\cdot 10^{-13}\, {\rm cm}$
(that's where the nucleon interaction becomes repulsive). We thus cover a volume $\sim r_0^3 \,A$ and thus have a radius
\be
R\sim r_0 \,A^{1/3} \, ,
\ee
(for our rough estimate we are not interested in factors of $\pi$), and a mass
\be
M\sim m\,A \, .
\ee
Now from the limit $R=2MG$ we obtain 
\be
A\sim \left(\frac{r_0}{2mG}\right)^{3/2} \sim 2.6\cdot 10^{57} \, .
\ee
This is the number of nucleons up to which we can fill our star before it gets unstable. In other words, the Schwarzschild radius
is proportional to the mass of the star and thus increases linearly in the number of nucleons $A$, while the radius increases with $A^{1/3}$; 
therefore, for $A$ smaller than the limit $A\sim 2.6\cdot 10^{57}$ the star is stable, while it collapses into a black hole
for nucleon numbers larger than this limit. We can plug the limit value for $A$ back into the radius and mass
of the star to obtain 
\be
R \sim 7\, {\rm km}  \, , \qquad M\sim 2.3\,M_\odot \, .
\ee 
Adding more nucleons would make the star too heavy in relation to its radius. We see that these values are not too far from the 
observed ones given in Sec.\ \ref{sec:what1}.

Besides giving an estimate for the baryon number in the star, we see from this simple exercise that general relativistic effects will 
be important because the Schwarzschild radius will be a significant fraction of the radius of the star. We can also estimate the 
gravitational energy of the star. To this end, we need the differential mass of the star at a given radius (i.e., 
the mass of a thin spherical layer)
\be \label{dm}
dm = \rho(r)\, dV \, ,
\ee
where $dV=4\pi r^2 dr$ is the volume of the thin spherical layer at radius $r$. For a rough estimate 
let us (unrealistically) assume a constant density $\rho(r)=\rho$ such that the mass $m(r)$ of the star up to a radius $r\le R$, 
is given by $m(r)=\frac{4\pi}{3}r^3\rho$. Then we obtain 
\be
E_{\rm grav} \simeq \int_0^R \frac{Gm(r)\,dm(r)}{r} \simeq \frac{3}{5}\frac{GM^2}{R}\simeq 0.12\, M \, ,
\ee
where we have used the above realistic values $M\simeq 1.4\,M_\odot$ and $R\simeq 10\,{\rm km}$. We thus see that the gravitational energy 
$E_{\rm grav}$ is more than 10\% of the mass of the star.
This suggests that for the mass-radius relation we need an equation that incorporates effects from general relativity. For simplicity, let
us first derive the equation that relates mass and radius without general relativistic effects and include them afterwards. 
We are looking for an equation that describes equilibrium between the gravitational force, seeking to compress the star, and the 
opposing force coming from the pressure of the matter inside the star. In the case of a compact star, this pressure is the Fermi pressure 
plus the pressure coming from the strong interactions of the nuclear or quark matter inside. The differential pressure $dP$ at a given radius 
$r$ is related to the gravitational force $dF$ via
\be \label{dP}
dP = \frac{dF}{4\pi r^2} \, ,
\ee
with 
\be \label{dF}
dF = -\frac{G m(r)\,dm}{r^2} \, .
\ee
The equation for the differential mass (\ref{dm}), together with Eq.\ (\ref{dP}) (into which we insert Eqs.\ (\ref{dm}) and (\ref{dF})), 
yields the two coupled differential 
equations,
\begin{subequations}\label{newton}
\bea
\frac{dm}{dr} &=& 4\pi r^2\epsilon(r) \, , \\  
\frac{dP}{dr} &=& -\frac{G\epsilon(r) m(r)}{r^2} \, . 
\eea
\end{subequations}
where we have expressed the mass density through the energy density $\epsilon(r)=\rho(r)$ (in units where $c=1$).
The second equation, which is easy to understand from elementary Newtonian physics, receives corrections from general relativity. 
It is beyond the scope of these lectures to derive these corrections. We simply quote the resulting equation,
\be \label{TOV}
\frac{dP}{dr} = -\frac{G\epsilon(r) m(r)}{r^2}\left[1+\frac{P(r)}{\epsilon(r)}\right]\left[1+\frac{4\pi r^3 P(r)}{m(r)}\right]
\left[1-\frac{2Gm(r)}{r}\right]^{-1} \, .
\ee
This equation is called Tolman-Oppenheimer-Volkov (TOV) equation and the derivation can be found for instance in Ref.\ \cite{shapiro2}. 
In order to solve
it, one first needs the energy density for a given pressure. Only then do we have a closed system of equations. This input is given from 
the microscopic physics which yields an equation of state in the form $P(\epsilon)$. We have thus found a first example how the microscopic
physics can potentially be ``observed'' from astrophysical data, namely from mass and radius of the star. We shall encounter many more of these 
examples. The equations of state for noninteracting nuclear and quark matter will be discussed in the subsequent sections.

For a given equation of state one needs two boundary conditions for the TOV equation. The first is obviously $m(r=0)=0$, the second
is a boundary value for the pressure in the center of the star, $P(r=0)=P_0$. Then, the solution of the equations will produce a mass and
pressure profile $m(r)$, $P(r)$ with the pressure going to zero at some point $r=R$, giving the radius of the star. The mass of the 
star is then read off at this point, $M=m(R)$. Doing this for varying initial pressures $P_0$ yields a curve $M(R)$ in the mass-radius plane, 
parametrized by $P_0$. This curve depends strongly on the chosen equation of state.

\section{Noninteracting nuclear matter}
\label{sec:freenuclear}

We start with a very simple system where we neglect all interactions. In this case, all we need is basic 
statistical physics and thermodynamics. The thermodynamic potential for the grand-canonical ensemble is 
given by 
\be
\Omega = E -\mu N -TS\, , 
\ee
with the energy $E$, chemical potential $\mu$, particle number $N$, temperature $T$ and entropy $S$. The pressure is then
\be \label{pressureTS}
P = -\frac{\Omega}{V} = -\epsilon + \mu n +Ts \, , 
\ee
where $V$ is the volume of the system. Number density $n= N/V$, energy density $\epsilon=E/V$, and entropy density $s=S/V$ are, for a 
fermionic system, given by 
\begin{subequations} \label{nes}
\bea
n &=& 2\int\frac{d^3{\bf k}}{(2\pi)^3} \,f_k \, , \\
\epsilon &=& 2\int\frac{d^3{\bf k}}{(2\pi)^3} \,E_k \,f_k \, , \\
s &=& -2\int\frac{d^3{\bf k}}{(2\pi)^3} \left[(1-f_k)\ln(1-f_k)+f_k\ln f_k\right] \, .
\eea
\end{subequations}
We shall first be interested in a system of neutrons ($n$), protons ($p$), and electrons ($e$), each giving a contribution 
to the pressure according to Eqs.\ (\ref{pressureTS}) and (\ref{nes}). Since they are spin-$\frac{1}{2}$ fermions, we have included 
a factor 2 for the two degenerate 
spin states. The Fermi distribution function is denoted by $f_k$,
\be
f_k \equiv \frac{1}{e^{(E_k-\mu)/T}+1} \, ,
\ee
and the single-particle energy is
\be
E_k = \sqrt{k^2+m^2} \, . 
\ee
Inserting number density, energy density, and entropy density into the pressure  (\ref{pressureTS}) yields 
\be \label{pressureTS1}
P = 2T\int\frac{d^3{\bf k}}{(2\pi)^3} \ln \left[1+e^{-(E_k-\mu)/T}\right] \, .
\ee 
This corresponds to the result obtained from field-theoretical methods in appendix \ref{app:fermions}, see Eq.\ (\ref{OV}). There 
also the antiparticle contribution is included, which can here, due to the large positive chemical potential, safely be neglected. 
One can easily check that
number density and entropy are obtained from the pressure (\ref{pressureTS1}) via the usual thermodynamic relations, i.e., 
by taking the derivatives with respect to $\mu$ and $T$. For the following we now take the limit $T=0$.
This is a good approximation since the temperature of a compact star is typically in the keV range and thus much smaller than 
the chemical potentials and masses of the nucleons.

For $T=0$ the Fermi distribution is a step function, $f_k=\Theta(k_F-k)$, and thus the $k$ integrals will
be cut off at the Fermi momentum $k_F$, i.e., 
\begin{subequations} \label{dens1eps1}
\bea
n &=& \frac{1}{\pi^2}\int_0^{k_F}dk\,k^2 = \frac{k_F^3}{3\pi^2} \, , \label{dens1}\\
\epsilon &=& \frac{1}{\pi^2}\int_0^{k_F}dk\,k^2 \sqrt{k^2+m^2} \non
&=& \frac{1}{8\pi^2}\left[(2k_F^3+m^2k_F)\sqrt{k_F^2+m^2}
-m^4\ln\frac{k_F+\sqrt{k_F^2+m^2}}{m}\right]  \, . \label{eps1} 
\eea
\end{subequations}
Then, with $\mu=\sqrt{k_F^2+m^2}$, the pressure is
\bea \label{pressure1}
P &=& \frac{1}{\pi^2} \int_0^{k_{F}} dk\,k^2(\mu - \sqrt{k^2+m^2}) \non
&=& \frac{1}{24\pi^2}\left[(2k_F^3-3m^2k_F)\sqrt{k_F^2+m^2}+3m^4
\ln\frac{k_F+\sqrt{k_F^2+m^2}}{m}\right] \, .
\eea
This can either be obtained by inserting Eqs.\ (\ref{dens1}) and (\ref{eps1}) into Eq.\ (\ref{pressureTS}) or, equivalently, by 
taking the $T=0$ limit in Eq.\ (\ref{pressureTS1}). For the latter one makes use of $\lim_{T\to 0} T \ln (1+e^{x/T})=x\,\Theta(x)$.  

For $n$, $p$, $e$ matter, the total pressure is 
\be
P = \frac{1}{\pi^2}\sum_{i=n,p,e} \int_0^{k_{F,i}} dk\,k^2(\mu_i - \sqrt{k^2+m_i^2}) \, .
\ee
The Fermi momenta can be thought of as variational parameters which have to be determined from maximizing the pressure, i.e., from the 
conditions
\be
\frac{\partial P}{\partial k_{F,i}} = 0 \, , \qquad i=n,p,e \, ,
\ee
which implies
\be \label{muis}
\mu_i = \sqrt{k_{F,i}^2+m_i^2} \, .
\ee
We have additional constraints on the Fermi momenta from the following two conditions. Firstly, we have to require the star to be 
electrically neutral,\footnote{In fact, a compact star has to be electrically neutral to a very high accuracy, 
as one can see from the following simple estimate. Suppose the star has an overall charge of $Z$ times the elementary charge, $Ze$, and
we consider the Coulomb repulsion of a test particle, say a proton, with mass $m$ and charge $e$ ($e$ having the same sign as the hypothetical 
overall charge of the star $Ze$). 
The Coulomb force, seeking to expel the test particle, has to be smaller than the gravitational force, seeking to keep the test particle 
within the star. 
This gives the condition
\be
\frac{(Ze)e}{R^2} \le \frac{GMm}{R^2} \, , 
\ee
with the mass $M$ and radius $R$ of the star. Even if we are generous with the limit on the right-hand side by assigning the upper
limit $M< Am$ to the mass of the star (if the star contains $A$ nucleons, its total mass will be less than $Am$ due to the 
gravitational binding energy), we will get a very restrictive limit on the overall charge. Namely, we find
\be
\frac{(Ze)e}{R^2} < \frac{GAm^2}{R^2} \qquad \Rightarrow \qquad Z < G\frac{m^2}{e^2}\,A \, . 
\ee
With the proton mass $m\sim 10^3\, {\rm MeV}$, the elementary charge $e^2\sim 10^{-1}$ (remember $\alpha=e^2/(4\pi)\simeq 1/137$), and the 
gravitational constant $G\sim 7\cdot 10^{-39}\,{\rm GeV}^{-2}$, we estimate 
\be
Z< 10^{-37}A \, , 
\ee
i.e., the average charge per nucleon has to be extremely small in order to ensure the stability of the star. Since we have found such an 
extremely small number, it is irrelevant for the argument whether we use a proton or an electron as a test particle.
The essence of this argument is the weakness of gravitation compared to the electromagnetic interactions: a tiny electric charge per
unit volume, distributed over the star, is sufficient to overcome the stability from gravity.
}, 
i.e., the densities of protons and electrons has to be equal,
\be
n_e = n_p \, . 
\ee
With Eq.\ (\ref{dens1}) this means
\be \label{kFekFp}
k_{F,e}=k_{F,p} \, .
\ee
Secondly, we require chemical equilibrium with respect to the weak processes
\begin{subequations} \label{betadecay}
\bea
n &\to& p + e + \bar{\nu}_e \, ,  \\
p+e&\to& n + \nu_e \,  .
\eea
\end{subequations}
The first of these processes is the usual $\beta$-{\it decay}, the second is sometimes called {\it inverse $\beta$-decay} or {\it electron capture}.
We shall assume that the neutrino chemical potential vanishes, $\mu_{\nu_e}=0$. This is equivalent to assuming that neutrinos and antineutrinos,
once created by the above processes, simply leave the system without further interaction. This
assumption is justified for compact stars since the neutrino mean free path is of the order of the size of the star or larger (except for the 
very early stages in the life of the star). Consequently, $\beta$-equilibrium, i.e., equilibrium with respect to the processes 
(\ref{betadecay}), translates into the following constraint for the chemical potentials,
\be
\mu_n = \mu_p + \mu_e \, .
\ee
Inserting Eq.\ (\ref{muis}) into this constraint yields
\be \label{sqrts}
\sqrt{k_{F,n}^2+m_n^2} =\sqrt{k_{F,p}^2+m_p^2} + \sqrt{k_{F,e}^2+m_e^2} \, .
\ee
We can eliminate the electron Fermi momentum with the help of Eq.\ (\ref{kFekFp}) and solve this equation to obtain the proton Fermi momentum 
as a function of the neutron Fermi momentum,
\be \label{kp}
k_{F,p}^2 = \frac{(k_{F,n}^2+m_n^2-m_e^2)^2 - 2(k_{F,n}^2+m_n^2+m_e^2)m_p^2 +m_p^4}{4(k_{F,n}^2+m_n^2)} \, .
\ee
To illustrate the physical meaning of this relation, let us consider some limit cases. First assume a vanishing proton contribution, 
$k_{F,p}=0$. Then the equation gives (which is most easily seen from Eq.\ (\ref{sqrts}))
\be
k_{F,n}^2=(m_p+m_e)^2-m_n^2 < 0\, .
\ee
This expression is negative because the neutron is slightly heavier than the electron and the proton together, 
$m_p\simeq 938.3\,{\rm MeV}$, $m_n\simeq 939.6\,{\rm MeV}$, $m_e\simeq 0.511\,{\rm MeV}$. Therefore, 
$k_{F,p}=0$ is impossible and there always has to be at least a small fraction of protons. Now let's assume $k_{F,n}=0$, which leads to
\be
k_{F,p}^2=\left(\frac{m_n^2+m_e^2-m_p^2}{2m_n}\right)^2-m_e^2 \simeq 1.4\,{\rm MeV}^2 \, .
\ee
This is the threshold below which there are no neutrons and the charge neutral system in $\beta$-equilibrium contains only protons and electrons
of equal number density. 

In general, we may consider a given baryon density $n_B=n_n+n_p$ to express the neutron Fermi momentum as 
\be
k_{F,n}=(3\pi^2 n_B-k_{F,p}^3)^{1/3} \, .
\ee
Inserting this into Eq.\ (\ref{kp}) yields an equation for $k_{F,p}$ as a function of the baryon density. In the ultrarelativistic limit, i.e., 
neglecting all masses, Eq.\ (\ref{kp}) obviously yields $k_{F,p}=k_{F,n}/2$ and thus $n_p=n_n/8$ or 
\be
n_p = \frac{n_B}{9} \, .
\ee
By solving Eq.\ (\ref{kp}) numerically one can check that this limit is approached from below for large baryon densities, i.e., 
in a compact star containing nuclear matter we deal with neutron-rich matter, which justifies the term neutron star. 

As a crude approximation we may thus consider the simple case of pure neutron matter. We also consider the nonrelativistic limit, 
$m_n\gg k_{F,n}$. In this case, the energy density (\ref{eps1}) and the pressure (\ref{pressure1}) become
\begin{subequations} \label{eps2pressure2}
\bea
\epsilon&\simeq& \frac{m_n^4}{3\pi^2}\left[\frac{k_{F,n}^3}{m_n^3}+{\cal O}\left(\frac{k_{F,n}^5}{m_n^5}\right)\right] \, , \\
P&\simeq& \frac{m_n^4}{15\pi^2}\left[\frac{k_{F,n}^5}{m_n^5}+{\cal O}\left(\frac{k_{F,n}^7}{m_n^7}\right)\right] \, .
\eea
\end{subequations}
(To see this, note that the ln term cancels the term linear in $k_F$ in the case of $\epsilon$, and the linear and cubic terms in 
$k_F$ in the case of $P$.)
Consequently, keeping the terms to lowest order in $k_{F,n}/m_n$,
\be \label{Pepsnonrel}
P(\epsilon)\simeq \left(\frac{3\pi^2}{m_n}\right)^{5/3}\frac{\epsilon^{5/3}}{15m_n\pi^2} \, .
\ee
We have thus found a particularly simple equation of state, where the pressure is given by a power of the energy density. 
The general (numerical) discussion of the equation of state, including protons and electrons, is left to the reader, see problem \ref{prob1}.

The next step to obtain the mass-radius relation of the star is to insert the equation of state into the TOV equation. 
The simplest case is a power-law behavior as in Eq.\ (\ref{Pepsnonrel}). The general form of such a so-called ``polytropic'' equation of state is
\be \label{polytrop}
P(\epsilon) = K\,\epsilon^\gamma \, . 
\ee
Using the Newtonian form of the mass-radius equations, Eqs.\ (\ref{newton}), this yields 
\begin{subequations}\label{TOVnewton}
\bea 
\frac{dm}{dr}&=&\frac{4\pi}{K^{1/\gamma}}\, r^2 P^{1/\gamma}(r) \, , \\
\frac{dP}{dr}&=&- \frac{G}{K^{1/\gamma}} \frac{P^{1/\gamma}(r)m(r)}{r^2} \, .
\eea
\end{subequations}
Even in this simplest example, we need to solve the equations numerically, see problem \ref{prob2}. 
The results of this problem show that the maximum mass reached within this model is about $M<0.7M_\odot$ which is well below
observed neutron star masses. (See also Refs.\ \cite{Balian:1999eb2,Silbar:2003wm2,Sagert:2005fw2} for a pedagogical introduction into the equation 
of state and mass-radius relation from solving the TOV equation.)
This small maximum mass is a consequence of the assumption of noninteracting nucleons. Taking into account interactions
will increase the maximum mass significantly.

\section{Noninteracting quark matter}
\label{sec:nonquark}

Whenever we talk about quark matter in these lectures we ignore the charm ($c$), bottom ($b$), and top ($t$) quarks. 
The quark chemical potential inside the star is at most of the order of 500 MeV and thus
much too small to create a population of these states. Therefore, we only consider at most three quark flavors,
namely up ($u$), down ($d$), and strange ($s$). We shall mostly neglect the masses of the $u$ and $d$ quarks;
their {\it current masses} are $m_u\simeq m_d \simeq 5\,{\rm MeV}\ll \mu \simeq (300 - 500)\, {\rm MeV}$. The mass of the 
strange quark, however, is not negligible. The current strange quark mass is $m_s\simeq 90\,{\rm MeV}$, and the density-dependent 
{\it constituent mass} can be significantly larger, making it non-negligible compared to the quark chemical potential.

If we consider free quarks, the energy density $\epsilon$, the number density $n$, and the pressure $P$, are obtained in the same way as 
demonstrated for nucleons in the previous subsection. We only have to remember that there are three colors for each quark flavor, 
$N_c=3$, i.e., the degeneracy factor for a
single quark flavor is $2 N_c = 6$, where the factor 2 counts the spin degrees of freedom. Consequently, for each quark flavor
$f=u,d,s$ we have at zero temperature (cf.\ Eqs.\ (\ref{dens1eps1}), (\ref{pressure1})),
\begin{subequations}
\bea
n_f &=& \frac{k_{F,f}^3}{\pi^2} \, , \\
\epsilon_f &=& \frac{3}{\pi^2}\int_0^{k_{F,f}} dk\,k^2\sqrt{k^2+m_f^2} \, , \\
P_f &=& \frac{3}{\pi^2}\int_0^{k_{F,f}} dk\,k^2\left(\mu_f-\sqrt{k^2+m_f^2}\right) \, .
\eea
\end{subequations}
Again, we need to impose equilibrium conditions with respect to the weak interactions. In the case of three-flavor 
quark matter, the relevant processes are the leptonic processes (including a neutrino or an antineutrino)
\begin{subequations} \label{proc3flavor}
\bea
d&\to&u+e+\bar{\nu}_e \, , \qquad s\to u + e + \bar{\nu}_e \, ,  \\
u+e&\to& d+\nu_e \, , \qquad u+e\to s+\nu_e \, , 
\eea
\end{subequations}
and the non-leptonic process
\be
s+u\leftrightarrow d+u \, .
\ee
These processes yield the following conditions for the quark and electron chemical potentials,
\be \label{quarkeq}
\mu_d = \mu_e +\mu_u \, , \qquad \mu_s = \mu_e + \mu_u \, .
\ee
(This automatically implies $\mu_d=\mu_s$.) The charge neutrality condition can be written in a general way as
\be \label{Q0}
\sum_{f=u,d,s} q_f n_f -n_e = 0 \, , 
\ee
with the electric quark charges
\be \label{quds}
q_u = \frac{2}{3} \, , \qquad q_d = q_s = -\frac{1}{3} \, , 
\ee
and the electron density $n_e$.

\subsection{Strange quark matter hypothesis}
\label{sec:strange}

Before computing the equation of state, we discuss the {\it strange quark matter hypothesis} within the so-called {\it bag model}. 
The bag model is a very crude phenomenological way to incorporate confinement into the description of quark matter. The effect of confinement 
is needed in particular if we compare quark matter with nuclear matter (which is ultimately what we want to do in this section). 
Put another way, although we speak of noninteracting quarks, we need to account for a specific -- in general very complicated -- aspect
of the interaction, namely confinement. 

To understand how the bag constant accounts for confinement, we compare the pressure of a noninteracting gas of massless 
pions with the pressure of a noninteracting gas of quarks and gluons at finite temperature and zero chemical potential. 
The pressure of a single bosonic degree of freedom at $\mu=0$ and at large temperatures compared to the mass of the boson is 
\be
P_{\rm boson} \simeq  -T\int\frac{d^3{\bf k}}{(2\pi)^3}\ln\left(1-e^{-k/T}\right) = \frac{\pi^2T^4}{90} \, . 
\ee
This is derived in appendix \ref{app:bosons} within thermal field theory, see Eq.\ (\ref{PressBos}). 
Analogously, a single fermionic degree of freedom gives (see Eq.\ (\ref{PressFer}) of appendix \ref{app:fermions})  
\be
P_{\rm fermion} \simeq  T\int\frac{d^3{\bf k}}{(2\pi)^3}\ln\left(1+e^{-k/T}\right) = \frac{7}{8}\frac{\pi^2T^4}{90} \, . 
\ee
Therefore, since there are three types of pions, their pressure is 
\be
P_\pi = 3\,\frac{\pi^2T^4}{90} \, .
\ee
This is a simple approximation for the pressure of the confined phase. In the deconfined phase, the degrees of freedom are gluons ($8\times 2$)
and quarks ($4N_c N_f =24$). Thus with $2\times 8+7/8\times 24 = 37$ we have
\be
P_{q,g} = 37\,\frac{\pi^2T^4}{90} - B \, , 
\ee
where the {\it bag constant} $B$ has been subtracted for the following reason.  
If $B$ were zero, the deconfined phase would have the larger pressure and thus would be preferred for all temperatures. We know
however, that at sufficiently small temperatures, the confined phase (that's the world we live in) must be preferred. 
This is phenomenologically accounted for by the bag constant B which acts like an energy penalty for the deconfined phase.
Without this penalty, at least in this very simply model description, the deconfined phase would be ``too favorable'' compared to 
what we observe. As a consequence, by including the bag constant there is certain critical temperature $T_c$ below which the confined phase 
is preferred, $P_{\pi}>P_{q,g}$, and above which the deconfined phase is preferred, $P_{q,g}>P_{\pi}$. 
This is indeed what one expects from QCD, where the deconfinement transition temperature is expected to be  
$T_c\simeq 170\,{\rm MeV}$. (As can be seen in the QCD phase diagram in Fig.\ \ref{figQCDpd}, this deconfinement transition is rather 
a crossover than a phase transition in the strict sense.)

In the context of compact stars we are not interested in such large temperatures. 
In this case, the chemical potential is large and the temperature practically zero. 
Nevertheless we compare nuclear (confined) with quark (deconfined) matter and thus have to include the bag constant in the
pressure and the free energy of quark matter, 
\begin{subequations} \label{bag1}
\bea
P+B &=& \sum_f P_f \, , \label{bag11}\\
\epsilon &=& \sum_f \epsilon_f + B \, . \label{bag12}
\eea
\end{subequations}
This phenomenological model of confinement is called the {\it bag model} \cite{Chodos:1974je2,Chodos:1974pn2} because 
the quarks are imagined to be confined in a bag. One can view the microscopic pressure $\sum_f P_f$ of the quarks to be counterbalanced by the 
pressure of the bag $B$ and an external pressure $P$. 

Equipped with the bag model, we can now explain the strange quark matter hypothesis. For simplicity we consider massless quarks. 
A nonzero strange quark mass will slightly change the results but is not important for the qualitative argument. 
We will also ignore electrons. They are not present in three-flavor massless quark matter at zero temperature. They are however required
in two-flavor quark matter to achieve electric neutrality. But also in this case their population is small enough to render their effect 
unimportant for the following argument.

With $m_f=0$ we simply have
\be
n_f = \frac{\mu_f^3}{\pi^2} \, , \qquad \epsilon_f = \frac{3\mu_f^4}{4\pi^2} \, , \qquad P_f=\frac{\mu_f^4}{4\pi^2} \, , 
\ee
which in particular implies
\be \label{Pe3}
P_f = \frac{\epsilon_f}{3} \, .
\ee
For the strange quark matter hypothesis we consider the energy $E$ per nucleon number $A$,
\be
\frac{E}{A} = \frac{\epsilon}{n_B} \, , 
\ee
where $n_B$ is the baryon number density, given in terms of the quark number densities as
\be
n_B = \frac{1}{3} \sum_f n_f \, , 
\ee
because a baryon contains $N_c=3$ quarks. At zero pressure, $P=0$, Eqs.\ (\ref{bag1}) and (\ref{Pe3}) imply $\epsilon=4B$ and thus
\be
\frac{E}{A} = \frac{4B}{n_B} \, .
\ee
We now apply this formula first to three-flavor quark matter (``strange quark matter''), then to two-flavor quark matter
of only up and down quarks. For strange quark matter, the neutrality constraint (\ref{Q0}) becomes   
\be
2n_u-n_d-n_s = 0 \, .
\ee
Together with the conditions from chemical equilibrium (\ref{quarkeq}) this implies
\be
\mu_u=\mu_d=\mu_s \equiv\mu\, .
\ee
We see that strange quark matter is particularly symmetric. The reason is that the electric charges of an up, down, and strange quark 
happen to add up to zero. Now with $n_B=\mu^3/\pi^2$ and 
\be
B=\sum_f P_f =\frac{3\mu^4}{4\pi^2} 
\ee
(still everything at $P=0$) we have 
\bea \label{EA3}
\left.\frac{E}{A}\right|_{N_f=3} &=& (4\pi^2)^{1/4}\,3^{3/4}\,B^{1/4} \simeq 5.714\,B^{1/4} \simeq 829\,{\rm MeV}\,B^{1/4}_{145} \, .
\eea
We have expressed $B^{1/4}$ in units of 145 MeV, $B_{145}^{1/4}\equiv B^{1/4}/(145\,{\rm MeV})$.

For two-flavor quark matter (neglecting the contribution of electrons), the charge neutrality condition is
\be
n_d=2n_u \, .
\ee
Hence,
\be
\mu_d=2^{1/3}\mu_u \, .
\ee
Then, with $n_B=\mu_u^3/\pi^2$ and 
\be
B=\sum_fP_f = \frac{(1+2^{4/3})\mu_u^4}{4\pi^2} \, ,
\ee
we find
\bea \label{EA2}
\left.\frac{E}{A}\right|_{N_f=2} &=& (4\pi^2)^{1/4}\,(1+2^{4/3})^{3/4}\,B^{1/4} \simeq 6.441\,B^{1/4} \simeq
 934\,{\rm MeV}\,B^{1/4}_{145} \, .
\eea
By comparing this to Eq.\ (\ref{EA3}) we see that two-flavor quark matter has a larger energy per baryon number than three-flavor quark matter. This
is a direct consequence of the Pauli principle: adding one particle species (and keeping the total number of particles fixed)
means opening a set of new available low-energy states that can be filled, thus lowering the total energy of the system.

We can now compare the results (\ref{EA3}) and (\ref{EA2}) with the energy per nucleon in nuclear matter. For pure neutron matter, it is
simply given by the neutron mass,
\be
\left.\frac{E}{A}\right|_{\rm neutrons} = m_n = 939.6\,{\rm MeV} \, .
\ee
For iron, $^{56}$Fe, it is 
\be \label{bindiron}
\left.\frac{E}{A}\right|_{{}^{56}{\rm Fe}} = \frac{56\,m_N-56\cdot 8.8\,{\rm MeV}}{56} = 930\, {\rm MeV} \, ,
\ee
with the nucleon mass $m_N=938.9\, {\rm MeV}$ and the binding energy per nucleon in iron of 8.8 MeV. 
Since we observe iron rather than deconfined quark matter, we know that 
\be
\left.\frac{E}{A}\right|_{{}^{56}{\rm Fe}}<\left.\frac{E}{A}\right|_{N_f=2} \quad  \Rightarrow \qquad  B^{1/4} > 144.4\, {\rm MeV} \, .
\ee
We have thus found a lower limit for the bag constant from the stability of iron with respect to two-flavor quark matter. 
Now what if the bag constant were only slightly larger than this lower limit? What if it were small enough for three-flavor quark matter
to have lower energy than iron? The condition for this would be
\be
\left.\frac{E}{A}\right|_{N_f=3}<\left.\frac{E}{A}\right|_{{}^{56}{\rm Fe}} \quad  \Rightarrow \qquad B^{1/4} < 162.8\, {\rm MeV} \, . 
\ee
This would imply that strange quark matter is absolutely stable (stable at $P=0$), 
while nuclear matter is metastable. This possibility, which would be
realized by a bag constant in the window $145\, {\rm MeV} < B^{1/4} < 162\, {\rm MeV}$, is called {\it strange quark matter hypothesis},
suggested by Bodmer \cite{Bodmer:1971we2} and Witten \cite{Witten:1984rs2}, see also \cite{Farhi:1984qu2}.

Note that the existence of ordinary nuclei does {\it not} rule out the strange quark matter hypothesis. The conversion of an ordinary
nucleus into strange quark matter requires the simultaneous conversion of many (roughly speaking $A$ many) $u$ and $d$ quarks into $s$ quarks.
Since this has to happen via the weak interaction, it is practically impossible. In other words, there is a huge energy
barrier between the metastable (if the hypothesis is true) state of nuclear matter and absolutely stable strange quark matter. This means
that strange quark matter has to be created in another way (``going around'' the barrier), 
by directly forming a quark-gluon plasma. This can for instance happen in 
a heavy-ion collision. Or, more importantly in our context, 
it may happen in the universe, giving rise to stars made entirely out of quark matter, so-called
{\it strange stars}.

Small ``nuggets'' of strange quark matter are called {\it strangelets} (a strange star would then in some sense simply be a huge strangelet). 
If a strangelet
is injected into an ordinary compact star (a neutron star), it would, assuming the strange quark matter hypothesis to be true, 
be able to ``eat up'' the nuclear matter, converting the neutron star into
a strange star. Note the difference between this transition and the above described impossible transition from ordinary nuclear matter to strange 
quark matter: once there is a sufficiently large absolutely stable strangelet, 
{\it successive} conversion of up and down quarks into strange quarks increase the size
of the strangelet; the energy barrier originating from the {\it simultaneous} creation of a large number of strange quarks now cannot 
cause the system to relax back into its original nuclear (metastable) state. This argument has important consequences. If there exist 
enough sizable strangelets in the universe to hit neutron stars, 
{\it every} neutron star would be converted into a strange star. In other words, the observation
of a single ordinary neutron star would rule out the strange quark matter hypothesis. Therefore, it is important to understand whether there are
enough strangelets around. It has been discussed recently in the literature that there may not be enough strangelets \cite{Bauswein:2008gx2}, 
in contrast to what was assumed before.

\subsection{Equation of state}
\label{sec:quarkeos}

\begin{figure}[t]
\begin{center}
\includegraphics[width=0.35\textwidth]{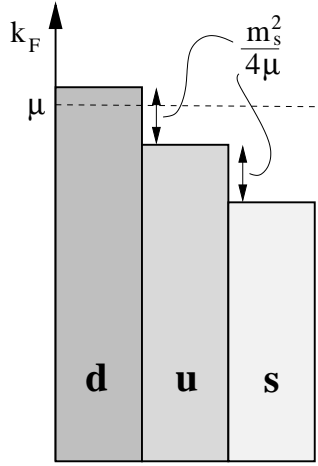}
\caption{Illustration of the Fermi momenta for neutral, unpaired quark matter in $\beta$-equilibrium with quark chemical potential $\mu$. 
The splitting of the Fermi momenta is due to the strange quark mass $m_s$ which is assumed to be small compared to $\mu$. 
}
\label{figuds}
\end{center}
\end{figure}

Next we derive the equation of state for strange quark matter. We include the effect of the strange quark mass to lowest order and also
include electrons. It is convenient to express the quark chemical potentials
in terms of an average quark chemical potential $\mu=(\mu_u+\mu_d+\mu_s)/3$ and the electron chemical potential $\mu_e$,
\begin{subequations}
\bea
\mu_u &=& \mu -\frac{2}{3}\mu_e \, , \\
\mu_d &=& \mu +\frac{1}{3}\mu_e \, , \\
\mu_s &=& \mu +\frac{1}{3}\mu_e \, .
\eea
\end{subequations}
Written in this form, the conditions from $\beta$-equilibrium (\ref{quarkeq}) are automatically fulfilled. Taking into
account the strange quark mass, the Fermi momenta for the approximately massless up and down quark and the massive strange quark are given by
\begin{subequations}
\bea
k_{F,u} &=& \mu_u \, , \\
k_{F,d} &=& \mu_d \, , \\
k_{F,s} &=& \sqrt{\mu_s^2-m_s^2} \, . \label{FermiKs}
\eea
\end{subequations}
The energy density and the pressure are
\begin{subequations}
\bea
\sum_{i=u,d,s,e}\epsilon_i &=& 
\frac{3\mu_u^4}{4\pi^2} + \frac{3\mu_d^4}{4\pi^2}+\frac{3}{\pi^2}\int_0^{k_{F,s}} dk\,k^2\sqrt{k^2+m_s^2} +\frac{\mu_e^4}{4\pi^2} \, , \\
\sum_{i=u,d,s,e}P_i &=& \frac{\mu_u^4}{4\pi^2} + \frac{\mu_d^4}{4\pi^2}+\frac{3}{\pi^2}\int_0^{k_{F,s}} dk\,k^2\left(\mu_s-\sqrt{k^2+m_s^2}\right)
+\frac{\mu_e^4}{12\pi^2} \, , 
\eea
\end{subequations}
where we have neglected the electron mass. 
The neutrality condition can now be written as
\be
0 = \frac{\partial}{\partial\mu_e} \sum_{i=u,d,s,e}P_i = -\frac{2}{3}n_u+\frac{1}{3}n_d+\frac{1}{3}n_s+n_e \, .
\ee
(Note that $\mu_e$ is defined as the chemical potential for {\it negative} electric charge.)
Solving this equation to lowest order in the strange quark mass yields
\be \label{muelowms}
\mu_e\simeq \frac{m_s^2}{4\mu} \, .
\ee
Consequently, the quark Fermi momenta become
\begin{subequations}
\bea
k_{F,u} &\simeq& \mu-\frac{m_s^2}{6\mu} \, , \\
k_{F,d} &\simeq& \mu+\frac{m_s^2}{12\mu} \, , \\
k_{F,s} &\simeq& \mu-\frac{5m_s^2}{12\mu} \, . 
\eea
\end{subequations}
We see that the Fermi momenta are split by an equal distance of $m_s^2/(4\mu)$, and $k_{F,s}<k_{F,u}<k_{F,d}$, see Fig.\ \ref{figuds}. 
The splitting and the order of the Fermi momenta can be understood from the following physical picture: 
start from the symmetric situation $m_s=\mu_e=0$. In this case, all 
quark flavors fill their Fermi spheres to a common Fermi momentum given by $\mu$, and the system is neutral. Now switch on the 
strange quark mass. This lowers the Fermi momentum of the strange quark according to Eq.\ (\ref{FermiKs}). Consequently, there are fewer
strange quarks in the system and thus there is a lack of negative charge. To counterbalance this missing negative charge, 
the system responds by switching on a chemical potential $\mu_e$. Because of $\beta$-equilibrium, the Fermi momenta 
of all quark flavors are rigidly coupled to this change. Electric neutrality is regained by lowering the up quark Fermi momentum and raising the 
down and strange quark Fermi momenta (which makes for the catchy phrase ``the Fermi momentum of the {\it down} goes {\it up}''). 
Since the strange quark Fermi momentum was already lowered by the finite mass, it is clear that
the resulting order is $k_{F,s}<k_{F,u}<k_{F,d}$. The electron contribution to the negative charge density is negligibly 
low, $n_e\propto \mu_e^3 \propto m_s^6/\mu^3$, while the contribution of the quarks due to the strange quark mass is proportional to $\mu m_s^2$.
The splitting of the Fermi momenta due to the effects of the strange quark mass, $\beta$-equilibrium, and electric neutrality is very important
in the context of {\it color superconductivity}. Since color superconductivity is usually based on Cooper pairing of quarks of different flavor, 
a mismatch in Fermi surfaces tends to disfavor this pairing. We shall discuss superconductivity in quark and nuclear matter in 
chapter \ref{sec:supersuper} and give a brief qualitative discussion of the consequences of Fermi surface splitting for color
superconductivity at the end of that chapter.   

Here we continue with unpaired quark matter and insert the result for $\mu_e$ (\ref{muelowms}) back into the energy
density and the pressure. Again keeping only terms to lowest order in the strange quark mass yields 
\begin{subequations}
\bea
\sum_{i}\epsilon_i &\simeq& \frac{9\mu^4}{4\pi^2} - \frac{3\mu^2m_s^2}{4\pi^2} \, , \\
\sum_{i}P_i &\simeq& \frac{3\mu^4}{4\pi^2} - \frac{3\mu^2m_s^2}{4\pi^2} \, .
\eea
\end{subequations}
Consequently,
\be
\sum_{i}\epsilon_i \simeq 3\sum_{i}P_i +\frac{3\mu^2m_s^2}{2\pi^2} \, .
\ee
With Eq.\ (\ref{bag11}) the pressure, including the bag constant, becomes
\be \label{Pquark}
P\simeq\frac{3\mu^4}{4\pi^2} - \frac{3\mu^2m_s^2}{4\pi^2} - B \, , 
\ee
and, expressing $P$ in terms of the energy density, we obtain with the help of Eq.\ (\ref{bag12})
\be
P \simeq \frac{\epsilon-4B}{3}-\frac{\mu^2m_s^2}{2\pi^2} \, .
\ee
This is the equation of state of noninteracting, unpaired strange quark matter within the bag model with strange quark mass corrections to 
lowest order.

\section{Mass-radius relation including interactions}
\label{sec:MRinter}

Let us briefly discuss the results for the mass-radius relation of a compact star for given equations of state for nuclear and
quark matter. Since the underlying calculations in general are complicated and have to be done on a computer, we only quote some results to 
illustrate the physical conclusions. So far we have only discussed the simplest cases of noninteracting matter. 
Interactions have a significant effect on both the equation of state and the mass-radius relation. We now discuss these effects briefly, 
only in the subsequent chapters shall we study the nature and details of these interactions (and discuss their relevance to 
other observables than the mass and the radius of the star). 

The maximum mass of a star for noninteracting nuclear matter is $\sim 0.7 M_\odot$ (see for instance Ref.\ \cite{Sagert:2005fw2} or solve
problem \ref{prob2}); including interactions increases the
mass to values well above $2M_\odot$. The significance of the equation of state and interactions for the maximum mass is easy to understand: if the 
pressure $P(\epsilon)$ for a given energy density $\epsilon$ is large, the system is able to sustain a large gravitational force that
seeks to compress it. Comparing two equations of state over a given energy density range, the one with the larger pressure (for all 
energy densities in the given range) is thus termed
stiff, the one with the smaller pressure is termed soft. Soft equations of state can sustain less gravitational force and thus 
lead to stars with lower maximum masses. In the case of noninteracting nuclear matter, it is only the Fermi pressure from the Pauli exclusion 
principle that prevents the star from the collapse. Interactions increase this pressure because the dominant effect in the case of nuclear
matter at the relevant densities is the short-range repulsion between the nucleons. 
Therefore, the maximum mass is significantly larger in this case. 

\begin{figure}[t]
\begin{center}
\includegraphics[width=0.6\textwidth]{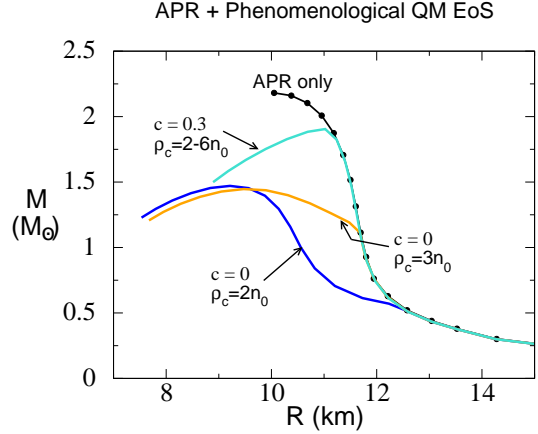}
\caption{Mass-radius plot from Ref.\ \cite{Alford:2004pf2} which shows 
the dependence of the mass-radius curve on the (uncertain) parameters of the quark matter equation of state in a hybrid star. 
We see that reasonable choices of the 
parameters lead to similar curves as for nuclear matter (here with the APR equation of state). In this plot, the 
transition density $\rho_c$ (in units of the nuclear ground state density $n_0$) between quark matter and nuclear matter has been 
used as a parameter, rather than the bag constant. From our discussion it is clear that one can be translated into the other. 
The coefficient $c$ describes QCD corrections to the quark Fermi momentum and thus
to the $\mu^4$ term in the pressure, see Eq.\ (\ref{Pquarkcorr}). 
}
\label{figMR1}
\end{center}
\end{figure}

\begin{figure}[h]
\begin{center}
\includegraphics[width=0.65\textwidth]{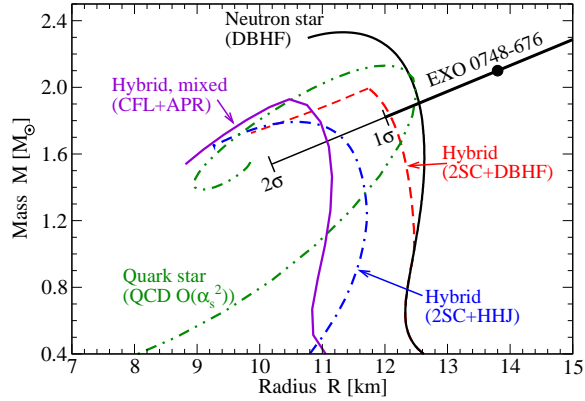}
\caption{Mass-radius plot from Ref.\ \cite{Alford:2006vz2}. A comparison of a neutron star, different hybrid stars,
and a quark star is shown, using several nuclear equations of state 
(DBHF, APR, HHJ) and several quark phases (CFL, 2SC). For more details and explanations of the various 
abbreviations, see Ref.\ \cite{Alford:2006vz2}. 
}
\label{figMR2}
\end{center}
\end{figure}

In Figs.\ \ref{figMR1} and \ref{figMR2} several models for the nuclear equation of state are applied to obtain maximum masses up
to $2.4\,M_\odot$. 
For the case of quark matter, we can understand some of the corrections through interactions in the following simple way. 
A generalization of the pressure (\ref{Pquark}) is
\be \label{Pquarkcorr}
P=\frac{3\mu^4}{4\pi^2}(1-c) - \frac{3\mu^2}{4\pi^2}(m_s^2-4\Delta^2) - B \, . 
\ee
This equation contains two corrections compared to Eq.\ (\ref{Pquark}). One is included in the coefficient $c$ and originates from the 
(leading order) correction of the Fermi momentum due to the QCD coupling $\alpha_s$, 
\be \label{Fermiliquid}
k_F = \mu\left(1-\frac{2\alpha_s}{3\pi}\right) \, , 
\ee
resulting in a correction of the $\mu^4$ term in the pressure with $c=2\alpha_s/\pi$. (This modification of the Fermi momentum 
will also become important in the context of neutrino emissivity in chapter \ref{sec:cooling}.) Higher order calculations suggest 
$c\gtrsim 0.3$ at densities
relevant for compact stars. However, the exact value of $c$ is unknown because perturbative calculations are not valid in the relevant 
density regime, cf.\ discussion in Sec.\ \ref{sec:what2}. Therefore, $c$ can only be treated as a parameter with values around 0.3, as done
for example in Fig.\ \ref{figMR1}. To get an idea about perturbative calculations beyond leading order in $\alpha_s$, you may 
consult the recent Ref.\ \cite{Kurkela:2009gj2}.


The second correction in Eq.\ (\ref{Pquarkcorr}) is the quantity $\Delta$. This is the energy gap arising from color superconductivity
whose microscopic origin we discuss in chapter \ref{sec:supersuper}.
It gives a correction to the $\mu^2$ term in the pressure. 
One might think that this correction is negligible compared to the $\mu^4$ term and the bag constant. However, it turns out that for reasonable
values of the bag constant these two terms largely cancel each other and the $\mu^2$ term becomes important. However, the effect of 
superconductivity is still hard to determine. Firstly, it would require a precise knowledge of the strange quark mass. Secondly, it turns
out that the maximum mass of a hybrid star is not very sensitive to the value of $m_s^2-4\Delta^2$ \cite{Alford:2004pf2}.

As a result of this discussion and the results in Figs.\ \ref{figMR1} and \ref{figMR2}, 
two points are important for the further contents of these lectures. Firstly, we should now be motivated to learn more about the nature
and the consequences of interactions in nuclear and quark matter.
Secondly, we have learned that, given our ignorance of the precise quantitative effects of the strong interaction and the uncertainty in 
astrophysical observations, the mass and the radius of the star are not sufficient to distinguish between a neutron star, a hybrid star,
and possibly a quark star. Therefore, we also have to take into account other observables
which are linked to the microscopic physics. While the equation of state is a bulk property, i.e., it is determined by the whole Fermi sea,
there are other phenomena which are only sensitive to the low-energy excitations at the Fermi surface. One class of such phenomena is given by  
transport properties. They can possibly be related to observables which are more restrictive than mass and radius for the question of the matter 
composition of the star. We shall discuss such observables in chapter \ref{sec:cooling} where we relate the cooling of the star to 
neutrino emissivity, and in chapter \ref{sec:discussion} where we qualitatively discuss other such observables.

\section*{Problems}
\addcontentsline{toc}{section}{Problems}

\begin{prob}
\label{prob1}
\textbf{Equation of state for noninteracting nuclear matter}\\
Find the full equation of state for noninteracting $n$, $p$, $e$ matter at $T=0$ numerically by plotting $P$ versus $\epsilon$. 
You should see the onset of neutrons 
and identify a region where the equation of state is well approximated by the power-law behavior of pure neutron matter in the nonrelativistic 
limit, Eq.\ (\ref{Pepsnonrel}).
\end{prob}

\begin{prob}
\label{prob2}
\textbf{Mass-radius relation}\\
(a) Solve Eqs.\ (\ref{TOVnewton}) numerically (for nonrelativistic pure neutron matter, i.e., $\gamma=5/3$) 
 and plot $m(r)$, $P(r)$ for a given value of the pressure $P_0=P(r=0)$. \\
(b) Use $P_0$ as a parameter to find the mass-radius relation $M(R)$. To this end, you need to do (a) for several values of $P_0$ and
find for each $P_0$ the radius $R$ at which $P(R)=0$ and the corresponding mass $M(R)$.\\
(c) You may incorporate general relativistic effects from the TOV equation (\ref{TOV}) 
and/or the full (numerical) equation of state for noninteracting nuclear matter from Problem \ref{prob1}.
\end{prob}

\chapter{Basic models and properties of dense nuclear matter}
\label{sec:internuclear}

There are numerous models to describe cold and dense interacting nuclear matter. Some of them have been used
to obtain the curves in Figs.\ \ref{figMR1} and \ref{figMR2}. From these curves we see that the models may differ significantly in their 
predictions of the properties of neutron stars and hybrid stars. The reason is that they all are extrapolated into a regime where there is
little theoretical control. In other words, for densities below the nuclear ground state density there are experimental data for instance from 
atomic nuclei or neutron scattering which serve to fit the parameters of the nuclear models unambiguously. However, it is very challenging 
to construct a model which reliably predicts the properties of nuclear matter for larger densities. Put another way, currently the 
only ``experiments'' in this density regime are astrophysical observations which themselves are naturally less controlled than 
experiments in the laboratory. Therefore, the state of the art in describing interacting nuclear matter at high densities is a competition between 
several models which all are prone to uncertainties. In these lectures we do not attempt to give an overview over 
these models. We rather focus on two basic models and discuss them in detail. The first is
the Walecka model and its extensions. The second is chiral perturbation theory, 
which is an effective model based on chiral symmetry of QCD and spontaneous breaking thereof in nuclear matter. 
We shall use it to discuss kaon condensation in nuclear matter. 

To put the following in the perspective of understanding QCD, we should keep in mind
that nucleons are ultimately built of quarks and gluons which are the fundamental degrees of freedom of the strong interactions. 
It is a highly nontrivial task to describe even the mass of a nucleon from quarks and gluons, let alone nuclear
interactions. An important tool for such a case is an effective theory which has non-fundamental degrees of 
freedom, baryons and mesons instead of quarks and gluons. An effective theory can in principle be obtained by taking the
low-energy limit of the underlying fundamental theory, in this case QCD. However, this procedure may turn out to be very difficult.
Therefore, one tries to ``guess'' an effective theory, for instance guided by symmetry principles. One obtains a theory with some unknown 
parameters which have to be fit, for instance to experimental results. Once the parameters are fitted, one may extrapolate the theory 
beyond the regime where the fit has been done. In our case, this will be the high-density region for which we have no experiments in the 
laboratory. There is of course no guarantee that this extrapolation works. Models for interacting nuclear matter at high densities 
have to be understood in this spirit. Of course, an upper density limit for the validity is the deconfinement phase transition to a phase
where quarks and gluons are the relevant degrees of freedom.
This limit density is not precisely known but may well be reached in compact stars.

\section{The Walecka model}
\label{sec:walecka}

The Walecka model contains nucleons which interact via the exchange of the scalar $\sigma$ meson and the vector $\omega$ meson.
The Lagrangian is
\be \label{LWal}
{\cal L} = {\cal L}_N + {\cal L}_{\sigma,\omega} + {\cal L}_I \, , 
\ee
Here, the free nucleon Lagrangian is
\be
{\cal L}_N = \bar{\psi}\left(i\gamma^\mu\partial_\mu-m_N+\mu\gamma^0\right)\psi \, ,
\ee
where $\bar{\psi}\equiv \psi^\dag\gamma^0$, and $\psi = \left(\begin{array}{c}\psi_n\\ \psi_p\end{array}\right)$ with the neutron
and proton spinors $\psi_n$ and $\psi_p$. For a basic discussion of the field-theoretical treatment of noninteracting fermions, in
particular the roles of finite temperature and chemical potential, see appendix \ref{app:fermions}. The free mesonic Lagrangian is
\be
{\cal L}_{\sigma,\omega} = \frac{1}{2}\left(\partial_\mu\sigma\partial^\mu\sigma -m_\sigma^2\sigma^2\right)-\frac{1}{4}\omega_{\mu\nu}\omega^{\mu\nu}
+\frac{1}{2}m_\omega^2\omega_\mu\omega^\mu \, , 
\ee
where $\omega_{\mu\nu}\equiv \partial_\mu\omega_\nu-\partial_\nu\omega_\mu$, and the interaction Lagrangian with Yukawa interactions
between the nucleons and the mesons is
\be
{\cal L}_I = g_\sigma\bar{\psi}\sigma\psi + g_\omega\bar{\psi}\gamma^\mu\omega_\mu\psi \, .
\ee
We shall consider isospin-symmetric matter, i.e., the masses and chemical potentials of protons and neutrons are assumed to be identical.
In general, $\mu$ is a matrix $\mu={\rm diag}(\mu_n,\mu_p)={\rm diag}(\mu_B+\mu_I,\mu_B-\mu_I)$ with the baryon and isospin chemical
potentials $\mu_B$ and $\mu_I$. Thus, in other words, we assume the 
isospin chemical potential to vanish. We can then simply denote $\mu\equiv \mu_B = \mu_n=\mu_p$. 
Also the interactions between the nucleons are assumed to be symmetric, i.e., the $nn$, $pp$, and $np$ 
interactions are identical. An isospin asymmetry in the interactions can be included by adding $\rho$-meson exchange. We will briefly discuss this
in Sec.\ \ref{sec:hyperons}. Also kaon condensation induces an asymmetry, discussed in Sec.\ \ref{sec:kaon}.

The parameters of the model are the masses and the coupling constants. The masses are
\be
m_N=939\, {\rm MeV} \, , \qquad m_\omega = 783\, {\rm MeV} \, , \qquad m_\sigma = (500-600)\, {\rm MeV} \, .
\ee
The $\sigma$ meson is in fact a broad resonance and thus we can only approximately assign a mass to this meson. Below we shall use 
$m_\sigma = 550\, {\rm MeV}$. The additional parameters are the coupling constants $g_\sigma$, $g_\omega$. We shall discuss below how
they are fixed. 

In order to compute the equation of state, we need to consider the partition function
\be
Z = \int {\cal D}\bar{\psi}{\cal D}\psi{\cal D}\sigma{\cal D}\omega \, \exp\int_X \, {\cal L} \, ,
\ee
where we abbreviated 
\be
\int_X \equiv \int_0^\beta d\tau\int d^3x \, , 
\ee
with the inverse temperature $\beta=1/T$. We shall allow for vacuum expectation values of the mesons. 
To this end, we write the meson fields as a sum of the condensate and fluctuations,
\begin{subequations}
\bea
\sigma &\to& \bar{\sigma} + \sigma \, , \\
\omega_\mu &\to& \bar{\omega}_0\delta_{0\mu} + \omega_\mu \, ,
\eea
\end{subequations}
as explained in appendix \ref{app:bosons} for a general bosonic field.
Now the simplest approximation is to neglect the fluctuations. This corresponds to the mean-field approximation. In this case 
the interaction between the nucleons and the mesons is simplified to a mesonic background, or mesonic mean field, which is 
seen by the nucleons. We can then simply drop all derivative terms of the mesons. As a consequence, the meson 
mean fields merely act as corrections to the nucleon mass and chemical potential, and we obtain the Lagrangian
\be
{\cal L} = \bar{\psi}\left(i\gamma^\mu\partial_\mu-m_N^*+\mu^*\gamma_0\right)\psi - \frac{1}{2}m_\sigma^2\bar{\sigma}^2
+\frac{1}{2}m_\omega^2\bar{\omega}_0^2 \, ,
\ee  
with 
\begin{subequations}
\bea
m_N^*&\equiv& m_N -g_\sigma \bar{\sigma} \, , \label{defmNstar}\\
\mu^*&\equiv& \mu -g_\omega \bar{\omega}_0 \, .
\eea
\end{subequations}
It is important to keep in mind that the actual chemical potential, associated with nucleon number, is $\mu$, not $\mu^*$. This becomes important
for the correct thermodynamic relations, see footnote before Eqs.\ (\ref{Peps}). The new effective ``chemical potential'' $\mu^*$ nevertheless
has physical meaning since it determines the Fermi energy as we shall see below.
 
The partition function now becomes
\be
Z = e^{\frac{V}{T}\left(-\frac{1}{2}m_\sigma^2\bar{\sigma}^2
+\frac{1}{2}m_\omega^2\bar{\omega}_0^2\right)}\,\int {\cal D}\bar{\psi}{\cal D}\psi \exp\int_X\bar{\psi}
\left(i\gamma^\mu\partial_\mu-m_N^*+\mu^*\gamma_0\right)\psi \, .
\ee
The evaluation of the free fermionic part (with modified mass and chemical potential) is now straightforward and is done in detail in 
appendix \ref{app:fermions}. Here we only repeat the most important steps.
One first introduces the Fourier transforms 
\be
\psi(X) = \frac{1}{\sqrt{V}}\sum_K e^{-iK\cdot X}\psi(K) \, , \qquad 
\bar{\psi}(X) = \frac{1}{\sqrt{V}}\sum_K e^{iK\cdot X}\bar{\psi}(K) \, .
\ee
Our conventions are $K=(-i\omega_n,{\bf k})$, $X=(-i\tau,{\bf x})$, and $K\cdot X=k_0x_0-{\bf k}\cdot{\bf x} 
= -(\omega_n\tau+{\bf k}\cdot{\bf x})$, with the fermionic Matsubara frequencies $\omega_n = (2n+1)\pi T$. Thus, after performing the $X$ 
integral in the exponent one obtains
\be
Z = e^{\frac{V}{T}\left(-\frac{1}{2}m_\sigma^2\bar{\sigma}^2+\frac{1}{2}m_\omega^2\bar{\omega}_0^2\right)}
\,\int {\cal D}\psi^\dag{\cal D}\psi \exp\left[-\sum_K \psi^\dag(K)\frac{G^{-1}(K)}{T}\psi(K)\right]
\,, 
\ee
with the inverse nucleon propagator
\be
G^{-1}(K) = -\gamma^\mu K_\mu-\gamma_0\mu^*+m_N^* \, .
\ee
Now using the standard formula for the functional integral over Grassmann variables one obtains
\be
Z = e^{\frac{V}{T}\left(-\frac{1}{2}m_\sigma^2\bar{\sigma}^2+\frac{1}{2}m_\omega^2\bar{\omega}_0^2\right)}\, {\rm det}\frac{G^{-1}(K)}{T} \, , 
\ee
where the determinant is taken over momentum space, Dirac space, and the (here trivial) neutron-proton space.
Consequently, 
\bea
\ln Z &=& \frac{V}{T}\left(-\frac{1}{2}m_\sigma^2\bar{\sigma}^2+\frac{1}{2}m_\omega^2\bar{\omega}_0^2\right)\non
&&\hspace{-1cm}+4V\int\frac{d^3{\bf k}}{(2\pi)^3}\left[\frac{E_k}{T}+\ln\left(1+e^{-(E_k-\mu^*)/T}\right)
+\ln\left(1+e^{-(E_k+\mu^*)/T}\right)\right]
\, ,
\eea
where we have performed the Matsubara sum and taken the thermodynamic limit, and where we have defined the single-nucleon
energy
\be
E_k = \sqrt{k^2+(m_N^*)^2} \, .
\ee
The pressure then becomes 
\be 
P = \frac{T}{V}\ln Z = -\frac{1}{2}m_\sigma^2\bar{\sigma}^2+\frac{1}{2}m_\omega^2\bar{\omega}_0^2 + P_N \, ,
\ee
with the nucleon pressure (after subtracting the vacuum part)
\be \label{PN}
P_N\equiv 4T\int\frac{d^3{\bf k}}{(2\pi)^3}\left[\ln\left(1+e^{-(E_k-\mu^*)/T}\right)+\ln\left(1+e^{-(E_k+\mu^*)/T}\right)\right] \, .
\ee
We have thus derived the fermionic pressure already used in chapter \ref{sec:massradius}, see Eq.\ (\ref{pressureTS1}), from thermal field theory. 
The factor 4 counts the two spin degrees of freedom and the two baryon degrees of freedom (proton and neutron). We also have obtained the 
contribution of antiparticles, for which $\mu^*\to -\mu^*$. 

The meson condensates have to be determined by maximizing the pressure. We obtain
\begin{subequations} \label{maxso}
\bea
0&=&\frac{\partial P}{\partial\bar{\sigma}} = -m_\sigma^2\bar{\sigma}-g_\sigma \frac{\partial P_N}{\partial m_N^*} \, , \\
0&=&\frac{\partial P}{\partial\bar{\omega}_0} = m_\omega^2\bar{\omega}_0-g_\omega \frac{\partial P_N}{\partial \mu^*} \, . 
\eea
\end{subequations}
In terms of the baryon and scalar densities 
\begin{subequations}
\bea
n_B &=& \langle \psi^\dag\psi \rangle = \frac{\partial P_N}{\partial \mu}=
 \frac{\partial P_N}{\partial \mu^*} = 4\sum_{e=\pm} e\,\int\frac{d^3{\bf k}}{(2\pi)^3}\frac{1}{e^{(E_k-e\mu^*)/T}+1} \, , \\
n_s &=& \langle \bar{\psi}\psi \rangle = -\frac{\partial P_N}{\partial m_N^*} = 4\sum_{e=\pm} \int\frac{d^3{\bf k}}{(2\pi)^3}\frac{m_N^*}{E_k}
\frac{1}{e^{(E_k-e\mu^*)/T}+1} \, , 
\eea
\end{subequations}
we can write the equations for the condensates (\ref{maxso}) as
\begin{subequations}
\bea
\bar{\sigma}&=&\frac{g_\sigma}{m_\sigma^2}\,n_s \, , \\
\bar{\omega}_0&=&\frac{g_\omega}{m_\omega^2}\,n_B \, . \label{baromega}
\eea
\end{subequations}
It is useful to rewrite the first of these equations as an equation for the corrected mass $m_N^*$ rather than for
the condensate $\bar{\sigma}$,
\be \label{mNstar}
m_N^* = m_N-\frac{g_\sigma^2}{m_\sigma^2}\,n_s \, ,
\ee
where we have used Eq.\ (\ref{defmNstar}).
We now take the zero-temperature limit, $T\ll m_N,\mu$, which is justified since 
the temperatures of interest are at most of the order of 10 MeV, while the baryon chemical potentials are above 1 GeV. 
The Fermi distribution function then becomes a step function. In particular, all antiparticle contributions
vanish. We obtain\footnote{One 
has to be careful with the thermodynamic relations in deriving the energy density (\ref{epsilon3}): 
remember that the actual chemical potential associated with baryon number $n_B$ is $\mu$, not $\mu^*$. This
means that the pressure at zero temperature can be written as $P=-\epsilon +\mu n_B$. The last
term of the pressure (term in square brackets on the right-hand side of Eq.\ (\ref{pressure3})) comes from a term of 
the structure $-\epsilon_0 +\mu^* n_B$, cf.\ for instance Eq.\ (\ref{pressure1}). 
With $\mu^*=\mu-g_\omega \bar{\omega}_0$ and the expression for $\bar{\omega}_0$ in Eq.\ (\ref{baromega})
we can write this as 
\bea
P&=&-\epsilon_0+\mu^* n_B+\frac{1}{2}\frac{g_\omega^2}{m_\omega^2}\,n_B^2-\frac{1}{2}\frac{g_\sigma^2}{m_\sigma^2}\,n_s^2  \non 
&=& -\left(\epsilon_0+\frac{1}{2}\frac{g_\omega^2}{m_\omega^2}\,n_B^2+\frac{1}{2}\frac{g_\sigma^2}{m_\sigma^2}\,n_s^2\right) +\mu n_B \, ,
\eea
from which we can read off the energy density 
\be
\epsilon= \epsilon_0+\frac{1}{2}\frac{g_\omega^2}{m_\omega^2}\,n_B^2+\frac{1}{2}\frac{g_\sigma^2}{m_\sigma^2}\,n_s^2\, , 
\ee
which yields Eq.\ (\ref{epsilon3}).} 
\begin{subequations} \label{Peps}
\bea 
P&=& \frac{1}{2}\frac{g_\omega^2}{m_\omega^2}\,n_B^2-\frac{1}{2}\frac{g_\sigma^2}{m_\sigma^2}\,n_s^2 \non
&&+\frac{1}{4\pi^2}\left[\left(\frac{2}{3}k_F^3-(m_N^*)^2k_F\right)E_F^* + (m_N^*)^4\ln\frac{k_F+E_F^*}{m_N^*}\right]
 \, , \label{pressure3} \\
\epsilon &=& \frac{1}{2}\frac{g_\omega^2}{m_\omega^2}\,n_B^2+\frac{1}{2}\frac{g_\sigma^2}{m_\sigma^2}\,n_s^2 \non
&&+\frac{1}{4\pi^2}\left[\left(2k_F^3+(m_N^*)^2k_F\right)E_F^* - (m_N^*)^4\ln\frac{k_F+E_F^*}{m_N^*}\right]
 \, , \label{epsilon3}
\eea
\end{subequations}
where we have defined the Fermi energy
\be
E_F^* = \mu^* = \sqrt{k_F^2+(m_N^*)^2} \, , 
\ee
and where the zero-temperature densities are
\begin{subequations}
\bea
n_B &=& \frac{2k_F^3}{3\pi^2} \,, \label{nBzero}\\
n_s &=& \frac{m_N^*}{\pi^2}\left[k_F E_F^*-(m_N^*)^2\ln\frac{k_F+E_F^*}{m_N^*}\right] \, .\label{nszero}
\eea
\end{subequations}
Pressure and energy density in Eqs.\ (\ref{Peps}) define the equation of state which has to be determined numerically. 
We may discuss the limits of small ($k_F\to 0$) and large ($k_F\to \infty$)
density analytically. For small density we find
\be
n_s\simeq \frac{2 k_F^3}{3\pi^2} = n_B\, ,
\ee
neglecting terms of the order of $k_F^5/(m_N^*)^2\ll k_F^3$ and higher.
Therefore, from Eq.\ (\ref{mNstar}) we conclude
\be
m_N^* \simeq m_N \, ,
\ee
where we have suppressed terms of the order of $k_F^3/m_\sigma^2\ll m_N$.
The pressure and the energy density are, within this small-density approximation, dominated by the nucleonic pressure $P_N$,
\be
P\simeq \frac{2k_F^5}{15\pi^2m_N} \, ,\qquad \epsilon \simeq \frac{2m_N k_F^3}{3\pi^2} \, . 
\ee
Comparing with Eqs.\ (\ref{eps2pressure2}) we see that we have exactly reproduced the noninteracting limit. This is no surprise
because the only effect of the interactions in the present approach is the modification of $\mu$ and $m_N$. 
In the small-density limit these effects are negligible and
we are back to the noninteracting result, where the equation of state has the form $P\propto \epsilon^{5/3}$. 

\begin{figure}[t]
\begin{center}
\includegraphics[width=0.65\textwidth]{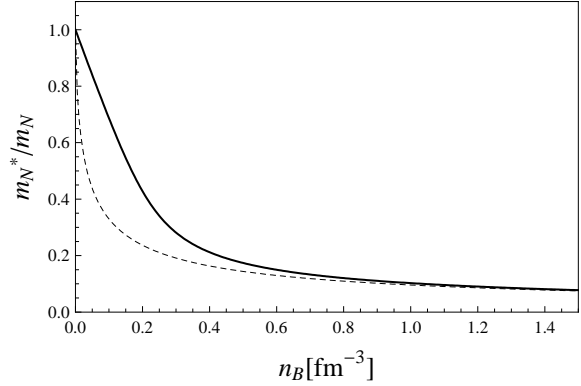}
\caption{Density-dependent effective nucleon mass $m_N^*$ at $T=0$ in the Walecka model in units of the zero-density mass $m_N$ and as a function
of the baryon density $n_B$. Solid line: full numerical result. Dashed line: high-density approximation from 
Eq.\ (\ref{mNstarhighn}). 
}
\label{figwalecka1}
\end{center}
\end{figure}

For large $k_F$, on the other hand, we have
\be
n_s \simeq \frac{m_N^*k_F^2}{\pi^2} \, , 
\ee
and thus 
\be \label{mNstarhighn}
m_N^*\simeq \frac{m_N}{1+\frac{g_\sigma^2k_F^2}{m_\sigma^2\pi^2}} \, .
\ee
We see that the effective nucleon mass goes to zero for large densities. For general values of the Fermi momentum, the effective mass has 
to be computed numerically from Eqs.\ (\ref{mNstar}) and (\ref{nszero}), see Fig.\ \ref{figwalecka1}. 

At large densities, the nucleonic pressure $P_N$ as well as the pressure from the scalar meson (which is proportional to $n_s^2/m_\sigma^2$)
behave like $k_F^4$. 
Therefore, the total pressure is dominated by the vector meson contribution which is proportional to $n_B^2/m_\omega^2$ and thus
behaves like $k_F^6$, 
\be
P\simeq \epsilon \simeq \frac{1}{2}\frac{g_\omega^2}{m_\omega^2}\,n_B^2 \, .
\ee
Consequently, the speed of sound approaches the speed of light at large densities,
\be
c_s^2 \equiv \frac{\partial P}{\partial \epsilon} \simeq 1 \, .
\ee
So far, our model cannot be used quantitatively since we have not yet fixed the numerical values of the coupling constants.
To do so one requires the model to reproduce the saturation density $n_0$ and the binding energy per nucleon at saturation $E_0$,  
\be \label{binding}
n_0 = 0.153\,{\rm fm}^{-3} \, , \qquad E_0 \equiv \left(\frac{\epsilon}{n_B}-m_N\right)_{n_B=n_0} = - 16.3\, {\rm MeV} \, .
\ee
Note the difference between the binding energy in (infinite) nuclear matter and the binding energy in finite nuclei. The latter is $-8.8\,{\rm MeV}$ 
for iron, see Eq.\ (\ref{bindiron}). 

\begin{figure}[t]
\begin{center}
\includegraphics[width=0.65\textwidth]{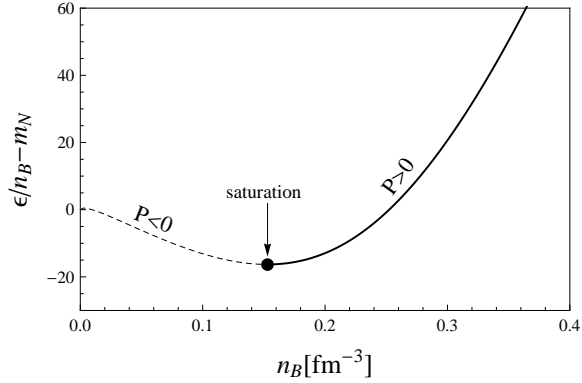}
\caption{Binding energy per nucleon at zero temperature in the Walecka model as a function of baryon density, obtained from computing the 
energy density with the density-modified nucleon mass. The binding energy has a minimum at which the pressure $P$ is zero, i.e., at this 
point nuclear matter is self-bound, and the corresponding density is called {\it saturation density}. 
The two parameters of the model, namely the coupling constants $g_\sigma$ and $g_\omega$, are fixed such that the binding energy per 
nucleon is $E_0=-16.3\,{\rm MeV}$ at the saturation density $n_0=0.153\,{\rm fm}^{-3}$. 
}
\label{figbinding}
\end{center}
\end{figure}

We leave it as an exercise to compute the coupling constants from the values (\ref{binding}), see problem \ref{prob3}. 
One obtains $g_\omega^2/(4\pi) = 14.717$ and $g_\sigma^2/(4\pi) = 9.537$. The result for the density-dependent binding energy with 
these values for the coupling constants is shown in Fig.\ \ref{figbinding}.
This figure shows that there is a finite density $n_0$ where the binding energy is minimal. This is a basic feature of nuclear matter which 
has to be reproduced by any physically meaningful model, reflecting the properties of the nuclear forces. It says in particular that 
if you add nucleons to a
large nucleus the density will stay approximately constant because there is a preferred distance between the nucleons that minimizes
the energy. We have implicitly made use of this fact in our estimate of the nucleon number in a neutron star at the beginning of 
chapter \ref{sec:massradius}. In the limit of infinite symmetric nuclear matter and ignoring the Coulomb forces, this density at which 
the binding energy is minimal is $n_0=0.153\,{\rm fm}^{-3}$. It is called {\it saturation density}. At the saturation density nuclear
matter is self-bound, i.e., it is stable at zero pressure. We have indicated in Fig.\ \ref{figbinding} 
that the minimum of the binding energy divides the
stable density regime with positive pressure from the unstable regime with negative pressure. The behavior of the pressure follows 
from the thermodynamic relation $P=-\frac{\partial E}{\partial V}$, where $E=\epsilon V$ is the energy and $V$ the volume, which implies
\be \label{PnBeps}
P = n_B^2\frac{\partial(\epsilon/n_B)}{\partial n_B} \, .
\ee
Consequently, at the minimum of $\epsilon/n_B$ as a function of $n_B$ the system has zero pressure. Moreover, we see that a {\it decrease} in the 
binding energy per baryon number upon {\it increasing} the baryon number leads to a negative pressure. (At very small densities, barely visible
in the plot, the energy also increases with density, i.e., $P>0$. This is the regime where the nucleons are too far apart to feel their 
attraction; the increasing energy is then a consequence of the increasing kinetic energy.) 
 
In our context of compact stars, the self-boundedness of nuclear matter implies that nuclear matter can
exist at the surface of the star where the pressure vanishes. In the interior, the gravitational pressure compresses the matter to densities
larger than $n_0$. As we see from the figure, this compressed matter, in turn, has itself positive pressure to counterbalance the pressure 
from gravity.  This is the reason why the high-density part of the curve in Fig.\ \ref{figbinding}  
is relevant for astrophysical applications. We shall see in the next subsection why the Walecka model in the simple form discussed here 
cannot be trusted for densities much larger than $n_0$ and how the model can be improved to yield predictions for the high-density regime.

\subsection{Including scalar interactions}

The Walecka model accommodates important aspects of nuclear matter such as 
the existence of a saturation density whose realistic value is reproduced upon fitting the parameters of the model. We have already discussed 
on general grounds that extrapolations to high densities are uncontrolled, and thus the Walecka model (and all similar models of this kind) 
have to be improved in an interplay with experimental observations, for example astrophysical data. 
But there is even a more obvious shortcoming of the simple version of the Walecka model discussed so far. 
Even at the saturation density it fails in its prediction for the {\it incompressibility} of nuclear matter which is defined as  
\be \label{defK2}
K\equiv k_F^2\frac{\partial^2(\epsilon/n_B)}{\partial k_F^2} \, . 
\ee
This quantity is a measure for the stiffness of nuclear matter. In some literature, $K$ is also called 
{\it compression modulus} or,
somewhat misleadingly, ``compressibility''. To see that a large value of $K$ corresponds to ``stiff'' matter, 
start from the usual thermodynamic definition for the compressibility $\chi$,
\be
\frac{1}{\chi} = n_B\frac{\partial P}{\partial n_B} = n_B^2\frac{\partial^2 \epsilon}{\partial n_B^2}\, .
\ee
This definition says that easily compressible (``soft'') matter has a small change in pressure upon changing the density. 
For the second equality we have used Eq.\ (\ref{PnBeps}).

On the other hand, from the definition (\ref{defK2}) we obtain
\bea \label{compress3}
K &=& k_F^2\frac{\partial^2(\epsilon/n_B)}{\partial n_B^2}\left(\frac{\partial n_B}{\partial k_F}\right)^2 = 
9 n_B^2\frac{\partial^2(\epsilon/n_B)}{\partial n_B^2} \non
&=& 9n_B \frac{\partial^2 \epsilon}{\partial n_B^2} + 18 \left(\frac{\epsilon}{n_B}-\frac{\partial\epsilon}{\partial n_B}\right) \, ,
\eea
where $\partial n_B/\partial k_F=3n_B/k_F$ (see Eq.\ (\ref{nBzero})) has been used. Now recall that in equilibrium, i.e., at the saturation density
where the pressure vanishes, $\epsilon/n_B$ as a function of $n_B$ has a minimum,
\be 
0=\left.\frac{\partial(\epsilon/n_B)}{\partial n_B}\right|_{n_B=n_0} = -\frac{1}{n_B}\left(\frac{\epsilon}{n_B}-
\frac{\partial\epsilon}{\partial n_B}\right)_{n_B=n_0}   \, .
\ee
Consequently, the second term on the right-hand side of Eq.\ (\ref{compress3}) vanishes at $n_B=n_0$ and the relation between $\chi$ and $K$ becomes
at saturation
\be
\frac{1}{\chi} = \frac{n_0 K}{9} \, ,
\ee
i.e., a large compressibility $\chi$ corresponds to a small incompressibility $K$, as it should be.
 
The calculation of the incompressibility in the given model yields $K\simeq 560\,{\rm MeV}$. 
This is more than twice as much as the experimentally inferred value. Also the nucleon mass itself can be determined experimentally and 
compared to the prediction of the model.
In total, there are thus four values which the model should reproduce. To improve the model, we add cubic and 
quartic scalar self-interactions of the form
\be
{\cal L}_{I,\sigma}=-\frac{b}{3} m_N(g_\sigma \sigma)^3-\frac{c}{4}(g_\sigma\sigma)^4 
\ee
to the Lagrangian (\ref{LWal}). Besides the phenomenological need of these terms, there is also a theoretical reason for their presence:
the model becomes renormalizable. With the self-interactions we have introduced two new dimensionless constants $b$ and $c$ which can be used, 
together with the two couplings 
$g_\sigma$, $g_\omega$ to fit four experimental values. Namely, the two from Eq.\ (\ref{binding}) plus the incompressibility and the {\it Landau mass}
\be
K \simeq 250\,{\rm MeV} \, , \qquad m_L = 0.83\,m_N \, .
\ee
The Landau mass is defined as
\be
m_L = \frac{k_F}{v_F} \, , 
\ee
where 
\be
v_F = \left.\frac{\partial E_k}{\partial k}\right|_{k=k_F} 
\ee
is the Fermi velocity. It is plausible that the Landau mass is experimentally more accessible than the mass parameter $m_N^*$ since it 
is an effective mass for fermions at the Fermi surface where all low-energy excitations are located. 

In the mean field approximation, it is easy to include the effect of the scalar self-interactions. The pressure becomes
\be
P=-\frac{1}{2}m_\sigma^2\bar{\sigma}^2-\frac{b}{3}m_N(g_\sigma\bar{\sigma})^3-\frac{c}{4}(g_\sigma\bar{\sigma})^4
+\frac{1}{2}m_\omega^2\bar{\omega}_0^2 +P_N \, , 
\ee
with $P_N$ defined in Eq.\ (\ref{PN}). The implicit equation for the effective nucleon mass (\ref{mNstar}) now receives contributions
from the additional terms and becomes
\be
m_N^* = m_N-\frac{g_\sigma^2}{m_\sigma^2}n_s +\frac{g_\sigma^2}{m_\sigma^2}\left[bm_N(m_N-m_N^*)^2+c(m_N-m_N^*)^3\right] \, .
\ee
To fit the four above mentioned values, one has to choose $g_\sigma^2/(4\pi) = 6.003$, $g_\omega^2/(4\pi) = 5.948$, $b=7.950\cdot 10^{-3}$, and
$c=6.952\cdot 10^{-4}$. The numerical evaluation of the binding energy is left as an exercise. 
\begin{figure}[t]
\begin{center}
\includegraphics[width=0.65\textwidth]{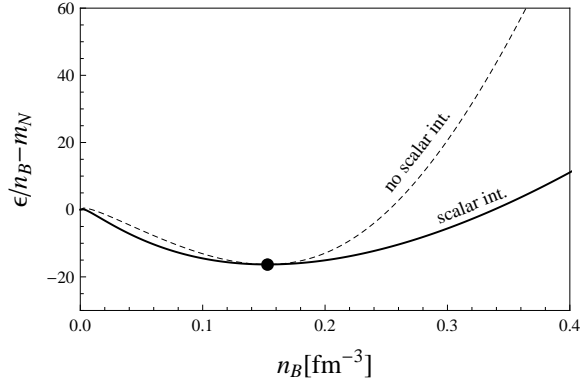}
\caption{Binding energy per nucleon as a function of density in the Walecka model, including cubic and quartic scalar self-interactions
(solid line). 
The four parameters of the model are fixed to the saturation density, the binding energy per nucleon at the saturation density, the 
incompressibility, 
and the Landau mass. For comparison, the dashed line shows the result from Fig.\ \ref{figbinding}, i.e., without scalar interactions.
The scalar interactions account for a much softer equation of state.
}
\label{figwalecka2}
\end{center}
\end{figure}
The result is plotted in Fig.\ \ref{figwalecka2}
and shows that the behavior at large densities has changed significantly compared to the case without scalar interactions. In particular, the 
lower value of the incompressibility goes along with a softer equation of state at large densities. In other words, if you choose a fixed 
binding energy on the vertical axis you find a larger baryon density after taking into account the scalar interactions. The matter has thus become 
easier to compress in the high-density regime, in accordance with a lower incompressibility. 
(See also discussion about stiff and soft equations of state in Sec.\ \ref{sec:MRinter}.)

\section{Hyperons}
\label{sec:hyperons}

In the interior of a compact star, densities can be as high as several times nuclear saturation density. Therefore, baryons with 
strangeness, {\it hyperons}, may occur (as well as muons). The lightest of these states are given by the baryon octet, see Table \ref{tableoctet}.
\begin{table*}[t]
\begin{center}
\begin{tabular}{|c||c|c|c|c|c|c|c|c|c|} 
\hline
& $p$ & $n$ & $\Lambda$ & $\;\;$$\Sigma^+$$\;\;$ &$\;\;$ $\Sigma^0$$\;\;$ &$\;\;$ $\Sigma^-$$\;\;$ & $\Xi^0$ & $\Xi^-$ \\ \hline\hline
$\;\;$ $m$ (MeV)$\;\;$ & \multicolumn{2}{c}{939}\vline  & $\;\;$1115$\;\;$ & \multicolumn{3}{c}{1190}\vline  & \multicolumn{2}{c}{1315}\vline  \\ \hline
$I_3$ &$\;\;$ 1/2$\;\;$ &$\;\;$ $-1/2$$\;\;$ & 0 & 1 & 0 & $-1$ &$\;\;$ 1/2$\;\;$ &$\;\;$ $-1/2$$\;\;$ \\ \hline
$Q$ & 1 & 0 & 0 & 1 & 0 & $-1$ & 0 & $-1$ \\ \hline
$S$ & \multicolumn{2}{c}{0}\vline  &\multicolumn{4}{c}{$-1$}\vline  &\multicolumn{2}{c}{$-2$}\vline \\ \hline
$J$ & \multicolumn{8}{c}{1/2} \vline \\ \hline
$\;\;$ quark content $\;\;$ & $uud$ & $udd$ & $uds$ & $uus$ & $uds$ & $dds$ & $uss$ & $dss$ \\ \hline
\end{tabular}
\end{center}
\caption{Mass, isospin, electric charge, strangeness, spin, and quark content for the spin-1/2 baryon octet.}
\label{tableoctet}
\end{table*}
It is rather straightforward to incorporate hyperons in the kind of model discussed above. Of course, the evaluation becomes
more laborious, and the model has many more parameters. Let us therefore briefly discuss the model with the hyperon
octet without going into too much detail. 

\begin{figure}[t]
\begin{center}
\includegraphics[width=0.65\textwidth,angle=-90]{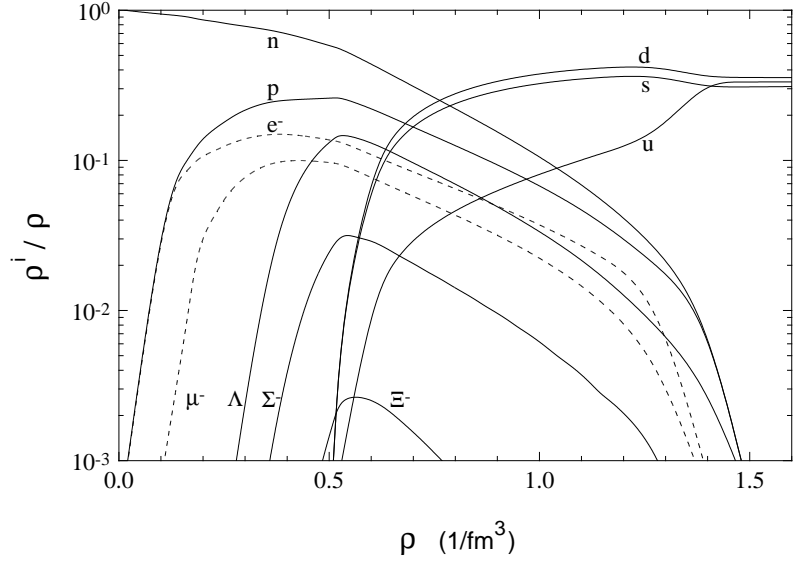}
\caption{Density fractions of baryons and leptons (and quarks, for a bag constant $B=250\,{\rm MeV}/{\rm fm}^{-3}$) 
as a function of the baryon number density. The figure is taken from Ref.\ \cite{Weber:2004kj3} where more details about the underlying 
calculation can be found.
For sufficiently large densities, hyperons and muons appear, and at densities of a few times nuclear ground state densities
their density fractions becomes comparable to the fractions of nucleons and electrons. There is a region of coexistence of 
deconfined quark matter and baryonic matter, see Sec.\ \ref{sec:mixed} for a discussion of these mixed phases. The curves shown 
here depend on the chosen models for nuclear and quark matter and the value of the bag constant.}
\label{fighyper}
\end{center}
\end{figure}

The interaction between the baryons is now extended by interactions mediated by the $\phi$ and $\rho$ vector mesons.  
(The $\phi$ meson has quark content $\bar{s}s$; the $\rho$ meson has the same quark content as a pion, i.e., it can be considered as an 
excited state of the pion.) The Lagrangian is
\bea
{\cal L} &=& \sum_j\bar{\psi}_j \left(i\gamma^\mu\partial_\mu-m_j+\mu_j\gamma_0+g_{\sigma j}\sigma 
-g_{\omega j}\gamma^\mu\omega_\mu-g_{\phi j}\gamma^\mu\phi_\mu - g_{\rho j}\gamma^\mu\rho_\mu^a\tau_a\right)\psi_j \non
&&+\,\frac{1}{2}\left(\partial^\mu\sigma\partial_\mu\sigma-m_\sigma^2\sigma^2\right)-\frac{b}{3} m_N(g_\sigma \sigma)^3-\frac{c}{4}(g_\sigma\sigma)^4
\non
&&-\,\frac{1}{4}\omega^{\mu\nu}\omega_{\mu\nu}+\frac{1}{2}m_\omega^2\omega^\mu\omega_\mu \non
&&-\,\frac{1}{4}\phi^{\mu\nu}\phi_{\mu\nu}+\frac{1}{2}m_\phi^2\phi^\mu\phi_\mu \non
&&-\,\frac{1}{4}\rho^{\mu\nu}_a\rho_{\mu\nu}^a+\frac{1}{2}m_\rho^2\rho^\mu_a\rho_\mu^a \, .
\eea
Here, $j$ runs over all eight baryons and $\tau_a$ are the isospin generators. In a compact star, we have to require chemical equilibrium with 
respect to the weak interactions. In the case of hyperons, the conditions are
\begin{subequations} \label{condhyp}
\bea
\mu_p &=& \mu_n-\mu_e \, , \qquad \mu_\Lambda = \mu_n \\
\mu_{\Sigma^+} &=& \mu_n-\mu_e \, , \qquad \mu_{\Sigma^0} = \mu_n \\
\mu_{\Sigma^-} &=& \mu_n+\mu_e \, , \qquad \mu_{\Xi^0} = \mu_n \\
\mu_{\Xi^-} &=& \mu_n+\mu_e \, ,
\eea
\end{subequations} 
and, including muons, $\mu_e=\mu_\mu$. The conditions (\ref{condhyp}) all come from weak processes which we have already discussed, see
Eqs.\ (\ref{proc3flavor}). For example the process $n\to \Sigma^++e+\bar{\nu}_e$, which gives rise to the condition 
$\mu_{\Sigma^+}=\mu_n-\mu_e$, can be understood from the elementary processes as
\be
\left.\begin{array}{c} u+e\to s+\nu_e \\ d\to u+e+\bar{\nu}_e \\ d\to u + e+\bar{\nu}_e \end{array}\right\} \qquad udd\to uus+e+\bar{\nu}_e \, .
\ee
Electric neutrality is given by the constraint
\be
n_p + n_{\Sigma^+} = n_e + n_\mu + n_{\Sigma^-}+n_{\Xi^-} \, .
\ee
We show the result of baryon and lepton density fractions in a model similar to the one discussed here in Fig.\ \ref{fighyper}.  

As a result of this rough discussion and the curves in the figure we learn that hyperons can be included in a rather straightforward 
extension of the
simple Walecka model and that hyperons do appear for sufficiently large densities. The physical reasons are that $(i)$ they 
{\it can} appear because the baryon chemical potential is large enough to provide energies larger than their mass, $(ii)$ they {\it do}
appear because $(a)$ the systems seeks to acquire neutrality and does so with electrons at low densities; if hyperons are available, 
electrons in high-energy states can be replaced by hyperons in low-energy states and $(b)$ the system seeks to 
become isospin symmetric; at low densities it is highly isospin asymmetric, and hyperons with nonzero isospin number provide a means to 
symmetrize the system.

\section{Kaon condensation}
\label{sec:kaon}

Another possible variant of dense nuclear matter, besides the occurrence of hyperons, is the condensation of mesons. Originally, pion condensation
was suggested \cite{Migdal:1973zm3}. Only many years later, it was realized that kaon condensation is possible in compact stars \cite{Kaplan:1986yq3}. 
This is somewhat 
surprising since kaons are much heavier than pions and thus pion condensation seems more likely. However, in the medium, the effective 
kaon mass becomes sufficiently small to allow for a kaon condensate. 

Kaon condensation is of interest for these lectures for several reasons. Besides being a variant of dense matter and thus relevant for the physics
of compact stars, its discussion requires the introduction of several important concepts in the theory of the strong interaction. It is thus
also interesting from a fundamental point of view. Moreover, we shall encounter kaon condensation again later in these lectures, when 
we discuss the quark-matter relatives of the kaon, see Sec.\ \ref{sec:CFLK0}. 

To explain kaon condensation, we will first have to say what a kaon is and will do so with the help
of chiral symmetry and spontaneous breaking thereof. Then, we will discuss chiral perturbation theory. This is one possible method
to study kaon condensation and has been used in the original work \cite{Kaplan:1986yq3}. For another approach, using models similar to 
the above discussed Walecka model, see for instance Ref.\ \cite{Ramos:2000dq3} and references therein. 
The evaluation of the chiral model has to be done numerically, so we will more or less only
be concerned with setting up and understanding the basic equations. As a modest goal, we will try to understand the onset of kaon condensation,
i.e., we will show how to compute the critical baryon density at which there is a second-order phase transition to the kaon-condensed phase.

\subsection{Chiral symmetry of QCD}
\label{sec:chiralsym}

Kaon condensation can be discussed in a low-energy effective theory, here chiral perturbation theory. This theory should describe
the fundamental theory, QCD, in the low-energy limit. In order to construct the theory, we need to understand the underlying symmetries.
The QCD Lagrangian is
\be
{\cal L}_{\rm QCD} = \bar{\psi}(i\gamma^\mu D_\mu + \mu\gamma_0 - M) \psi + {\cal L}_{\rm gluons}\, , 
\ee
with the quark spinor $\psi$ in color, flavor, and Dirac space, the mass matrix in flavor space
\be
M = \left(\begin{array}{ccc} m_u & 0 & 0 \\ 0 & m_d & 0 \\ 0 & 0 & m_s \end{array}\right) \, ,
\ee
and the covariant derivative $D_\mu = \partial_\mu-igT_a A_\mu^a$, where $T_a=\lambda_a/2$ ($a=1,\ldots 8$) are the generators of the color
gauge group $SU(3)_c$ with the Gell-Mann matrices $\lambda_a$, $A_\mu^a$ are the corresponding gauge fields, and $g$ is the strong coupling 
constant. The chemical potential $\mu$ is a diagonal matrix in flavor space. Without taking into account the weak interactions, each flavor 
is conserved and there are three independent chemical potentials. We have already seen in the previous sections that after taking into account
weak interactions there are only two chemical potentials, one for quark (baryon) number, and one for electric charge.

The purely gluonic contribution to the Lagrangian is given by 
\be 
{\cal L}_{\rm gluons} = -\frac{1}{4} G_a^{\mu\nu}G^a_{\mu\nu} \, ,
\ee
where $G_{\mu\nu}^a=\partial_\mu A_\nu^a - \partial_\nu A_\mu^a+g f^{abc}A_\mu^b A_\nu^c$ with the $SU(3)_c$ structure constants is the
gluon field strength tensor. Here we are not interested in this gluonic part, since we focus on the transformations of the 
fermion fields and the resulting symmetries of the Lagrangian. 
Also later, when we shall use QCD for explicit calculations, the gluonic part is negligible because we always work at very small temperatures
compared to the quark (or baryon) chemical potential. The interactions of the quarks via gluon exchange, included in the covariant derivative, 
is of course important; in Sec.\ \ref{sec:QCDgap} this interaction will be used on the microscopic level.

We now introduce the chirality projectors
\be
P_R = \frac{1+\gamma_5}{2} \, , \qquad P_L = \frac{1-\gamma_5}{2} \, .
\ee
They obey the identities
\be
P_{R/L}^2 = P_{R/L} \, , \qquad P_{R/L}^\dag = P_{R/L}\, , \qquad P_RP_L=0 \, , \qquad P_R+P_L=1 \, ,
\ee
i.e., they form a complete set of orthogonal projectors. (These identities are obvious with $\gamma_5^2=1$ and $\gamma_5^\dag = \gamma_5$.)
For a physical picture, remember that, for massless quarks, chirality eigenstates are also eigenstates of helicity. Therefore, in this
case, there is a one-to-one correspondence between chirality and the projection of the fermion momentum onto its spin.
We define left- and right-handed quark spinors by
\be
\psi_{R/L}\equiv P_{R/L}\psi \, , 
\ee
such that $\psi = P_R\psi +P_L\psi = \psi_R+\psi_L$. 
Then, using 
\be
\{\gamma_5,\gamma_\mu\} = 0 \, , 
\ee
we can write the Lagrangian as
\bea \label{LQCDchiral}
{\cal L}_{\rm QCD} &=& \bar{\psi}_R(i\gamma^\mu D_\mu +\mu\gamma_0)\psi_R + \bar{\psi}_L(i\gamma^\mu D_\mu +\mu\gamma_0) \psi_L \non
&& -\bar{\psi}_RM\psi_L-\bar{\psi}_LM\psi_R 
+{\cal L}_{\rm gluons}\, .
\eea
Let us first discuss the massless case, $M=0$. In this case, separate rotations of left- and right-handed spinors leave the 
Lagrangian invariant,
\be
\psi_R\to e^{i\phi_R^a t_a}\psi_R \, , \qquad \psi_L\to e^{i\phi_L^a t_a}\psi_L \, .
\ee
Since we are interested in three quark flavors, $t_a$ are the nine generators of the flavor group $U(3)$, 
$t_0={\bf 1}$ and $t_a=T_a$ ($a=1,\ldots 8$).
Consequently, the Lagrangian is invariant under $U(3)_L\times U(3)_R$. The corresponding Noether currents are
\be
J_{a,R/L}^\mu = \bar{\psi}_{R/L}\gamma^\mu t_a\psi_{R/L} \, .
\ee
They can be rewritten in terms of vector and axial-vector currents
\begin{subequations}
\bea
J_{a,V}^\mu &\equiv& J_{a,R}^\mu + J_{a,L}^\mu = \bar{\psi}\gamma^\mu t_a\psi \, , \\ 
J_{a,A}^\mu &\equiv& J_{a,R}^\mu - J_{a,L}^\mu = \bar{\psi}\gamma^\mu t_a \gamma_5 \psi \, .
\eea
\end{subequations}
To see this, note that $P_R\gamma_5 = P_R$ and $P_L\gamma_5 = -P_L$. In QCD the singlet axial-vector current is in general 
not conserved,
\be
\partial_\mu J_{0,A}^\mu = -\frac{g^2N_f}{16\pi^2} G_{\mu\nu}^a\tilde{G}^{\mu\nu}_a \, , 
\ee
where $\tilde{G}^{\mu\nu} = \frac{1}{2}\epsilon^{\mu\nu\sigma\rho} G_{\sigma\rho}$ is the dual field strength tensor. This is referred to as the 
{\it axial anomaly}.
We are left with the symmetry group $SU(3)_R\times SU(3)_L\times U(1)_V$. 
The vector symmetry $U(1)_V$ corresponds to baryon number conservation and is therefore also denoted as $U(1)_B$. 
The flavor symmetry group $SU(3)_R\times SU(3)_L$ is referred to as {\it chiral symmetry}. As we can see from Eq.\ (\ref{LQCDchiral}), 
nonzero masses break the chiral symmetry {\it explicitly}. They do not break the $U(1)_V$ symmetry, and for the special case $m_u=m_d=m_s$ 
the subgroup $SU(3)_{R+L}$ of simultaneous $R$ and $L$ rotations remains a symmetry of the Lagrangian. 

{\it Spontaneous} 
breaking of chiral symmetry is realized by a chiral condensate of the form $\langle\bar{\psi}_L\psi_R\rangle$. This is analogous to 
spontaneous symmetry breaking in simple models such as $\phi^4$ theory
(see for instance the discussion of Bose-Einstein condensation in appendix \ref{app:bosons}), or in a superconductor, or in the Higgs mechanism.
The chiral condensate is only invariant under simultaneous right- and left-handed rotations, i.e., the 
symmetry breaking pattern is
\be \label{breakchiral}
G\equiv SU(3)_R\times SU(3)_L \to H\equiv SU(3)_{R+L} \, .
\ee
As a comparison, in $\phi^4$ theory with a complex scalar field $\phi$, we have $G=U(1)$, $H={\bf 1}$, which gives rise to the 
familiar ``Mexican hat'' potential with a negative quadratic and a positive quartic term in $|\phi|$.   
Spontaneous breaking of a global symmetry goes along with massless Goldstone bosons. In the Mexican hat, there is one massless excitation
along the bottom of the Mexican hat, given by the angular component of the order parameter (while the radial component corresponds to a massive
mode). Here, the bottom of the Mexican hat is not just a one-dimensional line. It is rather given by the coset space $G/H$ (which 
is simply $U(1)$ in $\phi^4$ theory). This space has ${\rm dim}\,G-{\rm dim}\,H$ generators. Consequently, with ${\rm dim}\, G=8+8=16$ 
and ${\rm dim}\, H = 8$, there are 8 Goldstone modes. 
They are described by the $SU(3)$ matrix
\be \label{defU}
U = e^{i\theta_a \lambda_a/f_\pi} \, , 
\ee
with the pion decay constant $f_\pi\simeq 93\, {\rm MeV}$. 
The meson fields $\theta_a$ of the Goldstone octet are usually reparametrized as
\be \label{mesonoctet}
\theta_a \lambda_a =\left(\begin{array}{ccc} \displaystyle{\frac{\pi^0}{\sqrt{2}} + \frac{\eta}{\sqrt{6}}} & \pi^+ & K^+ \\ \pi^- &
\displaystyle{-\frac{\pi^0}{\sqrt{2}} + \frac{\eta}{\sqrt{6}}}  & K^0  \\ K^- & \bar{K}^0 & \displaystyle{-\sqrt{\frac{2}{3}}\eta} 
\end{array}\right) \, .
\ee
Since the rows (columns) of this matrix carry left-handed flavor (right-handed anti-flavor) labels, it is easy to read off the quark content of the 
various mesons, e.g., $K^+\sim \bar{s}u$, $\pi^+\sim \bar{d}u$ etc.  
According to its chiral structure, the chiral matrix transforms under a transformation $g=(g_L,g_R)\in G$ as
\be
U \to g_L U g_R^\dag \, .
\ee

\subsection{Chiral Lagrangian}

In the (unrealistic) case of vanishing quark masses, the chiral symmetry is an exact symmetry and the Goldstone bosons are exactly massless.
Exploiting the analogy to the Mexican hat potential, this means that the bottom of the Mexican hat is truly flat.  
Quark masses break the chiral symmetry explicitly. However, if the masses are small compared to the characteristic scale of chiral 
symmetry breaking $\Lambda \sim 4\pi f_\pi \sim 1\, {\rm GeV}$ we can still consider the chiral symmetry as approximate. 
The bottom of the Mexican hat then gets distorted
on a scale small compared to the deepness of the potential, and the Goldstone bosons acquire small masses. In this case it is more
appropriate to speak of {\it pseudo}-Goldstone bosons. One might still hope to describe the system at low energies
by an effective theory which is built on the underlying chiral symmetry, although this symmetry is strictly speaking broken. 
The mass matrix $M$, now nonvanishing, is required to transform just as the chiral field $U$, i.e., 
\be
M \to g_L M g_R^\dag \, .
\ee
We require the chiral Lagrangian to be invariant under $G$. The kinetic term and the mass term of the resulting effective theory are
\be \label{calLU}
{\cal L}_U = \frac{f_\pi^2}{4}\Tr[\partial_\mu U\partial^\mu U^\dag] + c\Tr[M^\dag U+M U^\dag] + \ldots \, , 
\ee
where the trace is taken over flavor space. The two constants $f_\pi$ and $c$ have to be fitted to experimental values, similarly
to the constants of the Walecka model. 
In principle, higher order terms in $U$ are allowed but shall be neglected here. 
Note that the Goldstone fields themselves appear in the exponent of the field $U$, i.e., they 
are already present to all orders.

In the context of compact stars, we do not only want to describe isolated mesons. We also need to include baryons and their interactions.
The baryon octet fields are given by the matrix 
\be
B =\left(\begin{array}{ccc} \displaystyle{\frac{\Sigma^0}{\sqrt{2}} + \frac{\Lambda}{\sqrt{6}}} & \Sigma^+ & p \\ \Sigma^- &
\displaystyle{-\frac{\Sigma^0}{\sqrt{2}} + \frac{\Lambda}{\sqrt{6}}}  & n  \\ \Xi^- & \Xi^0 & \displaystyle{-\sqrt{\frac{2}{3}}\Lambda} 
\end{array}\right) \, ,
\ee
which includes the proton $p$, the neutron $n$, and the hyperons from Table \ref{tableoctet}. A simple way to understand the structure 
of this matrix is as follows. Consider the baryons as composed of a diquark and a quark. 
The diquarks form an antitriplet, i.e., one can think of the columns of the matrix as labelled by $(\bar{u}, \bar{d},\bar{s})$ which 
corresponds to the quark content $(ds,us,ud)$. Then the rows are simply labelled by the flavors in the fundamental representation $(u,d,s)$, 
and one easily checks that this yields the quark content of the baryons as given in Table \ref{tableoctet}.     

The free baryon Lagrangian is
\be \label{calLB}
{\cal L}_B = \Tr[\bar{B}(i\gamma^\mu\partial_\mu-m_B)B] \, ,
\ee
where $m_B\simeq 1.2\, {\rm GeV}$ is the $SU(3)_L\times SU(3)_R$ symmetric baryon mass. 
To write down the interaction between baryons and the mesons it is convenient to decompose the 
chiral field into left- and right-handed fields,
\be
U = \xi_L\xi_R^\dag \, , 
\ee
where, without loss of generality, we may choose
\be
\xi\equiv \xi_L=\xi_R^\dag \, , 
\ee
such that 
\be
U=\xi^2 \, .
\ee
We now add the meson-baryon interaction terms \cite{Kaplan:1986yq3,Thorsson:1993bu3} (see also chapter 7 of Ref.\ \cite{thomas3} for more details), 
\bea \label{calLI}
{\cal L}_I &=& i\Tr[\bar{B}\gamma_\mu[J_V^\mu,B]] + D\,\Tr[\bar{B}\gamma_\mu\gamma_5\{J_A^\mu,B\}]+F\,\Tr[\bar{B}\gamma_\mu\gamma_5[J_A^\mu,B]] 
\non[2ex]
&& +a_1\Tr[B^\dag(\xi M\xi +\xi^\dag M^\dag \xi^\dag)B] + a_2\Tr[B^\dag B (\xi M\xi +\xi^\dag M^\dag \xi^\dag)] \non[2ex]
&&+ a_3 \Tr[B^\dag B]\Tr[M\xi^2+M^\dag(\xi^\dag)^2] \, ,
\eea
with the additional constants $D$, $F$, $a_1$, $a_2$, $a_3$, and the vector and axial-vector currents
\begin{subequations}\label{VAV}
\bea
J_V^\mu &=& \frac{1}{2}(\xi^\dag\partial^\mu\xi+\xi\partial^\mu\xi^\dag) \, , \\
J_A^\mu &=& \frac{i}{2}(\xi^\dag\partial^\mu\xi-\xi\partial^\mu\xi^\dag) \, .
\eea
\end{subequations}  
The Lagrangian is an expansion in $M/\Lambda$ and $\partial/\Lambda$ with the scale of chiral symmetry breaking $\Lambda$. 
Higher order terms in these parameters are omitted. 
In summary, we have the Lagrangian 
\be
{\cal L} = {\cal L}_U + {\cal L}_B + {\cal L}_I \, .
\ee
Later we shall also add electron and muon contributions, but they are simple and we ignore them for now to keep the notation brief.

\subsection{Kaon-nucleon matter}
\label{sec:kaonnucleon}

Since we expect (charged) kaon condensation in a compact star rather than any other meson condensation (possibly there is pion condensation)
let us for simplicity drop all mesons other than the kaons. We can then write 
\be
U=e^{iQ}=\cos Q +i\sin Q \, , 
\ee
with 
\be \label{Q4567}
Q = \sum_{a=4}^7 \phi_a \lambda_a = \left(\begin{array}{ccc} 0 & 0 & \phi_4 - i\phi_5 \\ 0 & 0 & \phi_6-i\phi_7 \\
\phi_4 + i\phi_5 & \phi_6+i\phi_7 & 0\end{array}\right) \, , 
\ee
where we have absorbed $f_\pi$ into the fields $\phi_a\equiv \theta_a/f_\pi$ such that the $\phi_a$'s are dimensionless. 
We can now compute a simple expression for the matrix $U$. To this end we first verify by explicit matrix multiplication 
\be \label{Q3Q1}
Q^3 = \phi^2 Q \, , 
\ee
where
\be \label{theta3}
\phi^2 \equiv \phi_4^2+\phi_5^2 + \phi_6^2+\phi_7^2 \, .
\ee
From Eq.\ (\ref{Q3Q1}) we obtain (for instance via complete induction) 
\be
Q^{2n} = \phi^{2(n-1)}Q^2 \, ,
\ee
for all $n\ge 1$, which can be used to compute
\bea \label{cosQ1}
\cos Q &=& 1-\left(\frac{Q^2}{2!}-\frac{Q^4}{4!}+\ldots \right) 
= 1-Q^2\left(\frac{1}{2!}-\frac{\phi^2}{4!}+\ldots \right)\non
&=& 1-\frac{Q^2}{\phi^2}(1-\cos\phi) \, , 
\eea
and
\bea \label{sinQ1}
\sin Q &=& Q\left(1-\frac{Q^2}{3!}+\frac{Q^4}{5!}-\ldots \right) 
=  Q\left(1-\frac{Q^2}{3!}+\frac{\phi^2 Q^2}{5!}-\ldots \right) \non
&=& Q - \frac{\phi^2 Q}{3!}+\frac{\phi^4 Q}{5!}-\ldots 
= \frac{Q}{\phi}\sin\phi \, . 
\eea
As a further simplification let us now drop the neutral kaon fields, $\phi_6=\phi_7=0$, because we expect charged kaon condensation.
Then, from Eqs.\ (\ref{cosQ1}) and (\ref{sinQ1}) we obtain  
\be
U = \left(\begin{array}{ccc} \cos\phi&0&\;\;i\,\displaystyle{\frac{\phi_4-i\phi_5}{\phi}}\,\sin\phi\;\; \\[1.5ex] 
0&1&0 \\[1.5ex] 
\;\;i\,\displaystyle{\frac{\phi_4+i\phi_5}{\phi}}\,\sin\phi\;\;&0&\cos\phi \end{array}\right) \, .
\ee
Now we interpret the fields $\phi_{4,5}$ as vacuum expectation values, $\phi_{4,5}\to \langle\phi_{4,5}\rangle$, and neglect the fluctuations 
around this background. The general procedure to describe Bose-Einstein condensation, including fluctuations, is explained in 
appendix \ref{app:bosons} for the $\phi^4$ model.
The condensates are assumed to be constant in space and to have the time dependence $\phi(t,{\bf x})\to\phi e^{-i\mu_K t}$, i.e., 
our ansatz is
\begin{subequations}
\bea
\langle K^-\rangle&=&\langle\phi_4\rangle+i\langle\phi_5\rangle = \phi e^{-i\mu_K t} \, , \\
\langle K^+\rangle&=&\langle\phi_4\rangle-i\langle\phi_5\rangle = \phi e^{i\mu_K t} \, . 
\eea
\end{subequations}
The real, constant (i.e., space-time independent) value of $\phi$ has to be determined later from minimizing the free energy; 
$\mu_K$ plays the role of a kaon chemical potential, as we shall see more explicitly below. More precisely, $\mu_K$ is the chemical 
potential for $K^-$ while $-\mu_K$ is the chemical potential for $K^+$. We thus arrive at
\be
U = \left(\begin{array}{ccc} \cos\phi&0&\;\;ie^{i\mu_K t}\sin\phi\;\; \\[1.5ex] 
0&1&0 \\ [1.5ex]
\;\;ie^{-i\mu_K t}\sin\phi\;\;&0&\cos\phi \end{array}\right) \, .
\ee
We are now prepared to evaluate ${\cal L}_U$ from Eq.\ (\ref{calLU}). We shall neglect the masses of the up and down quarks such 
that $M\simeq {\rm diag}(0,0,m_s)$. We also define the kaon mass
\be
m_K^2=\frac{2cm_s}{f_\pi^2} \, .  
\ee
Rather than $c$, we shall later use the kaon mass $m_K\simeq 494\, {\rm MeV}$ as a parameter of the model.
This yields 
\be
{\cal L}_U = -V(\phi)
\ee
with the tree-level potential 
\be
V(\phi) = -\frac{f_\pi^2\mu_K^2}{2}\sin^2\phi + m_K^2f_\pi^2(1-\cos\phi) \, ,
\ee
where we have subtracted the constant vacuum contribution $V(\phi=0)$.
This potential contains the kaon condensate to all orders. We shall work with this expression below, but it is instructive to expand it
up to fourth order in $\phi$, 
\be
V(\phi) \simeq \frac{m_K^2-\mu_K^2}{2}\,(f_\pi\phi)^2+\frac{4\mu_K^2-m_K^2}{24f_\pi^2} (f_\pi\phi)^4 \, .
\ee
This is the familiar expression from a $\phi^4$ model for the free energy of a Bose condensate with chemical potential  $\mu_K$, 
mass $m_K$, and effective coupling $(4\mu_K^2-m_K^2)/(6f_\pi^2)$, see for instance Eq.\ (\ref{Uphitree}) in the appendix. As expected, 
condensation occurs for $\mu_K^2>m_K^2$ because in this case the quadratic term is negative and the quartic term positive, i.e., we have
recovered the Mexican hat potential (where we have already picked one direction since $\phi$ is real).

For the baryonic Lagrangian we only keep the lightest baryons, the proton and the neutron. From Eq.\ (\ref{calLB}) we thus obtain
\be
{\cal L}_B = \bar{p}(i\gamma^\mu\partial_\mu-m_B+\gamma^0\mu_p)p+\bar{n}(i\gamma^\mu\partial_\mu-m_B+\gamma^0\mu_n)n \, ,
\ee
where we have added the proton and neutron chemical potentials $\mu_p$ and $\mu_n$.
For the interaction terms we need 
\be
\xi = \left(\begin{array}{ccc} \cos(\phi/2)&0&\;\;ie^{i\mu_K t}\sin(\phi/2)\;\; \\[1.5ex] 
0&1&0 \\[1.5ex] 
\;\;ie^{-i\mu_K t}\sin(\phi/2)\;\;&0&\cos(\phi/2) \end{array}\right) \, ,
\ee
which obviously fulfills $\xi^2=U$.
By inserting this into Eqs.\ (\ref{VAV}) we see that the spatial components of the currents vanish, ${\bf J}_V ={\bf J}_A = 0$ 
(since there is no spatial dependence in the condensate), and the temporal components are 
\begin{subequations}
\bea 
J_V^0 &=& i\mu_K\sin^2(\phi/2)\left(\begin{array}{ccc} -1&0&0\\0&0&0\\0&0&1 \end{array}\right) \,, \non 
J_A^0 &=& i\mu_K\cos(\phi/2)\sin(\phi/2)\left(\begin{array}{ccc} 0&0&-e^{i\mu_Kt}\\0&0&0\\e^{-i\mu_Kt}&0&0 \end{array}\right)\, .
\eea
\end{subequations}
Hence the various nonzero terms needed for ${\cal L}_I$ in Eq.\ (\ref{calLI}) become
\begin{subequations}
\bea
i\Tr[\bar{B}\gamma_0[J_V^0,B]] &=& \mu_K\,(2p^\dag p+n^\dag n)\sin^2(\phi/2)\,, \\[1.5ex] 
a_1\Tr[B^\dag(\xi M\xi +\xi^\dag M^\dag \xi^\dag)B] &=& -2a_1m_s\,p^\dag p\sin^2(\phi/2)\, , \\[1.5ex] 
a_2\Tr[B^\dag B (\xi M\xi +\xi^\dag M^\dag \xi^\dag)] &=& 2a_2m_s(p^\dag p+n^\dag n)\cos^2(\phi/2) \, , \\[1.5ex] 
a_3 \Tr[B^\dag B]\Tr[M\xi^2+M^\dag(\xi^\dag)^2] &=& 2a_3m_s(p^\dag p+n^\dag n)[1-2\sin^2(\phi/2)] \, .
\eea
\end{subequations}
It is left as an exercise to verify these results. Inserting this into Eq.\ (\ref{calLI}), and putting together the
contributions from the chiral field, the baryons, and the interactions between them, the total Lagrangian can be written as
\bea
{\cal L}&=& -V(\phi) +\bar{p}[i\gamma^\mu\partial_\mu-m_B+\gamma^0(\mu_p+\mu_p^*)]p \non
&& \hspace{1.2cm}+\bar{n}[i\gamma^\mu\partial_\mu-m_B+\gamma^0(\mu_n+\mu_n^*)]n \, . 
\eea
Similar to the Walecka model in Sec.\ \ref{sec:walecka}, the effect of the kaon condensate on the nucleons can be absorbed into an effective
chemical potential. In a slightly different notation than in Sec.\ \ref{sec:walecka} (where $\mu$ was absorbed into $\mu^*$),
we have kept the actual thermodynamic chemical potentials separate, and we have 
\begin{subequations}
\bea
\mu_p^* &=& 2(a_2+a_3)m_s+[2\mu_K-2(a_1+a_2+2a_3)m_s]\sin^2(\phi/2) \, , \\
\mu_n^* &=& 2(a_2+a_3)m_s+[\mu_K-2(a_2+2a_3)m_s] \sin^2(\phi/2) \, .
\eea
\end{subequations}
We can now, analogously to Sec.\ \ref{sec:walecka}, evaluate the partition function at $T=0$ to obtain the thermodynamic potential density
$\Omega=-T/V\,\ln Z$ which can be written as
\be \label{OmegaUE}
\Omega = V(\phi) + \epsilon_B -(\mu^*_n+\mu_n)n_n-(\mu^*_p+\mu_p)n_p \, , 
\ee
with the nucleon number densities $n_n$ and $n_p$, and the nucleon energy density 
\be
\epsilon_B  = 2\sum_{i=p,n}\int\frac{d^3{\bf k}}{(2\pi)^3}\sqrt{k^2+m_B^2}\;\Theta(k_{F,i}-k) \, , 
\ee
where $k_{F,i}$ are the respective Fermi momenta.

Before adding the lepton contribution we need to find the relations between the various chemical potentials through the 
conditions of chemical equilibrium. The leptonic processes including nucleons are 
\be
n\to p+\ell+\bar{\nu}_\ell \, , \qquad p+\ell \to n+\nu_\ell \, .
\ee  
Here, $\ell=e,\mu$ can either be an electron or a muon. We also have the purely leptonic processes,
\be
e\to\mu+\bar{\nu}_\mu+\nu_e \, , \qquad \mu\to e+\bar{\nu}_e+\nu_\mu \, , 
\ee
and the processes involving kaons,
\be
n\leftrightarrow p + K^- \, , \qquad e\leftrightarrow K^- +\nu_e \, .
\ee
These processes lead to the independent conditions
\be
\mu_e=\mu_K=\mu_\mu \, ,  \qquad \mu_n=\mu_p+\mu_e \, .
\ee
The system is thus characterized by two independent chemical potentials, say $\mu_e$ and $\mu_n$. 
 We implement the constraint $\mu_n=\mu_p+\mu_e$ 
by rewriting the terms containing the nucleon chemical potentials in the potential (\ref{OmegaUE}) as $\mu_nn_n+\mu_p n_p = \mu_n n_B-\mu_en_p$. 
Since we want to work at fixed $n_B=n_n+n_p$, 
we perform a Legendre transformation of $\Omega$ with respect to the variables $\mu_n$ and $\frac{\partial\Omega}{\partial \mu_n}=-n_B$. This
amounts to adding the term $\mu_nn_B$ to $\Omega$ which yields the relevant free energy for the baryons and the kaon condensate,
\be
\Omega_{B,K} = V(\phi) + \epsilon_B -\left[(\mu_p^*-\mu_e)x_p+(1-x_p)\mu_n^*\right]\,n_B \, .
\ee
Here we have introduced the proton fraction 
\be
x_p \equiv \frac{n_p}{n_B} \, ,
\ee
which has to be determined dynamically from minimizing the free energy. 
We can now add the lepton contributions  to arrive at 
\be 
\Omega_{B,K,\ell} = \Omega_{B,K} +\epsilon_e - \mu_e n_e + \Theta(\mu_e^2-m_\mu^2)(\epsilon_\mu-\mu_e n_\mu) \, , 
\ee
where we have used $\mu_e=\mu_\mu$, where $\epsilon_\ell$  are the lepton energy densities ($\ell=e,\mu$),
and where 
\be
n_e = \frac{\mu_e^3}{3\pi^2} \, ,\qquad n_\mu = \frac{(\mu_e^2-m_\mu^2)^{3/2}}{3\pi^2} 
\ee
are the corresponding lepton number densities. 
The step function in the muon contribution accounts for the fact that muons only appear if $\mu_e$ is larger than their 
mass $m_\mu=106\,{\rm MeV}$. On the relevant energy scale, electrons are massless to a very good approximation and thus are present for 
any nonzero $\mu_e$.  

For a given baryon number $n_B$, the variables of $\Omega_{B,K,\ell}$ 
are the proton fraction $x_p$, the kaon condensate $\phi$, and the chemical potential for (negative) electric
charge $\mu_e$. 
They are determined by minimizing the free energy with respect to $x_p$ and $\phi$ and by requiring charge neutrality,
\be
\frac{\partial\Omega_{B,K,\ell}}{\partial x_p} =\frac{\partial\Omega_{B,K,\ell}}{\partial \phi} =\frac{\partial\Omega_{B,K,\ell}}{\partial \mu_e} 
= 0 \, .
\ee
It is straightforward to compute the various derivatives, and after a few lines of algebra the result can be written as
\begin{subequations} \label{minimize123}
\bea
\mu_e &=& -\frac{1}{n_B\cos^2(\phi/2)}\frac{\partial \epsilon_B}{\partial x_p} -2a_1m_s\tan^2(\phi/2) \, , \\[1.5ex]
0&=& \cos\phi-\frac{m_K^2}{\mu_e^2}+\frac{n_B}{\mu_e^2f^2_\pi}\left[\frac{\mu_e}{2}(1+x_p)-(a_1x_p+a_2+2a_3)m_s\right] \, ,
\label{nonzeroK}\\[1.5ex]
0&=& f_\pi^2\mu_e\sin^2\phi-n_B\left[x_p\cos^2(\phi/2)-\sin^2(\phi/2)\right] \non
&& + n_e + n_\mu\Theta(\mu_e^2-m_\mu^2) \, . 
\eea
\end{subequations} 
The second equation has been obtained after dividing both sides by $\sin\phi$. 
This means that $\phi=0$ is always a solution and Eq.\ (\ref{nonzeroK}) is only valid
for nonvanishing condensates. In the third equation we recover the various contributions to the electric charge density: the
first term on the right-hand side is the pure contribution from the kaon condensate. It gives a positive contribution to the negative charge density
for $\mu_e>0$.
The second term on the right-hand side arises from the nucleons and their interactions with the kaon condensate. Only for $\phi=0$ does
it give the pure proton contribution $-n_p=-x_p n_B$. Finally, the other two terms are the expected contributions from the leptons. 

The onset of kaon condensation can be determined by 
setting $\phi=0$ in all three equations. This yields three equations which can be solved for $x_p^c$, $\mu_e^c$, and $n_B^c$, where 
$n_B^c$ is the critical density beyond which there is a condensate and $x_p^c$, $\mu_e^c$ the values of the proton fraction and the 
charge chemical potential at this density. Since Eq.\ (\ref{nonzeroK}) is only valid for $\phi\neq 0$, this has to be understood as approaching
$n_B^c$ from above. 

\begin{figure}[t]
\begin{center}
\includegraphics[width=0.75\textwidth]{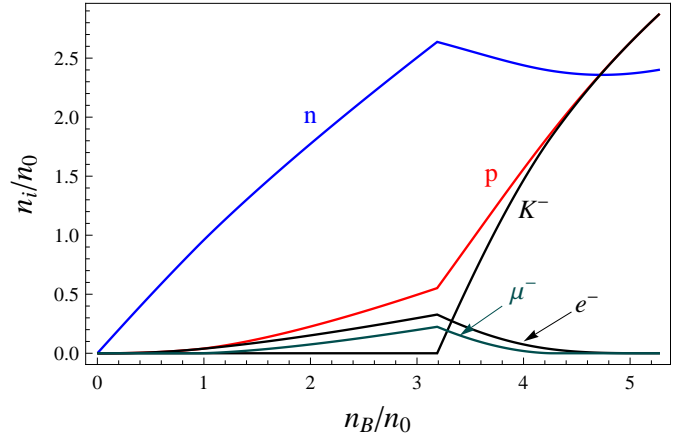}
\caption{Density fractions of neutrons ($n$), protons ($p$), electrons ($e^-$), muons ($\mu^-$), and the kaon condensate ($K^-$)
from Eqs.\ (\ref{minimize123}) with $\epsilon_B$ modified as given in Eqs.\ (\ref{modify1}), (\ref{modify2}). The parameters are 
(see Ref.\ \cite{Thorsson:1993bu3}) $a_1m_s=-67\,{\rm MeV}$, 
$a_2m_s=134\,{\rm MeV}$, $a_3m_s=-222\,{\rm MeV}$, $m_\mu=106\,{\rm MeV}$, $f_\pi=93\,{\rm MeV}$, 
$m_K=494\,{\rm MeV}$, $m_B=1200\,{\rm MeV}$. We see that for this parameter choice 
the onset of kaon condensation is at about three times nuclear saturation density, $n_B^c\simeq 3.2n_0$.
}
\label{figkaon}
\end{center}
\end{figure}

We leave the numerical evaluation of the critical density and the general evaluation for all $n_B$ as an exercise, see problem \ref{prob5}. 
An important modification, which we have neglected for simplicity, has to be taken into account for this evaluation. Namely, the 
energy density $\epsilon_B$ has to be modified due to interactions among
nucleons. It is beyond the scope of these lectures to derive this modification, see Ref.\ \cite{Thorsson:1993bu3} and references therein 
for more details. Here we simply quote this modification which is needed in order to get physically sensible results. 
One needs to use an expansion of $\epsilon_B$ around symmetric nuclear matter $x_p=1/2$ of the form
\be \label{modify1}
\epsilon_B \to \epsilon_0 + n_B(1-2x_p)^2S(u) \, , \qquad u \equiv \frac{n_B}{n_0} \, . 
\ee
Here $\epsilon_0$ is the energy density of symmetric nuclear matter, whose form is not relevant because we only need the derivative of $\epsilon_B$
with respect to $x_p$. The nuclear saturation density is denoted by $n_0$, and 
\be \label{modify2}
S(u) = (2^{2/3}-1)\frac{3}{5}\frac{(3\pi^2n_0/2)^{2/3}}{2m_B}\left[u^{2/3}-F(u)\right] + S_0F(u) 
\ee
is the nuclear symmetry energy (see Ref.\ \cite{Prakash:1988md3} for a discussion of the nuclear symmetry energy in the context of the 
maximum mass of neutron stars). For the numerical evaluation shown in Fig.\ \ref{figkaon}, the nuclear symmetry energy at 
the saturation point $S_0=30\,{\rm MeV}$ has been used, as well as the function $F(u)=u$. See caption of the figure 
for the choice of the other parameters.

\section{From hadronic to quark phases: possibility of a mixed phase}
\label{sec:mixed}

We have already mentioned the possibility of a hybrid star, i.e., a star with a quark matter core surrounded by nuclear matter. 
How does the interface between these two phases look? Is it a sharp interface or is there a shell in a hybrid star where the hadronic
and quark phases coexist in a mixed phase? If the former is true, there will be a jump in the density profile of the star, 
while the latter allows for a continuous change in density. 

Mixed phases are a very general phenomenon. In the context of compact stars, not only the mixed hadronic/quark matter phase is of relevance.
Also in the inner crust of a hybrid or neutron star one may find mixed phases. 
There one expects a neutron superfluid coexisting with a lattice of ions, i.e., a mixed 
phase of neutron matter and nuclei. In these lectures, we shall not discuss the properties of the crust of a compact star in detail. See
Sec.\ \ref{sec:misc} for a brief discussion and Ref.\ \cite{Chamel:2008ca3} for an extensive review. 
Other examples of mixed phases in different systems are liquid-gas mixtures or simply a solid, which is
a mixture of an electron gas and nuclear matter (sitting in the lattice of ions).  

\begin{figure}[t]
\begin{center}
\hbox{
\includegraphics[width=0.45\textwidth]{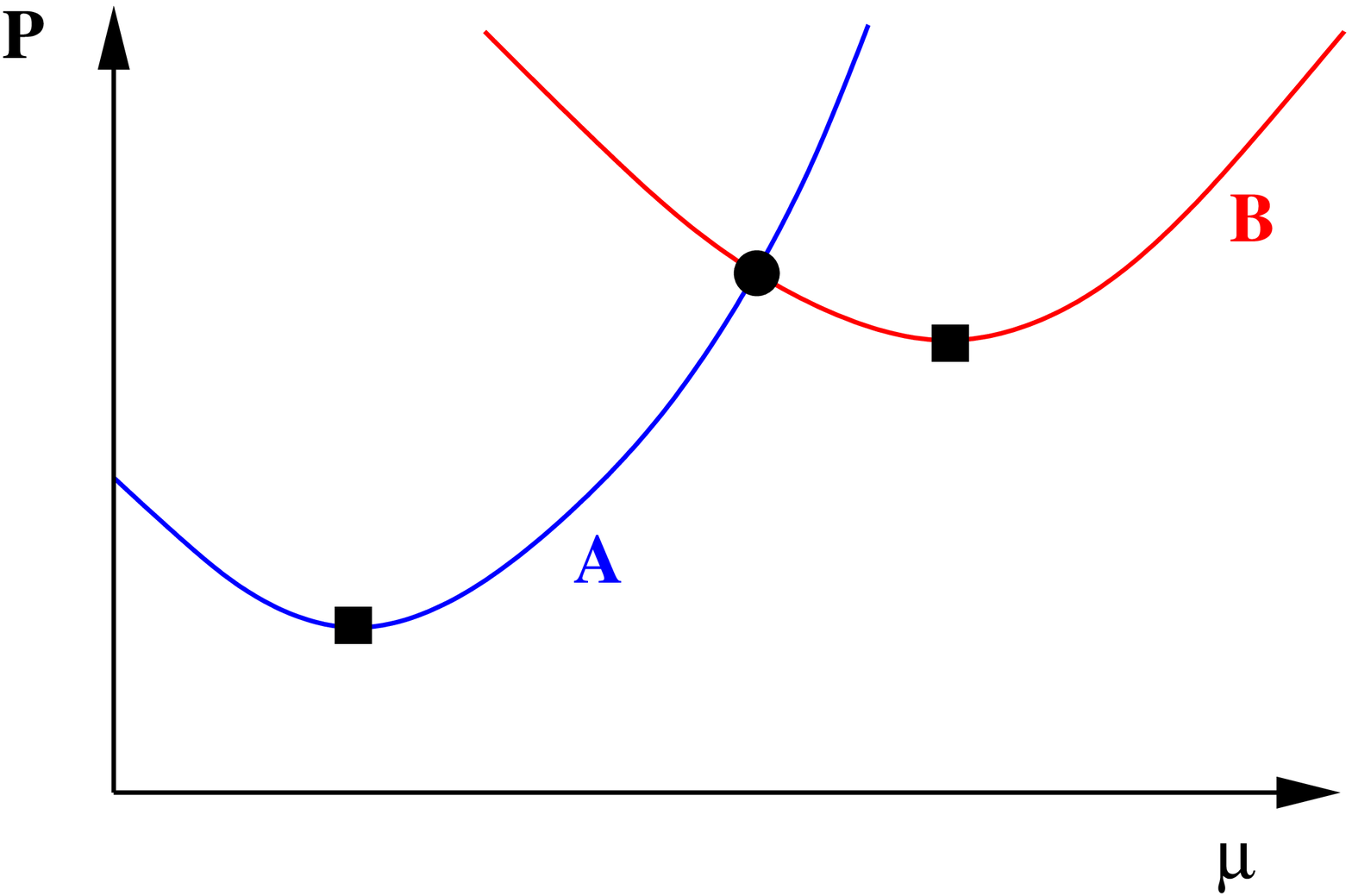}
\hspace{0.5cm}\includegraphics[width=0.45\textwidth]{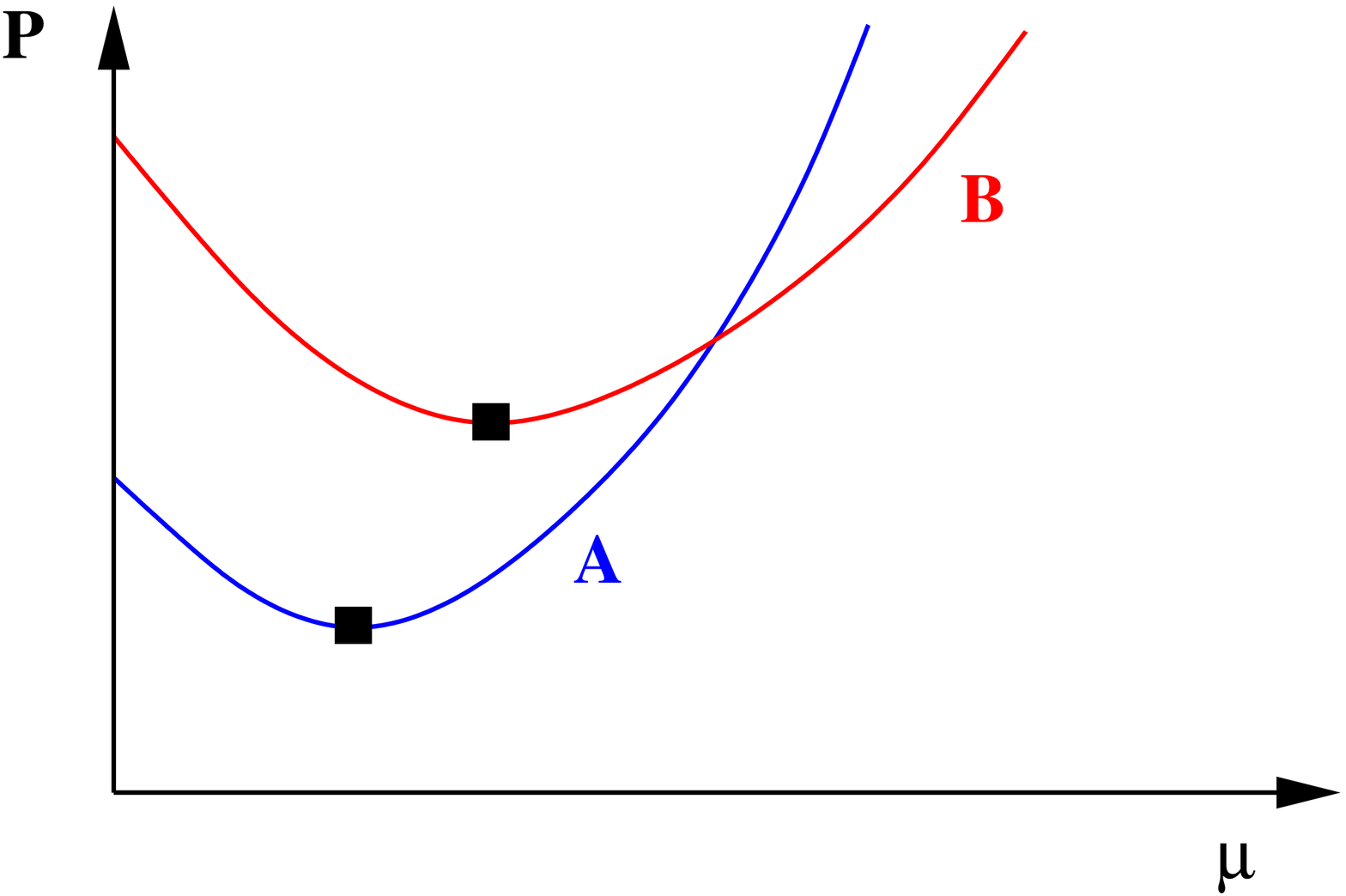}
}
\caption{Illustration of the possibility of a mixed phase. The pressure $P$ of two phases A and B is given by the 
respective curves as a function of a chemical potential $\mu$.  Note that $\partial P/\partial \mu$ has to increase with increasing $\mu$
(increasing $\mu$ cannot lead to a decrease of the corresponding charge; this would lead to an instability). Suppose $\mu$ is the electric charge 
chemical potential and we require charge neutrality. Then the squares mark the points at which a given phase is charge neutral. The circle
in the left panel marks a point where the two phases have equal pressure and opposite charge. Since this point has higher pressure than either
of the squares, this is the ground state (neglecting surface tension and Coulomb energy). 
In this state, phase A and B coexist and occupy different volume fractions, determined by the different slopes of the curves. In the
right panel, there is no point where both phases have equal pressure and opposite charges. Therefore, the square on the curve B is the ground state.
}
\label{figmixed}
\end{center}
\end{figure}

In Fig.\ \ref{figmixed} the possibility of a mixed phase is illustrated.
We see that the condition of charge neutrality plays an important role here. It is important that in a compact star charge neutrality 
is required globally, not locally. In other words, certain regions in the star may very well have a nonzero electric charge as long as other
regions have opposite charge to ensure an overall vanishing charge. 

It is plausible that such a mixed phase will have a crystalline structure. For instance, one phase may form spheres sitting at the points of a 
lattice which is immersed in the other phase. Other possibilities are rods or slabs \cite{Ravenhall:1983uh3}, 
such that the mixed phase looks like spaghetti or lasagna, 
wherefore astrophysicists have termed such phases {\it nuclear pasta}. In any case, if a mixed phase is possible because of a 
general argument such 
as given in Fig.\ \ref{figmixed}, this does not mean that it is indeed realized. One has to take into account Coulomb forces (which seek to 
break charged regions into smaller regions) and surface forces (which seek to minimize the surface and thus work in the opposite 
direction). We shall not discuss these forces quantitatively but rather give some general arguments about mixed phases.

We start from the simple picture that at small quark density (or quark chemical potential $\mu$) the hadronic phase is preferred and that
there is a first-order phase transition to the quark matter phase at some critical chemical potential. 
The question is whether there is a mixed phase between these two pure phases. The pressures of the two phases $P_h(\mu,\mu_e)$ and
$P_q(\mu,\mu_e)$ depend on the quark chemical potential and the charge chemical potential $\mu_e$ (we work at zero temperature). Phase coexistence
is possible when the pressures of the two phases are equal,
\be
P_h(\mu,\mu_e) = P_q(\mu,\mu_e) \, .
\ee
Now suppose the neutrality condition were {\it local} (which it isn't in our context). Then the charge must vanish in each phase separately,
\be
Q_h(\mu,\mu_e) = Q_q(\mu,\mu_e) =0 \, .
\ee
These two conditions yield $\mu_e$ for each phase separately as a function of $\mu$, $\mu_e^h(\mu)$ and $\mu_e^q(\mu)$. Consequently,
the condition of equal pressure,  
\be    
P_h(\mu,\mu_e^h(\mu)) = P_q(\mu,\mu_e^q(\mu)) \, ,
\ee
yields a unique $\mu$. Only at this $\mu$ do the phases coexist. This amounts to a sharp interface at a given value for the 
pressure, where on both sides of the interface the pure hadronic and the pure quark phases exist with different densities, i.e., 
there is a density jump in the profile of the star.

\begin{figure}[t]
\begin{center}
\includegraphics[width=\textwidth]{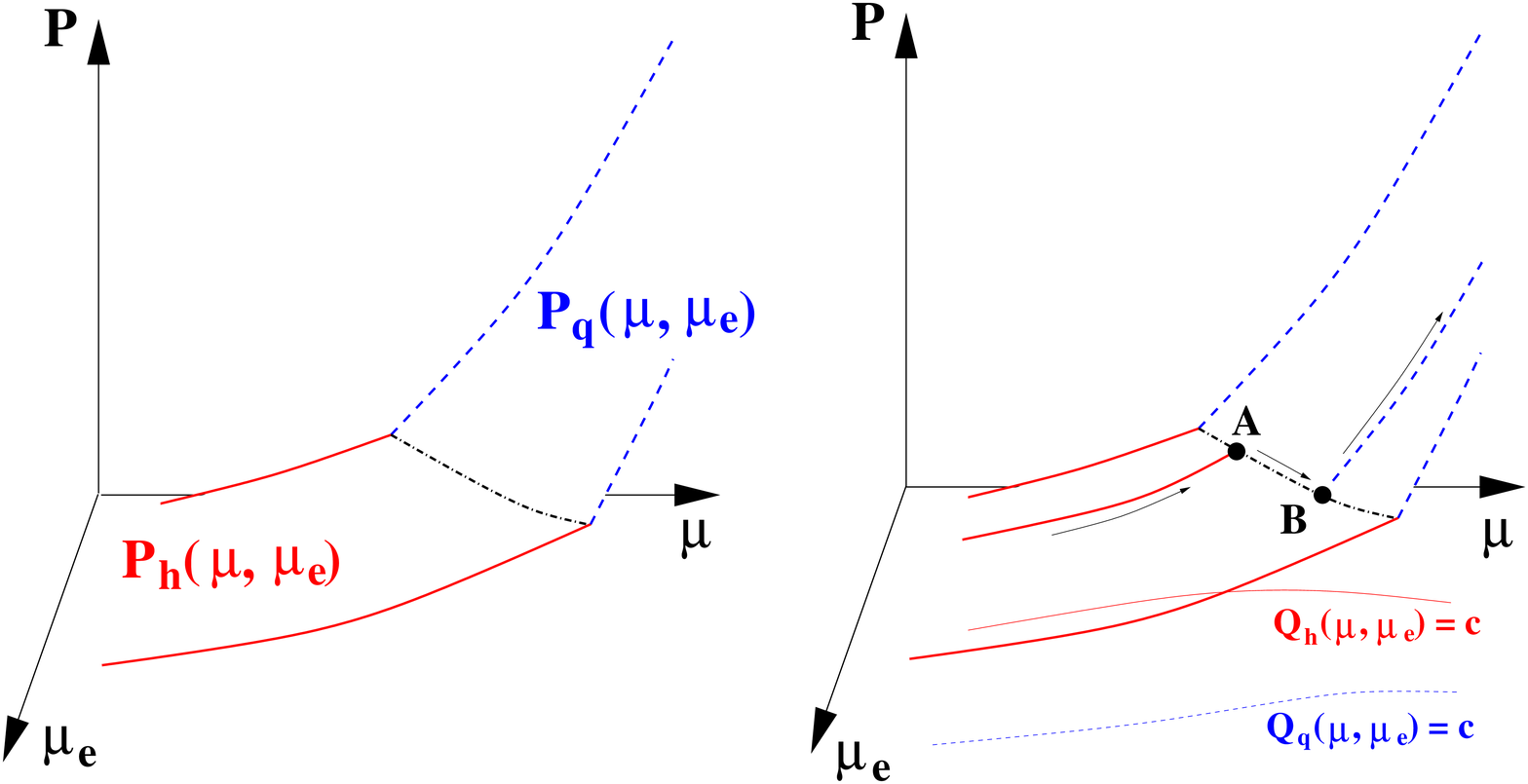}
\caption{Schematic picture of a hadron-quark mixed phase in a finite interval of $\mu$. Left panel: the pressures of the 
two phases define two surfaces parametrized by $\mu$ and $\mu_e$. The intersection of the two surfaces forms a line where coexistence 
of the two phases is possible. 
Right panel: the neutrality condition for each of the phases defines a curve in the $\mu$-$\mu_e$ plane, and thus
a curve on the respective surfaces (for illustrative purposes let the charge be nonzero -- denoted by $c$ -- 
since for zero charge there would have to be a valley of the pressure). A mixed phase may exist from A (where $\chi_q=0$) to B (where $\chi_q=1$),
provided that, for a given $\mu$, the pressure on this line is larger than the respective pressure on the neutrality curves of each phase. 
In this segment none of the phases is neutral separately, but they
may combine to a globally neutral phase. Note that the extra direction $\mu_e$ is crucial to have a finite segment along the 
$\mu$ axis where a mixed phase is possible. If the mixed phase is realized, the arrows indicate the ground state for increasing
values of $\mu$ (the pressure also has to increase along this line).  
}
\label{figmixed2}
\end{center}
\end{figure}
  
Now we impose the weaker (and realistic) condition of {\it global} charge neutrality. This means that in any mixed phase 
only the total charge has to vanish. We denote the quark volume fraction by 
\be
\chi_q \equiv \frac{V_q}{V_h+V_q} \;\; \in [0,1] \, ,
\ee
where $V_q$ and $V_h$ are the volumes occupied by the quark and hadron phases, respectively.
Then, neutrality reads
\be
(1-\chi_q)Q_h(\mu,\mu_e)+\chi_q Q_q(\mu,\mu_e) = 0 \, .
\ee
This yields a function $\mu_e(\chi_q,\mu)$ which is then inserted into the condition of equal pressure,
\be    
P_h(\mu,\mu_e(\chi_q,\mu)) = P_q(\mu,\mu_e(\chi_q,\mu)) \, .
\ee
The result is a chemical potential as a function of $\chi_q$, $\mu(\chi_q)$. Thus there is a finite interval on the $\mu$-axis where a mixed
phase is possible. We see that the looser condition of global charge neutrality allows for a shell with a mixed phase
in a hybrid star.
These formal arguments become more transparent in a geometric picture, see Fig.\ \ref{figmixed2}. 

\begin{figure}[t]
\begin{center}
\vspace{0.8cm}
\includegraphics[width=0.75\textwidth]{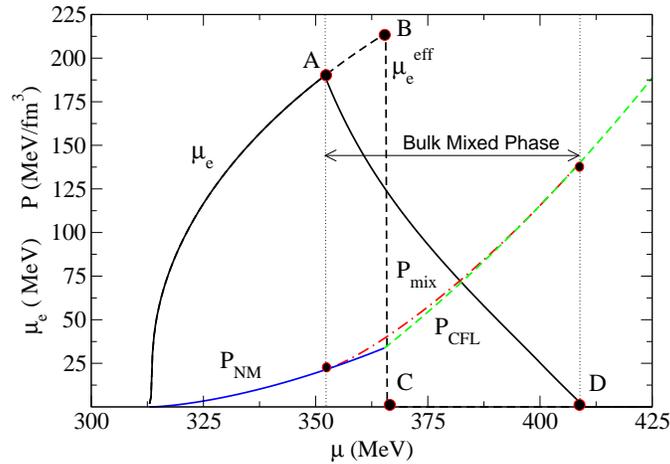}
\caption{Figure from Ref.\ \cite{Alford:2001zr3} showing the transition from nuclear matter (NM) to a mixed phase (mix) to a quark matter phase (CFL)
(color-flavor locking (CFL) is explained in Sec.\ \ref{sec:CFL}). 
In the mixed phase, $\mu_e$ is lowered in order to make the nuclear phase positively charged
and the CFL phase negatively charged. Taking into account Coulomb energy and surface energy shows that the $\mu$ interval for the mixed phase 
shrinks with increasing surface tension $\sigma$ until it completely disappears for $\sigma \gtrsim 40\, {\rm MeV}/{\rm fm}^2$. The 
exact value of $\sigma$ is not known but it is likely to be larger than that limit value such that a mixed phase appears unlikely. The limit 
value does not depend much on whether the mixed phase has spheres, rods, or slabs. 
}
\label{figmixed3}
\end{center}
\end{figure}

We shall not go into the details of an explicit calculation of the quark/hadron mixed phase because, even neglecting surface tension and 
Coulomb energy, this calculation eventually has to be performed numerically. Instead we show the result of such a calculation \cite{Alford:2001zr3} 
in  Fig.\ \ref{figmixed3} (cf.\ also Fig.\ \ref{fighyper} where we have already seen a mixed phase). One recovers the (projection of the) topology 
of Fig.\ \ref{figmixed2} in Fig.\ \ref{figmixed3}. The figure shows the mixed phase being the preferred phase in a certain $\mu$ interval 
without taking into account Coulomb energy and surface energy. In the complete calculation one finds that a relatively small surface energy is 
needed to destroy the mixed phase. It thus appears unlikely that a mixed phase of quarks and hadrons exists in a hybrid star.

\section*{Problems}
\addcontentsline{toc}{section}{Problems}

\begin{prob}
\label{prob3}
\textbf{Binding energy and saturation density in the Walecka model}\\
Solve Eq.\ (\ref{mNstar}) at zero temperature numerically for different values of the baryon density. Use the solution 
to compute the binding energy per nucleon and check that the values (\ref{binding}) are obtained upon using the values of the coupling
constants $g_\omega^2/(4\pi) = 14.717$, $g_\sigma^2/(4\pi) = 9.537$. In other words, reproduce the results from Fig.\ \ref{figbinding}. 
If you are a bit more ambitious you can also do it the other 
way around: set up and solve the two equations that are needed to determine the coupling constants from the conditions (\ref{binding}).
\end{prob}

\begin{prob}
\label{prob4}
\textbf{Walecka model with scalar interactions}\\
Reproduce the result of Fig.\ \ref{figwalecka2} numerically.
\end{prob}

\begin{prob}
\label{prob5}
\textbf{Onset of kaon condensation}\\
Solve equations (\ref{minimize123}) -- with the modifications given in Eqs.\ (\ref{modify1}) and (\ref{modify2}) -- numerically to 
determine the density fractions of nucleons, kaons, and leptons at $T=0$ as a function of baryon density. 
In particular, compute the critical baryon density for the onset of kaon condensation. See caption of Fig.\ \ref{figkaon} 
for the values of the parameters and compare your result to the plot in this figure.
\end{prob}

\chapter{Superconductivity and superfluidity in a compact star}
\label{sec:supersuper}

In our discussion of interacting nuclear matter we have so far ignored a very important physical effect. We have not included the possibility of 
superfluidity
and/or superconductivity, although we have briefly mentioned the effect of superconductivity on the equation of state of quark matter, 
see Sec. \ref{sec:MRinter}. 
In the following, we shall discuss these effects in more detail. But first let us recapitulate what superconductivity is. Once we have 
introduced the basic concept we shall see that it may appear in 
several variants in a compact star. And we will see that it is crucial for the understanding of transport properties of dense matter. 
And the transport properties of dense matter, in turn, are related to the phenomenology of the star. 

Consider a system of fermions at zero temperature with chemical potential $\mu$ and free energy
\be
\Omega = E -\mu N \, .
\ee
Now first suppose the fermions are noninteracting. Then, adding a fermion with energy $\mu$, i.e., at the Fermi surface, leaves the 
free energy $\Omega$ obviously
unchanged: the energy $E$ is increased by $\mu$, but the second term subtracts the same amount since we add $N=1$ fermion. 
Now let us switch on an arbitrarily small attractive interaction between the fermions. Then, by adding two fermions at the Fermi surface, 
we can actually 
lower the free energy because  the attractive interaction will lead to an energy gain from the binding energy. Therefore, the Fermi surface 
we have started with is unstable.
A new ground state is formed in which pairs of fermions are created at the Fermi surface. Since two fermions formally can be viewed as a 
boson, these fermion pairs will form a Bose condensate.\footnote{In fact, the fermions are correlated in momentum space, not in real space. 
Consequently, in the weak-coupling limit, the fermion pairs are not spatially separated bosons. The typical size of a pair
is rather larger than the mean distance between fermions. Therefore, one apparently has to be careful to describe the pairs as bosons. However, 
recent experiments with cold fermionic atoms show that there is no phase transition between the weak-coupling limit (where the pairs are wide 
spread) and the strong-coupling limit (where the pairs are actual difermions, i.e., bosons). This is the so called BCS-BEC crossover. 
This observation suggests in particular that it is not too bad to 
think of the fermion pairs as bosons even in the weak-coupling limit.} This formation of a condensate of 
fermion pairs due to an arbitrarily small interaction is called {\it Cooper's Theorem} and the fermion pairs are called {\it Cooper pairs}. 

This mechanism is completely general, i.e., it holds for arbitrary fermions with a Fermi surface as long as their interaction is attractive. 
It holds for electrons in a usual
superconductor, i.e., a metal or alloy, for $^3$He atoms in superfluid helium, for fermionic atoms in an optical trap etc. 
In our context, it can be applied to protons, neutrons, and 
quarks. Anticipating that the Cooper mechanism leads to {\it superfluidity} for neutral fermions and to {\it superconductivity} for charged 
fermions, we thus expect
$(i)$ neutron superfluidity, $(ii)$ proton superconductivity, and $(iii)$ quark superconductivity to be in principle possible in a compact star. 
Quarks are of course a bit more complicated since they not only carry electric charge but also color charge. Therefore, we need
to make more precise what we mean by quark superconductivity, see Sec.\ \ref{sec:CFL}.

\begin{figure}[t]
\begin{center}
\hspace{-0.3cm}\hbox{
\includegraphics[width=0.48\textwidth]{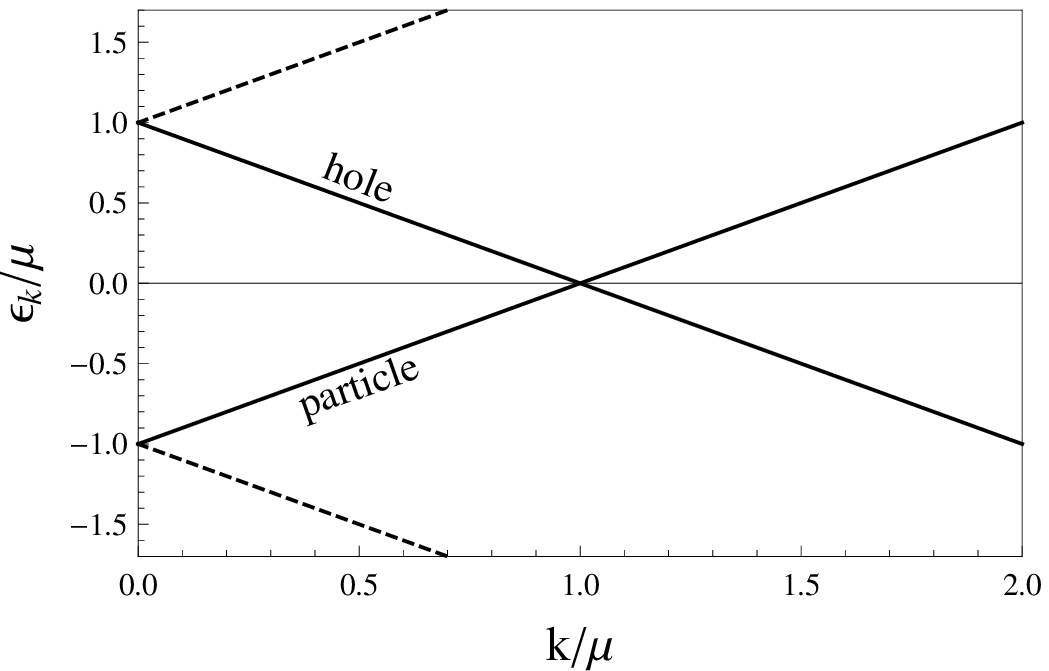}
\hspace{0.5cm}\includegraphics[width=0.48\textwidth]{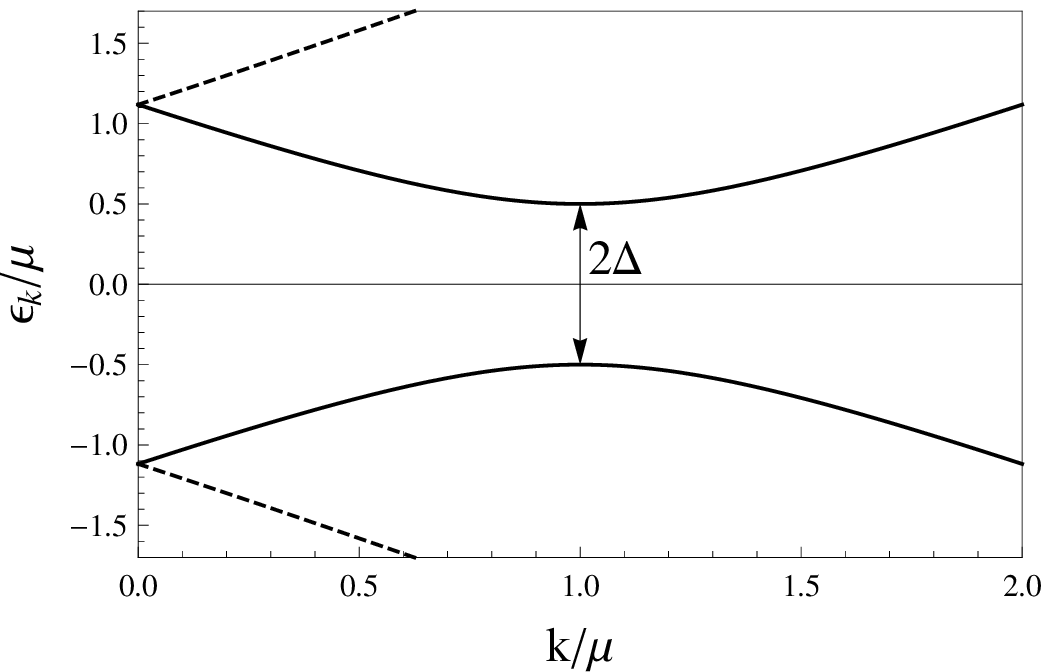}
}
\caption{Left panel: particle and hole excitations (solid lines) in a system of noninteracting ultrarelativistic fermions with chemical 
potential $\mu$.
The dashed lines are the antiparticle and antiparticle hole excitations. Right panel: quasiparticle excitations after switching on small interactions
which, via Cooper's Theorem, give rise to an energy gap $\Delta$ according to Eq.\ (\ref{excite}), here chosen to be $\Delta=0.5\mu$. 
What were pure particle and pure hole excitations
in the left panel have now become momentum-dependent mixtures of particles and holes. 
}
\label{figdispersion}
\end{center}
\end{figure}

Let us first stay on a very general level and discuss the basic consequences of Cooper pairing. A Cooper pair is held together by 
a sort of "binding energy" (although it is not a bound state), i.e., one needs a finite amount of energy to break a pair. Consequently, the 
single-particle dispersion relation acquires an energy gap $\Delta$,
\be \label{excite}
\epsilon_k = \sqrt{(E_k-\mu)^2+\Delta^2} \, ,
\ee  
with $E_k=\sqrt{k^2+m^2}$ as in the previous chapters. One might think that $\epsilon_k$ does not reproduce the 
usual dispersion $E_k-\mu$ for a vanishing gap, rather $\epsilon_k\to |E_k-\mu|$. This is no contradiction after taking into account
the fermion hole excitations, such that in the ungapped system $\epsilon_k=\pm(E_k-\mu)$ to which the $\Delta=0$ limit
of $\epsilon_k = \pm \sqrt{(E_k-\mu)^2+\Delta^2}$ is indeed equivalent.
The excitation described by Eq.\ (\ref{excite}) is also called {\it quasiparticle} since it contains the interaction of the 
original particles in an effective way. To excite a quasifermion in a superconductor, a finite amount of energy is needed, while a 
fermion at the Fermi surface of a noninteracting system can be excited by an infinitesimally small energy, see Fig.\ \ref{figdispersion}. 
The energy gap in the dispersion relation is responsible for most of the phenomenological properties of a superconductor. 
For instance, it gives rise
to the frictionless charge transport in an electronic superconductor, since (sufficiently low energy) scattering of electrons off 
phonons cannot excite a single-electron 
state. Or, in the context of superfluidity, the energy gap explains the frictionless flow in the same way. For quantitative predictions it is thus
crucial to compute the magnitude of $\Delta$. We shall perform this calculation within perturbative QCD for quarks in Sec.\ \ref{sec:QCDgap}.

The energy gap is in general a temperature-dependent quantity. It typically decreases with temperature and becomes zero at and above a certain
critical temperature $T_c$. This critical temperature indicates the phase transition from the superconducting to the non-superconducting phase,
as we shall demonstrate with the discontinuity of the specific heat in the following section. 
Since the onset of superconductivity or superfluidity is a phase transition, there must be a symmetry which is spontaneously broken below the 
critical temperature. In particular for quark matter, the symmetry breaking pattern is very useful to characterize the superconductor, see 
Sec.\ \ref{sec:CFL}.  

\section{Specific heat for isotropic and anisotropic superconductors}
\label{sec:cV}

As a example of the effect of $\Delta$ let us compute the specific heat of a superconductor.\footnote{More precisely, here we compute the
fermionic contribution to the specific heat. There may be light Goldstone modes which dominate the
specific heat at small temperatures. In this section we ignore such modes for the purpose of illustrating the effect of the fermionic energy gap.} 
The specific heat is easy to compute and shows characteristic
features of a superconductor. We start from the free energy of a superconductor made of fermions with 
two degenerate (spin-) degrees of freedom,
\be
\Omega = -2T\int\frac{d^3{\bf k}}{(2\pi)^3} \ln\left(1+e^{-\epsilon_k/T}\right) \, ,
\ee
where the quasiparticle energy $\epsilon_k$ is given by Eq.\ (\ref{excite}). We shall, for simplicity, consider massless fermions, $E_k=k$. 
The entropy (density) is given by the derivative 
with respect to the temperature (with respect to the {\it explicit} temperature dependence only, there is also an implicit temperature
dependence in $\Delta$) 
\be \label{s}
s  = -\frac{\partial\Omega}{\partial T} = -2 \int\frac{d^3{\bf k}}{(2\pi)^3} \left[(1-f_k)\ln(1-f_k) + f_k \ln f_k\right] \, .
\ee
with the Fermi distribution 
\be
f_k = \frac{1}{e^{\epsilon_k/T}+1} \, .
\ee
To derive Eq.\ (\ref{s}) one uses the identities
\be
\frac{\epsilon_k}{T} = \ln(1-f_k) -\ln f_k \, , \qquad \ln\left(1+e^{-\epsilon_k/T}\right) = -\ln(1-f_k)\, .
\ee
From the entropy we then compute the specific heat (at constant volume)
\be
c_V \equiv T\frac{\partial s}{\partial T} = 2\int\frac{d^3{\bf k}}{(2\pi)^3} \epsilon_k\frac{\partial f_k}{\partial T} \, .
\ee
For the temperature dependence of the gap we assume the following simple form,
\be
\Delta(T) = \Theta(T_c-T)\Delta_0\sqrt{1-\frac{T^2}{T_c^2}} \, , 
\ee
such that the zero-temperature gap is $\Delta_0$, the gap approaches  zero at $T=T_c$ and vanishes for all temperatures larger than 
$T_c$. Then, for $T<T_c$ we have
\be
\frac{\partial \Delta}{\partial T} = -\frac{\Delta_0^2}{T_c^2}\frac{T}{\Delta} \;\; \Rightarrow \;\; \frac{\partial\epsilon_k}{\partial T} 
= -\frac{T}{\epsilon_k}\frac{\Delta_0^2}{T_c^2} \;\;
\Rightarrow \;\; \frac{\partial f_k}{\partial T} = \frac{1}{\epsilon_k}\frac{e^{\epsilon_k/T}}{\left(e^{\epsilon_k/T}+1\right)^2}\left(\frac{\epsilon_k^2}{T^2}+\frac{\Delta_0^2}{T_c^2}\right)
\, , 
\ee
and consequently
\be \label{cV1}
c_V = 2\int \frac{d^3{\bf k}}{(2\pi)^3} \frac{e^{\epsilon_k/T}}{\left(e^{\epsilon_k/T}+1\right)^2}  \left(\frac{\epsilon_k^2}{T^2}+\frac{\Delta_0^2}{T_c^2}\right) \, .
\ee
We are only interested in temperatures much smaller than the chemical potential, $T\ll \mu$. Then, 
the main contribution comes from the Fermi surface, and we can approximate $dk\, k^2 \to \mu^2 dk$. We introduce the 
new variable $x=(k-\mu)/T$, and define
\be
\varphi \equiv \frac{\Delta}{T} \, .
\ee
This yields 
\be \label{cV2}
c_V \simeq \frac{\mu^2T}{\pi^2} \int_0^\infty dx\int_0^\pi d\theta\,\sin\theta \left(x^2+\varphi^2+\frac{\Delta_0^2}{T_c^2}\right)
\frac{e^{\sqrt{x^2+\varphi^2}}}{\left(e^{\sqrt{x^2+\varphi^2}}+1\right)^2} \, ,
\ee
where we have approximated the lower boundary by $-\mu/T\simeq -\infty$ and have used that the integrand is even in $x$ (which gives rise to the 
new integration boundaries $[0,\infty]$ and a factor 2). We have not yet performed the $\theta$ integral since we shall allow for 
anisotropic gaps. From this general expression we easily get the limit of a vanishing gap, 
$\varphi=\Delta_0=0$, i.e., the result for the non-superconducting state,
\be \label{cV0}
c_V^0 \simeq \frac{\mu^2 T}{\pi^2}\int_0^\infty dx\,\frac{x^2}{1+\cosh x} = \frac{\mu^2 T}{3}\, .
\ee
Before evaluating the specific heat in the superconducting phase at small temperatures, 
let us discuss the behavior of $c_V$ at the critical temperature. This is best done by looking at Eq.\ (\ref{cV1}). 
Approaching $T_c$ from above, $c_V$ is given by setting $\Delta_0$ and $\Delta(T)$ (appearing in $\epsilon_k$) to zero in that equation. 
In the superconducting phase, approaching $T_c$ from below, we only set $\Delta$ in $\epsilon_k$ to zero. Consequently, at $T_c$ there
is a jump in the specific heat which is given by
\bea
\Delta c_V  &=& 2\frac{\Delta_0^2}{T_c^2} \int \frac{d^3{\bf k}}{(2\pi)^3} \frac{e^{\epsilon_k/T}}{\left(e^{\epsilon_k/T}+1\right)^2}  \non
&\simeq& \frac{\Delta_0^2\mu^2}{\pi^2 T_c}\int_0^\infty dx\, \frac{1}{1+\cosh x} = \frac{\Delta_0^2\mu^2}{\pi^2 T_c} \, ,
\eea
where we have assumed the gap to be isotropic.
This jump is a typical signature for a second-order phase transition, since the specific heat is the second derivative of the thermodynamic 
potential.

Next we evaluate Eq.\ (\ref{cV2}) for temperatures much smaller than the gap, i.e., in the limit $\varphi\to \infty$. 
First we consider an isotropic gap. We can approximate
\be
\frac{e^{\sqrt{x^2+\varphi^2}}}{\left(e^{\sqrt{x^2+\varphi^2}}+1\right)^2} \simeq e^{-\sqrt{x^2+\varphi^2} }\simeq 
e^{-\varphi-\frac{x^2}{2\varphi}}\, .
\ee
Consequently, 
\bea
c_V &\simeq& 
\frac{2\mu^2 T}{\pi^2} e^{-\varphi} \left[\int_0^\infty dx\, x^2 e^{-\frac{x^2}{2\varphi}}+\left(\varphi^2+\frac{\Delta_0^2}{T_c^2}\right)
\int_0^\infty dx\,e^{-\frac{x^2}{2\varphi}}\right] \non
&\simeq& \frac{\sqrt{2}\mu^2T}{\pi^{3/2}}\varphi^{5/2}e^{-\varphi}\,,
\eea
where we used
\be
\int_0^\infty dx\,x^2 e^{-\frac{x^2}{2\varphi}} = \varphi^{3/2}\sqrt{\frac{\pi}{2}} \, , \qquad 
\int_0^\infty dx\,e^{-\frac{x^2}{2\varphi}} = \varphi^{1/2} \sqrt{\frac{\pi}{2}} \, .
\ee
The main result is that the specific heat is exponentially suppressed by the factor $e^{-\varphi}=e^{-\Delta/T}$ 
for temperatures much smaller than the gap. The suppression of the specific heat in a superconductor  
provides a good example to get some intuition for the properties of superconductors. 
To this end, note that the specific heat is a measure of how many degrees of freedom are available to store heat. 
A large number of degrees of freedom means a lot of "storage room" and thus a large specific heat. A small specific heat, 
such as for a superconductor at sufficiently small temperature, thus means there are 
very few states available. This is a direct consequence of the energy gap which obviously leads to a region in the energy spectrum with no 
allowed states. Only by increasing the temperature does the exponential suppression disappear because temperature provides the energy to 
populate states above the gap which in turn are then available to store thermal energy.

Next let us assume an anisotropic gap of the form
\be \label{pointnodes}
\Delta \to \Delta \sin \theta \, .
\ee
In a compact star, anisotropic gaps may be realized in neutron superfluidity and possibly for quark superconductivity. The reason 
is very different in the two kinds of matter: at large density, the $s$-wave interactions between neutrons become repulsive and thus 
only interactions in the
$p$-wave channel can lead to superfluidity (this is in contrast to protons which do form $s$-wave superconductors). In the 
case of quark matter, anisotropic gaps may occur due to a mismatch in Fermi momenta of the quarks that form Cooper pairs; anisotropies then
arise either because the mismatch allows only for pairing in certain directions in momentum space or because pairing occurs
in the spin-one channel which does not suffer from the mismatch. In either case, there
are several possibilities for the specific form of the angular dependence of the gap and it is not entirely clear which one is realized in the 
relevant density regime. For more details see for instance Refs.\ \cite{levyak14,levyak24} for nuclear matter and 
Ref.\ \cite{Schmitt:2004et4} for quark matter. 

With $\theta$ being the angle between the momentum and the $z$-axis, the form (\ref{pointnodes}) implies point-like nodes of the gap function at 
the north and south pole of the Fermi sphere. In other words, although there is a nonzero order parameter for superfluidity, there
are directions in momentum space where quasiparticles can be excited with infinitesimally small energy. 
For sufficiently small temperatures, these directions give the 
dominant contribution to the specific heat. Therefore, in the low-temperature approximation, we only integrate over angles in the vicinity 
of the nodes. 
We restrict the angular integration by requiring the quasiparticle energy (with respect to the Fermi surface)
to be at most of the order of the scale set by the temperature,
\be
\Delta_0 \sin\theta \lesssim \pi T \, , 
\ee
which, for small angles $\theta$ and small temperatures implies $\theta \lesssim \pi/\varphi$. Therefore, the specific heat becomes (note the factor 2 since we obtain
the same result for north and south pole)
\bea
c_V&\simeq& \frac{\mu^2T}{\pi^2}\int_0^\infty dx\,\frac{1}{1+\cosh x}\int_0^{\pi/\varphi} d\theta\, \theta (x^2+\varphi^2\theta^2) \non
&\simeq& \frac{5\pi^2}{4} \frac{\mu^2 T}{3}\frac{1}{\varphi^2} 
\eea
We see that instead of an exponential suppression we now get a power-law suppression $\propto (T/\Delta)^2$ of the specific heat compared 
to the non-superconducting result. In this sense, the specific heat measures how effectively the quasiparticle
excitations are suppressed by the gap. Our result shows that the dimensionality of the zero-energy excitations in momentum space translates into the 
temperature dependence of the specific heat: in the normal phase, there is a two-dimensional Fermi surface that contributes at $T=0$, while
for an isotropic gap, this Fermi surface is, simply speaking, gone. The anisotropic gap (\ref{pointnodes}) is an intermediate case, its 
suppression lies between the normal and the completely gapped phase. One may thus expect that between the zero-dimensional point nodes and the 
fully gapped spectrum there is another intermediate case, namely one-dimensional line nodes, see problem \ref{prob6}.
 
The low-temperature results for the specific heat are relevant for the physics of compact stars because the superconducting gap of 
either nucleonic superconductivity/superfluidity or quark superconductivity may well be much 
larger than the temperature of the star. In particular, 
the specific heat is important in the context of the cooling of the star, for example through neutrino emissivity $\epsilon_\nu$. 
With $\epsilon_\nu$ being the energy loss
per unit time and volume through neutrino emission (for example through the processes (\ref{betadecay}) in nuclear matter or the processes 
(\ref{proc3flavor}) in quark matter), the relation between $\epsilon_\nu$, $c_V$, and the change in temperature is 
\be
\epsilon_\nu(T) = -c_V(T)\frac{dT}{dt} \, .
\ee
(The minus sign is needed since a positive $\epsilon_\nu$ is an energy loss, i.e., the temperature will decrease, $dT/dt<0$.)
Integrating this relation from a time $t_0$ (with temperature $T(t_0)=T_0$) yields 
\be \label{tt0}
t-t_0	= -\int_{T_0}^T dT'\,\frac{c_V(T')}{\epsilon_\nu(T')} \, .
\ee
This shows that the ratio of the specific heat and the neutrino emissivity enters the cooling behavior of the star. Typically, 
for a given phase, the neutrino emissivity will exhibit 
a similar behavior as the specific heat. For instance, in a superconductor, the emissivity as well as the specific heat are exponentially 
suppressed in which case the subleading behavior becomes important. 
In a real compact star, however, there is most likely not just a single phase and the phase that dominates the behavior of the 
emissivity is not necessarily the one that dominates the specific heat. 

The neutrino emissivity is much more 
difficult to compute than the specific heat, and we devote a whole chapter to its discussion and to a detailed calculation 
for the case of quark matter, see chapter \ref{sec:cooling}.

\section{Color-flavor locked (CFL) quark matter}
\label{sec:CFL}

In our discussion of superconductivity and superfluidity in compact stars we first focus on a density regime where we can perform
rigorous calculations from first principles. This is the regime of asymptotically large densities, where we deal with weakly coupled, 
deconfined quark matter.\footnote{We shall not go into details of neutron superfluidity and proton superconductivity. For a detailed
review of these matters, see Ref.\ \cite{Lombardo:2000ec4}. A shorter discussion can be found for instance in Sec.\ 3.2
of Ref.\ \cite{Page:2006ud4}.} The quarks are weakly coupled due to {\it asymptotic freedom}, which says that the coupling of QCD
becomes weak for large exchanged momenta. For our purpose, the QCD coupling can be considered as a function of the quark chemical 
potential and becomes arbitrarily small for large chemical potentials. In other words, quarks at infinite chemical potential are free. 
Because of this important property of QCD we may use perturbative methods at high densities. The high-density region of the QCD 
phase diagram shown in Fig.\ \ref{figQCDpd} is therefore maybe the best understood regime of QCD. The other regimes in that phase diagram 
are more complicated: we have seen that for nuclear matter one usually relies on phenomenological models;
the high-temperature, small-density region, where the QCD coupling also becomes small, has subtle nonperturbative effects 
because of infrared degrees of freedom; first-principle QCD calculations via computer simulations ({\it lattice QCD}) are 
much more complicated than perturbative physics at high densities and are so far restricted to vanishing chemical potential. 

This possibility of understanding a region of the phase diagram rigorously from first principles is 
a good theoretical motivation to study ultra-dense quark matter. However, for our astrophysical purposes we need to point out 
that these studies are valid at densities much larger than expected
in compact stars. In a compact star, the quark chemical potential is at most of the order of $\mu \lesssim 500\, {\rm  MeV}$. The perturbative 
calculation of the energy gap $\Delta$, to be discussed in Sec.\ \ref{sec:QCDgap}, can be estimated to be reasonable at chemical potentials 
of the order of $\mu \gtrsim 10^8\, {\rm MeV}$ (!) Given this difference of many orders of magnitude, extrapolation of perturbative 
results down to compact star densities may seem bold. However, 
the (rough) quantitative agreement of these extrapolations with different approaches, using phenomenological models, gives us some 
confidence that the ultra-high density calculation may be of relevance for astrophysical calculations. Furthermore, we shall also apply 
general arguments, based on symmetries, which we can expect to hold even at moderate density where the coupling is strong. 
In summary, the following 
discussion, strictly speaking only valid for extreme densities, is of theoretical interest and may also give us insight into compact star physics.

At this point we may remember that we have already discussed the approach to compact star densities from the opposite side.
In the Walecka model of Sec.\ \ref{sec:walecka} 
we have constructed the model such that we have reproduced properties of nuclear matter at densities accessible in the laboratory. These
densities are {\it lower} than the ones in compact stars. We had to extrapolate up to higher densities to obtain predictions
of astrophysical relevance. Therefore, we learn that matter inside compact stars is quite hard to tackle; we have to approach it 
from different sides, and currently we do not have rigorous control over our approaches. This reflects the discussion begun in the introduction:
it shows that the question ``What is the matter composition inside a compact star?'' is, due to our lack of understanding of dense, 
strongly-interacting matter, not only an application of QCD but also relevant to understand QCD. 

From this somewhat philosophical discussion now back to superconductivity in quark matter. Cooper's Theorem tells us 
that an attractive interaction, however small it may be, 
leads to the formation of a quark Cooper pair condensate. At asymptotically high densities, this attractive interaction is 
provided by single-gluon exchange. We can formulate quark pairing in terms of representations of the color gauge group $SU(3)_c$,
\be
SU(3)_c: \qquad [{\bf 3}]_c\otimes [{\bf 3}]_c = [\bar{\bf 3}]^A_c \oplus [{\bf 6}]^S_c \, . 
\ee
On the left-hand side we have two quarks in the fundamental representation, i.e., two complex three-vectors since the number of colors is 
three, $N_c=3$. 
They interact in an antisymmetric ($A$) anti-triplet channel and a symmetric ($S$) sextet channel which are attractive and repulsive, 
respectively. The attractive channel thus provides an anti-triplet of diquarks which has (anti-)color charge. The attractiveness of this
channel can be understood for instance from the existence of baryons. Namely, in a simple picture a baryon contains a diquark in the  
$[\bar{\bf 3}]^A_c$ representation. If it is made of, say, a red and a green quark it has color anti-blue. The baryon is then color-neutralized 
by combining this anti-blue diquark with a blue quark. 

An obvious property of a quark Cooper pair is that it is color-charged. Therefore, it breaks the color symmetry $SU(3)_c$ spontaneously.
In analogy to electronic superconductors, which break the electromagnetic $U(1)_{\rm em}$, quark Cooper pairing is thus termed
{\it color superconductivity}. For an extensive review of color superconductivity see Ref.\ \cite{Alford:2007xm4}. The order parameter of color
superconductivity is the expectation value of the quark-quark two-point function $\langle\psi\psi\rangle$. The color structure of this object
has to be antisymmetric because the antisymmetric representation $[\bar{\bf 3}]^A_c$ is the attractive channel. 
The flavor structure is governed by the chiral symmetry group 
$SU(3)_R\times SU(3)_L$,\footnote{As already mentioned in the introduction, we neglect the heavy quark flavors although
in this section we consider asymptotically large densities. Since we are ultimately interested in extrapolating our results
down to compact star densities, we only take $u$, $d$, and $s$ quarks into account.} discussed in Sec.\ \ref{sec:chiralsym}. For now, we may 
consider these symmetries to be exact, since at the high densities we are working we may neglect all three quark masses compared to the 
chemical potential.
Each of these global $SU(3)$'s leads to the same representations as the color group,
\be \label{repflavor}
SU(3)_f: \qquad [{\bf 3}]_f\otimes [{\bf 3}]_f = [\bar{\bf 3}]^A_f \oplus [{\bf 6}]^S_f \, , 
\ee
with $f=L,R$. Since the overall wave function of the Cooper pair has to be antisymmetric and since pairing in the antisymmetric spin-zero channel 
is preferred, we need to pair in the flavor $[\bar{\bf 3}]^A_f$ channel. In other words, the color-flavor structure of the Cooper pair
is
\be
\langle\psi\psi\rangle \in [\bar{\bf 3}]^A_c \otimes [\bar{\bf 3}]^A_f \, .
\ee
More specifically, with $A,\alpha,\beta\le 3$ being color indices and $B,i,j\le 3$ being flavor indices, 
\be \label{psipsi}
\langle\psi_i^\alpha C\gamma_5 \psi_j^\beta \rangle \propto \epsilon^{\alpha\beta A}\epsilon_{ijB}\phi_A^B \, .
\ee
Here, we have added the Dirac structure with the charge-conjugation matrix $C\equiv i\gamma^2\gamma^0$, leading to even-parity, spin-singlet
pairing. The $3\times 3$ matrix $\phi$ now determines the specific color-flavor structure within the given antisymmetric representations.
This shows that there are in principle many different possible color-superconducting phases. They are distinguished by different 
pairing patterns, i.e., by which quark pairs with which other quark. (At asymptotically large densities, where the flavor symmetries are exact, 
many pairing patterns are equivalent by symmetry and only a few physically distinct phases exist.) In particular, one may construct phases in 
which some of the quarks are paired while some others are not. 

At high densities, the favored phase is the {\it color-flavor locked (CFL)} phase \cite{Alford:1998mk4}. We can characterize it by the 
following properties, 

\renewcommand{\labelenumi}{(\roman{enumi})}
\begin{enumerate}

\item The CFL order parameter is given by 
\be \label{propi}
\phi_A^B = \delta_A^B \;\; \Rightarrow \;\; \langle\psi_i^\alpha C\gamma_5 \psi_j^\beta \rangle \propto \epsilon^{\alpha\beta A}\epsilon_{ijA}\, .
\ee

\item  In the CFL phase, all quarks are paired with pairing pattern $rd-gu$, $bu-rs$, $bd-gs$, $ru-gd-bs$ (where $rd$ is a red down quark,
$gu$ a green up quark etc.), and there are 8 quasiparticles with 
gap $\Delta$ and 1 quasiparticle with gap $2\Delta$.

\item  The CFL phase has the following symmetry breaking pattern,
\be \label{propii}
SU(3)_c\times SU(3)_R \times SU(3)_L \times U(1)_B \to SU(3)_{c+L+R} \times \mathbb{Z}_2 \, .
\ee

\end{enumerate}
\renewcommand{\theenumi}{\arabic{enumi}}

These three properties are in fact equivalent. Before discussing their physical implications, many of which can be read off from properties
(ii) and  (iii),
let us show how the physical statement (ii) follows from the more abstract statement (i). To get a clear picture of the matrix 
structure of the order parameter, let us denote the
bases of the color and flavor antitriplets $[\bar{\bf 3}]^A_c$ and $[\bar{\bf 3}]^A_f$ by $(J^A)^{\alpha\beta} = -i\epsilon^{\alpha\beta A}$,  
$(I_B)_{ij} = -i\epsilon_{ijB}$. Then, we can write Eq.\ (\ref{propi}) as
\be \label{9times9}
\langle\psi C\gamma_5 \psi \rangle_{\rm CFL}\propto
{\bf J}\cdot {\bf I} = i\,\left(\begin{array}{ccc} 0 & -I_3 & I_2 \\ I_3 & 0 & -I_1 \\ -I_2 & I_1 & 0\end{array}\right) 
=\left(\begin{array}{ccccccccc} 0&0&0&0&-1&0&0&0&-1\\0&0&0&1&0&0&0&0&0\\0&0&0&0&0&0&1&0&0\\0&1&0&0&0&0&0&0&0\\-1&0&0&0&0&0&0&0&-1\\
0&0&0&0&0&0&0&1&0\\0&0&1&0&0&0&0&0&0\\0&0&0&0&0&1&0&0&0\\-1&0&0&0&-1&0&0&0&0\end{array}\right) \, .
\ee
This $9\times 9$ matrix is obviously symmetric, as required (the color-flavor structure is symmetric, giving overall antisymmetry through 
the antisymmetric Dirac structure). Its rows and columns are labelled with the nine quarks, $ru$, $rd$, $rs$,
$gu$, $gd$, $gs$, $bu$, $bd$, $bs$. A nonzero entry indicates that the corresponding quarks pair. We see that the matrix has a block 
structure with three $2\times 2$ blocks and one $3\times 3$ block. This leads to the pairing pattern given 
in point (ii). Note that this is a basis dependent statement. In particular, since the color symmetry is a gauge symmetry, 
$\langle\psi C\gamma_5\psi\rangle$ is a gauge variant object. The physically relevant statement, however, is the second part of point (ii) 
about the quasiparticle excitations.
This statement is gauge invariant. The gap structure is given by the eigenvalues of the square of the above $9\times 9$ matrix,
\be \label{exciteCFL}
\epsilon_{k,r} = \sqrt{(k-\mu)^2+\lambda_r \Delta^2} \, ,
\ee
where $\lambda_r$ are the eigenvalues of
\be \label{defL1}
L \equiv ({\bf J}\cdot{\bf I})^2 \, .
\ee
We shall prove the form of the quasiparticle excitations (\ref{exciteCFL}) in Sec.\ \ref{sec:QCDgap}. 
Here we simply compute the eigenvalues $\lambda_r$. They are 
given by the solutions of 
\be
{\rm det}(\lambda - L) = 0 \, .
\ee
This can be rewritten as
\be \label{exptrln}
0= \exp\left[\Tr\ln(\lambda-L)\right] = \exp\left[\Tr\left(\ln\lambda - \sum_{n=1}^\infty\frac{L^n}{n\lambda^n}\right)\right] \, .
\ee
We now have to compute $L^n$. First note that 
\be \label{JIL}
({\bf J}\cdot{\bf I})^{\alpha\beta}_{ij} = 
-\epsilon^{\alpha\beta A}\epsilon_{ijA} = -\delta_i^\alpha\delta_j^\beta + \delta_j^\alpha \delta_i^\beta \;\;
\Rightarrow \;\; L^{\alpha\beta}_{ij}= \delta^{\alpha\beta}\delta_{ij} + \delta_i^\alpha\delta_j^\beta \, .
\ee
This result can be used to compute 
\be \label{L54}
L^2 = 5L-4 \, .
\ee
Consequently, all powers of $L$ only have the matrix structures $L$ and ${\bf 1}$. 
Thus we make the ansatz
\be 
L^n = a_n L + b_n \, .
\ee
Multiplying both sides of this equation by $L$ and using Eq.\ (\ref{L54}) yields 
\be
a_{n+1}=5a_n+b_n \, , \qquad b_{n+1}=-4a_n \, .
\ee
These recursion relations can be solved with the ansatz $a_n=p^n$. This yields the equation $p^2=5p-4$ which is solved by $p_1=4$ and $p_2=1$.
Consequently, the general solution is the linear combination
\be
a_n = \alpha p_1^n +\beta p_2^n = 4^n\alpha + \beta \, .
\ee
From above we know $a_1=1$ and $a_2=5$ which yields $\alpha=-\beta=1/3$. Hence
\be
L^n = \frac{4^n-1}{3}\,L-\frac{4^n-4}{3} \, .
\ee
Inserting this into Eq.\ (\ref{exptrln}) yields
\be
0=\exp\left\{\Tr\left[\frac{L-1}{3}\ln(\lambda-4) - \frac{L-4}{3}\ln(\lambda-1)\right]\right\} \, .
\ee
Now we use $\Tr\,{\bf 1}=9$ and, from Eq.\ (\ref{JIL}), $\Tr \,L = 12$. Thus we have
\be
0=\exp\left[\ln(\lambda-4)+8\ln(\lambda-1)\right] = (\lambda-4)(\lambda-1)^8 \, .
\ee
Consequently, the eigenvalues of $L$ are 1 (8-fold) and 4 (1-fold). Physically speaking, together with Eq.\ (\ref{exciteCFL}) this means that
in the CFL phase 8 quasiparticle excitations have a gap $\Delta$ and 1 quasiparticle excitation has a gap $2\Delta$. This is the second part of 
point (ii). Of course, this discussion says nothing about the magnitude of $\Delta$, which has to be computed from the 
QCD gap equation, see subsequent section. 
We leave it as an exercise to show that (iii) follows from (i), see problem \ref{prob7}.

Points (ii) and (iii) reveal many important physical properties of the CFL state. Since these points are solely based on symmetry 
considerations, they are independent of the details of the interaction. Therefore, they can be expected to hold also at 
lower densities where perturbative QCD is not applicable. First, one may ask why CFL is the ground state and not
any other order parameter given by a different matrix $\phi_A^B$. The simple answer is that the CFL order parameter is the only one in which 
all quarks participate in pairing, as we have seen. All other possible order parameters leave several excitations ungapped. Therefore, the
CFL phase leads to the largest condensation energy and thus is the ground state at high densities (at lower densities the situation
is much more complicated). A more formal argument is that the CFL phase is the color superconductor with the largest residual symmetry group. 
It is thus a particularly symmetric state which also indicates that it is preferred over other color superconductors, although this is 
not a rigorous argument.  
 
From (iii) we read off the following properties of CFL,
\begin{itemize}
\item CFL breaks chiral symmetry. We see that the CFL symmetry breaking pattern (\ref{propii}) is, regarding chiral symmetry, the same as in 
Eq.\ (\ref{breakchiral}). However, the mechanisms are different. The latter is caused by a chiral condensate of the form 
$\langle \bar{\psi}_R\psi_L\rangle$, 
while the CFL condensate has the form $\langle \psi_R\psi_R\rangle$ (and the same with $R\to L$). At first sight, the CFL
condensate thus preserves the full chiral symmetry, i.e., apparently one can still do separate $L$ and $R$ rotations without
changing the ground state. However, the symmetry breaking occurs through the ``locking'' with color, i.e., 
in order to leave the order parameter invariant, a color rotation has to be undone by equal rotations in the left- and right-handed sectors. 
Although caused by different mechanisms, the two scenarios lead to similar physics. As for the usual chiral symmetry breaking, the CFL phase 
also has an octet of Goldstone modes. Since all fermions acquire an energy gap, these Goldstone modes become very important for the 
phenomenology of the CFL phase. Moreover, at lower densities, where the strange quark mass cannot be neglected, kaon condensation 
is expected in the CFL phase, not unlike its nuclear matter relative discussed in Sec.\ \ref{sec:kaon}. The kaon-condensed CFL phase is usually
called CFL-$K^0$ and will be discussed in the next subsection, Sec.\ \ref{sec:CFLK0}.

\item The color gauge group is completely broken. While spontaneous breaking of a global group leads to Goldstone bosons, spontaneous 
breaking of a gauge group leads to masses for the gauge bosons. Here, all gluons acquire a Meissner mass, just as the photon acquires a Meissner 
mass in an electronic superconductor. A nonzero Meissner mass for a gauge boson is the field-theoretical way of saying that there is a Meissner
effect, i.e., that the magnetic field can penetrate the superconductor only up to a certain penetration depth. The inverse of this penetration 
depth corresponds to the Meissner mass.  In the CFL phase, one linear combination of a gluon and the photon remains massless.
In other words, there is an unbroken $U(1)_{\tilde{Q}} \subseteq SU(3)_{c+L+R}$, generated by $\tilde{Q}$ which is a linear combination
of the original charge generator $Q$ and the eighth gluon generator $T_8$ (if you have done problem \ref{prob7} you can easily show this and determine
the exact form of the linear combination).
This phenomenon is also called {\it rotated electromagnetism}. Since the admixture of the gluon to the new gauge boson is small, one may say
that the CFL phase is a color superconductor but no electromagnetic superconductor. This is of relevance for compact stars since it implies
that the CFL phase does not expel magnetic fields. 

\item The CFL phase is a superfluid since it breaks the baryon number conservation group $U(1)_B$. This is important since this 
is an exact symmetry, even at lower densities where finite quark masses become important. Therefore, there is always one exactly massless 
Goldstone mode in the CFL phase.

\end{itemize}

\subsection{Kaon condensation in CFL quark matter}
\label{sec:CFLK0}

We have pointed out that chiral symmetry is not only broken in the hadronic phase, but also in CFL.
This is by itself an interesting fact since it means that in QCD chiral symmetry is spontaneously 
broken at very low and very high densities. How about
the region in between? This is unknown, but the possibility remains that chiral symmetry is, at small temperatures, broken for all
densities. Since the symmetry breaking patterns of nuclear matter and CFL are identical (note that in a neutron superfluid also the
$U(1)_B$ is broken), this implies that possibly there is no real phase transition at moderate densities and small densities
in the QCD phase diagram. In Fig.\ \ref{figQCDpd} this corresponds to the possibility that the ``non-CFL'' region is absent, at least at $T=0$.

Now let us use the chiral symmetry breaking of CFL for a concrete calculation.  
Since in the CFL phase all (quasi)fermions acquire energy gaps of at least $\Delta$  
-- whose magnitude we compute from first principles in Sec.\ \ref{sec:QCDgap} -- the physics of the CFL phase at temperatures 
smaller than $\Delta$ is determined by the pseudo-Goldstone modes associated to chiral symmetry breaking (and the exact Goldstone 
mode from breaking of $U(1)_B$ which we do not discuss here). As we discuss below, $\Delta$ can be expected to be of the order of 10 MeV at 
densities present in compact stars. This is large enough to make fermionic excitations in a possible CFL phase in a star essentially
irrelevant. Therefore, for astrophysical applications, the discussion of the physical properties of the Goldstone modes
is crucial. 

In the context of kaon condensation in nuclear matter, Sec.\ \ref{sec:kaon}, we have used an effective theory for the chiral field $U$ and its 
interactions with nucleons. Also for the mesons in CFL we can write down such a theory. In this case, the chiral field is  given by
\be \label{chiralsigma}
\Sigma = \phi_L^\dag \phi_R \, ,
\ee
where $\phi_L$ and $\phi_R$ are the $3\times 3$ matrix order parameters in the left- and right-handed sector. In our above discussion 
we have not distinguished between $\phi_L$ and $\phi_R$ since in ``pure'' CFL we have $\phi_L=\phi_R={\bf 1}$. 
For unitary $3\times 3$ matrices $\phi_L$ and $\phi_R$, $\Sigma$ is unitary, 
$\Sigma\in U(3)$. It thus contains 9 degrees of freedom, one of which one usually ignores since it corresponds to the $\eta'$ which is heavy 
due to the explicitly broken $U(1)_A$. Eight degrees of freedom remain, $\Sigma\in SU(3)$, and we can identify them as pions, kaons 
etc.\ just like in hadronic matter, see Eq.\ (\ref{mesonoctet}). Despite the similarities, there is an important difference to hadronic matter: as 
one can see from the definition of the chiral field (\ref{chiralsigma}), a meson in CFL is composed of two fermions and two fermion holes 
(each $\phi$ in Eq.\ (\ref{chiralsigma}) represents a diquark). For example, a neutral kaon  should be viewed as an excitation 
$K^0\sim \bar{u}\bar{s}du$. Note that this ``CFL kaon'' has the same quantum numbers as the ``usual kaon'', composed of a particle and
an antiparticle, $K^0\sim \bar{s}d$. Hence, if you want to construct a CFL kaon from a usual kaon you need to replace 
$s\to \bar{d}\bar{u}$ and $d\to \bar{u}\bar{s}$. This identification reflects the {\it anti}-triplet representation in Eq.\ (\ref{repflavor}).
As a consequence, the meson masses in CFL are ordered inversely compared to the usual mesons. To see this, first note  
that the quark flavors $(u,d,s)$, ordered with increasing mass, $m_u<m_d<m_s$, have the anti-triplet counterpart 
$(\bar{d}\bar{s}, \bar{u}\bar{s},\bar{u}\bar{d})$. Here the masses (squared) have become ordered in the opposite way, $m_dm_s>m_um_s>m_um_d$. 
Therefore, in nuclear matter (and ignoring finite density effects), $m_{\pi^0}<m_{K^0}$ because $m_{\pi^0} \propto m_u+m_d$
and $m_{K^0} \propto m_s+m_d$, whereas in CFL $m_{K^0}<m_{\pi^0}$ because $m_{K^0}^2\propto m_u m_d+m_um_s$ and $m_{\pi^0}^2\propto m_d m_s+m_um_s$.
We shall verify the form of the kaon mass in CFL below within the effective theory.

The effective Lagrangian for mesons in CFL is given by 
\be \label{calLSigma}
{\cal L} = \frac{f_\pi^2}{4}\Tr[\nabla_0\Sigma\nabla_0\Sigma^\dag -v_\pi^2\partial_i\Sigma\partial_i\Sigma^\dag]+\frac{af_\pi^2}{2}{\rm det}\,M\,
\Tr[M^{-1}(\Sigma+\Sigma^\dag)] \, , 
\ee
with 
\be \label{covA}
\nabla_0\Sigma \equiv \partial_0\Sigma + i[A,\Sigma] \, , \qquad A\equiv -\frac{M^2}{2\mu} \, , 
\ee
where $M={\rm diag}(m_u,m_d,m_s)$ is the quark mass matrix. The matrix $A$ enters the theory as the temporal component of a gauge field; 
it plays the role of an effective chemical potential for the field $\Sigma$. We shall see below how this translates into effective 
chemical potentials for the neutral and charged kaons.

The original works where this Lagrangian has been proposed are 
Refs.\ \cite{Son:1999cm4,Bedaque:2001je4}. There you can find detailed explanations about the structure of the effective Lagrangian and its
differences to the effective meson Lagrangian for hadronic matter (\ref{calLU}). Comparing with Eq.\ (\ref{calLU}) we see that 
in CFL we do not have a term linear in the quark masses, rather only quadratic, $M^{-1} {\rm det}\, M \propto m^2$
(and higher even powers which we have neglected). We also have different coefficients in front of the temporal and spatial part of the 
kinetic term, 
originating from the breaking of Lorentz invariance in a medium, $v_\pi=1/\sqrt{3}$. As for the hadronic phase, there are two constants
$f_\pi$ and $a$. This reminds us of the nature of effective theories like the ones given by Eqs.\ (\ref{calLU}) and (\ref{calLSigma}):
they are expected to give at least a qualitatively correct description even beyond the regime where the theory can be tested experimentally
or from first-principle calculations. 
The reason is that they are almost entirely determined by symmetries. Only the coefficients have to be taken from the experiment or
an underlying microscopic theory. The former is done in the effective theory of the hadronic phase. The latter, namely
fixing the constants $f_\pi$ and $a$ from perturbative QCD, is done for the effective theory of CFL. In particular, 
one can expect that, if CFL is the ground state of dense quark matter at densities relevant for compact stars, the effective
theory is a powerful tool to compute the phenomenology of a potential quark core of the star.

Although terms of higher order in the fields and the mass matrix have already been neglected in Eq.\ (\ref{calLSigma}), the Lagrangian 
still looks complicate. The meson fields $\theta_a$ appear in the exponent of $\Sigma$,
\be \label{Sithla}
\Sigma = e^{i\theta_a \lambda_a/f_\pi} \, , 
\ee
with the Gell-Mann matrices $\lambda_a$, and thus they appear to all orders even in the given truncated theory. Let us first rewrite the 
Lagrangian by abbreviating $Q\equiv \theta_a \lambda_a/f_\pi$ such that 
\be
\Sigma = e^{iQ} = \cos Q + i\sin Q \, .
\ee
Then, the various terms of the Lagrangian become 
\begin{subequations}
\bea
\Tr[\partial_0 \Sigma\partial_0 \Sigma^\dag] &=& \Tr[(\partial_0\cos Q)^2 +(\partial_0\sin Q)^2] \, ,  \\
\Tr[\partial_i\Sigma\partial_i\Sigma^\dag] &=& \Tr[(\nabla\cos Q)^2 +(\nabla\sin Q)^2] \, , \\  
\Tr\left[[A,\Sigma][A,\Sigma]^\dag\right] &=& 2\Tr[A^2 - (A\cos Q)^2-(A\sin Q)^2] \, , \\
i\Tr[-\partial_0\Sigma[A,\Sigma]^\dag + [A,\Sigma]\partial_0\Sigma^\dag] &=& 2i\Tr[(\partial_0\cos Q)[A,\cos Q] \non
&& \hspace{1cm} +(\partial_0\sin Q)[A,\sin Q]] \, , 
\eea
\end{subequations}
and thus 
\bea \label{Lsincos}
{\cal L} &=& \frac{f_\pi^2}{2}\Tr[A^2- (A\cos Q)^2-(A\sin Q)^2 + 2a({\rm det}\, M)M^{-1}\cos Q] \non 
&&+\frac{f_\pi^2}{4}\Tr[(\partial_0\cos Q)^2+(\partial_0\sin Q)^2-v_\pi^2[(\nabla\cos Q)^2 +(\nabla\sin Q)^2]] \non
&&+i\frac{f_\pi^2}{2}\Tr[(\partial_0\cos Q)[A,\cos Q]+(\partial_0\sin Q)[A,\sin Q] ] \, .
\eea
Let us first interpret $Q$ as a constant background, i.e., as the meson condensate, and neglect the fluctuations.
This will allow us to compute the values of the various condensates at zero temperature. 
In general, all mesons may condense and the parameters of the theory determine which of the 
condensates becomes nonzero. We recall from the above discussion that we expect the kaons, not the pions, to be the lightest mesons in CFL. 
Therefore, let us simplify $Q$ by setting all fields except the kaon fields to zero,
\be \label{Qkaon}
Q = \sum_{a=4}^7 \phi_a \lambda_a = \left(\begin{array}{ccc} 0 & 0 & \phi_4 - i\phi_5 \\ 0 & 0 & \phi_6-i\phi_7 \\
\phi_4 + i\phi_5 & \phi_6+i\phi_7 & 0\end{array}\right) \, , 
\ee
with the dimensionless condensates $\phi_a\equiv \theta_a/f_\pi$. 
With this ansatz we shall be able to 
construct a zero-temperature phase diagram that contains regions of no condensates, charged kaon condensates, neutral kaon condensates, and 
possibly coexistence of both. This is exactly the same ansatz as we have made in Sec.\ \ref{sec:kaonnucleon} 
for kaon condensation in nuclear matter, 
see Eq.\ (\ref{Q4567}). We can thus follow the steps below Eq.\ (\ref{Q4567}) to obtain 
\bea
\cos Q &=& 1-\frac{Q^2}{\phi^2}(1-\cos\phi) \, , \label{cosQ}
\eea
and
\bea
\sin Q &=& \frac{Q}{\phi}\sin\phi \, , \label{sinQ}
\eea
where 
\be
\phi^2 \equiv \phi_4^2+\phi_5^2+\phi_6^2+\phi_7^2 \, .
\ee
Since we assume our condensates to be constant in time and space, only the first line of the Lagrangian (\ref{Lsincos}) survives. 
The tree-level zero-temperature free energy is the negative of this Lagrangian and becomes 
\bea \label{LzeroT}
U &=& \frac{f_\pi^2}{2}\Tr\left[2\frac{1-\cos\phi}{\phi^2}\left(a({\rm det}\,M)M^{-1} Q^2-A^2Q^2\right)\right. \non
&& \left. +\frac{(1-\cos\phi)^2}{\phi^4} (AQ^2)^2
+\frac{\sin^2\phi}{\phi^2}(AQ)^2\right] \, ,
\eea
where we have subtracted the ``vacuum'' contribution 
\be
U_{\rm CFL} = U(\Sigma={\bf 1}) = -f_\pi^2a\,{\rm det}\,M \,\Tr[M^{-1}] \, ,
\ee
such that the state without kaon condensates, i.e., the pure CFL state has free energy $U=0$.  
With the definitions of the matrices $A$ and $Q$ in Eqs.\ (\ref{covA}) and (\ref{Qkaon}),
the notations 
\be
\phi_{K^+}^2\equiv \phi_4^2+\phi_5^2 \,, \qquad \phi_{K^0}^2\equiv \phi_6^2+\phi_7^2 \, , 
\ee
and abbreviating $A={\rm diag}(a_1,a_2,a_3)$, the various traces are
\begin{subequations} \label{traces}
\bea
{\rm Tr}[a({\rm det} M)M^{-1}Q^2-A^2Q^2]&=& (m^2_{K^+}-\mu^2_{K^+})\phi^2_{K^+}+
(m^2_{K^0}-\mu^2_{K^0})\phi^2_{K^0} \non
&& -2a_3(a_1\phi^2_{K^+}+a_2\phi^2_{K^0}) \, , \\
{\rm Tr}[(AQ^2)^2]&=&(a_1\phi^2_{K^+}+a_2\phi^2_{K^0})^2+a_3^2\phi^4 \, , \\
{\rm Tr}[(AQ)^2]&=& 2a_3(a_1\phi^2_{K^+}+a_2\phi^2_{K^0}) \, , 
\eea
\end{subequations}
where we have defined the kaon chemical potentials and masses
\begin{subequations} \label{mumK}
\bea
\mu_{K^+} &\equiv& \frac{m_s^2-m_u^2}{2\mu} \, , \qquad \mu_{K^0} \equiv \frac{m_s^2-m_d^2}{2\mu} \, , \\
m_{K^+}^2 &\equiv& am_d(m_s+m_u) \, , \qquad m_{K^0}^2 \equiv am_u(m_s+m_d) \, .
\eea
\end{subequations}
It will become clear below that these quantities really act as masses and chemical potentials for the kaons. 
For simplicity we have omitted the electric charge chemical potential in the Lagrangian which would have appeared in $\mu_{K^+}$ as an additional
contribution. Inserting Eqs.\ (\ref{traces}) into Eq.\ (\ref{LzeroT}), we can write the free energy as
\bea \label{Utt}
\frac{U(\phi_1,\phi_2)}{f_\pi^2}  &=& (1-\cos\phi)\left[(m^2_1-\mu^2_1)
\frac{\phi^2_1}{\phi^2} + (m^2_2-\mu^2_2)
\frac{\phi^2_2}{\phi^2}\right]  \non
&& +\frac{1}{2}(1-\cos\phi)^2\left(\mu_1\frac{\phi^2_1}{\phi^2}+\mu_2
\frac{\phi^2_2}{\phi^2}\right)^2 \, .
\eea
Here and in the following we use, for notational convenience, the subscript 1 for $K^+$ and 2 for $K^0$. 
To understand the expression for the free energy we consider the limit case of small condensates, $\theta_i\ll f_\pi$, i.e., 
$\phi_i=\theta_i/f_\pi\ll 1$ for $i=K^+,K^0$. Then we can expand $U(\phi_1,\phi_2)$ up to fourth order in the condensates to obtain
\bea \label{Utt2}
U(\theta_1^2,\theta_2^2) &\simeq& \frac{m_1^2-\mu_1^2}{2}\theta_1^2 + \frac{m_2^2-\mu_2^2}{2}\theta_2^2 \non
&& + \frac{\beta_1}{4}\theta_1^4 + \frac{\beta_2}{4}\theta_2^4  + \frac{\alpha}{4}\theta_1^2\theta_2^2 \, , 
\eea
with 
\be \label{defalphabeta}
\beta_i \equiv \frac{4\mu_i^2-m_i^2}{6f_\pi^2} \quad (i=K^+,K^0) \, ,  \qquad \alpha\equiv \frac{\beta_1+\beta_2}{2}
-\frac{(\mu_1-\mu_2)^2}{4f_\pi^2} \, .
\ee
We have thus reduced the effective theory to a two-component $\phi^4$ theory, cf.\ Eq.\ (\ref{Uphitree}) in appendix \ref{app:bosons}, 
with effective coupling constants $\beta_i$ for the self-coupling of the kaons and an effective coupling constant $\alpha$ for the interaction 
between charged and neutral kaons. 

\begin{figure}[t]
\begin{center}
\includegraphics[width=0.7\textwidth]{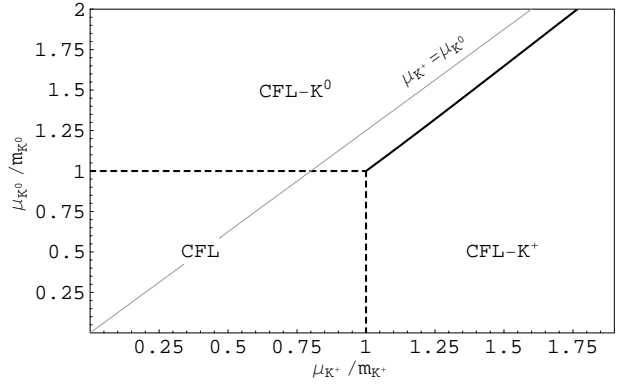}
\caption{
Zero-temperature phase diagram for kaon condensation in the $\mu_{K^+}$-$\mu_{K^0}$-plane. 
No condensation occurs if the chemical potential is smaller than the meson mass. Coexistence of the two 
condensates is only possible along the (solid) line that separates the CFL-$K^0$ from the CFL-$K^+$ phase. 
This line is given by Eq.\ (\ref{curve}) and marks a first order phase transition. For large chemical 
potentials, it approaches the line $\mu_{K^+}=\mu_{K^0}$. The (dashed) lines separating 
either of 
the two meson-condensed phases from the pure CFL phase are second order phase transition lines. In the 
condensed phases, the condensate and free energy are given by Eqs.\ (\ref{costheta}) and 
(\ref{freeenergy}), respectively.}
\label{figmu1mu20}
\end{center}
\end{figure}

We may come back to the full free energy (\ref{Utt}) to find the ground state of the system for arbitrary chemical 
potentials $\mu_1$, $\mu_2$. To this end, one has to minimize the free energy through the equations
\be \label{saddle}
\frac{\partial U}{\partial\phi_1}=\frac{\partial U}{\partial\phi_2}=0 \, .
\ee
By construction, the free energy of the CFL state without kaon condensation, $\phi_1=\phi_2=0$, is given by $U=0$. 
If one of the condensates vanishes, say $\phi_2=0$, one of the equations (\ref{saddle}) is 
automatically fulfilled, and the other one becomes
\be
0=\frac{1}{f_\pi^2}\left.\frac{\partial U}{\partial\phi_1}\right|_{\phi_2=0} = \sin\phi_1\left(m_1^2-\mu_1^2\cos\phi_1\right) \, . 
\ee
This has a nontrivial solution for $m_1^2<\mu_1^2$,   
\be \label{costheta}
\cos \phi_1 = \left\{\begin{array}{cc} 1 & \;\;{\rm for}\;\;\; m_1^2>\mu_1^2 \\[2ex] \displaystyle{\frac{m_1^2}{\mu_1^2}} 
& \;\;{\rm for}\;\;\; m_1^2<\mu_1^2 \end{array}\right. \, , 
\ee
and the free energy density becomes 
\be \label{freeenergy}
U(\phi_2=0)= \left\{\begin{array}{cc} 0 & \;\;{\rm for}\;\;\; m_1^2>\mu_1^2 \\[2ex] 
-\displaystyle{\frac{f_\pi^2(m_1^2-\mu_1^2)^2}{2\mu_1^2}} & \;\;{\rm for}\;\;\; m_1^2<\mu_1^2 \end{array}\right. \, .
\ee
By symmetry, we find the same solution for $\phi_2$ if we set $\phi_1=0$. 
Equating the free energies of the two phases $\phi_1=0$, $\phi_2\neq 0$ and 
$\phi_1\neq 0$, $\phi_2= 0$
one finds the condition for coexistence of two condensates,
\be \label{curve}
\mu_2^2(\mu_1^2 - m_1^2)^2 =\mu_1^2(\mu_2^2 - m_2^2)^2 \, .
\ee
This condition can also be obtained by assuming two nonvanishing condensates in Eqs.\ (\ref{saddle}).
As a result we obtain the phase diagram shown in Fig.\ \ref{figmu1mu20}, where we restrict ourselves to 
$\mu_1,\mu_2>0$ without loss of generality. 

What are the values of the kaon chemical potentials in the real world? In other words, where in the phase diagram of Fig.\ \ref{figmu1mu20}
does a compact star sit? Let us first see whether in a star we can expect the kaon chemical potentials to be larger than their mass, i.e., 
whether kaon condensation is possible. As discussed above, for quantitative predictions of the effective theory 
we rely on the results for the constants $f_\pi$ and $a$ at asymptotically large densities and their extrapolation down to densities in a 
compact star. This extrapolation yields
\be \label{fpia}
f_\pi^2 = \frac{21-8\ln 2}{18}\frac{\mu^2}{2\pi^2} \simeq (100\,{\rm MeV})^2 \, , \qquad a=\frac{3\Delta^2}{\pi^2 f_\pi^2} \simeq 0.03 \, ,
\ee
where we used a quark chemical potential $\mu\simeq 500\, {\rm MeV}$ and a fermionic energy gap $\Delta\simeq 30\, {\rm MeV}$. 
Then, from Eqs.\ (\ref{mumK}) we conclude that both kaon masses are of the order of 
$m_{K^+}\simeq m_{K^0} \simeq (a\,m_{\rm light}\,m_s)^{1/2}\simeq 5\, {\rm MeV}$, where we
used a quark mass for $u$ and $d$ quarks $m_{\rm light}\simeq 5\, {\rm MeV}$ and a strange quark mass $m_s\simeq 150\, {\rm MeV}$. The kaon 
chemical potential then is $\mu_{K^+}\simeq\mu_{K^0}\simeq m_s^2/(2\mu)\simeq 20\, {\rm MeV}$. This suggests that the interior of the star sits 
outside the rectangle given by the dashed lines in Fig.\ \ref{figmu1mu20}, i.e., if there is a color-flavor locked core in a compact star it 
is likely 
to be kaon-condensed CFL matter rather than ``pure'' CFL matter (we shall confirm this conclusion below for nonzero temperatures). 
Does this matter contain a charged or a neutral kaon condensate? 
Firstly, the slightly heavier $d$ quark compared to the $u$ quark makes the $K^+$  slightly heavier than the $K^0$. This 
asymmetry is taken into account in Fig.\ \ref{figmu1mu20}. Moreover, the electric charge of a potential $K^+$ condensate would require the presence 
of electrons to neutralize the system, which further disfavors the charged kaon condensate. 
We thus expect the CFL-$K^0$ phase to be the most likely meson-condensed phase in CFL.

As a second application of the effective theory for mesons in CFL let us compute an estimate of the critical temperature of (neutral) 
kaon condensation. This is important to answer the question: if there is CFL matter in a compact star and if there 
is kaon condensation at zero temperature, at which temperature (i.e., at which point in the life of the star) does condensation set in? 

The full temperature-dependent theory defined by the effective Lagrangian is very complicated. We therefore expand 
the Lagrangian (\ref{Lsincos}) up to fourth order in the matrix-valued field $Q$ to obtain
\bea \label{LAQ}
{\cal L} &=& \frac{f_\pi^2}{2}\Tr\left[(A^2-a({\rm det}\,M)M^{-1})\left(Q^2-\frac{A^2Q^4}{12}\right)-(AQ)^2-\frac{(AQ^2)^2}{4}
+\frac{(AQ^2)Q^2}{3}\right] \non
&& + \frac{f_\pi^2}{4}\Tr\left[(\partial_0 Q)^2-v_\pi^2(\nabla Q)^2+2i(\partial_0 Q)[A,Q]\right] \, , 
\eea
where we have neglected terms of fourth order in $Q$ which contain derivatives such as $\Tr[(Q\partial_0Q)^2]$ etc, and where we have
dropped the contribution constant in $Q$, which serves to normalize the free energy of the pure CFL state to zero, see remark below 
Eq.\ (\ref{LzeroT}).
Next, one has to separate the condensate from the fluctuations, as demonstrated in appendix \ref{app:bosons}. 
The resulting Lagrangian has the same structure as given in the appendix for the $\phi^4$ model, see Eq.\ (\ref{L234}): a tree-level
potential; terms of second order in the fluctuations which define the tree-level propagator; terms cubic in the fluctuations which 
correspond to interactions due to the presence of the condensate; and finally terms quartic in the fluctuations. 
Here we do not discuss the explicit structure of these terms in general, for details you may consult Ref.\ \cite{Alford:2007qa4}.
We rather restrict ourselves again to the kaon degrees of freedom. As a further simplification, we set the charged kaon condensate 
to zero, $\theta_4=\theta_5=0$ (but keep the charged kaon fluctuations). 
This is motivated by the above discussion about the more favorable neutral kaon condensate. 
For the neutral kaon condensate we choose, without loss of generality, a direction in the degeneracy space of the condensate by 
setting $\theta_7=0$, and we denote $\theta\equiv \theta_{K^0}=\theta_6$. The tree-level potential 
from Eq.\ (\ref{Utt2}) then simply becomes
\be \label{Utheta}
U(\theta) = \frac{m_2^2-\mu_2^2}{2}\theta^2 + \frac{\beta_2}{4}\theta^4   \, . 
\ee
The kaon sector of the inverse tree-level propagator is block diagonal, 
\be \label{treeprop}
D_0^{-1} = \left(\begin{array}{cc}D_{01}^{-1} & 0 \\ 0 & D_{02}^{-1} \end{array}\right) \,  , 
\ee
where
\begin{subequations} \label{S01S02}
\bea
D_{01}^{-1} &=& \left(\begin{array}{cc} -K^2 + m_1^2-\mu_1^2+\alpha\theta^2  & 
-2i\mu_1k_0 \\[1ex] 2i\mu_1k_0 &
-K^2+ m_1^2-\mu_1^2+\alpha\theta^2  \end{array}\right) \, , \\ \non
D_{02}^{-1} &=& \left(\begin{array}{cc} -K^2 + m_2^2-\mu_2^2 +3\beta_2\theta^2  & 
-2i\mu_2k_0 \\[1ex] 
2i\mu_2k_0 &
-K^2 + m_2^2-\mu_2^2 +\beta_2\theta^2 \end{array}\right) \, ,
\eea
\end{subequations}
with the abbreviation $K^2\equiv k_0^2-v_\pi^2k^2$. The verification of this form of the kaon tree-level propagator
is left as an exercise, see problem \ref{prob8}.
Analogously to the calculation in the appendix, we obtain the kaon dispersion relations. They are given by the poles of the propagator $D_0$,
which are the zeros of the determinant of the inverse propagator $D_0^{-1}$. 
The dispersion for the charged kaon is
\be \label{epsKp}
\epsilon_1^\pm(k) = \sqrt{v_\pi^2k^2+m_1^2+\alpha\theta^2}\mp\mu_1 \, .
\ee
We see that the $K^0$ condensate gives a contribution to the mass of the $K^+$. For the neutral kaon we obtain  
\be \label{epsK0}
\epsilon_2^\pm(k) = \sqrt{E_k^2+\mu_2^2\mp\sqrt{4\mu_2^2E_k^2+\delta M^4}} \, , 
\ee
where
\be
E_k \equiv \sqrt{v_\pi^2k^2+m_2^2+2\beta_2\theta^2} \, , \qquad \delta M^2 = \beta_2 \theta^2 \, .
\ee
Since kaon condensation breaks a global symmetry of the system, namely the $U(1)$ associated to conservation of strangeness, 
we expect a Goldstone mode. (Notice the two-fold condensation process: due to the condensation of quark Cooper pairs, chiral symmetry is broken
and pseudo-Goldstone bosons appear in the system; on top of that, these pseudo-Goldstone modes -- here the neutral kaons -- condense themselves,
breaking the global symmetry further and giving rise to another Goldstone mode.)
This mode is expected to be gapless.\footnote{Due to the weak interactions this mode acquires a small energy 
gap in the keV range which we neglect here.} To check this expectation, we first compute the condensate from 
the tree-level potential (\ref{Utheta}). The nontrivial minimum of this potential is 
\be \label{thetazeroT}
\theta^2 = \frac{\mu_2^2-m_2^2}{\beta_2} \, .
\ee
This implies $4\mu_2^2E_{k=0}^2+\delta M^4=(3\mu_2^2-m_2^2)^2$ and 
$E_{k=0}^2+\mu_2^2 = 3\mu_2^2-m_2^2$ which we can insert into the kaon dispersion (\ref{epsK0}). The result is 
\be
\epsilon_2^+(k=0) = 0 \, ,
\ee
confirming the existence of a gapless mode. 

Following the calculation in the appendix, we can immediately write down the thermodynamic potential at finite temperature,
\be
\Omega = U(\theta) + T\sum_{i=1,2}\sum_{e=\pm} \int \frac{d^3{\bf k}}{(2\pi)^3} \ln\left(1-e^{-\epsilon_i^e/T}\right) \, .
\ee
In order to extract an estimate for the critical temperature, we expand the potential for large $T$, 
\bea \label{OmegaU}
\Omega &\simeq& U(\theta) - \frac{2\pi^2}{45v_\pi^3}T^4 + \left(\frac{\alpha+2\beta_2}{12v_\pi^3}\theta^2 + 
\frac{m_1^2+m_2^2-2(\mu_1^2+\mu_2^2)}{12v_\pi^3}\right) \, T^2 + \ldots \non
&=& \left(\frac{m_2^2-\mu_2^2}{2}+\frac{\alpha+2\beta_2}{12v_\pi^3}T^2\right)\theta^2+\frac{\beta_2}{4}\theta^4 \non
&&- \frac{2\pi^2}{45v_\pi^3}T^4 +\frac{m_1^2+m_2^2-2(\mu_1^2+\mu_2^2)}{12v_\pi^3}\, T^2 + \ldots 
\eea
The $T^4$ term is easy to obtain and has also been discussed in appendix \ref{app:bosons}. For the $T^2$ term we have neglected $\delta M$ in the
neutral kaon dispersions (\ref{epsK0}). Then they assume the same form as the ones for the charged kaons (\ref{epsKp}) and we can 
use the expansion for the pressure of a noninteracting Bose gas, see for instance the appendix of Ref.\ \cite{kapusta4}.

We have arrived at a potential with terms constant, quadratic, and quartic in $\theta$. Since we assume the existence of a condensate at $T=0$,
we have $\mu_2>m_2$, which we have argued to be realistic for densities in compact stars. Therefore, the quartic term is always positive, while
the quadratic term starts from a negative value at $T=0$ and becomes positive for sufficiently large temperatures. 
Consequently, the nontrivial solution for the condensate ceases to exist when the coefficient in front of the 
quadratic term vanishes. This yields the condition for the critical temperature which we thus estimate to be
\be \label{TcK0}
T_c^2 \simeq 6 v_\pi^3 \frac{\mu_2^2-m_2^2}{\alpha+2\beta_2} \, .
\ee
With the definitions (\ref{defalphabeta}) we can express $T_c$ as a function of kaon chemical potentials and masses.\footnote{Notice that
for $\alpha+2\beta_2<0$ the critical temperature formally becomes imaginary, 
i.e., the condensate apparently ``refuses'' to melt. This situation cannot occur for realistic parameters
in our case but is an interesting theoretical possibility. See appendix C in Ref.\ \cite{Alford:2007qa4} and references therein for more 
information.}
Before we interpret the result we point out a problem of the current approach. We have seen that at zero temperature, with $\theta(T=0)$ given 
by Eq.\ (\ref{thetazeroT}), we have $\epsilon_2^+(k=0)=0$. At finite temperature we expect the condensate to melt, 
i.e., $\theta(T)<\theta(T=0)$ for all $T$. 
In this case, however, the excitation of the Goldstone mode (which should remain gapless for all $T<T_c$ due to the Goldstone theorem)
becomes imaginary if written in the form (\ref{epsK0}). This is clearly unphysical and due to the approximation we have made. 
The solution to this problem is to set up a more elaborate approximation scheme which evaluates the thermal kaon masses self-consistently. 
This is beyond the scope of these lectures, see Ref.\ \cite{Alford:2007qa4} for such a treatment. 

It turns out that our
estimate of the critical temperature coincides with the self-consistent calculation. We can therefore use Eq.\ (\ref{TcK0})
for a physical conclusion.
With the definition of the effective coupling constants $\alpha$ and $\beta_2$ in Eq.\ (\ref{defalphabeta}) 
and the approximate numbers for the kaon chemical potentials and masses discussed below Eq.\ (\ref{fpia}) we obtain $T_c\simeq 60\,{\rm MeV}$.
This is of the order of or even larger than the critical temperature $T_c^{\rm CFL}$ for CFL itself. We do not aim to compute the 
critical temperature of CFL in these lectures. We simply give the (mean-field) result,  
\be \label{TcCFL}
T_c^{\rm CFL} \simeq 2^{1/3}\cdot 0.57\Delta \, , 
\ee
where $\Delta$ is the zero-temperature gap. This relation differs by a prefactor of
order one from the relation obtained from the usual Bardeen-Cooper-Schrieffer (BCS) theory, $T_c\simeq 0.57\Delta$; see remark below 
Eq.\ (\ref{twogapstr}) for the origin of this prefactor. For our present purpose it is sufficient to notice 
that the critical temperature in a superconductor is typically of the same order as the zero-temperature gap. 
Since $\Delta$ is also of the order of tens of MeV, we may apparently conclude that the kaon condensate does not melt before the CFL phase itself
melts. However, we need to remember that our effective theory
is only valid for temperatures smaller than the gap $\Delta$. Therefore, the estimated critical temperature for kaon condensation 
is close to or beyond the limit of validity of our effective description. Nevertheless, as a tentative conclusion we can say that 
as soon as quark matter is cold enough to be in the CFL state, we also expect it to be cold enough for kaon condensation, provided that the 
parameters are such that kaon condensation is present at zero temperature. In other words, upon decreasing the temperature, one encounters
the transition from unpaired quark matter to CFL-$K^0$, not from unpaired quark matter to CFL and then to CFL-$K^0$.
The critical temperature we have found is
larger than all temperatures we are interested in for compact star applications. Therefore, we have learned that the temperature inside a 
compact star is, for all times in the life of the star, sufficiently low for the CFL-$K^0$ phase.

\section{Color-superconducting gap from QCD}
\label{sec:QCDgap}

Let us now go through a true QCD calculation from first principles. Our goal is to compute the gap $\Delta$ 
with perturbative methods. As explained above, this calculation can be expected to be strictly valid only at densities much larger than present 
in compact stars. 

In the theoretical treatment of superconductivity one introduces charge-conjugate
fermions, which can be thought of as hole degrees of freedom. A hole is left in the Fermi sea if you remove a fermion. 
One might thus say that introducing fermion holes leads to an overcounting of the degrees of freedom because if the theory knows about all 
fermions it also knows about where a fermion is missing. And indeed, we have formally doubled the degrees of freedom. However, since
in a superconductor quasiparticles are mixtures of fermions and fermion holes, this is a necessary extension of the theory.
The fermion spinors become spinors in the so-called {\it Nambu-Gorkov space} and the fermion propagator becomes a $2\times 2$ matrix in this 
space. The Cooper 
pair condensate is taken into account in the off-diagonal elements of this propagator, i.e., it couples fermions with holes.
The inverse tree-level propagator in Nambu-Gorkov space is
\be
S_0^{-1} = \left(\begin{array}{cc} [G_0^+]^{-1} & 0 \\ 0 & [G_0^-]^{-1} \end{array}\right) \, ,
\ee
with the inverse tree-level fermion and charge-conjugate fermion propagators
\be \label{G0inv}
[G_0^\pm]^{-1} = \gamma^\mu K_\mu \pm \mu\gamma_0 = \sum_{e=\pm}[k_0\pm(\mu-ek)]\gamma_0\Lambda_k^{\pm e} \, , 
\ee
where 
\be
\Lambda_k^{\pm e} \equiv \frac{1}{2}\left(1\pm e\gamma_0\vg\cdot\uk\right)
\ee
are projectors onto positive and negative energy states. Since our QCD calculation applies to asymptotically large densities, we
can safely neglect all quark masses. See appendix \ref{app:fermions} for a derivation of the tree-level fermion propagator and its
representation in terms of energy projectors. From Eq.\ (\ref{G0inv}) we immediately get the tree-level propagators
\be \label{G0}
G_0^\pm = \sum_{e=\pm}\frac{\Lambda_k^{\pm e}\gamma_0}{k_0\pm(\mu-ek)} \, . 
\ee
The full inverse propagator $S^{-1}$ is obtained from a Dyson-Schwinger equation
\be \label{DS}
S^{-1} = S_0^{-1} + \Sigma \, ,
\ee
with the self-energy 
\be \label{SigmaPhi}
\Sigma \simeq  \left(\begin{array}{cc} 0 & \Phi^- \\ \Phi^+ & 0  \end{array}\right) \, .
\ee
In principle, $\Sigma$ also has nonvanishing diagonal elements which we neglect here. The off-diagonal elements contain the 
gap function $\Delta(K)$,
\be \label{Phipm}
\Phi^+(K) = \Delta(K) {\cal M}\gamma_5 \, , \qquad \Phi^-(K) = -\Delta(K) {\cal M}^\dag \gamma_5 \, , 
\ee
where ${\cal M}$ specifies the color-flavor structure of the color-superconducting phase; in the CFL phase ${\cal M}={\bf J}\cdot{\bf I}$, see
Eq.\ (\ref{9times9}).
From the Dyson-Schwinger equation (\ref{DS}) we obtain the inverse propagator, which we formally invert to obtain the propagator,
\be \label{SFG}
S = \left(\begin{array}{cc} G^+ & F^- \\ F^+  & G^-  \end{array}\right) \, ,
\ee
with
\begin{subequations}
\bea
G^\pm &=& \left([G_0^\pm]^{-1}-\Phi^\mp G_0^\mp \Phi^\pm\right)^{-1} \, ,  \label{normalGpm}\\
F^\pm &=& -G_0^\mp \Phi^\pm G^\pm \, . \label{anomalous}
\eea
\end{subequations}
The off-diagonal elements $F^\pm$ are termed {\it anomalous propagators}. They are typical for all superconductors, 
see for example Ref.\ \cite{fetter4}. From their structure 
(\ref{anomalous}) we see that they describe the propagation of a charge-conjugate fermion that is converted into a  fermion through 
the condensate (or vice versa). One can thus think of the condensate as a reservoir of fermions and holes, and the quasiparticles are
not just single fermions but superpositions of states with fermion number $\ldots,-5,-3,-1,1,3,5,\ldots$.

Inserting Eqs.\ (\ref{G0inv}), (\ref{G0}), and (\ref{Phipm}) into Eq.\ (\ref{normalGpm}),  we compute the diagonal elements of the propagator 
(for simplicity we
assume ${\cal M}^\dag = {\cal M}$ which is true in the CFL phase, but may not be true in other phases),
\bea
G^\pm = \left\{\sum_{e=\pm}\left[k_0\pm(\mu-ek)-\frac{\Delta^2 L}{k_0\mp(\mu-ek)}\right]\Lambda_k^{\mp e}\gamma_0\right\}^{-1} \, , 
\eea
with $L={\cal M}^2$, as defined for the CFL phase in Eq.\ (\ref{defL1}). Now we write $L$ in its spectral representation,
\be
L = \sum_{r=1,2} \lambda_r {\cal P}_r \, ,
\ee
with $\lambda_r$ being the eigenvalues of $L$, $\lambda_1=1$, $\lambda_2=4$, and ${\cal P}_r$ the projectors onto the corresponding eigenstates,
\be
{\cal P}_1 = -\frac{L-4}{3} \, , \qquad {\cal P}_2 = \frac{L-1}{3} \, .
\ee
Obviously, these projectors are complete, ${\cal P}_1+{\cal P}_2={\bf 1}$; they are also orthogonal, ${\cal P}_1{\cal P}_2=0$, as one can see
with the help of Eq.\ (\ref{L54}). We obtain 
\bea \label{Gpm}
G^\pm &=& \left\{\sum_{e,r}\left[k_0\pm(\mu-ek)-\frac{\lambda_r\Delta^2}{k_0\mp(\mu-ek)}\right]{\cal P}_r \Lambda_k^{\mp e}\gamma_0\right\}^{-1} \non
&=& \sum_{e,r}\left[k_0\pm(\mu-ek)-\frac{\lambda_r\Delta^2}{k_0\mp(\mu-ek)}\right]^{-1}{\cal P}_r \gamma_0\Lambda_k^{\mp e} \non
&=& [G_0^\mp]^{-1} \sum_{e,r}\frac{{\cal P}_r\Lambda_k^{\mp e}}{k_0^2 - (\epsilon_{k,r}^e)^2} \, , 
\eea
with 
\be
\epsilon_{k,r}^e = \sqrt{(ek-\mu)^2+\lambda_r \Delta^2} \, .
\ee
The poles of the propagator are $k_0=\pm \epsilon_{k,r}^e$, i.e., $\epsilon_{k,r}^e$ are the dispersion relations of the quasiparticles ($e=+$) and
quasiantiparticles ($e=-$). 
We have thus confirmed Eq.\ (\ref{exciteCFL}), in particular we now understand why the eigenvalues of $L$ appear in the excitation
energies. Note that the structure of the dispersion relations is thus determined entirely by the color-flavor (and Dirac) structure of the order
parameter, and thus ultimately by the symmetry breaking pattern. Only the calculation of the magnitude of $\Delta$ goes beyond simple symmetry 
considerations and depends on the form of the interaction between the fermions.

\begin{figure}[t]
\begin{center}
\includegraphics[width=0.75\textwidth]{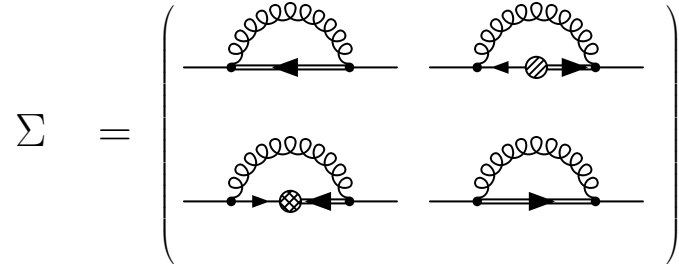}\\[2ex]
\caption{Diagrammatic representation of the one-loop self-energy in Nambu-Gorkov space. Curly lines are gluon propagators,
double lines correspond to $G^+$ (left-pointing arrow) and $G^-$ (right-pointing arrow), single lines to $G_0^+$ (left-pointing arrow) and 
$G_0^-$ (right-pointing arrow), and the circles are the gap matrices $\Phi^+$ (cross-hatched) and $\Phi^-$ (hatched). The vertices
have the form $g\gamma^\mu T_a$ with the QCD coupling $g$. 
}
\label{figsigmaNG}
\end{center}
\end{figure}

\begin{figure}[h]
\begin{center}
\includegraphics[width=0.5\textwidth]{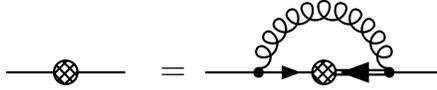}
\caption{Diagrammatic representation of the gap equation which arises as follows. On the one hand, the one-loop self-energy is given 
by cutting a fermion line in the corresponding two-loop diagram of the effective action. 
In Nambu-Gorkov space, this yields the matrix of four diagrams
shown in Fig.\ \ref{figsigmaNG}. On the other hand, the self-energy is given by Eq.\ (\ref{SigmaPhi}). Equating these
two matrices leads to the gap equation in the off-diagonal elements. The algebraic form of the gap 
equation is given in Eq.\ (\ref{gapgeneral}). It is a self-consistent equation for 
$\Phi^+$ (equivalently, one may solve the equation for $\Phi^-$), and thus for the gap function $\Delta(K)$.  
}
\label{figgapeq}
\end{center}
\end{figure}

Using the result (\ref{Gpm}) for $G^\pm$ and Eq.\ (\ref{anomalous}), one easily obtains the anomalous propagators, 
\be
F^\pm = \pm\Delta {\cal M}\gamma_5 \sum_{e,r} \frac{{\cal P}_r\Lambda_k^{\mp e}}{k_0^2 - (\epsilon_{k,r}^e)^2} \, .
\ee
The gap equation is a self-consistent equation for the off-diagonal elements of the self-energy $\Sigma$. 
We shall not discuss the detailed derivation of the gap equation (see Sec.\ IV.A in Ref.\ \cite{Alford:2007xm4} for this derivation). 
The gap equation reads 
\be \label{gapgeneral}
\Phi^+(K) = g^2\frac{T}{V}\sum_Q \gamma^\mu T_a^T F^+(Q) \gamma^\nu T_b D_{\mu\nu}^{ab}(K-Q) \, ,
\ee
where $g$ is the QCD coupling constant, which will be our expansion parameter, where $D_{\mu\nu}^{ab}$ is the gluon propagator, and
where $T_a=\lambda_a/2$ $(a=1,\ldots, 8$) with the Gell-Mann matrices $\lambda_a$.
In Figs.\ \ref{figsigmaNG} and \ref{figgapeq} we show the self-energy and the gap equation diagrammatically.


The first step is to transform the matrix equation (\ref{gapgeneral}) into an equation for the scalar gap function $\Delta(K)$.
To this end, we multiply both sides of the gap equation with $\gamma_5{\cal M}\Lambda_k^+$ from the right and  take the trace on both sides. 
Furthermore, we neglect the antiparticle contribution $e=-$ (and denote $\epsilon_{k,r}\equiv \epsilon_{k,r}^+$) and use the fact
that the gluon propagator can be taken to be diagonal in color space, $D_{\mu\nu}^{ab} = \delta^{ab}D_{\mu\nu}$. This yields
\bea
\Delta(K) &=& \frac{g^2}{24}\frac{T}{V}\sum_Q\sum_r\frac{\Delta(Q)}{q_0^2-\epsilon_{q,r}^2}
\Tr[\gamma^\mu\gamma_5\Lambda_q^-\gamma^\nu\gamma_5\Lambda_k^+]\,\Tr[T_a^T{\cal M}{\cal P}_rT_a{\cal M}]D_{\mu\nu}(P)  \non
&=& -\frac{g^2}{3}\frac{T}{V}\sum_Q\left[\frac{2}{3}\frac{\Delta(Q)}{q_0^2-\epsilon_{q,1}^2}+\frac{1}{3}\frac{\Delta(Q)}{q_0^2-\epsilon_{q,2}^2}
\right] \non[1.5ex]
&&\times \Tr[\gamma^\mu\gamma_5\Lambda_q^-\gamma^\nu\gamma_5\Lambda_k^+]\,D_{\mu\nu}(P) \, , 
\eea
where we abbreviated $P\equiv K-Q$, and where we have used the results for the color-flavor traces
\be
\Tr[T_a^T{\cal M}{\cal P}_1T_a{\cal M}] = 2\,\Tr[T_a^T{\cal M}{\cal P}_2T_a{\cal M}] = -\frac{16}{3} \, .
\ee
It is left as an exercise to verify these traces. With the gluon propagator in Coulomb gauge,
\be
D_{00}(P) = D_\ell(P) \, , \qquad D_{0i}(P)=0 \, , \qquad D_{ij}=(\delta_{ij}-\hat{p}_i\hat{p}_j)D_t(P) \, , 
\ee
where $D_\ell$ and $D_t$ are the longitudinal and transverse components, we have
\bea \label{twogapstr}
\Delta(K) &=& \frac{g^2}{3}\frac{T}{V}\sum_Q\left[\frac{2}{3}\frac{\Delta(Q)}{q_0^2-\epsilon_{q,1}^2}
+\frac{1}{3}\frac{\Delta(Q)}{q_0^2-\epsilon_{q,2}^2}\right]\non[1.5ex]
&& \times \left[(1+\uq\cdot\uk)D_\ell(P)-2(1-\up\cdot\uq\;\up\cdot\uk)D_t(P)\right] \, .
\eea
Again, it is left as an exercise to verify this result by performing the trace in Dirac space.
The two terms on the right-hand side arising from $\epsilon_{q,1}$ and $\epsilon_{q,2}$ are due to the two-gap structure
of CFL. Let us for simplicity ignore this structure in the following, i.e., we replace $\epsilon_{q,2}$ by $\epsilon_{q,1}$ (for more
details about the QCD gap equation for CFL, see Ref.\ \cite{Shovkovy:1999mr4}). This 
simplification does not change the main result which is the dependence of the gap on the QCD coupling $g$. The
two-gap structure has a nontrivial effect for instance on the relation between the critical temperature and the zero-temperature
gap, see Eq.\ (\ref{TcCFL}). In fact the $2^{1/3}$ in that equation is actually $(\lambda_1^{2/3}\lambda_2^{1/3})^{1/2}$ where the 
exponents 2/3 and 1/3 are the prefactors in front of the two fractions in Eq.\ (\ref{twogapstr}). 
 
For the sake of brevity, let us now skip a few steps in the calculation. One inserts the specific form of the longitudinal and transverse gluon 
propagators (in the so-called hard-dense loop approximation), performs the Matsubara sum and the angular integral. 
Details of all these steps can be found for instance in 
Ref.\ \cite{Pisarski:1999tv4}, and one obtains 
\be \label{Deltak}
\Delta_k \simeq \frac{g^2}{24\pi^2}\int_{\mu-\delta}^{\mu+\delta} dq\,\frac{\Delta_q}{\epsilon_q}\tanh\frac{\epsilon_q}{2T}
\left(\ln\frac{4\mu^2}{3m_g^2}+\ln\frac{4\mu^2}{M^2}+\frac{1}{3}\ln\frac{M^2}{|\epsilon_q^2-\epsilon_k^2|}\right) \, .
\ee
Here, the three terms in parentheses arise from static electric gluons, non-static magnetic gluons, and (Landau-damped) soft
magnetic gluons, respectively. The last of these terms is responsible for the leading behavior of the gap which will turn
out to be different from the usual BCS behavior in electronic superconductors. The reason is the existence of a long-range interaction
mediated by the magnetic gluons in QCD for which there is no analogue in the interaction of electrons in a metal. 
We have defined $m_g^2\equiv N_fg^2\mu^2/(6\pi^2)$ ($N_f$ being the number of flavors), and $M^2\equiv (3\pi/4)m_g^2$, and we have 
restricted the momentum integral to a small vicinity around the Fermi surface, $q\in [\mu-\delta,\mu+\delta]$ ($\delta\ll\mu$), 
where we expect the gap function $\Delta_q$ to be peaked. The three logarithms can be combined to obtain
\be 
\Delta_k = \bar{g}^2\int_0^\delta d(q-\mu) \frac{\Delta_q}{\epsilon_q}
\frac{1}{2}\ln\frac{b^2\mu^2}{|\epsilon_q^2-\epsilon_k^2|} \, ,
\ee
with 
\be
\bar{g}\equiv \frac{g}{3\sqrt{2}\pi} \, , \qquad b\equiv 256\pi^4\left(\frac{2}{N_f g^2}\right)^{5/2} \, ,
\ee
and where we have taken the zero-temperature limit $\tanh\frac{\epsilon_q}{2T}\to 1$. The logarithm can be approximated by
\be
\frac{1}{2}\ln\frac{b^2\mu^2}{|\epsilon_q^2-\epsilon_k^2|}\simeq \Theta(k-q)\ln\frac{b\mu}{\epsilon_k}+\Theta(q-k)\ln\frac{b\mu}{\epsilon_q} \, .
\ee
Moreover, we define the new integration variable
\be
y\equiv \bar{g}\ln\frac{2b\mu}{q-\mu+\epsilon_q} \, , 
\ee
and abbreviate
\be
x\equiv \bar{g}\ln\frac{2b\mu}{k-\mu+\epsilon_k} \, , \qquad x^*\equiv \bar{g}\ln\frac{2b\mu}{\Delta} \, , \qquad x_0\equiv 
\bar{g}\ln\frac{b\mu}{\delta} \, , 
\ee
where $\Delta$ is the zero-temperature value of the gap at the Fermi surface, $\Delta\equiv \Delta_{q=\mu}$. We have
\be
dy = -\frac{\bar{g}}{\epsilon_q}d(q-\mu) \, , \qquad \epsilon_q = b\mu e^{-y/\bar{g}}\left[1+\frac{\Delta_q^2}{(q-\mu+\epsilon_q)^2}\right] \, .
\ee
With the latter relation we approximate $\ln(b\mu/\epsilon_q)\simeq y/\bar{g}$, $\ln(b\mu/\epsilon_k)\simeq x/\bar{g}$ to obtain
\be \label{gapeqxy}
\Delta(x) = x\int_x^{x^*}dy\, \Delta(y) + \int_{x_0}^x dy\, y\,\Delta(y) \, .
\ee
We can rewrite this integral equation as a second-order differential equation,
\be
\frac{d\Delta}{dx} = \int_x^{x^*} dy \, \Delta(y) \quad \Rightarrow \qquad \frac{d^2\Delta}{dx^2} = -\Delta(x) \, .
\ee
This equation is solved by 
\be \label{solutiongap}
\Delta(x) = \Delta \cos(x^*-x) \, ,
\ee
such that the value of the gap at the Fermi surface (which corresponds to $x=x^*$) is $\Delta$, and such that the first derivative
of the gap at the Fermi surface vanishes, since the gap peaks at the Fermi surface. To compute the value of the gap at the Fermi 
surface, we insert the solution (\ref{solutiongap}) back into the gap equation (\ref{gapeqxy}) and consider the point $x=x^*$,
\bea
\Delta &=& \Delta\int_{x_0}^{x^*} dy\,y\,\cos(x^*-y) = \Delta\left[\cos(x^*-y)-y\sin(x^*-y)\right]_{y=x_0}^{y=x^*} \non 
&=& \Delta\left[1-\cos(x^*-x_0)+x_0\sin(x^*-x_0)\right] \, .
\eea
Since $x_0$ is of order $\bar{g}$, we approximate $\cos(x^*-x_0) = \cos x^*\cos x_0+\sin x^*\sin x_0 \simeq \cos x^* + x_0\sin x^*$, 
$\sin(x^*-x_0) = \sin x^*\cos x_0-\cos x^*\sin x_0\simeq \sin x^* -x_0\cos x^*$, and thus
\be
\Delta \simeq \Delta (1-\cos x^*) \, .
\ee
Hence, $\cos x^* \simeq 0$ and thus
\be \label{Delta1}
\Delta = 2b\mu \exp\left(-\frac{3\pi^2}{\sqrt{2}g}\right) \, .
\ee
This important result, first derived in Ref.\ \cite{Son:1998uk4}, shows that the color-superconducting gap is parametrically enhanced 
compared to the BCS gap in conventional superconductors. In BCS theory there is a contact interaction instead of gluon exchange, and the 
resulting gap equation has the form 
\be
\Delta \propto g^2\int_0^\delta d(q-\mu)\frac{\Delta}{\epsilon_q} \, .
\ee
Here the gap does not depend on momentum and one obtains $\Delta\propto \exp(-{\rm const}/g^2)$, i.e., the coupling appears quadratic in the 
denominator of the exponential. This is in contrast to the color-superconducting gap (\ref{Delta1}) where the coupling appears linear 
in the denominator of the exponential. As mentioned above, this is due to the long-range interaction from magnetic gluons.
For more details and a more general solution of the QCD gap equation see Sec.\ IV in Ref.\ \cite{Alford:2007xm4} and references therein.

\begin{figure}[t]
\begin{center}
\includegraphics[width=0.65\textwidth]{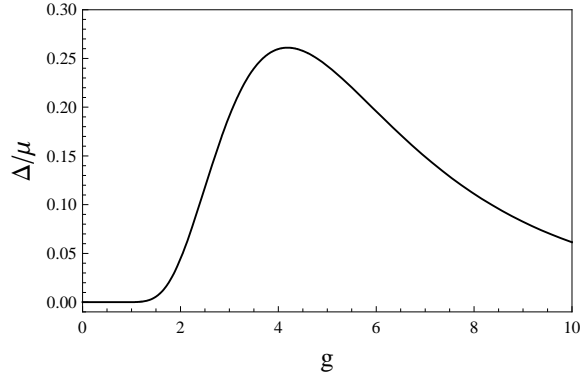}
\caption{Color-superconducting gap $\Delta$ over quark chemical potential $\mu$ as a function of the QCD coupling $g$. The 
curve shows the result from Eq.\ (\ref{Delta1}) with $N_f=2$, predicting a weak-coupling behavior $\Delta/\mu \propto \exp(-{\rm const}/g)$. The 
values of $\Delta/\mu$ for large coupling is a simple (and in principle unreliable) extrapolation of the weak-coupling result.
}
\label{figQCDgap}
\end{center}
\end{figure}

The solution of the QCD gap equation is a weak-coupling result and thus only valid at very large chemical potentials where the QCD coupling is 
sufficiently small. It is nevertheless interesting to extrapolate this result to larger couplings. Of course one
should keep in mind that this extrapolation has no theoretical justification. We show the gap as a function of the coupling 
in Fig.\ \ref{figQCDgap}. We see the exponentially small gap at small coupling and observe a maximum of the gap at a coupling of about 
$g\simeq 4.2$. For compact stars we make the following rough estimate. According to the two-loop $\beta$-function (which should 
not be taken too seriously at these low densities), the coupling at $\mu = 400\,{\rm MeV}$ is $g\simeq 3.5$. 
From Fig.\ \ref{figQCDgap} we then read off
$\Delta\simeq 80\,{\rm MeV}$. However in our derivation of the result we have ignored a subleading effect which yields an additional prefactor
$\simeq 0.2$. Therefore, we can estimate the color-superconducting gap for compact star densities to be of the order of 
$\Delta \sim 10\, {\rm MeV}$. 

This result suggests that the critical temperature of color superconductivity is also of the order of $T_c \sim 10\, {\rm MeV}$, 
cf.\ Eq.\ (\ref{TcCFL}). Remember that compact stars have temperatures well below that value (only in the very early stages of the 
life of the star, temperatures around $10\,{\rm MeV}$ are reached). This suggests that color superconductors are viable
candidates for the matter inside the star. More precisely, if there is deconfined quark matter inside the star, it is very likely that
it is in a color-superconducting state.

We conclude this chapter about color superconductivity by noticing that, besides the strong-coupling nature, other interesting questions 
arise at lower densities. We have seen in Sec.\ \ref{sec:quarkeos}, that in unpaired quark matter the Fermi momenta of 
up, down, and strange quarks split apart, see Fig.\ \ref{figuds}. This is due to the nonzero strange quark mass and the conditions of neutrality and 
weak equilibrium. In our discussion of superconductivity we have always assumed that the fermions that form Cooper pairs have identical 
Fermi momenta. This is true in the region of asymptotically large densities where the strange quark mass can be neglected. It is 
not true, however, at lower densities. The different Fermi momenta rather impose a ``stress'' on the pairing.\footnote{Cooper pairing
with mismatched Fermi momenta is an interesting general phenomenon and not only relevant for quark matter, but also in condensed matter 
physics and atomic physics. See for instance Ref.\ \cite{zwierlein4} where mismatched pairing of fermionic atoms is investigated experimentally
in an optical trap.} It is a quantitative
question whether the pairing gap is large enough to overcome this stress. Roughly speaking, if the gap is larger than the mismatch in 
Fermi momenta, the usual pairing is still possible. It is therefore conceivable that the CFL phase persists down to densities where 
the transition to hadronic matter takes place. If the gap is too small, however, or the mismatch too large, Cooper pairing in the 
conventional way is not possible anymore. There are several versions of unconventional pairing which may take over and constitute one or several
phases between the CFL phase at high densities and hadronic matter. Some of them break rotational invariance and may lead to nodes
of the gap in certain directions in momentum space as discussed in the context of the specific heat in Sec.\ \ref{sec:cV}. Others even break 
translational invariance and exhibit crystalline structures. All of the unconventional phases have in common that there is less, and less
symmetric, pairing than in the CFL phase. There is less pairing because the CFL phase is the only color superconductor where all quarks  are gapped
in all directions in momentum space. There is less symmetric pairing because the CFL phase is the color superconductor 
with the largest residual symmetry group. In the phase diagram of Fig.\ \ref{figQCDpd} all color superconductors other than CFL are 
collectively denoted by non-CFL. From what we just said it is clear that this region of the phase diagram may either be completely absent or, if
present, may itself contain several phase transition lines separating different color superconductors. More details about stressed pairing
in quark matter and unconventional color superconductors can be found in Ref.\ \cite{Alford:2007xm4}. 

In summary, we emphasize that not only the strong-coupling nature but 
also the less symmetric situation (due to the finite strange quark mass) complicates our understanding of quark matter in compact 
stars. This supports the theme of these lectures that we need to compute properties of candidate phases and check them for 
their compatibility with astrophysical observations. In the following section we shall turn to one of these properties, namely the 
neutrino emissivity.

\section*{Problems}
\addcontentsline{toc}{section}{Problems}

\begin{prob}
\label{prob6}
\textbf{Specific heat for anisotropic superfluid}\\
Compute the low-temperature behavior of the specific heat for 
a gap function with line nodes, i.e., instead of Eq.\ (\ref{pointnodes}), take $\Delta\to \Delta |\cos\theta|$ and apply analogous 
approximations as for the case of point nodes.
\end{prob}

\begin{prob}
\label{prob7}
\textbf{Symmetries of CFL}\\
Show that from the structure of the CFL order parameter given in Eq.\ (\ref{propi}) it follows that the CFL
symmetry breaking pattern is given by Eq.\ (\ref{propii}). Hints: it is sufficient to treat the chiral group $SU(3)_L\times
SU(3)_R$ as one single flavor group $SU(3)_f$. A color-flavor transformation $(U,V)\in SU(3)_c\times SU(3)_f$ with $U=\exp(i\phi_a^c T_a)$, 
$V=\exp(i\phi^f_a T_a)$ acts on the 
order parameter as $(U,V)({\bf J}\cdot {\bf I}) = (UJ^A U^T)\,(VI_A V^T)$. One then has to show that only $SU(3)_{c+f}$ transformations 
leave the order parameter invariant.
\end{prob}

\begin{prob}
\label{prob8}
\textbf{Kaon propagator}\\
Derive the inverse tree-level propagator for neutral and charged kaons 
given in Eqs.\ (\ref{treeprop}) and (\ref{S01S02}) from the Lagrangian (\ref{LAQ}).

\end{prob}

\chapter{Neutrino emissivity and cooling of the star} 
\label{sec:cooling}

We have seen in Sec.\ \ref{sec:massradius} that measuring mass and radius of a compact star is not sufficient to 
deduce the matter composition inside the star; it is neither conclusive for a distinction  
between nuclear matter and quark matter nor between unpaired quark matter and color-superconducting quark matter. We now turn to an observable
which is more sensitive to the microscopic properties of dense matter, namely the temperature of the star. More precisely, its cooling curve,
i.e., the temperature as a function of the age of the star. Approximately one minute after the star is born, the temperature has cooled
below 1 MeV and the star becomes transparent for neutrinos. Consequently, neutrinos (and antineutrinos) which are produced in the star can 
leave the system and carry away energy. Neutrino emission is thus the dominant cooling mechanism of a compact star in about the first million
years of its life. After that, photon emission takes over. We shall not be concerned with this late regime here. 

A very detailed review about neutrino emissivity in nuclear matter is Ref.\ \cite{Yakovlev:2000jp5}. If you are interested in a shorter review, also 
discussing quark matter, I recommend Ref.\ \cite{Page:2005fq5}. 
Before turning to the microscopic calculation of the neutrino emissivity $\epsilon_\nu$, 
let us discuss its importance for the cooling curves. First of all, 
as already discussed briefly in Sec.\ \ref{sec:cV} it is not only the emissivity which is important for the cooling. Once you know how much energy
per time and volume is carried away, you need to know how this affects the temperature of the star. Hence you also need to know
the specific heat. The specific heat $c_V$ is a thermodynamic quantity and thus much easier to compute than the neutrino emissivity. We have 
done so in Sec.\ \ref{sec:cV} and have seen that superconductivity has a huge effect on $c_V$, namely, due to the energy gap, $c_V$ is 
exponentially suppressed at sufficiently 
small temperatures. We shall see that superconductivity has a similar effect on the neutrino emissivity. Besides $\epsilon_\nu$ and $c_V$, also 
the heat conductivity is important for the cooling behavior. Most forms of dense matter are very good heat conductors, such that the star becomes 
isothermal. As a consequence, in a realistic star which may have layers of different phases of dense matter, cooling tends to be
dominated by the phase with the highest emissivity and the phase with the highest specific heat. 

\section{Urca processes in nuclear matter}

\begin{figure}[t]
\begin{center}
\includegraphics[width=0.85\textwidth]{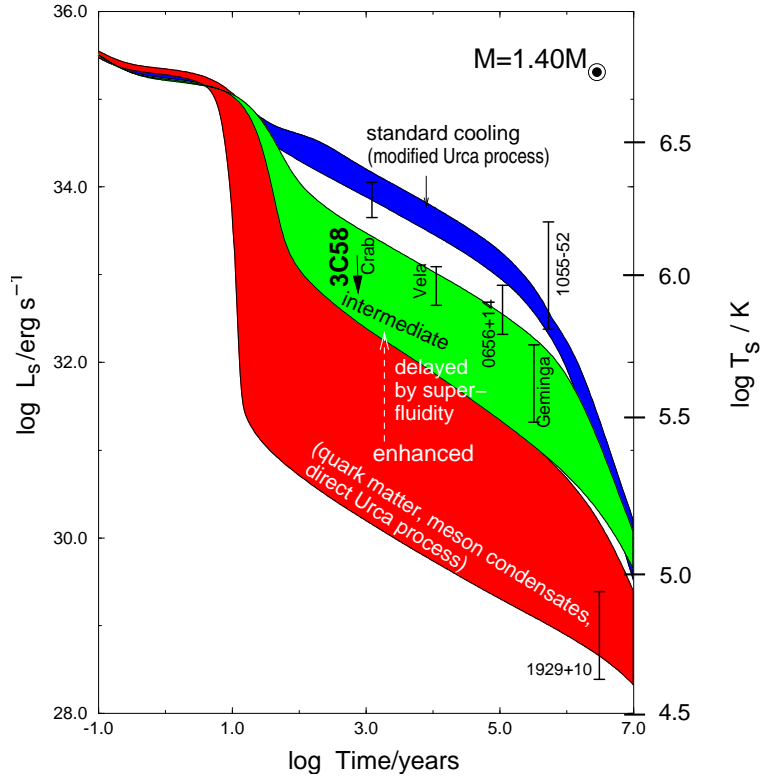}
\caption{Effective surface temperature $T_s$ and luminosity $L_s$ vs.\ age of compact stars, taken from Ref.\ \cite{Weber:2004kj5}. 
Observed values are compared with different cooling scenarios, each represented by a band that reflects the large uncertainties in 
the microscopic calculations.
}
\label{figcooling}
\end{center}
\end{figure}

In Fig.\ \ref{figcooling} we show some data and schematic comparison with calculations for the cooling curves. We see that there are different 
classes of processes which lead to significantly different cooling scenarios. The most efficient process is the so-called
{\it direct Urca process} which leads to a very fast cooling.\footnote{This process is as efficient in sucking energy out of the star as the 
{\it Casino de Urca} in Rio de Janeiro is in sucking money out of the pockets of the gamblers. Hence the name.} 
In nuclear matter, the direct Urca processes are
\be \label{urca}
n\to p+e+\bar{\nu}_e \,, \qquad p+e\to n+\nu_e \, .
\ee
We have discussed these processes in the context of $\beta$-equilibrium, where they serve to establish the relation $\mu_p+\mu_e=\mu_n$, 
assuming that neutrinos and antineutrinos escape from the star, $\mu_\nu=0$. Here we are interested in the question how both processes 
contribute to the neutrino emissivity. Since it 
does not matter for the energy balance whether neutrinos or antineutrinos are emitted, both processes contribute -- in chemical equilibrium --
equally to the emissivity. For the neutron, proton, and electron, 
the dominant contribution in momentum space to the processes comes from the momenta close to the Fermi momentum. The neutrino momentum is of the 
order of the temperature $T$ which can be neglected compared to the Fermi momenta. Therefore, momentum conservation for both processes in 
Eq.\ (\ref{urca}) reads
\be
{\bf k}_{F,n} = {\bf k}_{F,p} + {\bf k}_{F,e} \, .
\ee
In other words, the Fermi momenta ${\bf k}_{F,n}$, ${\bf k}_{F,p}$, and ${\bf k}_{F,e}$ must form a triangle. For this triangle to exist,
the triangle inequality has to be fulfilled,
\be
k_{F,n} <  k_{F,p} + k_{F,e} \, .
\ee
We know that in a neutral system we have $k_{F,p} = k_{F,e}$, and thus the triangle inequality becomes 
\be
k_{F,n} < 2k_{F,p} \, .
\ee
Consequently, with $n_i\propto k_{F,i}^3$ ($i=n,p$),
\be
n_n < 8 n_p \;\; \Rightarrow \;\; \frac{n_p}{n_B}>\frac{1}{9} \, ,
\ee
i.e., the proton fraction has to be larger than 11\%. We have seen in Sec.\ \ref{sec:freenuclear} that this is not the case for 
noninteracting nuclear matter. 
Interactions can change this, especially for very large densities. At lower densities, this means that the direct Urca process 
is strongly suppressed in nuclear matter. 

This brings us to a second class of processes which are less efficient than the direct Urca process, but may be the most efficient ones to emit
neutrinos when the direct Urca process is suppressed. Momentum conservation 
can be fulfilled by adding a spectator neutron or proton. This is the so-called {\it modified Urca process}, 
\bea
N+n\to N+p+e+\bar{\nu}_e \, , \qquad N+p+e\to N+n+\nu_e \, ,  \qquad N=n,p \, .
\eea
As can be seen from Fig.\ \ref{figcooling}, this process typically results in a much slower cooling. The cooling is thus very sensitive to the proton
fraction of nuclear matter, especially around the threshold of 11\%. In other words, this sensitivity provides a good check on the equation of state.
Phenomenological models with equations of state which predict the proton fraction to be above this threshold can be excluded since the star would
cool too fast. 
There are several other neutrino emissivity processes 
in nuclear matter which we shall not discuss here. Some of these processes happen only with superconducting protons and superfluid neutrons,
and are due to constant formation of Cooper pairs. 

\section{Direct Urca process in quark matter}
\label{sec:dUrca}

The direct Urca processes in quark matter are 
\begin{subequations}\label{urcaquark}
\bea 
d&\to& u+e+\bar{\nu}_e \,, \qquad u+e\to d+\nu_e \,, \label{urcaud}\\
s&\to& u+e+\bar{\nu}_e \,, \qquad u+e\to s+\nu_e \, . \label{urcaus}
\eea
\end{subequations}
These processes obviously require the availability of single quarks. If quarks are paired in Cooper pairs one first has to break a pair.
This costs energy. Therefore, in a phase where all quarks are paired (gapped), such as the CFL phase, we can expect the direct Urca process 
to be strongly suppressed. As for the specific heat, we expect an exponential suppression at temperatures small compared to the gap
(at larger temperatures, but still below the color-superconducting phase transition, thermal energy is available to break the pairs).
Recall that the gap is of the order of 10 MeV, and the temperature of the star is
well below that. Therefore, the exponential suppression $\exp(-\Delta/T)$ forbids any sizable effect of the Urca process. Other 
processes coming from Goldstone modes dominate the neutrino emissivity in the CFL phase \cite{Jaikumar:2002vg5}. 
However, their contribution is much lower than 
that of the unsuppressed direct Urca process. Therefore, if the core of a hybrid star is made of CFL quark matter, with outer layers of
nuclear matter where any kind of Urca process is possible, the cooling properties are utterly dominated by these outer layers. 

We have briefly discussed that at lower densities the CFL phase may not be the ground state anymore. Any other color-superconducting phase 
will have ungapped modes.\footnote{A possible exception is the {\it color-spin locked phase} which has Cooper pairs with total angular momentum
one and which we do not discuss here.} The simplest example is the so-called {\it 2SC phase} where all blue and strange quarks are ungapped while the
others are gapped. There are more complicated candidate phases with ungapped modes only in certain directions in momentum space.
In any case, the neutrino emissivity of these phases will be dominated by these ungapped modes, and thus will be comparable to 
the emissivity of unpaired quark matter. We are thus interested in the neutrino emissivity of unpaired quark matter. To be a bit more
ambitious, let us discuss the emissivity in the 2SC phase. From this calculation we will obtain the result for the unpaired 
phase ``for free'' because of the unpaired modes in the 2SC phase. Furthermore, we learn something about computing reaction rates in 
a superconductor which show some interesting features. And also we will see in an actual calculation why the emissivity of the 
gapped modes is exponentially suppressed. In other words, the goal of this section will be to understand
\begin{itemize}
\item the role of the Cooper pair condensate and the energy gap on the Urca process (we shall estimate this qualitatively)

\item the result of the emissivity of unpaired (ultrarelativistic) quark matter (we shall compute this quantitatively).
\end{itemize} 
All we shall need from the 2SC phase is the propagator. From Eq.\ (\ref{Gpm}) we know that the general form of the propagator can be written as  
\be \label{G2SC}
G^\pm = \gamma^0\Lambda_k^\mp\sum_r {\cal P}_r \frac{k_0\mp(\mu-k)}{k_0^2-\epsilon_{k,r}^2} \, ,
\ee
where we used Eq.\ (\ref{G0}) and where we dropped the antiparticle contribution. Note that this form of the propagator assumes that 
all flavor chemical potentials are the same. For the neutrino emissivity we need to drop this assumption. 
The order parameter in the 2SC phase is characterized by 
$\phi_A^B=\delta_{A3}\delta^{B3}$ where $\phi$ is the color-flavor matrix from Eq.\ (\ref{psipsi}). For simplicity, we drop the strange quarks and 
consider only a two-flavor system of up and down quarks.\footnote{The weak interaction between $u$ and $s$ quarks is suppressed compared
to the one between $u$ and $d$ quarks due to the Cabibbo angle. However, the finite strange quark mass may partially compensate this effect 
because it leads to a larger phase space for the Urca process. Here in these lectures we do not want to deal with these complications and thus
simply consider a system of massless up and down quarks, and thus only the processes (\ref{urcaud}).} Then, the color-flavor structure
of the gap matrix is
\be
{\cal M} = \tau_2 J_3 \, , 
\ee
with the second Pauli matrix $\tau_2$ in flavor space and $J_3$ in color space, as defined above Eq.\ (\ref{9times9}). The 
color-flavor structure of the 2SC phase is much easier to deal with than the one of the CFL phase 
because color and flavor matrices factorize. Since $\tau_2^2={\bf 1}$, we have
${\cal M}^2 = J_3^2$, whose eigenvalues are $\lambda_1=1$ (4-fold) and $\lambda_2=0$ (2-fold). This is the formal way of saying that in the 
2SC phase quarks of one color, say blue, remain ungapped. The projectors onto the corresponding eigenspaces in color-flavor space are
\be
{\cal P}_1 = J_3^2 \, , \qquad {\cal P}_2 = {\bf 1} - J_3^2 \, .
\ee
They are trivial in flavor space and project onto red and green quarks (which are gapped) and blue quarks (which are ungapped), respectively. 

In a neutral two-flavor system, up and down chemical potentials are different, namely 
$\mu_u+\mu_e=\mu_d$, where $\mu_e$ turns out to be nonzero due to the neutrality constraint. The generalization of the propagator (\ref{G2SC}) 
to this case can be written in terms of the flavor components (see problem \ref{prob9})
\begin{subequations}\label{GuGd}
\bea
G^\pm_u &=& \gamma^0\Lambda_k^\mp \sum_r \frac{k_0\mp(\mu_u-k)}{(k_0\mp\delta\mu)^2-\epsilon_{k,r}^2} {\cal P}_r \, , \\ 
G^\pm_d &=& \gamma^0\Lambda_k^\mp \sum_r \frac{k_0\mp(\mu_d-k)}{(k_0\pm\delta\mu)^2-\epsilon_{k,r}^2} {\cal P}_r \, , 
\eea
\end{subequations}
with 
\bea \label{GuGd1}
\epsilon_{k,r} \equiv \sqrt{(\bar{\mu}-k)^2+\lambda_r\Delta^2} \, , \qquad \delta\mu\equiv \frac{\mu_d-\mu_u}{2} \, , \qquad 
\bar{\mu}\equiv \frac{\mu_d+\mu_u}{2} \, ,
\eea
and $\lambda_r$, ${\cal P}_r$ as above (${\cal P}_r$ now being only matrices in color space since the flavor components are written 
separately). This structure of the propagator and the resulting quasiparticle dispersion relations are interesting on their own, since 
they describe Cooper 
pairing with a mismatch in Fermi momenta, as discussed at the end of Sec.\ \ref{sec:QCDgap}. However, in the present context of 
neutrino emissivity, we are only interested in the qualitative features of the gapped modes. Thus we shall ignore this complicated structure 
of the propagator and temporarily set $\mu_u=\mu_d$.
Only when we compute the emissivity from the unpaired modes we shall reinstate the difference in up and down chemical potentials.

\begin{figure}[t]
\begin{center}
\includegraphics[width=0.95\textwidth]{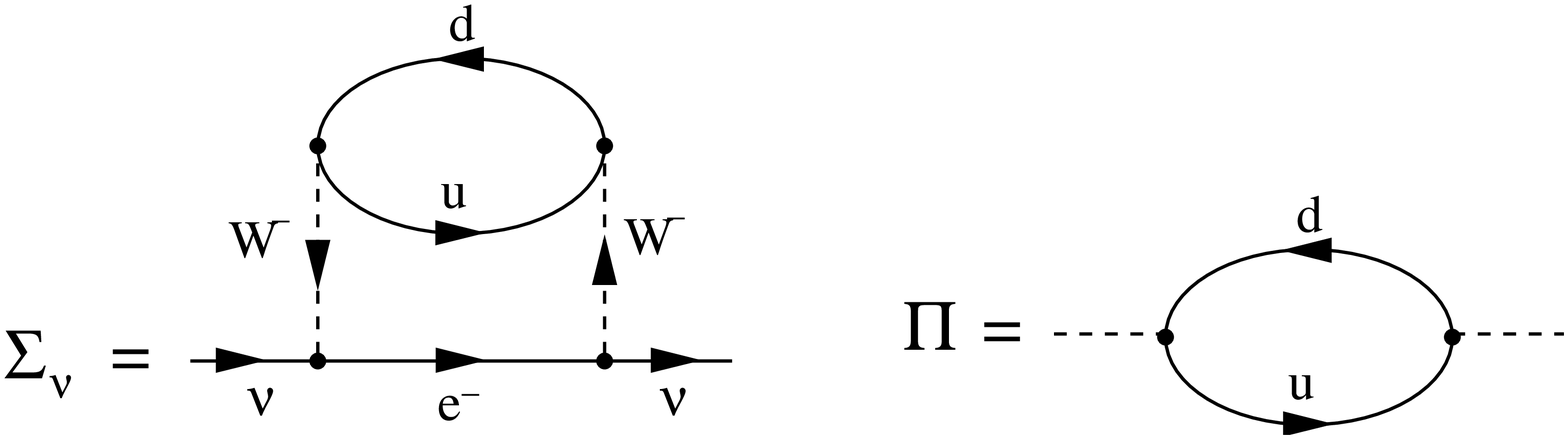}
\caption{Neutrino self-energy $\Sigma_\nu$ and $W$-boson polarization tensor $\Pi$ needed for the neutrino emissivity from the quark Urca
process. 
}
\label{figsigma}
\end{center}
\end{figure}

Next we need to set up the equation that determines the neutrino emissivity. One possible formalism is the finite temperature real-time formalism.
We shall not explain this formalism but refer the reader for more details to the textbooks \cite{kapusta5} and \cite{lebellac5}. For 
our purpose it is enough to know that the real-time formalism can be used for nonequilibrium calculations. Therefore it is 
well suited for transport properties and neutrino emissivity. Since these properties are always close-to-equilibrium properties, one often
simply uses an equilibrium formalism, such as the imaginary-time formalism, and adds whatever is needed as a small out-of-equilibrium feature
by hand. In the real-time formalism we can start from the kinetic equation
\be \label{kinetic}
i\frac{\partial}{\partial t}{\rm Tr}[\gamma_0 G_\nu^<(P_\nu)]=-\Tr[G_\nu^>(P_\nu)\Sigma_\nu^<(P_\nu)-\Sigma_\nu^>(P_\nu)G_\nu^<(P_\nu)] \, , 
\ee 
where $G_\nu^>$ and $G_\nu^<$ are the so-called ``greater'' and ``lesser'' neutrino propagators, and $P_\nu$ is the neutrino four-momentum. 
The greater and lesser propagators are obtained from the retarded propagator in the same way as given in Eqs.\ (\ref{Pigreatless}) 
for the case of the $W$-boson polarization tensor.  
The trace in Eq.\ (\ref{kinetic}) is taken over Dirac space. The two terms on the right-hand side correspond to the two directions of both 
processes (\ref{urcaud}), i.e., there is a neutrino gain term from $d\to u + e + \bar{\nu}_e$, $u+e\to d+\nu_e$, and a neutrino loss
term from $u + e + \bar{\nu}_e\to d$, $d+\nu_e\to u + e$. Since neutrinos, once created, simply leave the system, only the gain terms,
namely the directions given in Eq.\ (\ref{urcaud}), contribute. The neutrino self-energies
$\Sigma_\nu^{<,>}$ are given by the diagram in Fig.\ \ref{figsigma}. The present formalism amounts to cutting this diagram. 
The figure shows that a cut through the internal $u$, $d$, and $e$ lines produces two diagrams which represent the Urca process.
One part of the neutrino self-energies are the $W$-boson polarization tensors $\Pi^{<,>}$, as shown diagrammatically in Fig.\ \ref{figsigma}. 
They are defined through the 
imaginary part of the retarded polarization tensor ${\rm Im}\,\Pi_R$,
\begin{subequations} \label{Pigreatless}
\bea
\Pi^>(Q) &=& -2i[1+f_B(q_0)]{\rm Im}\,\Pi_R(Q) \, , \\
\Pi^<(Q) &=& -2if_B(q_0){\rm Im}\,\Pi_R(Q) \, , 
\eea
\end{subequations}
with the Bose distribution function $f_B(x) = 1/(e^{x/T}-1)$. 
We shall discuss the calculation of ${\rm Im}\,\Pi_R$ in detail below. The kinetic equation (\ref{kinetic})
becomes
\bea \label{df}
\frac{\partial}{\partial t} f_\nu(t,{\bf p}_\nu) &=& \frac{G_F^2}{8} \int\frac{d^3{\bf p}_e}{(2\pi)^3 p_\nu p_e} 
L_{\lambda\sigma}({\bf p}_e,{\bf p}_\nu) \non[1.5ex]
&& \times f_F(p_e-\mu_e)f_B(p_\nu+\mu_e-p_e){\rm Im}\,\Pi_R^{\lambda\sigma}(Q) \, , 
\eea
where $f_F(x) = 1/(e^{x/T}+1)$ is the Fermi distribution function, where, due to four-momentum conservation,
\be \label{Qcons}
Q = (p_e-p_\nu-\mu_e, {\bf p}_e-{\bf p}_\nu) \, , 
\ee
and where
\begin{equation} \label{defL}
L^{\lambda\sigma}({\bf p}_e,{\bf p}_\nu)\equiv\mbox{Tr}\left[(\g_0p_e-\vg\cdot{\bf p}_e)\, 
\gamma^\sigma (1-\gamma^5)(\g_0p_\nu-\vg\cdot{\bf p}_\nu)\, \gamma^\lambda
(1-\gamma^5)\right].
\end{equation}
(In this section, Lorentz indices are denoted $\lambda, \sigma,\ldots $ in order to avoid confusion with the subscript $\nu$ which indicates
neutrino quantities.)
If you are interested in the details of the derivation of Eq.\ (\ref{df}) or more details about the real-time formalism, 
see Ref.\ \cite{Schmitt:2005wg5} and references therein. In this reference the neutrino emissivity is computed in the same 
formalism; however, for anisotropic phases, which leads to more complicated calculations than we shall present here. The following is equally 
understandable if you simply start with Eq.\ (\ref{df}) whose features are physically plausible as we explain now.
 
The left-hand side of Eq.\ (\ref{df}) is the change of the neutrino occupation number in time. It is related to the emissivity by
\be \label{defeps}
\e_\nu \equiv 
 -2\frac{\partial}{\partial t}\int\frac{d^3{\bf p}_\nu}{(2\pi)^3}\,p_\nu\,
f_\nu (t, {\bf p}_\nu) \, ,
\ee
where the factor 2 accounts for the contribution from antineutrinos. The neutrino emissivity is thus the change in energy per unit time
and volume. 
Our task is to compute the right-hand side of Eq.\ (\ref{df}) and integrate over the neutrino momentum according to Eq.\ (\ref{defeps}) 
to obtain $\epsilon_\nu$. To understand the right-hand side of Eq.\ (\ref{df}) first note that the vertex $\Gamma^\mu$ for the processes 
$d\leftrightarrow u+W^-$ and $e\leftrightarrow \nu +W^-$ is given by 
\be 
\Gamma^\mu = -\frac{e}{2\sqrt{2}\sin\theta_W}\gamma^\mu(1-\gamma^5) \, , 
\ee
with the Weinberg angle $\theta_W$. (For the process $d\leftrightarrow u+W^-$ there is an additional factor $V_{ud}$ from the 
Cabibbo-Kobayashi-Maskawa (CKM) matrix; however, \mbox{$V_{ud}\simeq 1$.)} The $W$-boson propagators can be approximated by the inverse 
$W$-boson mass squared $M_W^2$ since
all momenta we are interested in are much smaller than this mass $M_W\simeq 80\,{\rm GeV}$. Thus, pulling out the constant factors of the 
vertices in the $W$-boson polarization tensor, we obtain the overall factor $G_F^2$ with the Fermi coupling constant
\be \label{FermiCoupling}
G_F = \frac{\sqrt{2}e^2}{8M_W^2\sin^2\theta_W} = 1.16637\cdot 10^{-11} \, {\rm MeV}^{-2} \, .
\ee
The additional factors in the trace of Eq.\ (\ref{defL}) come from the electron and neutrino propagators. And finally, the distribution functions
in Eq.\ (\ref{df}) belong to the electron and the $W$-boson. Eventually, the Bose distribution of the $W$ will drop out since
the $W$-boson polarization tensor will turn out to be $\propto f_B^{-1}$, see below. This makes sense because the 
$W$ does not appear in the initial or final state of the process we are interested in.

\subsection{$W$-boson polarization tensor}
 
Next we need to compute ${\rm Im}\,\Pi_R^{\lambda\sigma}$ for which we first compute
\be
\Pi^{\lambda\sigma}(Q) = \frac{T}{V}\sum_K\Tr[\Gamma_-^\lambda S(K)\Gamma_+^\sigma S(P)] \, , 
\ee
where the trace is taken over Dirac, color, flavor, and Nambu-Gorkov space. We have defined $P\equiv K+Q$; $K$ and $P$ will play the role of the 
$u$ and $d$ quark momentum, respectively. The weak vertices in Nambu-Gorkov space are
\be
\Gamma^\lambda_{\pm} = \left(\begin{array}{@{\extracolsep{2mm}}cc}
\gamma^\lambda(1-\gamma^5)\,\tau_{\pm} & 0 \\
0 & -\gamma^\lambda (1+\gamma^5)\,\tau_{\mp}
\end{array}\right) \, ,
\label{vert}
\ee
where $\tau_{\pm}\equiv(\tau_1\pm i\tau_2)/2$ are matrices in flavor 
space, constructed from the Pauli matrices $\tau_1$, $\tau_2$. They take care of the fact that a $u$ and a $d$ quark interact at the vertices.
Recall that, for notational convenience,
we have pulled out the constants of the weak vertices already and
absorbed them in the overall factor $G_F^2$. With the quark propagator $S$ from Eq.\ (\ref{SFG}), the trace over 
Nambu-Gorkov space yields
\bea \label{NGTr}
\Pi^{\lambda\sigma}(Q) &=& \frac{T}{V}\sum_K\Big\{\Tr\left[\gamma^\lambda(1-\gamma^5)\tau_- G^+(K)\gamma^\sigma(1-\gamma^5)\tau_+G^+(P)\right]\non
&&+ \Tr\left[\gamma^\lambda(1+\gamma^5)\tau_+ G^-(K)\gamma^\sigma(1+\gamma^5)\tau_-G^-(P)\right] \non
&&-\,\Tr\left[\gamma^\lambda(1-\gamma^5)\tau_- F^-(K)\gamma^\sigma(1+\gamma^5)\tau_-F^+(P)\right] \non
&&-\Tr\left[\gamma^\lambda(1+\gamma^5)\tau_+ F^+(K)\gamma^\sigma(1-\gamma^5)\tau_+F^-(P)\right]\Big\} \, . 
\eea
We see that there is a contribution from the anomalous propagators $F^\pm$. 
\begin{figure}[t]
\begin{center}
\includegraphics[width=0.75\textwidth]{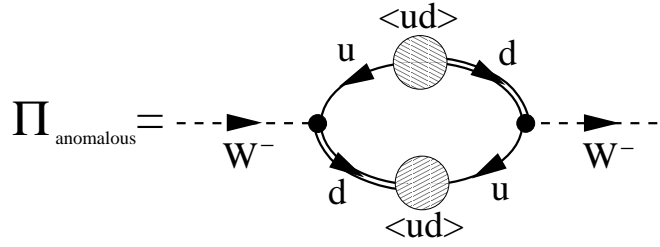}
\caption{Anomalous contribution to the $W$-boson polarization tensor $\Pi$. The loop consists of two anomalous fermion propagators, 
according to Eq.\ (\ref{anomalous}). The lower propagator consists of a full fermion propagator (double line), the condensate (hatched circle),
and a charge-conjugate free propagator (single line in opposite direction), and analogously for the upper one. Electric charge conservation at the 
weak vertices determines the flavor content of each line. As a consequence, one reads off that the condensate acts as a reservoir that 
can convert a $u$ quark into a $d$ quark hole and vice versa.   
}
\label{figanomalous}
\end{center}
\end{figure}
The corresponding diagram in Fig.\ \ref{figanomalous} is an instructive example for processes in a superconductor which 
are only possible due to the Cooper pair condensate, see explanation in the caption of the figure. The anomalous contribution is thus 
only present for the gapped modes. We shall
ignore it here for simplicity (it is smaller than the contribution from the normal propagators, but not negligibly small \cite{Jaikumar:2005hy5}). 
This leaves us with the first two traces in Eq.\ (\ref{NGTr}) which are the contribution 
of the normal propagators (they of course also contain the superconducting gap). It turns out that both traces are identical
which we use without explicit proof. We thus continue simply with twice the first term,
\bea
\Pi^{\lambda\sigma}(Q) &\simeq& 2\frac{T}{V}\sum_K\sum_{r,s}\Tr\left[\gamma^\lambda(1-\gamma^5)\tau_-\gamma_0{\cal P}_r\Lambda_k^-
\gamma^\sigma(1-\gamma^5)\tau_+ \gamma_0{\cal P}_s\Lambda_p^-\right] \non
&& \times \frac{k_0-(\mu-k)}{k_0^2-\epsilon_{k,r}^2}
\frac{p_0-(\mu-p)}{p_0^2-\epsilon_{p,s}^2} \, ,
\eea
where we have inserted the propagator (\ref{G2SC}). (Recall that we have set $\mu_u=\mu_d$ temporarily to avoid complications; 
this is sufficient to discuss the effects of superconductivity qualitatively, but eventually we shall reinstate the difference in $\mu_u$ and $\mu_d$
to compute the result for unpaired quark matter. In principle, for the 2SC phase we would have to use the propagators given in Eq.\ (\ref{GuGd})). 
The color-flavor traces are
\begin{subequations}
\bea 
\Tr[\tau_-{\cal P}_1\tau_+{\cal P}_1] &=& 2 \, , \label{Tr11}\\
\Tr[\tau_-{\cal P}_1\tau_+{\cal P}_2] &=& 0 \, , \\
\Tr[\tau_-{\cal P}_2\tau_+{\cal P}_1] &=& 0 \, , \\
\Tr[\tau_-{\cal P}_2\tau_+{\cal P}_2] &=& 1 \, . 
\eea
\end{subequations}
Recalling that ${\cal P}_1$ projects onto the gapped red and green quarks and ${\cal P}_2$ onto the ungapped blue quarks, this is easy to interpret:
the weak interaction cannot change colors. Therefore, the quark loop in the polarization tensor -- see right diagram in Fig.\ \ref{figsigma} --
contains an up quark and a down quark of the same color. They are either both gapped (then they are red or green, hence the result 2 in 
Eq.\ (\ref{Tr11})), or they are both ungapped (then they are blue). There is no term involving one gapped and one ungapped quark. 
We thus get two contributions, 
\bea \label{Pi12}
\Pi^{\lambda\sigma}(Q) &\simeq& 2\frac{T}{V}\sum_K{\cal T}^{\lambda\sigma}(\uk,\up)\non 
&&\hspace{-0.6cm}\times \left[2\frac{k_0-(\mu-k)}{k_0^2-\epsilon_{k,1}^2}
\frac{p_0-(\mu-p)}{p_0^2-\epsilon_{p,1}^2} + \frac{k_0-(\mu-k)}{k_0^2-\epsilon_{k,2}^2}
\frac{p_0-(\mu-p)}{p_0^2-\epsilon_{p,2}^2}\right] \, ,
\eea
where we abbreviated the Dirac trace
\be \label{defcalT}
{\cal T}^{\lambda\sigma}(\uk,\up)\equiv \Tr\left[\gamma^\lambda(1-\gamma^5)\gamma_0\Lambda_k^-
\gamma^\sigma(1-\gamma^5)\gamma_0\Lambda_p^-\right] \, .
\ee
We notice that the second contribution in Eq.\ (\ref{Pi12}) is obtained from the first upon setting $\Delta=0$. Thus, for notational 
convenience, let us simply continue with one color degree of freedom, say the first term without the factor 2, and
denote $\epsilon_k\equiv \epsilon_{k,1}$. In the end it is then straightforward to get the full result. 

Next one has to perform the sum over the fermionic Matsubara frequencies. 
This technique is discussed in detail for a simple example in appendix \ref{app:matsufermions}. Here we need the more complicated result 
from problem \ref{prob12},
\bea
&& T\sum_{k_0}\frac{k_0-(\mu-k)}{k_0^2-\epsilon_k^2}\frac{p_0-(\mu-p)}{p_0^2-\epsilon_p^2} \non
&& = -\frac{1}{4\epsilon_k\epsilon_p}\sum_{e_1,e_2}
\frac{[\epsilon_k +e_1(\mu-k)][\epsilon_p+e_2(\mu-p)]}{q_0-e_1\epsilon_k+e_2\epsilon_p}
\frac{f_F(-e_1\epsilon_k)f_F(e_2\epsilon_p)}{f_B(-e_1\epsilon_k+e_2\epsilon_p)} \, .
\eea
(Remember $P=Q+K$.)
We comment on the physical meaning of the sum over the signs $e_1,e_2=\pm$ below.
To obtain the retarded polarization tensor, we need to replace $q_0 \to q_0-i\eta$. Then, the imaginary part is obtained by using the identity
\be
\lim_{\eta\to 0^+} \frac{1}{x\pm i\eta} = {\cal P}\frac{1}{x}\mp i\pi\delta(x) \, ,
\ee
where ${\cal P}$ denotes the principal value.
This yields
\bea \label{ImPifinal}
{\rm Im}\,\Pi_R^{\lambda\sigma}(Q) &\simeq& -2\pi\sum_{e_1e_2}\int\frac{d^3{\bf k}}{(2\pi)^3}{\cal T}^{\lambda\sigma}(\uk,\up)B_k^{e_1}B_p^{e_2} \non
&& \times \frac{f_F(-e_1\epsilon_k)f_F(e_2\epsilon_p)}{f_B(-e_1\epsilon_k+e_2\epsilon_p)}\,\delta(q_0-e_1\epsilon_k+e_2\epsilon_p) \, ,
\eea
where we have defined the {\it Bogoliubov coefficients}
\be
B_k^e\equiv \frac{1}{2}\left(1+e\frac{\mu-k}{\epsilon_k}\right) \, . 
\ee
These coefficients appear in the theory of any kind of superconductor or superfluid, see for instance Ref.\ \cite{fetter5}.
Inserting the result (\ref{ImPifinal}) back into Eq.\ (\ref{df}) yields
\bea \label{df1}
\frac{\partial}{\partial t} f_\nu(t,{\bf p}_\nu) &=& -\frac{\pi G_F^2}{4} \sum_{e_1e_2}\int\frac{d^3{\bf p}_ed^3{\bf k}}{(2\pi)^3(2\pi)^3 p_\nu p_e} 
L_{\lambda\sigma}({\bf p}_e,{\bf p}_\nu){\cal T}^{\lambda\sigma}(\uk,\up) B_k^{e_1}B_p^{e_2} \non
&& \times \,f_F(p_e-\mu_e)f_F(-e_1\epsilon_k)f_F(e_2\epsilon_p)\delta(q_0-e_1\epsilon_k+e_2\epsilon_p) \, . 
\eea
As expected, the Bose distribution from Eq.\ (\ref{df}) cancels with the denominator from Eq.\ (\ref{ImPifinal}) since on the one hand 
$q_0=p_e-p_\nu-\mu_e$ according to Eq.\ (\ref{Qcons}) and on the other hand $q_0=e_1\epsilon_k-e_2\epsilon_p$ according to the $\delta$-function.

\subsection{Effect of superconductivity on Urca process}

Eq.\ (\ref{df1}) describes the change in the neutrino occupation number due to the process $u+e\to d+\nu_e$. 
(The other relevant process $d\to u + e +\bar{\nu}_e$ yields the 
same result and is taken into account by the factor 2 in Eq.\ (\ref{defeps}).) For this process one expects Fermi distributions of the form 
$f_ef_u(1-f_d)$, the factors $f_e$ and $f_u$ standing 
for the incoming fermions, and the factor $1-f_d$ standing for the outgoing fermion (for the neutrino, $f_\nu\simeq 0$). So what is the meaning of 
the sum over $e_1$, $e_2$? With $f(-x)=1-f(x)$ it seems that all combinations $f_ef_uf_d$, $f_ef_u(1-f_d)$, $f_e(1-f_u)f_d$,
and $f_e(1-f_u)(1-f_d)$ appear. In other words, also processes where both the up and down quark are created or annihilated apparently 
give a contribution. More precisely, the quasiparticles, which are mixtures of up and down particles and holes, are allowed to 
appear on either side of the reaction. This is an interesting property of a superconductor or superfluid where particle number conservation 
is spontaneously broken and particles can be created from or deposited into the condensate. 

To see explicitly that in the unpaired phase only one of the four subprocesses survives, 
let us define the new Bogoliubov coefficients and the new dispersion relations
\be \label{newbogol}
\tilde{B}_k^e \equiv \frac{1}{2}\left(1+e\frac{k-\mu}{\tilde{\epsilon}_k}\right) \, , \qquad \tilde{\epsilon}_k\equiv {\rm sgn}(k-\mu)\epsilon_k \, .
\ee
Then we use that for any function $F$ we have
\be
\sum_e\int_0^\infty dk \, B_k^e F(e\epsilon_k) = \sum_e\int_0^\infty dk \, \tilde{B}_k^e F(-e\tilde{\epsilon}_k) \, .
\ee
This reformulation is useful to understand the mixing of particles and holes, which is manifest in the Bogoliubov coefficients. 
Had we taken the limit $\Delta\to 0$ with the original formulation in $B_k^e$, $\epsilon_k$, we would have obtained the excitation
energy $\epsilon_k=|k-\mu|$ which describes a hole for $k<\mu$ and a particle for $k>\mu$. The more conventional excitation
$\epsilon_k=k-\mu$ which describes a particle for all $k$ is only obtained as a limit using $\tilde{B}_k^e$, $\tilde{\epsilon}_k$
(both formulations are of course physically equivalent).
Now, since in the unpaired phase $\tilde{B}_k^+=1$, $\tilde{B}_k^-=0$, we see that only the subprocess with $e_1=e_2=1$ survives 
in the unpaired phase. The other three subprocesses are only possible in the superconducting phase. 

The general result in the superconducting phase has to be computed numerically. Here we proceed with a discussion of the  
behavior at temperatures much smaller than the gap, $T\ll\Delta$. The neutrino emissivity is obtained by integrating Eq.\ (\ref{df1}) over
the neutrino momentum according to Eq.\ (\ref{defeps}). For the purpose of a simple estimate we may consider the expression
\bea \label{estimate1}
\epsilon_\nu &\sim&  \sum_{e_1,e_2=\pm}
\int_{v,x,y} \, 
\left(e^{v+e_1\sqrt{y^2+\varphi^2}-e_2\sqrt{x^2+\varphi^2}} + 1\right)^{-1} \non
&&\times \left(e^{-e_1\sqrt{y^2+\varphi^2}} + 1\right)^{-1}\,
\left(e^{e_2\sqrt{x^2+\varphi^2}} + 1\right)^{-1} \, ,
\eea
where we have abbreviated
\be
\varphi\equiv \frac{\Delta}{T} \, , \qquad \int_{v,x,y} \equiv \int_0^\infty dv \, v^3 \, \int_0^\infty dx \,\int_0^\infty dy \, ,
\ee
and introduced the new dimensionless variables
\be \label{xyv}
x=\frac{p-\mu_d}{T} \, , \qquad y=\frac{k-\mu_u}{T} \, , \qquad v=\frac{p_\nu}{T} \, . 
\ee
The integration over the electron momentum has been rewritten as an integration over the $d$-quark momentum. 
We shall discuss the calculation more explicitly for the case of unpaired quark matter below. Especially the angular integral, i.e., 
the phase space for the process, needs to be considered in detail. For now we are only interested in the suppression due to the gap.
In the integrand of Eq.\ (\ref{estimate1}) one recovers the distribution functions for the electron, the $u$-quark, and the $d$-quark. 
We may now perform the sum over $e_1$ and $e_2$ and approximate $e^{\sqrt{x^2+\varphi^2}}\gg 1$ and 
$e^{\sqrt{y^2+\varphi^2}}\gg 1$, since $\varphi\to\infty$ for small temperatures. 
Then the four terms, in the order $(e_1,e_2)= (+,+), (-,-), (-,+), (+,-)$, become 
\bea
\label{int-sum}
\epsilon_\nu &\sim& 
\int_{v,x,y} \, 
\Bigg(
\frac{1}{e^{\sqrt{x^2+\varphi^2}}+e^{v+\sqrt{y^2+\varphi^2}}}
+\frac{1}{e^{v+\sqrt{x^2+\varphi^2}}+e^{\sqrt{y^2+\varphi^2}}}\non
&& +\frac{1}{e^{v}+e^{\sqrt{x^2+\varphi^2}+\sqrt{y^2+\varphi^2}}}
+\frac{1}{e^{v+\sqrt{x^2+\varphi^2}+\sqrt{y^2+\varphi^2}}}\Bigg)\, .
\eea
The terms where $e_1$, $e_2$ assume different signs, i.e., the third and fourth term, yield contributions of the order of 
$e^{-2\varphi}$. They are thus even stronger suppressed than the first two terms which are identical and yield
contributions proportional to $e^{-\varphi}$,
\bea
\int_{v,x,y}\, \frac{1}{e^{\sqrt{x^2+\varphi^2}}+e^{v+\sqrt{y^2+\varphi^2}}} &\simeq& \int_{v,x,y} \,
\frac{e^{-\varphi}}{e^{x^2/(2\varphi)}+e^{v+y^2/(2\varphi)}}\non
&=& 2\varphi e^{-\varphi}
\int_{v,x,y} \, \frac{1}{e^{x^2}+e^{v+y^2}} \simeq 21.27\,\varphi e^{-\varphi}\, .
\label{I3}
\eea
In the last step we have performed the remaining integral numerically which yields a numerical factor, unimportant for our present purpose.
The main result is the expected exponential suppression of 
the neutrino emissivity for case of gapped $u$ and $d$ quarks, $\epsilon_\nu\propto e^{-\Delta/T}$ for $T\ll \Delta$. The full
numerical solution, also taking into account the temperature dependence of the gap $\Delta$, shows that this approximation is valid up to 
temperatures of about $T\lesssim T_c/3$ where $T_c$ is the critical temperature of superconductivity.

\subsection{Result for unpaired quark matter}

\begin{figure}[t]
\begin{center}
\includegraphics[width=0.75\textwidth]{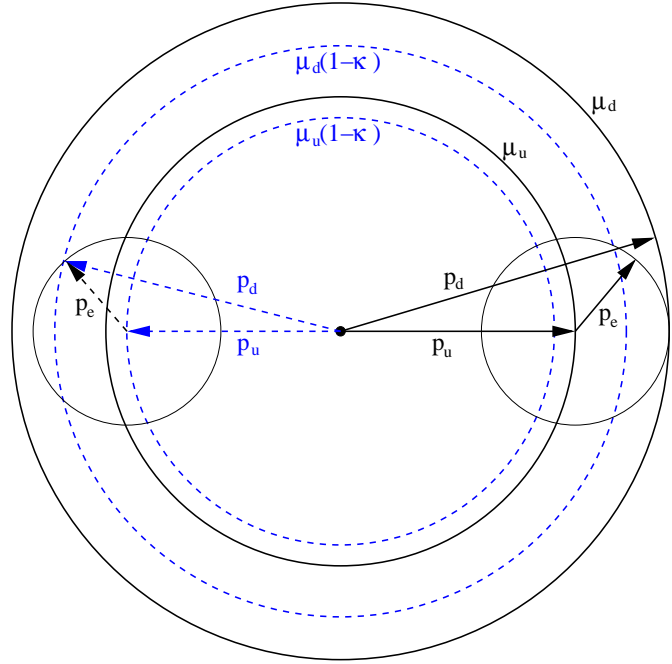}
\caption{Illustration of how Fermi liquid effects from the strong interaction open up the phase space for the direct Urca
process in unpaired quark matter. Right-hand side (solid Fermi momenta): 
without Fermi liquid corrections, the Fermi momenta of the ultrarelativistic quarks are
given by $p_{F,u}=\mu_u$, $p_{F,d}=\mu_d$. Start with the momentum of the up-quark, ${\bf p}_u$. The circle with center at its tip 
indicates possible endpoints of the electron momentum ${\bf p}_e$. Since $p_{F,e}=\mu_e$ and $\mu_u+\mu_e=\mu_d$ ($\beta$-equilibrium), one
cannot form a triangle with ${\bf p}_u$, ${\bf p}_e$ and the down-quark momentum ${\bf p}_d$, unless one chooses the three vectors to be collinear. 
In this case, the triangle collapses to a line and the phase space for the Urca process vanishes. Note that the neutrino momentum $p_\nu\sim T$
is negligibly small on the scale of the figure. Left-hand side (dashed Fermi momenta): the strong interaction changes the quark Fermi momenta to 
$p_{F,u}\simeq\mu_u(1-\kappa)$, $p_{F,d}\simeq\mu_d(1-\kappa)$ with $\kappa=2\alpha_s/(3\pi)$. In other words, both Fermi momenta are reduced, but 
the down-quark Fermi momentum is reduced by a larger absolute amount. Since the electron Fermi momentum is not changed, a finite region in phase
space opens up. The resulting triangle has a fixed angle between up- and down-quark momenta given by the values of the chemical potentials and 
$\kappa$.  
}
\label{figphasespace}
\end{center}
\end{figure}

With the help of the new Bogoliubov coefficients (\ref{newbogol}) we can easily take the limit of unpaired quarks. 
For an explicit calculation of the emissivity for this case we need the following ingredients. First we need to perform the 
remaining traces in Dirac space and do the contraction over Lorentz indices. This is done in problem \ref{prob10} with the result
\be
L_{\lambda\sigma}({\bf p}_e,{\bf p}_\nu){\cal T}^{\lambda\sigma}(\uk,\up) = 64(p_e-{\bf p}_e\cdot\uk)(p_\nu-{\bf p}_\nu\cdot\up) \, .
\ee
Next we observe that the result for the right-hand side of Eq.\ (\ref{df1}) would be zero without further corrections: we have to take into
account so-called Fermi liquid corrections which are induced by the strong interaction. We have mentioned these corrections briefly in 
Sec.\ \ref{sec:MRinter}, see Eq.\ (\ref{Fermiliquid}). To lowest order in the strong coupling constant $\alpha_s$ -- 
which is related to the coupling $g$ from Sec.\ \ref{sec:QCDgap} by $\alpha_s=g^2/(4\pi)$ -- we have
\be
p_{F,u/d} = \mu_{u/d}(1-\kappa) \, , \qquad \kappa\equiv \frac{2\alpha_s}{3\pi} \, .
\ee
We illustrate in Fig.\ \ref{figphasespace} how these corrections open up the phase space for the direct Urca process. As a consequence, there
is a fixed angle $\theta_{ud}$ between the $u$ and $d$ quarks, and the $\delta$-function in Eq.\ (\ref{df1}) can be approximated by
\be
\delta(p_e-p_\nu+k-p) \simeq \frac{\mu_e}{\mu_u\mu_d}\delta(\cos\theta_{ud}-\cos\theta_0) \, , \qquad 
\cos\theta_0 \equiv 1-\kappa\frac{\mu_e^2}{\mu_u\mu_d}
\, .
\ee
(We have reinstated the different chemical potentials $\mu_u$, $\mu_d$.) 
We denote the angle between the neutrino and the $d$ quark by $\theta_{\nu d}$ and approximate the factor 
\be
(p_e-{\bf p}_e\cdot\uk)(p_\nu-{\bf p}_\nu\cdot\up)\simeq 2\mu_ep_\nu\kappa(1-\cos\theta_{\nu d}) \, . 
\ee
This factor vanishes for the case of collinear scattering. 
The $\alpha_s$ effect renders it nonzero, hence this factor and in consequence the total neutrino emissivity is proportional to 
$\alpha_s$.  Putting all this together and 
changing the integration variable from ${\bf p}_e$ to the $d$ quark momentum ${\bf p}$ yields 
\bea \label{df2}
\frac{\partial}{\partial t} f_\nu(t,{\bf p}_\nu) &=& -64 G_F^2\alpha_s \mu_e\mu_d\mu_u\int\frac{dp\,d\Omega_p}{(2\pi)^3}
\int\frac{dk\,d\Omega_k}{(2\pi)^3}(1-\cos\theta_{\nu d}) \non 
&& \hspace{-1cm}\times \, \delta(\cos\theta_{ud}-\cos\theta_0)\,f_F(p_e-\mu_e)f_F(k-\mu_u)[1-f_F(p-\mu_d)] \, .
\eea
Since we have taken only one color degree of freedom from Eq.\ (\ref{Pi12}), we have reinstated a factor $N_c=3$.
Next we introduce the dimensionless variables $x$, $y$, $v$ from Eq.\ (\ref{xyv}), and with the definition (\ref{defeps}) of the total 
neutrino emissivity we obtain
\bea
\epsilon_\nu &=& 128\alpha_sG_F^2\mu_e\mu_u\mu_d T^6\int\frac{d\Omega_{p_\nu}}{(2\pi)^3}\int\frac{d\Omega_p}{(2\pi)^3}
\int\frac{d\Omega_k}{(2\pi)^3}\non[1.5ex]
&&\times (1-\cos\theta_{\nu d}) \delta(\cos\theta_{ud}-\cos\theta_0) \non[1.5ex]
&&\times \int_0^\infty dv\,v^3\int_{-\infty}^{\infty}dx\int_{-\infty}^{\infty}dy\,f_F(v+x-y)f_F(y)[1-f_F(x)] \, .
\eea
Here we have approximated the lower boundaries by $-\mu_{u/d}/T\simeq\infty$. With the integral
\be
\int_0^\infty dv\,v^3\int_{-\infty}^{\infty}dx\int_{-\infty}^{\infty}dy\,f_F(v+x-y)f_F(y)[1-f_F(x)] = \frac{457}{5040}\pi^6 \, , 
\ee
and the (trivial) angular integral
\be
\int\frac{d\Omega_{p_\nu}}{(2\pi)^3}\int\frac{d\Omega_p}{(2\pi)^3}
\int\frac{d\Omega_k}{(2\pi)^3}(1-\cos\theta_{\nu d})\delta(\cos\theta_{ud}-\cos\theta_0) = \frac{1}{16\pi^6} \, , 
\ee
we obtain the final result
\be \label{emitfinal}
\epsilon_\nu \simeq \frac{457}{630}\alpha_sG_F^2\mu_e\mu_u\mu_d T^6 \, .
\ee
This result has first been computed by Iwamoto in 1980 \cite{Iwamoto:1980eb5}. 

\section{Cooling with quark direct Urca process}

\begin{figure}[t]
\begin{center}
\includegraphics[width=0.65\textwidth]{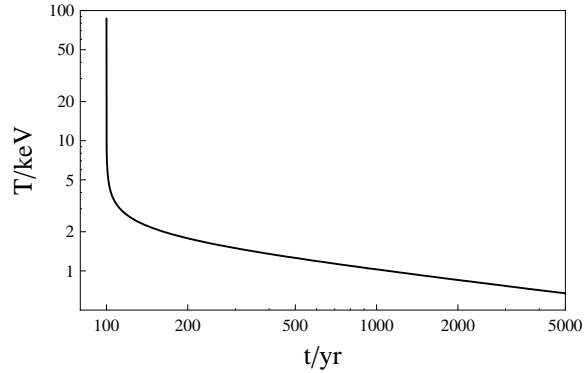}
\caption{Cooling curve from the direct Urca process in two-flavor, unpaired, ultrarelativistic quark matter, see Eq.\ (\ref{Tt}).
}
\label{figdUrca}
\end{center}
\end{figure}

From the result for the neutrino emissivity we can now get a simple cooling curve for unpaired quark matter. 
Of course we shall ignore a lot of details 
of realistic stars. The result will simply show how a chunk of unpaired two-flavor 
quark matter cools via the direct Urca process.
Nevertheless, the result is very illustrative and shows that the direct Urca process is indeed an efficient cooling mechanism.
We use Eq.\ (\ref{tt0}), which relates the temperature as a function of time to the emissivity and the specific heat. For the emissivity we
use the result (\ref{emitfinal}). For the specific heat, recall the result (\ref{cV0}) which is valid for two fermionic degrees of freedom, 
taking into account spin; we thus have to multiply this result by the number of colors and add up the contributions of $u$ and $d$ quarks,
\be
c_V = (\mu_u^2 +\mu_d^2) T \, .
\ee
Then, performing the integration in Eq.\ (\ref{tt0}) yields
\be \label{Tt}
T(t) = \frac{T_0\tau^{1/4}}{(t-t_0+\tau)^{1/4}} \, , 
\ee
where we have defined
\be
\tau = \frac{315}{914}\frac{\mu_u^2+\mu_d^2}{\alpha_s G_F^2\mu_e\mu_u\mu_d}\frac{1}{T_0^4} \, .
\ee
To get an estimate for this characteristic time scale, we assume $\mu_d=500\, {\rm MeV}$, $\mu_u = 400\,{\rm MeV}$, $\mu_e=100\,{\rm MeV}$,
$\alpha_s=1$, an initial temperature of $T_0=100\,{\rm keV}$ at an initial time $t_0=100\,{\rm yr}$, 
and use the value of the Fermi coupling (\ref{FermiCoupling}) to obtain
\be
\tau\simeq 10^{-5}\,{\rm yr} \simeq 5\,{\rm min} \, .
\ee
This is a very short time compared to the astrophysical time scales we are interested in. 
The function $T(t)$ is plotted in Fig.\ \ref{figdUrca}. We see the rapid drop
in temperature on a time scale of minutes down to a few keV. We thus recover the shape of the direct Urca cooling from Fig.\ \ref{figcooling}.
For late times $t\gg t_0$, we have $T(t)\propto t^{-1/4}$.

\section*{Problems}
\addcontentsline{toc}{section}{Problems}

\begin{prob}
\label{prob9}
\textbf{2SC propagator}\\
Show that for the case of different flavor chemical potentials the fermion propagator of the 2SC phase is given by 
Eqs.\ (\ref{GuGd}) and (\ref{GuGd1}).
\end{prob}

\begin{prob}
\label{prob10}
\textbf{Trace over Dirac space}\\
Show that 
\be
L_{\lambda\sigma}({\bf p}_e,{\bf p}_\nu){\cal T}^{\lambda\sigma}(\uk,\up) = 64(p_e-{\bf p}_e\cdot\uk)(p_\nu-{\bf p}_\nu\cdot\up) \, ,
\ee
with $L_{\lambda\sigma}({\bf p}_e,{\bf p}_\nu)$ and ${\cal T}^{\lambda\sigma}(\uk,\up)$ defined in Eqs.\ (\ref{defL}) and (\ref{defcalT}), 
respectively.
\end{prob}

\chapter{Discussion}
\label{sec:discussion}

Let us summarize what we have learned about compact stars and dense matter, having in mind
the two questions we have formulated in the preface. In addition, let us also list a few things which would in principle 
have fitted into these lectures
topic-wise. We haven't discussed them in the main part because either I found them not suitable for a concise, and yet pedagogical,
introduction or because they are simply beyond the scope of these lectures, such as some of the theoretical approaches to dense matter listed
at the end of Sec.\ \ref{sec:misc}. And, well, some selection has to be made, so for some of the following points there is no good reason 
why they appear here and not in the main part. The volume of the main part is chosen such that it should conveniently fit into a 
one-semester course, maybe dropping one 
or two of the more specialized subsections. In Sec.\ \ref{sec:misc} I will give some selected references where interested readers can find more 
information about the questions we haven't addressed in the main part. 

\section{What we have discussed}

\begin{itemize}

\item{\it Astrophysical observables and their relation to microscopic physics.} 
The first thing you should have learned in these lectures is in which sense  
compact stars are laboratories for the understanding of dense matter. 
The experiments we can do in this laboratory are less controlled as for example tabletop experiments in condensed matter physics. This means
we cannot always measure the quantities we would like to know, or at least not to an accuracy we would need for our purposes.
And it means that it is often impossible to switch on or off certain unwanted effects at will, which 
would be desirable to extract an exact value for a given quantity. For instance we would ideally like to have a precise look into the interior
of a compact star, but these kind of observations will always be indirect at best since the information we get from the interior
is filtered through the surface and the atmosphere of the star. However, in spite of these restrictions (which, to some extent, have been and will be
overcome through improvements in observational technology), we have seen that our observational data of compact stars can be closely 
linked to the properties of dense matter. Examples we have discussed in detail are the mass-radius relation which is related to the equation of state
and the cooling curve which is related to the neutrino emissivity and the specific heat.

\item{\it Theoretical approaches to dense matter.} 
We have emphasized at several points in these lectures that the density regime which 
is of interest for the physics of compact stars is very difficult to tackle. After all, this difficulty led us to consider 
compact stars not only as an application of QCD but also as an important means to understand QCD. The main reason for the 
theoretical difficulty is the strong-coupling nature of QCD. We have discussed attempts to approach the relevant 
density regime from two sides, coming from lower and higher density. 

First, we have discussed nuclear matter, 
for which we have solid knowledge at low densities. This knowledge
is strongly built upon experimental data. In principle, even a single nucleon is theoretically a very complicated object if considered from 
first principles. First-principle calculations, at least for sufficiently simple properties of nucleons, are possible in 
computer simulations, but effective theories remain an important tool to describe nuclear matter, and they work well (by construction) at 
sufficiently low densities. One of the basic examples we have discussed is the Walecka model. However, finding the correct description of 
nuclear matter at high density is a challenge, and 
astrophysical data can be used to rule out or confirm certain models. 

Second, we have discussed QCD from first principles in the context of deconfined quark matter. This approach is
rigorous at asymptotically high density and therefore is interesting on its own right. We have discussed that it predicts the CFL state.
Whether CFL persists down to densities relevant for compact stars is unknown. We have discussed that, to get a rough idea about 
the low-density region, one may simply extrapolate the rigorous results. But this of course stretches the results beyond their range
of validity. We have also introduced a more powerful approach to deduce intermediate-density properties from the high-density calculations. 
This approach relies on the symmetries of the CFL state. Building on these symmetries, one can construct an effective theory which 
can give us at least qualitative insight into the properties of CFL at lower densities, although this approach cannot tell us whether CFL is indeed
the ground state of matter at densities present in the core of a compact star. 

\end{itemize}

\section{What we could have, but haven't, discussed}
\label{sec:misc}

\begin{itemize}

\item{\it $r$-modes -- bulk/shear viscosity.}
We have said little about the rotation of a compact star except for stating that it can rotate very fast, up to almost a thousand times per 
second. For the purpose of our lecture, however, the rotation frequency is a very interesting observable because it is 
sensitive to the microscopic physics. One of the reasons is as follows. 

Certain non-radial oscillatory modes of a rotating star, 
in particular the so-called 
$r$-modes,\footnote{Oscillatory modes of compact stars are classified according to their restoring force. In the case of $r$-modes, this is the 
Coriolis force.}
are generically unstable with respect to gravitational radiation. The reason can be understood in a rather simple argument. Consider the situation 
where the star rotates {\it counterclockwise}, seen from the polar view, and where an observer in the co-rotating frame sees non-radial 
oscillations which 
propagate {\it clockwise}. These modes lower the total angular momentum of the star, i.e., if the star's angular momentum is positive, the 
oscillations have negative angular momentum. Now assume that these propagating modes are seen from a distant observer to move counterclockwise, 
i.e., they are ``dragged'' by the star's rotation or, in other words, their angular velocity in the co-rotating frame is smaller in magnitude
than the angular velocity of the star, seen from a distant observer. The pulsations now couple to gravitational radiation. The emitted radiation
has positive angular momentum since a distant observer sees the pulsations move counterclockwise. Consequently, the total angular momentum of
the star must be lowered. This, however, means that the angular momentum of the oscillations, which is already negative, is {\it increased} in 
magnitude (becomes more negative). Therefore, the emission of gravitational radiation tends to increase the amplitude of the pulsation which in turn 
leads to a stronger gravitational radiation etc. This is the $r$-mode instability. Note that the rotation of the star is crucial for this
argument. In a non-rotating star, the effect of gravitational radiation is dissipative, i.e., the non-radial oscillations would be damped.
For a nice pedagogical introduction into this general relativistic effect see Ref.\ \cite{Lindblom:2000jw6}. 

\begin{figure}[t]
\begin{center}
\includegraphics[width=0.65\textwidth]{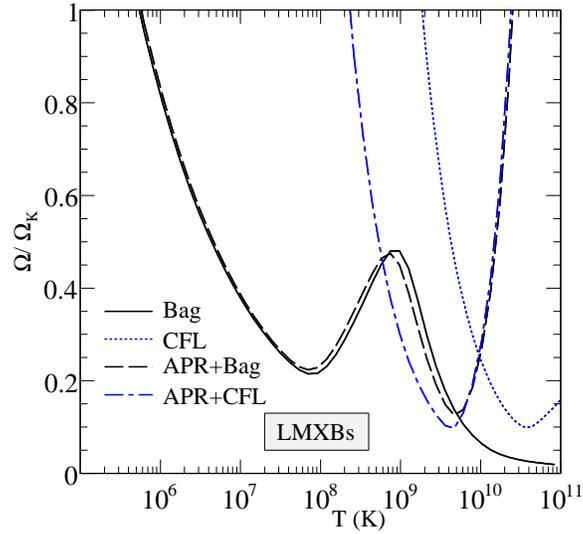}
\caption{Critical rotation frequency (normalized to the {\it Kepler frequency} $\Omega_K$, the upper limit for the rotation frequency beyond which
the star would start shedding mass from its equator) as a function of temperature for hybrid and quark stars. If a star is 
put somewhere above the respective
curves, the $r$-mode instability will set in and the star will spin down quickly.  
``APR'' stands for a certain nuclear equation of state, ``Bag'' denotes unpaired quark matter in the bag model
and the box labelled LMXB indicates the location of observed {\it low-mass X-ray binaries}. Within the given calculation they are located in 
a stable region for both hybrid and quark stars. For more explanations and details see Ref.\ \cite{Jaikumar:2008kh6}
where this figure is taken from.
}
\label{figviscous}
\end{center}
\end{figure}

The energy loss from gravitational radiation due to the $r$-mode instability makes the star spin down drastically and quickly. 
Consequently, the observation of sufficiently high 
rotation frequencies implies that some mechanism must be at work to avoid the instability. The above argument for the instability
is generic for all rotating perfect fluid stars. If there is dissipation, i.e., if the matter inside the star has a nonzero viscosity,
the instability can be damped. Put differently, in order to rotate fast the star has to be viscous. This 
statement seems paradoxical at first sight but makes sense with the above explanation. In Fig.\ \ref{figviscous} we show an example for 
critical frequencies of hybrid and quark stars, derived from viscosity calculations. 

In hydrodynamics, there are two kinds of viscosity, 
shear and bulk viscosity.\footnote{In the case of a superfluid, there are in fact several bulk viscosities.} Bulk viscosity describes
dissipation for the case of volume expansion or compression while shear viscosity is relevant for shear forces. Both kinds of viscosities
are relevant for the damping of the $r$-mode instability, typically they act in different temperature regimes, bulk viscosity at rather large, 
shear viscosity at rather small temperatures. What is the microscopic
physics behind the viscosity? Let us explain this for the case of the bulk viscosity. 

Imagine a chunk of nuclear or quark matter 
in thermal and chemical equilibrium in a volume $V_0$. Now we compress and expand this volume periodically, $V(t)=V_0+\delta V_0 \cos\omega t$. 
In the astrophysical setting, these will be local volume oscillations where $\omega$ is typically of the order of the rotation frequency 
of the star. Through the volume change the matter gets out of thermal and, possibly, chemical equilibrium. The latter may happen if the 
matter is composed of different components whose chemical potentials react differently on a density change. An example is unpaired quark matter
with massless up and down quarks and massive strange quarks. The system now seeks to reequilibrate. For instance, if the compression has
increased the down quark chemical potential compared to the strange quark chemical potential (in chemical equilibrium they are equal), 
the system reacts by producing strange quarks, for instance via the process $u+d\to u+s$. If it does so 
on the same time scale as the external oscillation, there can be sizable dissipation (think of compressing a spring which changes its 
spring constant during the process; you will not get back the work you have put in). Consequently, the calculation of the bulk viscosity 
requires the calculation of the rate of processes such as $u+d\to u+s$ which indeed turns out to be the dominant process for the bulk 
viscosity in unpaired quark matter.
Other processes which contribute are leptonic processes, such as the direct Urca process we have discussed in the context of  
neutrino emissivity in Sec.\ \ref{sec:dUrca}. It is important to note that again the weak processes are the relevant ones. In principle,
also strong processes contribute to the bulk viscosity since they reequilibrate the system thermally. However, they do so on time
scales much smaller than the external oscillation. Therefore, the system reequilibrates basically instantaneously during the compression process
and no energy is dissipated. These arguments also show that the bulk viscosity is a function of the (external) frequency. Maximum 
bulk viscosity is obtained when the rate of the respective microscopic process (which is a function of temperature) is closest to this frequency. 
Hence, it may well be that for a certain temperature regime a superconducting state has {\it larger} bulk viscosity than a non-superconducting 
state. This may sound counterintuitive but note that the (partial) suppression of the rate of the microscopic process by $\exp(-\Delta/T)$
may actually help the viscosity if it brings the rate closer to the external frequency. 
See for instance Sec.\ VII in Ref.\ \cite{Alford:2007xm6} for a brief review about viscosity in quark matter phases, and 
Refs.\ \cite{Madsen:1992sx6,Alford:2006gy6} for examples
of detailed microscopic calculations of bulk viscosity in quark matter.

\item{\it Magnetic fields.} 
We have mentioned in the introduction that compact stars can have huge magnetic fields, the highest 
magnetic fields measured for
the surface of a star (then called {\it magnetar}) are about 10$^{15}$ G. The first question one might ask is what the origin of these
magnetic fields is. The conventional explanation is that they are inherited from the star's progenitor, a giant star that has exploded in 
a supernova.
While the magnetic flux is conserved in this process, the magnetic field is greatly enhanced because the magnetic field lines are confined in 
a much smaller region after the explosion. 

Other questions regarding the magnetic field concern their interplay with dense matter. 
We have learned that nuclear matter can contain superconducting protons. Protons form a type-II superconductor where the magnetic field
is confined into flux tubes. Since at the same time the rotating neutron superfluid forms vortices, a complicated picture
emerges, where arrays of flux tubes and vortices intertwine each other. Their dynamics is complicated and relevant for instance 
for the observed precession times of the star, see for instance Ref.\ \cite{Link:2003hq6}. This issue is also related to {\it pulsar glitches}, see below. 

In the main part we have only touched the interplay of color superconductors with magnetic fields. 
We have stated without calculation that the CFL phase is not an 
electromagnetic superconductor, i.e., a magnetic field can penetrate CFL matter. More precisely, Cooper pairs in CFL are neutral
with respect to a certain mixture of the photon and one of the gluons. Because of the smallness of the electromagnetic coupling compared to the 
strong coupling, the gluon admixture is small and the new gauge boson is called ``rotated photon''. There are color superconductors which 
do expel magnetic fields, for instance the color-spin-locked (CSL) phase. In this case, Cooper pairs are formed of quarks with the same flavor, and
a Cooper pair carries total spin one (instead of zero in the CFL phase). The CSL phase is an electromagnetic superconductor. It is
of type I, i.e., expels magnetic fields completely. For a short review about spin-one color superconductors in compact stars and their
effect on magnetic fields see Ref.\ \cite{Aguilera:2006xv6}.

Magnetic fields also play a role in the cooling of the star since they have an effect on the heat transport, resulting
in an anisotropic surface temperature, see Ref.\ \cite{Page:2005fq6} and references therein. An extensive review about magnetic fields in compact 
stars is Ref.\ \cite{Harding:2006qn6}.

\item{\it Crust of the star.} 
The crust of the star is a very important ingredient for the understanding of observations. In the conventional picture of a neutron star 
there is an outer
crust with an ion lattice, and an inner crust with a neutron (super)fluid immersed in this lattice. This crust typically 
has a thickness of about 1 km. A lot about the crust can be found in Refs.\ \cite{haensel6,Chamel:2008ca6}.
In our discussion of neutron stars vs.\ hybrid stars vs.\ quark stars it is important that the crust provides a crucial distinction 
between an ordinary neutron star 
(or a hybrid star) and a quark star. How does the crust of a quark star look? Several scenarios have been suggested. First suppose that the
surface of a quark star exhibits an abrupt transition from strange quark matter to the vacuum. This is possible under the assumption of 
the strange quark matter hypothesis we discussed in Sec.\ \ref{sec:strange}, because, if the hypothesis is true, strange quark matter is stable 
at zero pressure. ``Abrupt'' means that the density drops to zero on a length scale of about 1 fm, given by the
typical length scale of the strong interaction. Now recall that (unpaired) three-flavor quark matter contains electrons. They interact with 
quark matter through the electromagnetic interaction, therefore their surface will be smeared (several hundred fm) compared to the sharp 
surface of the quark matter. As a consequence, an outward-pointing electric field develops (i.e., at the surface 
positively charged test particles are 
accelerated away from the center of the star). This electric field can support a thin layer of positively charged ions, separated from the 
quark matter by a layer of electrons. Hence a ``normal'' crust for a quark star is conceivable, consisting of an ion lattice. 
In contrast to the crust of a neutron star, such a crust of a quark star would be 
very thin, at most of the order of 100 m. See Ref.\ \cite{Alcock:1986hz6} for more details about this picture of the surface of a quark star 
(and for other properties of quark stars). This picture may be challenged by the possibility of a mixed phase at the surface of the star. 
Here, mixed phase refers to a crystalline structure of strangelets immersed in a sea of electrons. In this case, there would be no electric
field and thus no possibility for a ``normal'' crust. The quark matter would rather have its own crystalline crust. Estimates in 
Ref.\ \cite{Alford:2006bx6} show that it is unlikely that such a mixed phase is formed once surface tension is taken into account. 
In any case, a rigid crust, if at all present, will be much thinner in a quark star than in a neutron star or a hybrid star.

This difference is relevant in the context of ``magnetar seismology''. Quasi-periodic oscillations observed in the aftermath of 
X-ray bursts from magnetars can be related to typical oscillation frequencies of the crust. In other words, ``star quakes''
have significantly different properties depending on whether one assumes the star to be a neutron star or a quark star. 
In fact, the ordinary crust explains the data quite well while the crust of a quark star seems to be incompatible with the observed 
phenomenology \cite{Watts:2006hk6}.

\item {\it Pulsar glitches.} 
Pulsar glitches are an interesting phenomenon related to the rotation frequency, the crust (more precisely the crystalline 
structure of the crust), and superfluidity. For spinning-down pulsars one observes sudden spin-ups, i.e., in the overall trend
of a decreasing rotation frequency, the frequency increases in irregular intervals significantly on a very short time scale. 
This is conventionally explained through superfluid vortices in the neutron superfluid 
that pin at the lattice sites of the inner crust \cite{Anderson:1975zze6}. 

To understand this statement and the consequences for glitches, we recall the following
property of superfluids. A superfluid, be it superfluid helium, superfluid neutron matter, or any other superfluid, is irrotational
in the sense that the superfluid velocity has vanishing curl. Therefore, if the superfluid is rotated it develops regions where the 
order parameter vanishes, i.e., where it becomes a normal fluid.\footnote{It is instructive to view this phenomenon in analogy to 
a type-II superconductor. There, a magnetic field (if sufficiently large but not too large) penetrates the superconductor through 
flux tubes. It partially destroys superconductivity, i.e., in the center of the 
flux tubes the order parameter is zero. Hence the analogy is superfluid -- superconductor; angular momentum -- magnetic field; vortices --
flux tubes.}  The angular momentum is then ``stored'' in these regions which 
are called vortices. An array of vortices, which are ``strings'' in the direction of the angular momentum, is formed with the total 
angular momentum of the superfluid being proportional to the density of vortices (because each vortex carries one quantum of circulation).
Consequently, if the rotation frequency decreases, the array of vortices becomes sparser, i.e., the vortices move apart. 

The next 
ingredient in the glitch mechanism is the pinning of the vortices at the lattice of nuclei in the inner crust. Generally speaking,
there is an effective interaction between the vortices and the nuclei, resulting in a certain path of the vortex string through the lattice which
minimizes the free
energy of the system. You may think of this preferred configuration as follows. Superfluidity, i.e., neutron Cooper pairing, lowers the free energy 
of the system. Therefore, the system may want to put the vortices, where there is no Cooper pairing, through the lattice sites because they 
are not superfluid anyway. Otherwise, i.e., by putting them between the lattice sites, one loses pairing energy. The actual  
details of the pinning mechanism are complicated and, depending on the density, the preferred path of the vortices may in fact  
be between the sites, in contrast to the above intuitive argument. However, this does not matter for our argument for the mechanism of 
glitches:
 
In a rotating neutron star, the neutron vortices pin at the 
lattice of the inner crust. Now the star spins down. On the one hand, the vortices ``want'' to move apart. On the other hand, there is an effective
pinning force which keeps them at there sites. Hence, for a while they will not move which implies that the superfluid 
(the vortex array) is spinning faster than the rest of the star. At some point, when the tension is sufficiently large, the vortices
will un-pin, move apart and thus release their angular momentum which spins up in particular the surface of the star whose rotation
is observed. Then, they re-pin and the process starts again. 

An alternative scenario, where nuclear matter is replaced by quark matter has been suggested \cite{Mannarelli:2007bs6}. 
In our discussion of the CFL phase we have seen that quark matter can be a superfluid. This means that one of the conditions for
the mechanism of vortex pinning is fulfilled. The second condition, a sufficiently rigid lattice, may be provided 
by one of the unconventional color-superconducting phases which are possible in the case of mismatched Fermi momenta, 
see discussion at the end of in Sec.\ \ref{sec:QCDgap}. Some of these phases indeed exhibit a crystalline
structure. Such a quark crystal is of very different nature than the ion lattice because it is the energy gap from superconductivity
which varies periodically in space, giving rise to crystals characterized by surfaces where the gap vanishes. 
It remains to be seen in the future which of these scenarios passes all 
observational constraints and can explain the pulsar glitches or if there is a yet unknown mechanism for these curious spin-ups.

\item {\it Other theoretical approaches to dense matter.}
What are the alternatives to understand QCD at large, but not asymptotically large, densities? {\it Lattice QCD}, i.e., solving 
QCD by brute force on a computer, is by now a powerful tool for strong-coupling phenomena at zero chemical potential. However, at nonzero 
chemical potential, one encounters the so-called {\it sign problem} which renders lattice calculations unfeasible. Progress has been made
to extend lattice calculations to small chemical potentials, more precisely to small values of $\mu/T$. But calculations
at large $\mu$ and small $T$, as needed for compact stars, are currently not within reach. See Ref.\ \cite{Stephanov:2007fk6} 
for a non-technical recent 
overview article about lattice QCD, in particular its contributions to the QCD phase diagram and about the sign problem; you may also try 
Ref.\ \cite{Schmidt:2007jg6}.

Because of the problems of lattice calculations at finite chemical potential one has to rely on model calculations or on
extrapolations similar to the ones discussed in these lectures. One model for quark matter we have not discussed is the Nambu-Jona-Lasinio (NJL)
model. This model does not contain gluons and describes the interaction between quarks by an effective pointlike interaction. It has been used 
to compute the QCD phase diagram at intermediate densities. Since the result depends strongly on the parameters of the model, it 
should be taken as an indicator for how the phase diagram might look, not as an accurate prediction. Due to its simplicity it is
widely used and can indeed give some interesting results which serve as a guideline for the understanding of QCD. For an extensive review about
the NJL model in dense quark matter see Ref.\ \cite{Buballa:2003qv6}; for an application of the NJL model in the context of compact stars,
see for instance Ref.\ \cite{Blaschke:2005uj6}.

Finally, we point out that arguments for large numbers of colors $N_c$ 
may be applied to gain some insight to QCD where $N_c=3$. In particular, it has been argued that at $N_c=\infty$ an interesting
novel phase, termed {\it quarkyonic matter}, populates the $T$-$\mu$ phase diagram \cite{McLerran:2007qj6}. The (yet unsolved) problem is to 
find out whether this 
phase, or some modification of it, survives for $N_c=3$. More generally speaking, the large $N_c$ approach is another approach where 
calculations can be performed in a regime 
where everything is under rigorous control. From these rigorous results one then tries to get closer to the regime one is interested in. In this 
sense, this approach is not unlike the perturbative approach. In view of the possible, but not at all obvious,
relevance of large-$N_c$ physics to $N_c=3$ physics, one can also apply the duality of certain string theories to field theories similar
to QCD, based on the so-called {\it AdS/CFT correspondence}. For pedagogical reviews see Refs.\ \cite{Peeters:2007ab6,Gubser:2009md6}.  
This somewhat speculative but popular approach to QCD has recently been pursued especially for large-$T$, small-$\mu$ physics, 
but is, in certain variants, also suited for the physics at finite chemical potential.   

\end{itemize}

%% file: appendix2.tex
%
%
%

\appendix

\chapter{Basics of quantum field theory at finite temperature and chemical potential}

Many of the discussions in the main part of these lectures rely on field-theoretical methods, in particular on quantum field 
theory at finite temperature and chemical potential. One purpose of the following basic discussion is therefore to explain how a chemical 
potential is introduced in quantum field theory. We shall also discuss how finite temperature enters the formalism, 
although for most quantities we discuss in these lecture notes we consider the zero-temperature limit, which is 
a good approximation for our purposes. For instance in the discussion of the Walecka model, Sec.\ \ref{sec:walecka}, we give the 
finite-temperature expressions, based on appendix \ref{app:fermions}, before we set $T=0$ in the physical discussion. In other 
parts, we do keep $T\neq 0$ in our results, for instance when we are interested in the cooling behavior of dense matter, see chapter 
\ref{sec:cooling}. 

We shall start with the Lagrangian for a complex bosonic field and derive the partition function in the path integral formalism, taking
into account Bose-Einstein condensation. This part is particularly useful for our treatment of kaon condensation in CFL quark matter, see
Sec.\ \ref{sec:CFLK0}. We shall in particular see how bosonic Matsubara frequencies are introduced and how the summation over these is
performed with the help of contour integration in the complex frequency plane. In the second part of this appendix we shall then 
discuss the analogous derivation for fermions.

\section{Bosonic field}
\label{app:bosons} 

We start from the Lagrangian
\be
{\cal L}_0 = \partial_\mu\varphi^*\partial^\mu\varphi -m^2|\varphi|^2 -\lambda |\varphi|^4 \, ,
\ee
with a complex scalar field $\varphi$ with mass $m$ and coupling constant $\lambda$.
We shall first show how a chemical potential $\mu$ is introduced. This will lead to a new Lagrangian ${\cal L}$, wherefore we have denoted
the Lagrangian without chemical potential by ${\cal L}_0$. The chemical potential $\mu$ must be associated with a conserved charge.
We thus need to identify the conserved current. From Noether's theorem we know that the conserved current 
is related to the symmetry of the Lagrangian. We see that ${\cal L}_0$ is invariant under $U(1)$ rotations of the field,
\be
\varphi \to e^{-i\alpha} \varphi \, .
\ee
This yields the Noether current
\be
j^\mu = \frac{\partial{\cal L}_0}{\partial(\partial_\mu\varphi)}\,\frac{\delta\varphi}{\delta\alpha}+
\frac{\partial{\cal L}_0}{\partial(\partial_\mu\varphi^*)}\,\frac{\delta\varphi^*}{\delta\alpha}
= i(\varphi^*\partial^\mu\varphi-\varphi\partial^\mu\varphi^*) \, , 
\ee
with $\partial_\mu j^\mu = 0$, and the conserved charge (density) is 
\be \label{j0}
j^0 = i(\varphi^*\partial^0\varphi-\varphi\partial^0\varphi^*) \, .
\ee
In the following we want to see how the chemical potential associated to $j^0$ enters the Lagrangian. The partition function for a scalar field 
is 
\bea \label{part1}
Z&=& {\rm Tr}\, e^{-\beta(\hat{H}-\mu \hat{N})} \non 
&=& \int {\cal D}\pi\int_{\rm periodic}{\cal D}\varphi\,\exp\left[-\int_X \left({\cal H}-\mu{\cal N}
-i\pi\partial_\tau\varphi\right)\right] \, .
\eea
This equation should remind you that the partition function can be written in the operator formalism in terms of the Hamiltonian $\hat{H}$ 
and the charge operator $\hat{N}$, or, 
as we shall use here, in terms of 
a functional integral over $\varphi$ and the conjugate momentum $\pi$, with the Hamiltonian ${\cal H}$ and the charge density ${\cal N}=j^0$.  
We have abbreviated the space-time integration by 
\be
\int_X\equiv \int_0^\beta d\tau\int d^3 x \, , 
\ee
where the integration over ``imaginary time'' $\tau=it$ goes from 0 to the inverse temperature $\beta=1/T$. In the 
following, the four-vector in position space is denoted by 
\be
X\equiv (t,{\bf x}) = (-i\tau,{\bf x}) \, .
\ee
The term ``periodic'' for the $\varphi$ integral in Eq.\ (\ref{part1}) means that 
all fields $\varphi$ over which we integrate have to be periodic in the imaginary time direction, 
$\varphi(0,{\bf x})=\varphi(\beta,{\bf x})$. This is essentially a consequence of the trace operation in the first line of Eq.\ (\ref{part1}):
the partition function is formally reminiscent of a sum over transition amplitudes which have the same initial and final states at ``times''
0 and $\beta$.

Let us, for convenience, introduce the two real fields $\varphi_1$, $\varphi_2$,
\be \label{phi12}
\varphi=\frac{1}{\sqrt{2}}(\varphi_1+i\varphi_2) \, .
\ee
Then, the Lagrangian becomes
\be
{\cal L}_0 = \frac{1}{2}\left[\partial_\mu\varphi_1\partial^\mu\varphi_1+\partial_\mu\varphi_2\partial^\mu\varphi_2-m^2(\varphi_1^2+\varphi_2^2)
-\frac{\lambda}{2}(\varphi_1^2+\varphi_2^2)^2\right] \, .
\ee
The conjugate momenta are
\be \label{pi12}
\pi_i = \frac{\partial{\cal L}_0}{\partial(\partial_0\varphi_i)} = \partial^0\varphi_i \, , \qquad i=1,2 \, .
\ee
Consequently, with $j^0 =  \varphi_2\pi_1-\varphi_1\pi_2$, which follows from Eqs.\ (\ref{j0}), (\ref{phi12}), and (\ref{pi12}), 
we have 
\bea
{\cal H}-\mu{\cal N} &=& \pi_1\partial_0\varphi_1+\pi_2\partial_0\varphi_2 - {\cal L}_0 -\mu{\cal N} \non
&=& \frac{1}{2}\left[\pi_1^2+\pi_2^2+(\nabla\varphi_1)^2+(\nabla\varphi_2)^2+m^2(\varphi_1^2+\varphi_2^2)\right] \non
&& -\mu(\varphi_2\pi_1-\varphi_1\pi_2) \, .
\eea
The integration over the conjugate momenta $\pi_1$, $\pi_2$ can be separated from the integration over
the fields $\varphi_1$, $\varphi_2$ after introducing the shifted momenta
\be
\tilde{\pi}_1 \equiv \pi_1-\partial_0\varphi_1-\mu\varphi_2 \, , \qquad \tilde{\pi}_2 \equiv \pi_2-\partial_0\varphi_2+\mu\varphi_1 \, .
\ee
This yields 
\be\label{pidphi}
\pi_1\partial_0\varphi_1+\pi_2\partial_0\varphi_2 - {\cal H}+ \mu{\cal N} = -\frac{1}{2}(\tilde{\pi}_1^2+\tilde{\pi}_2^2)+{\cal L} \, , 
\ee
where the new Lagrangian ${\cal L}$ now includes the chemical potential,
\bea  \label{nowmu}
{\cal L} &=& \frac{1}{2}\Big[\partial_\mu\varphi_1\partial^\mu\varphi_1+\partial_\mu\varphi_2\partial^\mu\varphi_2
+2\mu(\varphi_2\partial_0\varphi_1-\varphi_1\partial_0\varphi_2) \non
&&+(\mu^2-m^2)(\varphi_1^2+\varphi_2^2)-\frac{\lambda}{2}(\varphi_1^2+\varphi_2^2)^2\Big] \, .
\eea
Thus we see that the chemical potential produces, besides the expected term $\mu j^0$, the additional term
$\mu^2(\varphi_1^2+\varphi^2_2)/2$. This is due to the momentum-dependence of $j^0$. In terms of the complex field $\varphi$, the Lagrangian reads
\be
{\cal L}=|(\partial_0-i\mu)\varphi|^2-|\nabla\varphi|^2-m^2|\varphi|^2 -\lambda|\varphi|^4 \, ,
\ee
which shows that the chemical potential looks like the temporal component of a gauge field. We can now insert Eq.\ (\ref{pidphi}) into the partition 
function (\ref{part1}). The integration over conjugate momenta and over fields factorize, and the momentum integral yields an irrelevant constant
$N$,
such that we can write  
\be \label{part2}
Z = N\int_{\rm periodic}{\cal D} \varphi_1 {\cal D}\varphi_2 \, \exp \int_X {\cal L} \, .
\ee
In order to take into account Bose-Einstein condensation, we divide the field into a constant background field and fluctuations around this
background, $\varphi_i\to\phi_i+\varphi_i$. A nonzero condensate $\phi_1+i\phi_2$ picks a direction in the $U(1)$ degeneracy space and thus 
breaks the symmetry spontaneously. We can choose $\phi_2=0$ and thus may denote $\phi\equiv \phi_1$. Then, the Lagrangian (\ref{nowmu}) becomes
\be \label{L234}
{\cal L} = -U(\phi^2) + {\cal L}^{(2)} + {\cal L}^{(3)} + {\cal L}^{(4)} , 
\ee
with the tree-level potential 
\be \label{Uphitree}
U(\phi^2) = \frac{m^2-\mu^2}{2}\,\phi^2  + \frac{\lambda}{4}\,\phi^4 \, , 
\ee
and terms of second, third, and fourth order in the fluctuations,
\begin{subequations}
\bea
{\cal L}^{(2)} &=& -\frac{1}{2}\left[-\partial_\mu\varphi_1\partial^\mu\varphi_1 -\partial_\mu\varphi_2\partial^\mu\varphi_2 
-2\mu(\varphi_2\partial_0\varphi_1-\varphi_1\partial_0\varphi_2) \right.\non
&& \left. +\,(m^2-\mu^2)(\varphi_1^2+\varphi_2^2)+3\lambda \phi^2 \varphi_1^2+
\lambda \phi^2 \varphi_2^2 \right] \, , \\
{\cal L}^{(3)} &=& -\lambda \phi \varphi_1 (\varphi_1^2+\varphi_2^2) \, , \label{Lcubic}\\
{\cal L}^{(4)} &=& -\frac{\lambda}{4}(\varphi_1^2+\varphi_2^2)^2 \, .\label{quarticphi4}
\eea
\end{subequations}
We have omitted the linear terms since they do not contribute to the functional integral. Note that the cubic interactions are induced by the 
condensate.

In this appendix we are only interested in the tree-level contributions $U(\phi^2)$ and ${\cal L}^{(2)}$ 
in order to explain the basic calculation of the partition function for 
the simplest case. We therefore shall ignore the cubic and quartic contributions ${\cal L}^{(3)}$ and ${\cal L}^{(4)}$. We introduce the 
Fourier transforms of the fluctuation fields via 
\be \label{fourier}
\varphi(X) = \frac{1}{\sqrt{TV}}\sum_K e^{-iK\cdot X}\varphi(K) =\frac{1}{\sqrt{TV}}\sum_K e^{i(\omega_n\tau+{\bf k}\cdot{\bf x})}\varphi(K) \, ,
\ee
with the four-momentum 
\be 
K\equiv (k_0,{\bf k}) = (-i\omega_n,{\bf k}) \, ,
\ee
and with the Minkowski scalar product $K\cdot X = k_0 x_0-{\bf k}\cdot{\bf x} = -(\tau\omega_n + {\bf k}\cdot {\bf x})$. (Although for convenience 
we have defined the time components with a factor $i$ and thus can use Minkowski notation, the scalar product is essentially Euclidean.)
The normalization is chosen such that the Fourier-transformed fields $\varphi(K)$ are dimensionless. 
The 0-component of the four-momentum is given by the {\it Matsubara frequency} $\omega_n$. To fulfill the
periodicity requirement $\varphi(0,{\bf x})=\varphi(\beta,{\bf x})$ we need $e^{i\omega_n\beta}=1$, i.e., $\omega_n\beta$ 
has to be an integer multiple of $2\pi$, or
\be
\omega_n = 2\pi n T \, , \qquad n\in \mathbb{Z} \, .
\ee
With the Fourier transform (\ref{fourier}), and 
\be
\int_X e^{iK\cdot X} = \frac{V}{T}\delta_{K,0} \, , 
\ee
we have 
\be \label{actionmu}
\int_X{\cal L}^{(2)}= -\frac{1}{2}\sum_K(\varphi_1(-K),\varphi_2(-K))\frac{D_0^{-1}(K)}{T^2}\left(
\begin{array}{c}\varphi_1(K) \\ \varphi_2(K)\end{array}\right) \, ,
\ee
with the free inverse propagator in momentum space 
\be
D_0^{-1}(K)=\left(\begin{array}{cc} -K^2+m^2+3\lambda\phi^2 -\mu^2 & -2i\mu k_0 \\ 2i\mu k_0 & -K^2+m^2+\lambda\phi^2 -\mu^2 \end{array}\right) \, .
\ee
With Eqs.\ (\ref{part2}), (\ref{actionmu}) and using that $\varphi(K) = \varphi^*(-K)$ (because $\varphi(X)$ is real) we can write the 
tree-level thermodynamic potential as 
\bea \label{OmegaV}
\frac{\Omega}{V} &=& -\frac{T}{V}\ln Z \non
&=& U(\phi^2) -\frac{T}{V} \ln \int {\cal D}\varphi_1{\cal D}\varphi_2
\exp\left[-\frac{1}{2}\sum_K(\varphi_1(-K),\varphi_2(-K))\frac{D_0^{-1}(K)}{T^2}\left(
\begin{array}{c}\varphi_1(K) \\ \varphi_2(K)\end{array}\right)\right] \non
&=& U(\phi^2) + \frac{T}{2V}\ln{\rm det}\frac{D_0^{-1}(K)}{T^2} \, ,
\eea
where the determinant is taken over $2\times 2$ space and momentum space.
Here we have used the general formula
\be \label{gaussbosons}
\int d^D x \,e^{-\frac{1}{2}{\bf x}\cdot \hat{A}{\bf x}} = (2\pi)^{D/2}({\rm det}\,\hat{A})^{-1/2} \, , 
\ee
for a Hermitian, positive definite matrix $\hat{A}$, which is a generalization of the one-dimensional Gaussian integral 
\be \label{gauss}
\int_{-\infty}^\infty dx\,e^{-\frac{1}{2}\alpha x^2} = \sqrt{\frac{2\pi}{\alpha}} \, .
\ee
To further evaluate the thermodynamic potential, we first define the tree-level masses
\begin{subequations} \label{m12}
\bea
m_1^2&\equiv& m^2+3\lambda\phi^2 \, , \\
m_2^2&\equiv& m^2+\lambda\phi^2 \, . 
\eea
\end{subequations}
Then, we obtain 
\bea \label{lndet}
\ln{\rm det} \frac{D_0^{-1}(K)}{T^2} &=& \ln\prod_K\frac{1}{T^4}[(-K^2+m_1^2-\mu^2)(-K^2+m_2^2-\mu^2)-4\mu^2k_0^2] \non
&=& \ln\prod_K\frac{1}{T^4}[(\epsilon_k^+)^2-k_0^2][(\epsilon_k^-)^2-k_0^2] \non
&=&  \sum_K \left[\ln\frac{(\epsilon_k^+)^2-k_0^2}{T^2}+\ln\frac{(\epsilon_k^-)^2-k_0^2}{T^2}\right] \, , 
\eea
where we defined the quasiparticle energies 
\be \label{epspm}
\epsilon_k^\pm = \sqrt{E_k^2+\mu^2\mp \sqrt{4\mu^2 E_k^2+\delta M^4}} \, ,  
\ee
with 
\be
E_k \equiv \sqrt{k^2+M^2} \, , \quad M^2 \equiv \frac{m_1^2+m_2^2}{2} = m^2+2\lambda\phi^2 \, , \quad 
\delta M^2 \equiv \frac{m_1^2-m_2^2}{2} = \lambda\phi^2 \, .
\ee
Even at tree-level, the quasiparticle energies (\ref{epspm}) look complicated, but become simple in the noninteracting limit,
\be
\lambda=0:\qquad \epsilon^\pm_k = \sqrt{k^2+m^2}\mp\mu \, , 
\ee
and for vanishing chemical potential,
\be
\mu =0:\qquad \epsilon^\pm_k=\sqrt{k^2 + m^2_{2/1}} \, .
\ee
Further properties of these quasiparticle energies are discussed in the context of kaon condensation in CFL, see Sec.\ \ref{sec:CFLK0}. 
Next, we perform the sum over Matsubara frequencies in Eq.\ (\ref{lndet}). We use the result 
\be \label{matsubos}
\sum_{n} \ln \frac{\omega_n^2+\epsilon_k^2}{T^2} = \frac{\epsilon_k}{T}+2\ln \left(1-e^{-\epsilon_k/T}\right)+{\rm const} \, ,
\ee
for a real number $\epsilon_k$, and where ``const'' is a temperature-independent constant. Before we prove this result 
via contour integration in the complex plane, we use it to compute the final result for the tree-level thermodynamic potential. 
We insert Eq.\ (\ref{matsubos}) into Eq.\ (\ref{lndet}), the result into Eq.\ (\ref{OmegaV}), and take the thermodynamic limit to obtain 
\be 
\frac{\Omega}{V} = U(\phi^2) + T\int\frac{d^3{\bf k}}{(2\pi)^3}\left[\frac{\epsilon_k^++\epsilon_k^-}{2T}+\ln\left(1-e^{-\epsilon_k^+/T}\right)
+\ln\left(1-e^{-\epsilon_k^-/T}\right)\right] \, .
\ee  
From this expression we can for instance compute the pressure $P=-\Omega/V$. The first term in the integrand yields an infinite contribution
which however is temperature-independent. We may thus use a renormalization such that the vacuum pressure vanishes. 
Then, for sufficiently large temperatures, where in particular $\phi=0$, particles
and antiparticles yield the same contribution and we obtain 
\be \label{PressBos}
P \simeq  -2\frac{T^4}{2\pi^2}\int_0^\infty dx\,x^2\ln\left(1-e^{-x}\right)
= 2\,\frac{\pi^2 T^4}{90} \, .
\ee

\subsection{Summation over bosonic Matsubara frequencies}

Here we prove Eq.\ (\ref{matsubos}) via contour integration in the complex frequency plane. Especially for more complicated Matsubara sums
this is a very useful technique as can be seen by applying the following method to the Matsubara sums in problems \ref{prob11} and \ref{prob12}. 

First, in order to get rid of the logarithm, we write
\be \label{original}
\sum_{n} \ln \frac{\omega_n^2+\epsilon_k^2}{T^2} =  \int_1^{(\epsilon_k/T)^2} dx^2 \sum_n \frac{1}{(2n\pi)^2+x^2} + \sum_n \ln[1+(2n\pi)^2] \, .
\ee
We now perform the sum in the integrand which, denoting $\epsilon_k\equiv Tx$, we write as 
a contour integral,
\be \label{contour1}
T\sum_n\frac{1}{\omega_n^2+\epsilon_k^2} = -\frac{1}{2\pi i}\oint_C d\omega\,\frac{1}{\omega^2-\epsilon_k^2}\frac{1}{2}\coth\frac{\omega}{2T}
\, .
\ee
The second identity follows from the residue theorem, 
\be
\frac{1}{2\pi i}\oint_C dz\,f(z) = \sum_n \left. {\rm Res}\,f(z)\right|_{z=z_n} \, , 
\ee
where $z_n$ are the poles of $f(z)$ in the area enclosed by the contour $C$. If we can write the function $f$ as $f(z)=\varphi(z)/\psi(z)$, 
with analytic functions $\varphi(z)$, $\psi(z)$, the residues are 
\be
\left. {\rm Res}\,f(z)\right|_{z=z_n} = \frac{\varphi(z_n)}{\psi'(z_n)} \, .
\ee
The contour $C$ in Eq.\ (\ref{contour1}) is chosen such that it encloses all poles of  $\coth[\omega/(2T)]$ and none of 
$1/(\omega^2-\epsilon_k^2)$. The poles of $\coth[\omega/(2T)]$ are given by $e^{\omega/2T}-e^{-\omega/2T}=0$, i.e.,
they are on the imaginary axis, $\omega=i\omega_n$ with the Matsubara frequencies $\omega_n$. In the above notation
with the functions $\varphi$ and $\psi$,
\bea
\varphi(\omega) &=& \frac{1}{2}\frac{e^{\omega/(2T)}+e^{-\omega/(2T)}}{\omega^2-\epsilon_k^2} \, , \qquad 
\psi(\omega) =e^{\omega/(2T)}-e^{-\omega/(2T)} \, , \non
&\Rightarrow& \frac{\varphi(i\omega_n)}{\psi'(i\omega_n)} = -T\frac{1}{\omega_n^2+\epsilon_k^2} \, ,
\eea
from which Eq.\ (\ref{contour1}) follows immediately.
Next, we may deform the contour $C$ 
(which consists of infinitely many circles surrounding the poles) and obtain
\bea
T\sum_n\frac{1}{\omega_n^2+\epsilon_k^2} &=& -\frac{1}{2\pi i}\int_{-i\infty+\eta}^{i\infty+\eta} d\omega\,
\frac{1}{\omega^2-\epsilon_k^2}\frac{1}{2}\coth\frac{\omega}{2T} \non
&& -\frac{1}{2\pi i}\int_{i\infty-\eta}^{-i\infty-\eta} d\omega\,
\frac{1}{\omega^2-\epsilon_k^2}\frac{1}{2}\coth\frac{\omega}{2T} \non
&=& -\frac{1}{2\pi i}\int_{-i\infty+\eta}^{i\infty+\eta} d\omega\,
\frac{1}{\omega^2-\epsilon_k^2}\coth\frac{\omega}{2T} \, , 
\eea
where we have changed the integration variable $\omega\to -\omega$ in the second integral.
We now use the residue theorem a second time: we can close the contour in the positive half-plane at infinity
and pick up the pole at $\omega=\epsilon_k$,
\bea \label{oneloopsum}
T\sum_n\frac{1}{\omega_n^2+\epsilon_k^2} &=& \frac{1}{2\epsilon_k}\coth\frac{\epsilon_k}{2T} 
= \frac{1}{2\epsilon_k}[1+2f_B(\epsilon_k)] \, , 
\eea
(note the minus sign from clockwise contour integration). Here, 
\be
f_B(\epsilon)\equiv \frac{1}{e^{\epsilon/T}-1} 
\ee
is the Bose distribution function. We have thus found
\be
\frac{1}{T}\sum_n \frac{1}{(2n\pi)^2+x^2} = \frac{1}{Tx}\left(\frac{1}{2}+\frac{1}{e^x-1}\right) \, .
\ee
Now we insert the result into the original expression (\ref{original}) and integrate over $x^2$ 
to obtain (with const denoting $T$-independent constants)
\bea
\sum_{n} \ln \frac{\omega_n^2+\epsilon_k^2}{T^2} &=& \int_1^{(\epsilon_k/T)^2} dx^2\frac{1}{x}\left(\frac{1}{2}+\frac{1}{e^x-1}\right) +{\rm const}
\non &=& \frac{\epsilon_k}{T}+2\ln \left(1-e^{-\epsilon_k/T}\right)+{\rm const} \, ,
\eea
which is the result we wanted to prove.

\section{Fermionic field}
\label{app:fermions}

To describe a system of non-interacting fermions with mass $m$ we start with the Lagrangian
\be
{\cal L}_0 = \overline{\psi}\left(i\gamma^\mu\partial_\mu-m\right)\psi \, , 
\ee
where $\overline{\psi}=\psi^\dag\gamma^0$. As for the bosons we are interested in adding a chemical potential to this Lagrangian. 
To this end, we determine the conserved current as above, i.e., we first identify the global symmetry of the Lagrangian which is given by the 
transformation $\psi\to e^{-i\alpha}\psi$. The conserved current is 
\be
j^\mu=\frac{\partial{\cal L}_0}{\partial(\partial_\mu\psi)}\,\frac{\delta\psi}{\delta\alpha} = \overline{\psi}\gamma^\mu\psi \, , 
\ee
which yields the conserved charge (density)  
\be
j^0 = \psi^\dag\psi \, .
\ee
The conjugate momentum is
\be
\pi = \frac{\partial{\cal L}_0}{\partial(\partial_0\psi)} = i\psi^\dag \, .
\ee
This means that in the case of fermions we need to treat $\psi$ and $\psi^\dag$ as independent variables. The partition function 
for fermions is
\bea \label{partfermi}
Z&=& {\rm Tr}\, e^{-\beta(\hat{H}-\mu \hat{N})} \non 
&=& \int_{\rm antiperiodic} {\cal D}\psi^\dag {\cal D}\psi\,\exp\left[-\int_X \left({\cal H}-\mu{\cal N}
-i\pi\partial_\tau\psi\right)\right] \, .
\eea
This has to be compared to the analogous expression for bosons, Eq.\ (\ref{part1}). Recall that the periodicity of the bosonic fields 
is a consequence of taking the trace in the operator formalism. In other words, the partition function in the path integral formalism
can be derived from a transition amplitude with identical initial and final states. In the case of fermions, the fields in the path integral 
are Grassmann variables, as a consequence of the anticommutation relations of creation and annihilation operators. In this 
case, the trace involves a transition amplitude where initial and final states differ by a sign. Therefore, in the fermionic 
partition function the integration is over antiperiodic fields $\psi(0,{\bf x})=-\psi(\beta,{\bf x})$ and $\psi^\dag(0,{\bf x})
=-\psi^\dag(\beta,{\bf x})$. 

With the Hamiltonian 
\be \label{Hfermion}
{\cal H} = \pi\partial_0\psi - {\cal L}_0 = \overline{\psi}(i\bm{\gamma}\cdot \nabla +m)\psi \, ,  
\ee
(here and in the following we mean by the scalar product $\bm{\gamma}\cdot \nabla$ the product where the Dirac matrices 
appear with a lower index $\gamma_i$) we thus obtain 
\be
Z = \int_{\rm antiperiodic}{\cal D}\psi^\dag {\cal D}\psi\,\exp\left[\int_X \overline{\psi}\left(-\gamma^0\partial_\tau-i\bm{\gamma}\cdot\nabla
+\gamma^0\mu-m\right)\psi\right] \, .
\ee
In this case we cannot separate the $\pi\sim \psi^\dag$ integration from the $\psi$ integration. Remember that, in the bosonic case, this
led to a new Lagrangian which contained the chemical potential not just in the term $j^0\mu$. Here, the Lagrangian with chemical potential
simply is
\be
{\cal L} = \bar{\psi}(i\gamma^\mu\partial_\mu+\gamma^0\mu-m)\psi \, .
\ee
Note that again the chemical potential enters just like the temporal component of a gauge field that couples to the fermions.
Analogously to the bosonic case, we introduce the (dimensionless) Fourier-transformed fields 
\be
\psi(X) = \frac{1}{\sqrt{V}}\sum_Ke^{-iK\cdot X}\psi(K) \, , \qquad  
\overline{\psi}(X) = \frac{1}{\sqrt{V}}\sum_Ke^{iK\cdot X}\overline{\psi}(K) \, , 
\ee
(note the different dimensionality of fields compared to bosons; here the 
field $\psi(X)$ in position space has mass dimension 3/2). Again we denote $k_0= -i\omega_n$ such that 
$K\cdot X = -(\omega_n\tau +{\bf k}\cdot{\bf x})$. 
Now antiperiodicity, $\psi(0,{\bf x}) = -\psi(\beta,{\bf x})$, implies $e^{i\omega_n\beta}=-1$ and thus 
the fermionic Matsubara frequencies are
\be
\omega_n = (2n+1)\pi T \, , \qquad n\in \mathbb{Z} \, .
\ee
With the Fourier decomposition we find  
\be
\int_X\overline{\psi}\left(-\gamma^0\partial_\tau-i\bm{\gamma}\cdot\nabla
+\gamma^0\mu-m\right)\psi = -\sum_K \psi^\dag(K) \frac{G_0^{-1}(K)}{T} \psi(K) \, , 
\ee
where the free inverse fermion propagator in momentum space is
\be \label{noprojectors}
G_0^{-1}(K) = -\gamma^\mu K_\mu -\gamma^0\mu +m \, .
\ee
Although not needed for the rest of the calculation in this appendix, 
let us introduce a useful form of the inverse propagator in terms of 
energy projectors. This form is convenient for more involved calculations such as done in chapters \ref{sec:supersuper} 
and \ref{sec:cooling}. Equivalently to Eq.\ (\ref{noprojectors}) we can write 
\be \label{prop1}
G_0^{-1}(K) = -\sum_{e=\pm}(k_0+\mu-eE_k)\gamma^0\Lambda_{\bf k}^e \, , 
\ee
where $E_k=\sqrt{k^2+m^2}$, and where the projectors onto positive and negative energy states are given by
\be
\Lambda_{\bf k}^e\equiv \frac{1}{2}\left(1+e\gamma^0\frac{\bm{\gamma}\cdot {\bf k}+m}{E_k}\right) \, .
\ee
These (Hermitian) projectors are complete and orthogonal,
\be \label{lambda}
\Lambda_{\bf k}^+ + \Lambda_{\bf k}^- = 1 \,, \qquad \Lambda_{\bf k}^e\Lambda_{\bf k}^{e'}=\delta_{e,e'} \Lambda_{\bf k}^e\, .
\ee
The first property is trivial to see, the second follows with $\{\gamma^0,\gamma^i\}=0$ which follows from the general 
anticommutation property $\{\gamma^\mu,\gamma^\nu\}=2g^{\mu\nu}$, and with $(\bm{\gamma}\cdot{\bf k})^2=-k^2$.

From the form of the inverse propagator (\ref{prop1}) we can immediately read off the propagator itself,
\be \label{prop2}
G_0(K) = -\sum_{e=\pm}\frac{\Lambda_{\bf k}^e\gamma^0}{k_0+\mu-eE_k} \, .
\ee
With the properties (\ref{lambda}) one easily checks that $G_0^{-1}G_0=1$. One can also rewrite (\ref{prop2}) as
\be \label{prop20}
G_0(K) = \frac{-\gamma^\mu K_\mu-\gamma^0\mu-m}{(k_0+\mu)^2-E_k^2} \, .
\ee
Let us now come back to the calculation of the partition function. For the functional integration we use
\be 
\int \prod_k^N d\eta_k^\dag d\eta_k \exp\left(-\sum_{i,j}^N\eta_i^\dag D_{ij}\eta_j\right) = {\rm det}\, D \, .
\ee
Note the difference of this integration over Grassmann variables $\eta^\dag$, $\eta$ to the corresponding formula for bosons (\ref{gaussbosons}).
We obtain for the partition function 
\bea
Z &=& {\rm det} \,\frac{G_0^{-1}(K)}{T} = {\rm det}\frac{1}{T}\left(\begin{array}{cc} -(k_0+\mu)+m & -\bm{\sigma}\cdot{\bf k} \\ 
\bm{\sigma}\cdot{\bf k} & (k_0+\mu)+m \end{array}\right) \, ,
\eea
where the determinant is taken over Dirac space and momentum space, and where $\sigma_1, \sigma_2, \sigma_3$ are the Pauli matrices.
We can use the general formula
\be
{\rm det}\left(\begin{array}{cc} A & B \\ C & D \end{array}\right) = {\rm det}(AD-BD^{-1}CD) \, ,
\ee
for matrices $A$, $B$, $C$, $D$ with $D$ invertible, to get
\be
\ln Z = \sum_K\ln\left(\frac{E_k^2-(k_0+\mu)^2}{T^2}\right)^2 \, , 
\ee
where we have used $(\bm{\sigma}\cdot{\bf k})^2 = k^2$.
With $k_0=-i\omega_n$ we can write this as
\bea \label{lnZfermi}
\ln Z &=& \sum_K\ln\left(\frac{E_k^2+(\omega_n+i\mu)^2}{T^2}\right)^2 \non
&=& \sum_K\left(\ln\frac{E_k^2+(\omega_n+i\mu)^2}{T^2}+\ln\frac{E_k^2+(-\omega_n+i\mu)^2}{T^2}\right) \non
&=& \sum_K\left(\ln\frac{\omega_n^2+(E_k-\mu)^2}{T^2}+\ln\frac{\omega_n^2+(E_k+\mu)^2}{T^2}\right) \, , 
\eea
where, in the second term of the second line, we have replaced $\omega_n$ by $-\omega_n$ which does not change the result since we sum over all 
$n\in \mathbb{Z}$. The third line can be easily checked by multiplying out all terms. 

Next we need to perform the sum over fermionic Matsubara frequencies. This is similar to the bosonic case and yields
\be \label{fermimatsu}
\sum_n\ln\frac{\omega_n^2+\epsilon_k^2}{T^2} = \frac{\epsilon_k}{T}+2\ln\left(1+e^{-\epsilon_k/T}\right) +{\rm const} \, .
\ee
Using this result to evaluate Eq.\ (\ref{lnZfermi}) and taking the thermodynamic limit yields the thermodynamic potential $\Omega=-T\ln Z$, 
\be \label{OV}
\frac{\Omega}{V} = -2\int\frac{d^3{\bf k}}{(2\pi)^3}\left[E_k+T\ln\left(1+e^{-(E_k-\mu)/T}\right)+T\ln\left(1+e^{-(E_k+\mu)/T}\right)
\right]  \, .
\ee
The overall factor 2 accounts for the two spin states of the spin-1/2 fermion. Together with the particle/antiparticle 
degrees of freedom we recover all four degrees of freedom of the Dirac spinor. Again we conclude this section by computing 
the pressure for large temperatures,
\be \label{PressFer}
P\simeq 4 \frac{T^4}{2 \pi^2}\int_0^\infty dx\, x^2\ln\left(1+e^{-x}\right) = 4\cdot \frac{7}{8} \frac{\pi^2T^2}{90} \, .
\ee
Comparing with the bosonic pressure (\ref{PressBos}) we see that for large $T$ a single fermionic degree of freedom contributes 7/8 times as much to
the thermal pressure as a single bosonic degree of freedom.  

\subsection{Summation over fermionic Matsubara frequencies}
\label{app:matsufermions}

It remains to prove Eq.\ (\ref{fermimatsu}) by summing over fermionic Matsubara frequencies. 
As for the bosonic case, we write
\be \label{startwith}
\sum_n\ln\frac{\omega_n^2+\epsilon_k^2}{T^2} = \int_1^{(\epsilon_k/T)^2} dx^2\sum_n\frac{1}{(2n+1)^2\pi^2+x^2}+\sum_n\ln[1+(2n+1)^2\pi^2] \, .
\ee
This time, we need to use the tanh instead of the coth when we write the sum in terms of a contour integral,
\be \label{contourfermi}
T\sum_n\frac{1}{\omega_n^2+\epsilon_k^2} = - \frac{1}{2\pi i}\oint_C d\omega\,\frac{1}{\omega^2-\epsilon_k^2}
\frac{1}{2}\tanh\frac{\omega}{2T} \, .
\ee
(We have denoted $\epsilon_k\equiv xT$.)
The poles of $\tanh[\omega/(2T)]$ are given by the zeros of $e^{\omega/(2T)}+e^{-\omega/(2T)}$, i.e., they are located at $i$ times 
the fermionic Matsubara frequencies, 
$\omega = i\omega_n$. The contour $C$ encloses these poles and none of the poles of $1/(\omega^2-\epsilon_k^2)$. 
Then, with the residue theorem and with 
\begin{subequations}
\bea
\left.\left(e^{\omega/(2T)}-e^{-\omega/(2T)}\right)\right|_{\omega=i\omega_n} &=& 2i(-1)^n \, , \\ 
\left.\frac{d}{d\omega}\left(e^{\omega/(2T)}+e^{-\omega/(2T)}\right)\right|_{\omega=i\omega_n} &=& \frac{i(-1)^n}{T} \, , 
\eea
\end{subequations}
one confirms Eq.\ (\ref{contourfermi}). We can now close the contour in the positive half-plane to obtain
\bea
T\sum_n\frac{1}{\omega_n^2+\epsilon_k^2} &=& -\frac{1}{2\pi i}\int_{-i\infty+\eta}^{i\infty+\eta} d\omega\,\frac{1}{\omega^2-\epsilon_k^2}
\tanh\frac{\omega}{2T} \non 
&=& \frac{1}{2\epsilon_k}\tanh\frac{\epsilon_k}{2T} = \frac{1}{2\epsilon_k}[1-2f_F(\e_k)] \, , 
\eea
where 
\be
f_F(\epsilon) \equiv \frac{1}{e^{\epsilon/T}+1}  
\ee
is the Fermi distribution function. Inserting this result into Eq.\ (\ref{startwith}) yields 
\bea \label{resultmatsu2}
\sum_n\ln\frac{\omega_n^2+\epsilon_k^2}{T^2} 
&=&\int_1^{(\epsilon_k/T)^2} dx^2 \frac{1}{x}\left(\frac{1}{2}-\frac{1}{e^x+1}\right) +{\rm const} \non
&=& \frac{\epsilon_k}{T}+2\ln\left(1+e^{-\epsilon_k/T}\right) +{\rm const} \, ,
\eea
which proves Eq.\ (\ref{fermimatsu}).

\section*{Problems}
\addcontentsline{toc}{section}{Problems}

\begin{prob}
\label{prob11}
\textbf{Matsubara sum for boson loop}\\
\bigskip
Show via contour integration that 
\be
T\sum_{k_0}\frac{1}{(k_0^2-\epsilon_1^2)[(p_0-k_0)^2-\epsilon_2^2]} = -\sum_{e_1,e_2=\pm}\frac{e_1e_2}{4\epsilon_1\epsilon_2}
\frac{1+f_B(e_1\epsilon_1)+f_B(e_2\epsilon_2)}{p_0-e_1\epsilon_1-e_2\epsilon_2} \, ,
\ee
with $k_0=-i\omega_n$, $p_0=-i\omega_m$ bosonic Matsubara frequencies, and $\epsilon_1,\epsilon_2>0$. 
\end{prob}

\begin{prob}
\label{prob12}
\textbf{Matsubara sum for fermion loop}\\
Prove via contour integration the following result for the summation over fermionic Matsubara frequencies,
\bea \label{ex4}
&&T\sum_{k_0} \frac{(k_0+\xi_1)(k_0+q_0+\xi_2)}{(k_0^2-\epsilon_1^2)[(k_0+q_0)^2-\epsilon_2^2]} \non 
&&= -\frac{1}{4\epsilon_1\epsilon_2}
\sum_{e_1,e_2=\pm}\frac{(\epsilon_1-e_1\xi_1)(\epsilon_2-e_2\xi_2)}{q_0-e_1\epsilon_1+e_2\epsilon_2}\frac{f_F(-e_1\epsilon_1)f_F(e_2\epsilon_2)}{f_B(-e_1\epsilon_1+e_2\epsilon_2)} \, ,
\eea
where $k_0=-i\omega_n$ with fermionic Matsubara frequencies $\omega_n$, and $q_0=-i\omega_m$ with bosonic Matsubara frequencies $\omega_m$, and 
where $\xi_1,\xi_2,\epsilon_1,\epsilon_2>0$ are real numbers.
The result of this problem is used in the calculation of the neutrino emissivity in chapter \ref{sec:cooling}.
\end{prob}

%% file: glossary2.tex
%
%

\Extrachap{Glossary}

\runinhead{2SC phase} 
Color superconductor in which strange quarks and quarks of one color remain unpaired. Because of the asymmetry induced by the 
strange quark mass, viable candidate for the ground state of quark matter at moderate chemical potential. 
In these lectures we discuss the 2SC phase in the context of neutrino emissivity, to illustrate the effect of both paired and unpaired quarks. 

\runinhead{AdS/CFT correspondence} 
Theoretical tool not discussed in these lectures, but an interesting approach to tackle QCD at strong coupling. 
The idea is that -- relatively simple -- calculations in the gravity approximation of a certain string theory 
provide results for the -- otherwise hard to access -- strong coupling limit of a corresponding (``dual'')
field theory. The problem is that currently no gravity dual of QCD is known.

\runinhead{anomalous propagator} 
Technically speaking, off-diagonal components of the propagator in Nambu-Gorkov space; nonzero in the 
case of a superconductor or a superfluid. More physically speaking, anomalous propagators describe a fermion which is, via the Cooper
pair condensate, converted into a fermion hole. 

\runinhead{asymptotic freedom} 
Important property of QCD which says that the running coupling constant of QCD becomes small for large exchanged momenta. For our context
this means that quarks at large densities, where the distance between them is small and hence the exchanged momentum large, are weakly
interacting; quarks at infinite density are free. In compact stars, however, the density is large, but by no means asymptotically large.

\runinhead{axial anomaly} Non-conservation of the axial current in QCD. In our context of (moderately) dense  
matter originating mainly from instantons which are certain semi-classical gauge field configurations. Leads to an explicit 
breaking of the axial $U(1)_A$, which is a subgroup of the chiral group, and thus gives a large mass to the $\eta'$.  

\runinhead{bag model (MIT bag model)} 
Simple model to take into account confinement. Via the bag constant, an energy penalty is introduced by hand for the deconfined phase.
The model amounts to the picture of a hadron as a bag which confines the quarks; the bag exerts an external pressure on the quarks, 
given by the bag constant. In our astrophysical context, the bag model is a simple way to compare free energies of dense quark matter and 
dense nuclear matter. 

\runinhead{BCS theory} Original theory for electronic superconductors, developed in 1957 by Bardeen, Cooper, and Schrieffer. 
Many concepts and approximations
can be adopted for nuclear and quark matter. In color-superconducting quark matter, an important difference to BCS theory is the
parametric dependence of the pairing gap on the coupling constant due to long-range interactions via magnetic gluons. 

\runinhead{$\beta$-decay} Process due to the weak interaction of the form $n\to p + e + \bar{\nu}_e$ in nuclear matter and 
$d \to u + e + \bar{\nu}_e$ in quark matter. Relevant in these lectures for two reasons: firstly, equilibrium with respect to this process 
($\beta$-equilibrium) yields important constraints for the chemical potentials and secondly, this process contributes to the neutrino
emissivity which in turn is responsible for the cooling of a compact star.  

\runinhead{Bogoliubov coefficients} Momentum-dependent coefficients in the theory of superconductivity and superfluidity which 
characterize the mixing of fermions and fermion holes due to Cooper pair condensation. In these lectures, the Bogoliubov coefficients
arise naturally in the calculation of the neutrino emissivity in color-superconducting quark matter. 

\runinhead{Cabibbo-Kobayashi-Maskawa (CKM) matrix} Matrix that characterizes the relative strength of the weak interaction for 
different quark flavors. In these lectures relevant for the calculation of the neutrino emissivity
in quark matter. 

\runinhead{chiral symmetry} 
For massless quarks, QCD possesses a global symmetry for right- and left-handed quarks separately, called chiral symmetry. This 
symmetry can be spontaneously broken, giving rise to Goldstone modes. These Goldstone modes (or pseudo-Goldstone modes in the case of nonzero 
quark masses) are for instance pions and kaons. In these lectures we discuss kaon condensation in nuclear and quark matter (in quark
matter, chiral symmetry is spontaneously broken in the CFL phase). 

\runinhead{color superconductivity} 
Cooper pair formation and condensation in cold and dense quark matter, analogous to electronic superconductivity in metals. If quark
matter is present in compact stars, it can be expected to be a color superconductor.  

\runinhead{color-flavor locking (CFL)} 
Ground state of three-flavor quark matter at asympto\-tically large densities. Particularly symmetric color superconductor where the 
order parameter is invariant only under simultaneous color and flavor transformations. May persist down to densities where the 
hadronic phase takes over or may be replaced before this transition
by a different color superconductor because of the effects of the strange quark mass.

\runinhead{compact star} 
Very dense astrophysical object with a mass close to the sun's mass and a radius of about ten kilometers. The term shows our ignorance
of the exact composition of these objects. They may be neutron stars, hybrid stars, or quark stars. In a more general terminology, compact star also
is used to include white dwarfs and black holes, neither of which are the subject of these lectures.

\runinhead{constituent quark mass} Quark mass including the quark's interaction energy in a baryon, such that the sum of the 
three constituent quark masses adds up to the baryon mass. More generally, in dense matter the density-dependent ``constituent'' 
quark mass includes any finite-density effects. Can be hundreds of MeV larger than the current quark mass. 

\runinhead{Cooper pairs} Microscopic explanation for superfluidity and superconductivity within BCS theory. 
Arise from an instability of the Fermi surface in the presence of an arbitrarily small interaction. In compact stars, there are possibly
Cooper pairs of neutrons, protons, hyperons and/or quarks.  

\runinhead{crust} Outer, km thick, layer of a neutron star or hybrid star. Composed of ordinary nuclei which form a crystalline
structure and which, upon increasing the density and thus going further inside the star, become more and more neutron rich. In the inner
crust a neutron superfluid is immersed in the lattice of nuclei. Quark stars have, if at all, much thinner crusts.  

\runinhead{current quark mass} Quark mass without effects from the interactions with other quarks and gluons, see also {\it constituent quark mass}.
Since interactions become weak at asymptotically large densities (much larger than densities in compact stars), 
current and constituent quark masses become identical in this limit.

\runinhead{dense matter}
In these lectures, dense matter means matter at densities of a few times nuclear ground state density, as expected in the interior of 
compact stars. Governed by the strong interaction, and 
thus very difficult to describe theoretically. We discuss several theoretical concepts and sometimes have to escape to lower or even higher 
densities, just to make life simpler. 


\runinhead{equation of state} 
Relation between the pressure and the energy density for a given form of dense matter. In our context, the equation of state 
determines, together with the TOV equation, the mass-radius relation of a compact star. In particular, a stiff (soft) equation of state
allows for a large (small) maximum mass.

\runinhead{Goldstone boson} Massless boson arising from spontaneous symmetry breaking of a global symmetry. The only exact 
(i.e., truly massless) Goldstone boson in dense matter is the one associated to superfluidity, i.e., to the breaking
of baryon number conservation. Such a mode exists in a nuclear superfluid as well as in the color-flavor locked phase.  

\runinhead{hybrid star} 
Compact star with a quark matter core and a nuclear mantle. Most likely scenario to find quark matter in a compact star. 

\runinhead{hyperon} 
Baryon with nonzero strangeness. Hyperons may occur in hadronic matter at sufficiently large densities. In these lectures only
discussed briefly, in the context of Walecka-like models. 

\runinhead{incompressibility} Thermodynamic property of nuclear matter at the saturation density, sometimes also called compression modulus.
Can be (at least indirectly and approximately) determined in the experiment and thus yields a value that can, among other quantities,
be used to fit the parameters of theoretical models, in these lectures the coupling constants of the Walecka model with scalar interactions.

\runinhead{kaon condensation} 
Possible example of Bose-Einstein condensation in a compact star. May appear at sufficiently large densities. Is possible not
only in nuclear matter, but also in quark matter, where kaons exist in the CFL phase. These kaons carry the same quantum numbers as the 
usual kaons, however are made of two quarks and two quark holes.

\runinhead{Kepler frequency} Absolute upper limit for the rotation frequency of compact stars beyond which mass shedding at the equator
sets in. Given by the equality of the centrifugal and gravitational forces (more precisely, the general relativistic version thereof).
For typical compact stars in the ${\rm ms}^{-1}$ regime, i.e., for some pulsars actually observed rotation frequencies are not too 
far from that limit. Below that limit stars can suffer from other rotational instabilities, for instance the $r$-mode instability.

\runinhead{Landau mass} Effective mass of (nonrelativistic) fermions at the Fermi surface, in the framework of Landau's
Fermi liquid theory. In these lectures, the Landau mass for nucleons is mentioned in the context of the Walecka model where its
experimental value serves to fit the parameters of the model.

\runinhead{lattice QCD} 
QCD on the computer. Powerful method to perform calculations from first principles. Not discussed in these lectures, mostly because 
lattice QCD is currently unable to provide results at large chemical potential and small temperature because of the so-called sign problem.
 
\runinhead{Low-mass X-ray binary (LMXB)} System of two stars, where a pulsar is accreting matter from its companion which 
has a mass typically smaller than one solar mass (as opposed to high-mass X-ray binaries where the companion has a mass larger 
than about ten solar masses).
Measured rotation frequencies of pulsars in LMXBs are mentioned in our brief discussion of the $r$-mode instability of rotating compact stars.

\runinhead{magnetar} 
Compact star with unusually large magnetic field, up to $10^{15}\, {\rm G}$ at the surface and possibly larger in the interior.

\runinhead{Matsubara frequency} 
In thermal field theory, the time direction in Minkowski space becomes imaginary and compact, giving rise
to Euclidean space with discrete energies, given by the Matsubara frequencies. In these lectures we mostly consider the 
zero-temperature limit, but in some instances we have to perform a sum over Matsubara frequencies. 

\runinhead{mixed phase} Coexistence of two (or more) phases which occupy certain volume fractions
-- for instance bubbles of one phase immersed in the other phase -- in a given total volume. In our context, global charge neutrality, 
as opposed to local charge neutrality, allows for mixed phases for instance of nuclei and nuclear matter or quark and hadronic matter. 
These phases may be disfavored by large surface energy costs.   

\runinhead{Nambu-Gorkov space} Contains Nambu-Gorkov spinors which arise from a doubling of the fermionic degrees of freedom in the theoretical 
description of superconductors and superfluids. Allows to introduce Cooper pairing in the off-diagonal elements of the Nambu-Gorkov propagators.
See also {\it anomalous propagators}.

\runinhead{Nambu-Jona-Lasinio (NJL) model} Phenomenological model, not discussed in these lectures, where the QCD interaction 
between quarks is replaced by 
a point-like four-quark interaction. Since it has attraction in the same channels as QCD, this model is frequently used to describe
color-superconducting quark matter at moderate densities.   

\runinhead{neutron star} Compact star made of neutron-rich nuclear matter. In some literature the term neutron star is 
also used to include the possibility of a quark matter core. Mostly, also in these lectures, these stars are called hybrid stars.

\runinhead{nuclear pasta} Mixed phase of ordinary nuclei (ions) and nuclear matter, typically found in the inner cores of neutron 
stars. Because different geometries can be realized -- spheres, rods, slabs, the latter two reminiscent of spaghetti or lasagna -- 
these phases have been termed
nuclear pasta. In these lectures we discuss the possibility of mixed phases of quark and hadronic matter.

\runinhead{pion condensation} Bose-Einstein condensation of pions in nuclear matter. Although pions are lighter than kaons in the vacuum, 
kaon condensation seems to be more likely in dense nuclear matter. Therefore, in these lectures, kaon condensation, not pion condensation,
is discussed.

\runinhead{pseudo-Goldstone boson} Less impressive brother of the Goldstone boson, arising from spontaneous breaking of a global symmetry
which is broken explicitly by a small amount (small compared to the scale of the spontaneous breaking). Light, but not exactly massless. 
Dense matter is full of pseudo-Goldstone modes, for instance
mesons in nuclear matter or color-flavor-locked quark matter, 
arising from the spontaneous breaking of chiral symmetry which is explicitly broken by quark masses.  

\runinhead{pulsar} Star whose radiation is observed in periodic pulses. Pulsars are rotating compact stars with large magnetic fields; 
their apparent pulsation is due to the alignment of the radiation in a beam along the magnetic axis. 
When the magnetic axis is different from the rotation axis, the beam may point towards the earth periodically, 
just as the light of a lighthouse flashes periodically when you observe it from the beach. 

\runinhead{pulsar glitch} 
Sudden spin-up of a rotating compact star. Not discussed in detail in these lectures but very interesting phenomenon since closely 
related to the microscopic physics, presumably to crystalline structures and vortices in the star. 

\runinhead{QCD phase diagram} 
Collection of equilibrium states of QCD, typically depicted in the plane of quark (or baryon) chemical potential and temperature. We roughly know 
where compact stars sit in this diagram, but we do not know the phase(s) that occupy this region of the diagram. These lectures
are about exploring this unknown territory.

\runinhead{quantum chromodynamics (QCD)} Theory of the strong interaction. Governs the physics that determines the ground state of dense matter
present in a compact star. In these lectures we perform one explicit calculation in QCD and discuss several effective approaches to this
very elegant, but for most practical purposes very difficult, theory.  

\runinhead{quarkyonic matter} 
Form of dense matter covering a large portion of the QCD phase diagram for the case of asymptotically large number of colors. 
Not discussed in these lectures because there are only three colors in the real world. However, it is a viable option that 
a small region of quarkyonic matter survives and thus becomes also important for compact stars. 

\runinhead{quasiparticle} Term originally used in condensed matter physics and carried over to dense QCD matter. Absorbs interactions
of the original particles into effective new particles. For instance, quasiparticles in a superconductor are gapped due to the 
attractive interaction between the original particles.  

\runinhead{r-modes} Non-radial oscillations of a star with the Coriolis force as the restoring force. Interesting for 
dense matter physics because they grow unstable in a pulsar unless the matter inside the star is sufficiently viscous.  

\runinhead{rotated electromagnetism} 
Effect in some color superconductors which is responsible for them being no electromagnetic superconductors. Therefore important for
the physics of compact stars since magnetic fields penetrate these color superconductors. Technically speaking, rotated electromagnetism
refers to a gauge boson which is a mixture of a gluon and the photon.

\runinhead{saturation density} Density at which the binding energy is minimized, here always used in the context of nuclear matter
for which the saturation density is approximately 0.15 baryons per ${\rm fm}^3$ and the corresponding binding energy per nucleon is about 
16 MeV. 

\runinhead{sign problem} Problem of QCD lattice calculations at finite values of the baryon chemical potential. For finite
chemical potential, the action, more precisely the quark determinant in the functional integral of the partition function, loses
its positivity and even becomes complex. This makes the probabilistic sampling method (``Monte Carlo method''), on which lattice QCD
is based, unfeasible. In our context this means that currently there is no input from lattice calculations to the properties of dense matter.

\runinhead{strange quark matter hypothesis} Hypothesis that strange quark matter, not nuclear matter, is the ground state at zero 
pressure. The hypothesis does not contradict our existence since, even if the hypothesis is true, the transition from nuclear matter, made
of $u$ and $d$ quarks, to strange quark matter is essentially forbidden. We discuss that, within the bag model, the strange quark matter
hypothesis is true if the bag constant is between a lower bound (since we know that ordinary nuclear matter is stable with respect to 
two-flavor quark matter) and an upper limit (beyond which nuclear matter is absolutely stable).    

\runinhead{strange star (quark star)} 
Compact star made entirely out of quark matter, thus the most radical scenario for quark matter in compact stars.  

\runinhead{strangelet} Small nugget of strange quark matter. Stretching the original meaning a bit -- well, from femtometers
to kilometers -- a strange star is a huge strangelet. Relevant for us in the context of the strange quark matter hypothesis: 
since strangelets would convert neutron stars into strange stars, the unambiguous observation of a single neutron star would 
invalidate the strange quark matter hypothesis, provided that there are enough strangelets in the cosmos to hit neutron stars.

\runinhead{supernova} Compact stars are expected to be born in (type II) supernova explosions, where a giant star, after burning
its nuclear fuel, undergoes a gravitational collapse. The energy of the explosion is mostly released in the form of neutrinos.
The theoretical description of supernovae requires very complicated hydrodynamical simulations.

\runinhead{Tolman-Oppenheimer-Volkov (TOV) equation} Differential equation from general relativity 
for the mass, pressure, and energy density as functions of the distance from the center of the star. In connection with the equation of state,
which relates energy density and pressure, used to compute the mass-radius relation for a compact star.  

\runinhead{unpaired quark matter} Term used for (dense) quark matter which does not form Cooper pairs and thus is no color superconductor. 
Since dense quark matter is expected to be some kind of color superconductor, completely unpaired dense quark matter is unlikely to exist. 
Therefore mostly used for reference calculations or when, for the computed quantity, it is a good approximation to paired quark matter.

\runinhead{Urca process} 
Most efficient process for neutrino emission, and thus for the cooling of the star. In quark matter the direct Urca process
is given by $u+e\to d +\nu_e$ and variants thereof. We compute the emission rate of this process in detail in these lectures. 
In the modified Urca process, a spectator particle is added which increases the available phase space.  

\runinhead{viscosity (bulk/shear)} Transport coefficients of nuclear or quark matter relevant in particular in the context of rotation and 
oscillation of the star. Requires microscopic calculation of processes typically governed by the weak interaction. Not discussed in
detail in these lectures. See also {\it r-modes}. 

\runinhead{Walecka model} Phenomenological model for interacting nuclear matter, based on Yukawa couplings of the 
nucleons with the $\sigma$ and $\omega$ meson. Used for extrapolation to 
large densities after fitting the parameters of the model at saturation density. Discussed in these lectures as a basic example 
for numerous more complicated nuclear models of similar kind.

\runinhead{white dwarf} Dense star with a mass of about the sun's mass and radius of a few thousand kilometers, which makes it  
less dense than a neutron star. Composed of nuclei immersed in a degenerate electron gas.

%% file: densematter2.bbl
\begin{thebibliography}{99}


\bibitem{weber1}
F.\ Weber, Pulsars as astrophysical laboratories for nuclear and particle physics, CRC Press, Boca Raton, FL (1999).

\bibitem{glendenning1}
N.K. Glendenning, Compact Stars: Nuclear Physics, Particle Physics, and General Relativity, Springer, Heidelberg (2000).

\bibitem{haensel1}
P.\ Haensel, A.Y.\ Potekhin, D.G.\ Yakovlev, Neutron Stars, Springer, Heidelberg (2007).

\bibitem{Page:2006ud1}
  D.~Page and S.~Reddy, Dense Matter in Compact Stars: Theoretical Developments and Observational Constraints,
  Ann.\ Rev.\ Nucl.\ Part.\ Sci.\  {\bf 56}, 327 (2006)
  [arXiv:astro-ph/0608360].

\bibitem{Madsen:1998uh1}
  J.~Madsen, Physics and astrophysics of strange quark matter,
  Lect.\ Notes Phys.\  {\bf 516}, 162 (1999)
  [arXiv:astro-ph/9809032].

\bibitem{Weber:2004kj1}
  F.~Weber, Strange quark matter and compact stars,
  Prog.\ Part.\ Nucl.\ Phys.\  {\bf 54}, 193 (2005)
  [arXiv:astro-ph/0407155].


\bibitem{Lattimer:2006xb1}
  J.~M.~Lattimer and M.~Prakash,
  Neutron Star Observations: Prognosis for Equation of State Constraints,
  Phys.\ Rept.\  {\bf 442}, 109 (2007)
  [arXiv:astro-ph/0612440].

\bibitem{Alford:2007xm1}
  M.~G.~Alford, A.~Schmitt, K.~Rajagopal and T.~Sch\"{a}fer, Color superconductivity in dense quark matter,
  Rev.\ Mod.\ Phys.\  {\bf 80}, 1455 (2008)
  [arXiv:0709.4635 [hep-ph]].

\bibitem{kapusta1}
J.I.\ Kapusta, C.\ Gale, Finite-temperature field theory: Principles and Applications, Cambridge Univ.\ Press, New York (2006).

\bibitem{lebellac1}
M.\ Le Bellac, Thermal Field Theory, Cambridge Univ.\ Press, Cambridge (2000).

\end{thebibliography}

\begin{thebibliography}{99}

\bibitem{shapiro2}
S.L.\ Shapiro and S.A.\ Teukolsky, 
Black Holes, White Dwarfs and Neutron Stars: The Physics of Compact Objects, Wiley, New York (1983).

\bibitem{Balian:1999eb2}
  R.~Balian and J.~P.~Blaizot, Stars and statistical physics: a teaching experience, 
  arXiv:cond-mat/9909291.

\bibitem{Silbar:2003wm2}
  R.~R.~Silbar and S.~Reddy, Neutron Stars for Undergraduates,
  Am.\ J.\ Phys.\  {\bf 72}, 892 (2004)
  [Erratum-ibid.\  {\bf 73}, 286 (2005)]
  [arXiv:nucl-th/0309041].

\bibitem{Sagert:2005fw2}
  I.~Sagert, M.~Hempel, C.~Greiner and J.~Schaffner-Bielich, Compact Stars for Undergraduates,
  Eur.\ J.\ Phys.\  {\bf 27}, 577 (2006)
  [arXiv:astro-ph/0506417].

\bibitem{Chodos:1974je2}
  A.~Chodos, R.~L.~Jaffe, K.~Johnson, C.~B.~Thorn and V.~F.~Weisskopf, A New Extended Model Of Hadrons,
  Phys.\ Rev.\  D {\bf 9}, 3471 (1974).

\bibitem{Chodos:1974pn2}
  A.~Chodos, R.~L.~Jaffe, K.~Johnson and C.~B.~Thorn, Baryon Structure In The Bag Theory,
  Phys.\ Rev.\  D {\bf 10}, 2599 (1974).

\bibitem{Bodmer:1971we2}
  A.~R.~Bodmer, Collapsed nuclei,
  Phys.\ Rev.\  D {\bf 4}, 1601 (1971).

\bibitem{Witten:1984rs2}
  E.~Witten, Cosmic Separation Of Phases,
  Phys.\ Rev.\  D {\bf 30}, 272 (1984).

\bibitem{Farhi:1984qu2}
  E.~Farhi and R.~L.~Jaffe, Strange Matter,
  Phys.\ Rev.\  D {\bf 30}, 2379 (1984).

\bibitem{Bauswein:2008gx2}
  A.~Bauswein {\it et al.},
  Mass Ejection by Strange Star Mergers and Observational Implications,
  Phys.\ Rev.\ Lett.\  {\bf 103}, 011101 (2009)
  [arXiv:0812.4248 [astro-ph]].

\bibitem{Kurkela:2009gj2}
  A.~Kurkela, P.~Romatschke and A.~Vuorinen,
  Cold Quark Matter,
  Phys.\ Rev.\  D {\bf 81}, 105021 (2010)
  [arXiv:0912.1856 [hep-ph]].

\bibitem{Alford:2004pf2}
  M.~Alford, M.~Braby, M.~W.~Paris and S.~Reddy, Hybrid stars that masquerade as neutron stars,
  Astrophys.\ J.\  {\bf 629}, 969 (2005)
  [arXiv:nucl-th/0411016].

\bibitem{Alford:2006vz2}
  M.~Alford, D.~Blaschke, A.~Drago, T.~Klahn, G.~Pagliara and J.~Schaffner-Bielich, Quark matter in compact stars?,
  Nature {\bf 445}, E7 (2007)
  [arXiv:astro-ph/0606524].

\end{thebibliography}

\begin{thebibliography}{99}

\bibitem{Weber:2004kj3}
  F.~Weber, Strange quark matter and compact stars,
  Prog.\ Part.\ Nucl.\ Phys.\  {\bf 54}, 193 (2005)
  [arXiv:astro-ph/0407155].

\bibitem{Migdal:1973zm3}
  A.~B.~Migdal, $\pi$ condensation in nuclear matter,
  Phys.\ Rev.\ Lett.\  {\bf 31}, 257 (1973).

\bibitem{Kaplan:1986yq3}
  D.~B.~Kaplan and A.~E.~Nelson, Strange Goings on in Dense Nucleonic Matter,
  Phys.\ Lett.\  B {\bf 175}, 57 (1986).


\bibitem{Ramos:2000dq3}
  A.~Ramos, J.~Schaffner-Bielich and J.~Wambach, Kaon condensation in neutron stars,
  Lect.\ Notes Phys.\  {\bf 578}, 175 (2001)
  [arXiv:nucl-th/0011003].

\bibitem{Thorsson:1993bu3}
  V.~Thorsson, M.~Prakash and J.~M.~Lattimer, Composition, structure and evolution of neutron stars with kaon condensates,
  Nucl.\ Phys.\  A {\bf 572}, 693 (1994)
  [Erratum-ibid.\  A {\bf 574}, 851 (1994)]
  [arXiv:nucl-th/9305006].

\bibitem{thomas3}
A.W.\ Thomas, W.\ Weise, The Structure of the Nucleon, Wiley, Berlin, 2001.

\bibitem{Prakash:1988md3}
  M.~Prakash, T.~L.~Ainsworth and J.~M.~Lattimer,
  Equation of state and the maximum mass of neutron stars,
  Phys.\ Rev.\ Lett.\  {\bf 61}, 2518 (1988).

\bibitem{Chamel:2008ca3}
  N.~Chamel and P.~Haensel,
  Physics of Neutron Star Crusts,
  Living Rev.\ Rel.\  {\bf 11}, 10 (2008)
  [arXiv:0812.3955 [astro-ph]].

\bibitem{Ravenhall:1983uh3}
  D.~G.~Ravenhall, C.~J.~Pethick and J.~R.~Wilson, Structure Of Matter Below Nuclear Saturation Density,
  Phys.\ Rev.\ Lett.\  {\bf 50}, 2066 (1983).

\bibitem{Alford:2001zr3}
  M.~G.~Alford, K.~Rajagopal, S.~Reddy and F.~Wilczek, The minimal CFL-nuclear interface,
  Phys.\ Rev.\  D {\bf 64}, 074017 (2001)
  [arXiv:hep-ph/0105009].

\end{thebibliography}

\begin{thebibliography}{99}

\bibitem{levyak14}
  K.P.\ Levenfish, D.G.\ Yakovlev,
  Specific heat of neutron star cores with superfluid nucleons,
  Astron.\ Rep.\ {\bf 38}, 247 (1994).

\bibitem{levyak24}
  K.P.\ Levenfish, D.G.\ Yakovlev, 
  Suppression of neutrino energy losses in reactions of direct urca processes by superfluidity in neutron star nuclei,
  Astron.\ Lett.\ {\bf 20}, 43 (1994). 

\bibitem{Schmitt:2004et4}
  A.~Schmitt, The ground state in a spin-one color superconductor,
  Phys.\ Rev.\  D {\bf 71}, 054016 (2005)
  [arXiv:nucl-th/0412033].

\bibitem{Lombardo:2000ec4}
  U.~Lombardo and H.~J.~Schulze, Superfluidity in Neutron Star Matter,
  Lect.\ Notes Phys.\  {\bf 578}, 30 (2001)
  [arXiv:astro-ph/0012209].

\bibitem{Page:2006ud4}
  D.~Page and S.~Reddy, Dense Matter in Compact Stars: Theoretical Developments and Observational Constraints,
  Ann.\ Rev.\ Nucl.\ Part.\ Sci.\  {\bf 56}, 327 (2006)
  [arXiv:astro-ph/0608360].

\bibitem{Alford:2007xm4}
  M.~G.~Alford, A.~Schmitt, K.~Rajagopal and T.~Sch\"{a}fer, Color superconductivity in dense quark matter,
  Rev.\ Mod.\ Phys.\  {\bf 80}, 1455 (2008)
  [arXiv:0709.4635 [hep-ph]].

\bibitem{Alford:1998mk4}
  M.~G.~Alford, K.~Rajagopal and F.~Wilczek, Color-flavor locking and chiral symmetry breaking in high density {QCD},
  Nucl.\ Phys.\  B {\bf 537}, 443 (1999)
  [arXiv:hep-ph/9804403].

\bibitem{Son:1999cm4}
  D.~T.~Son and M.~A.~Stephanov,
  Inverse meson mass ordering in color-flavor-locking phase of high  density
  QCD,
  Phys.\ Rev.\  D {\bf 61}, 074012 (2000)
  [arXiv:hep-ph/9910491].

\bibitem{Bedaque:2001je4}
  P.~F.~Bedaque and T.~Sch\"{a}fer,
  High Density Quark Matter under Stress,
  Nucl.\ Phys.\  A {\bf 697}, 802 (2002)
  [arXiv:hep-ph/0105150].

\bibitem{Alford:2007qa4}
  M.~G.~Alford, M.~Braby and A.~Schmitt,
  Critical temperature for kaon condensation in color-flavor locked quark
  matter,
  J.\ Phys.\ G {\bf 35}, 025002 (2008)
  [arXiv:0707.2389 [nucl-th]].

\bibitem{kapusta4}
J.I.\ Kapusta, C.\ Gale, Finite-temperature field theory: Principles and Applications, Cambridge Univ.\ Press, New York (2006).

\bibitem{fetter4}
A.\ Fetter, J.D.\ Walecka, Quantum Theory of Many-Particle Systems, McGraw Hill, New York (1971). 

\bibitem{Pisarski:1999tv4}
  R.~D.~Pisarski and D.~H.~Rischke, Color superconductivity in weak coupling,
  Phys.\ Rev.\  D {\bf 61}, 074017 (2000)
  [arXiv:nucl-th/9910056].

\bibitem{Shovkovy:1999mr4}
  I.~A.~Shovkovy and L.~C.~R.~Wijewardhana,
  On gap equations and color flavor locking in cold dense QCD with three massless flavors,
  Phys.\ Lett.\  B {\bf 470}, 189 (1999)
  [arXiv:hep-ph/9910225].

\bibitem{Son:1998uk4}
  D.~T.~Son,
  Superconductivity by long-range color magnetic interaction in  high-density
  quark matter,
  Phys.\ Rev.\  D {\bf 59}, 094019 (1999)
  [arXiv:hep-ph/9812287].

\bibitem{zwierlein4}
M.W.\ Zwierlein, A.\ Schirotzek, C.H.\ Schunck, W.\ Ketterle,
Fermionic Superfluidity with Imbalanced Spin Populations and the Quantum Phase Transition to the Normal State,
Science 311 (5760), 492-496 (2006) [arXiv:cond-mat/0511197].


\end{thebibliography}

\begin{thebibliography}{99}

\bibitem{Yakovlev:2000jp5}
  D.~G.~Yakovlev, A.~D.~Kaminker, O.~Y.~Gnedin and P.~Haensel, Neutrino emission from neutron stars,
  Phys.\ Rept.\  {\bf 354}, 1 (2001)
  [arXiv:astro-ph/0012122].

\bibitem{Page:2005fq5}
  D.~Page, U.~Geppert and F.~Weber, The Cooling of Compact Stars,
  Nucl.\ Phys.\  A {\bf 777}, 497 (2006)
  [arXiv:astro-ph/0508056].

\bibitem{Weber:2004kj5}
  F.~Weber, Strange quark matter and compact stars,
  Prog.\ Part.\ Nucl.\ Phys.\  {\bf 54}, 193 (2005)
  [arXiv:astro-ph/0407155].

\bibitem{Jaikumar:2002vg5}
  P.~Jaikumar, M.~Prakash and T.~Sch\"{a}fer,
  Neutrino emission from Goldstone modes in dense quark matter,
  Phys.\ Rev.\  D {\bf 66}, 063003 (2002)
  [arXiv:astro-ph/0203088].

\bibitem{kapusta5}
J.I.\ Kapusta, C.\ Gale, Finite-temperature field theory: Principles and Applications, Cambridge Univ.\ Press, New York (2006).

\bibitem{lebellac5}
M.\ Le Bellac, Thermal Field Theory, Cambridge Univ.\ Press, Cambridge (2000).


\bibitem{Schmitt:2005wg5}
  A.~Schmitt, I.~A.~Shovkovy and Q.~Wang, Neutrino emission and cooling rates of spin-one color superconductors,
  Phys.\ Rev.\  D {\bf 73}, 034012 (2006)
  [arXiv:hep-ph/0510347].

\bibitem{Jaikumar:2005hy5}
  P.~Jaikumar, C.~D.~Roberts and A.~Sedrakian,
  Direct Urca neutrino rate in colour superconducting quark matter,
  Phys.\ Rev.\  C {\bf 73}, 042801 (2006)
  [arXiv:nucl-th/0509093].

\bibitem{fetter5}
A.\ Fetter, J.D.\ Walecka, Quantum Theory of Many-Particle Systems, McGraw Hill, New York (1971). 

\bibitem{Iwamoto:1980eb5}
  N.~Iwamoto, Quark Beta Decay And The Cooling Of Neutron Stars,
  Phys.\ Rev.\ Lett.\  {\bf 44}, 1637 (1980).

\end{thebibliography}

\begin{thebibliography}{99}

\bibitem{Lindblom:2000jw6}
  L.~Lindblom, Neutron star pulsations and instabilities,
  arXiv:astro-ph/0101136.

\bibitem{Jaikumar:2008kh6}
  P.~Jaikumar, G.~Rupak and A.~W.~Steiner,
  Viscous damping of r-mode oscillations in compact stars with quark
  matter,
  Phys.\ Rev.\  D {\bf 78}, 123007 (2008)
  [arXiv:0806.1005 [nucl-th]].

\bibitem{Alford:2007xm6}
  M.~G.~Alford, A.~Schmitt, K.~Rajagopal and T.~Sch\"{a}fer, Color superconductivity in dense quark matter,
  Rev.\ Mod.\ Phys.\  {\bf 80}, 1455 (2008)
  [arXiv:0709.4635 [hep-ph]].

\bibitem{Madsen:1992sx6}
  J.~Madsen,
  Bulk Viscosity Of Strange Dark Matter, Damping Of Quark Star Vibration, And
  The Maximum Rotation Rate Of Pulsars,
  Phys.\ Rev.\  D {\bf 46}, 3290 (1992).

\bibitem{Alford:2006gy6}
  M.~G.~Alford and A.~Schmitt, 
  Bulk viscosity in 2SC quark matter,
  J.\ Phys.\ G {\bf 34}, 67 (2007)
  [arXiv:nucl-th/0608019].

\bibitem{Link:2003hq6}
  B.~Link, Constraining Hadronic Superfluidity with Neutron Star Precession,
  Phys.\ Rev.\ Lett.\  {\bf 91}, 101101 (2003)
  [arXiv:astro-ph/0302441].

\bibitem{Aguilera:2006xv6}
  D.~N.~Aguilera, Spin-one color superconductivity in compact stars? An analysis within NJL-type models,
  Astrophys.\ Space Sci.\  {\bf 308}, 443 (2007)
  [arXiv:hep-ph/0608041].

\bibitem{Page:2005fq6}
  D.~Page, U.~Geppert and F.~Weber, The Cooling of Compact Stars,
  Nucl.\ Phys.\  A {\bf 777}, 497 (2006)
  [arXiv:astro-ph/0508056].

\bibitem{Harding:2006qn6}
  A.~K.~Harding and D.~Lai, Physics of Strongly Magnetized Neutron Stars,
  Rept.\ Prog.\ Phys.\  {\bf 69}, 2631 (2006)
  [arXiv:astro-ph/0606674].

\bibitem{haensel6}
P.\ Haensel, A.Y.\ Potekhin, D.G.\ Yakovlev, Neutron Stars, Springer, Heidelberg (2007).

\bibitem{Chamel:2008ca6}
  N.~Chamel and P.~Haensel,
  Physics of Neutron Star Crusts,
  Living Rev.\ Rel.\  {\bf 11}, 10 (2008)
  [arXiv:0812.3955 [astro-ph]].

\bibitem{Alcock:1986hz6}
  C.~Alcock, E.~Farhi and A.~Olinto, Strange stars,
  Astrophys.\ J.\  {\bf 310}, 261 (1986).

\bibitem{Alford:2006bx6}
  M.~G.~Alford, K.~Rajagopal, S.~Reddy and A.~W.~Steiner, The stability of strange star crusts and strangelets,
  Phys.\ Rev.\  D {\bf 73}, 114016 (2006)
  [arXiv:hep-ph/0604134].

\bibitem{Watts:2006hk6}
  A.~L.~Watts and S.~Reddy, Magnetar oscillations pose challenges for strange stars,
  Mon.\ Not.\ Roy.\ Astron.\ Soc.\  {\bf 379}, L63 (2007)
  [arXiv:astro-ph/0609364].

\bibitem{Anderson:1975zze6}
  P.~W.~Anderson and N.~Itoh,
  Pulsar glitches and restlessness as a hard superfluidity phenomenon,
  Nature {\bf 256}, 25 (1975).

\bibitem{Mannarelli:2007bs6}
  M.~Mannarelli, K.~Rajagopal and R.~Sharma,
  The rigidity of crystalline color superconducting quark matter,
  Phys.\ Rev.\  D {\bf 76}, 074026 (2007)
  [arXiv:hep-ph/0702021].

\bibitem{Stephanov:2007fk6}
  M.~A.~Stephanov, QCD phase diagram: An overview,
  PoS {\bf LAT2006}, 024 (2006)
  [arXiv:hep-lat/0701002].

\bibitem{Schmidt:2007jg6}
  C.~Schmidt, QCD thermodynamics at zero and non-zero density,
  PoS C {\bf POD2006}, 002 (2006)
  [arXiv:hep-lat/0701019].

\bibitem{Buballa:2003qv6}
  M.~Buballa, NJL model analysis of quark matter at large density,
  Phys.\ Rept.\  {\bf 407}, 205 (2005)
  [arXiv:hep-ph/0402234].

\bibitem{Blaschke:2005uj6}
  D.~Blaschke, S.~Fredriksson, H.~Grigorian, A.~M.~Oztas and F.~Sandin,
  The phase diagram of three-flavor quark matter under compact star
  constraints,
  Phys.\ Rev.\  D {\bf 72}, 065020 (2005)
  [arXiv:hep-ph/0503194].

\bibitem{McLerran:2007qj6}
  L.~McLerran and R.~D.~Pisarski,
  Phases of Cold, Dense Quarks at Large $N_c$,
  Nucl.\ Phys.\  A {\bf 796}, 83 (2007)
  [arXiv:0706.2191 [hep-ph]].

\bibitem{Peeters:2007ab6}
  K.~Peeters and M.~Zamaklar,
  The string/gauge theory correspondence in QCD,
  Eur.\ Phys.\ J.\ ST {\bf 152}, 113 (2007)
  [arXiv:0708.1502 [hep-ph]].

\bibitem{Gubser:2009md6}
  S.~S.~Gubser and A.~Karch,
  From gauge-string duality to strong interactions: a Pedestrian's Guide,
  Ann.\ Rev.\ Nucl.\ Part.\ Sci.\  {\bf 59}, 145 (2009)
  [arXiv:0901.0935 [hep-th]].


\end{thebibliography}
